\documentclass[11pt,a4paper]{article}
\usepackage{jheppub}

\usepackage{bbm,bm,float,mathrsfs,placeins,epstopdf,multicol,siunitx,color,colortbl,multirow,hhline}

\allowdisplaybreaks
\epsfclipon
\input{epsf.sty}
\input epsf.tex

\def\sin{{\rm sin}}
\def\cos{{\rm cos}}
\def\tan{{\rm tan}}

\DeclareMathOperator\arctanh{arctanh}
\definecolor{lightblue}{rgb}{0.68,0.85,0.9}
\definecolor{celadon}{rgb}{0.67,0.88,0.69}
\definecolor{corn}{rgb}{0.98,0.93,0.36}

\subheader{Companion paper}
\title{A companion to ``Knot invariants and M-theory I'': proofs and derivations} 
\author{Ver\'{o}nica Errasti D\'{i}ez}
\affiliation{Ernest Rutherford Physics Department, McGill University, \\
3600 rue University, Montr\'{e}al, Qu\'{e}bec, Canada H3A 2T8}
\emailAdd{vediez@physics.mcgill.ca}

\date{\today}

\abstract{
We construct two distinct yet related M-theory models that provide suitable frameworks for the study of knot invariants.
We then focus on the four-dimensional gauge theory that follows from appropriately compactifying one of these M-theory models.
We show that this theory has indeed all required properties to host knots.
Our analysis provides a unifying picture of the various recent works that attempt an understanding of knot invariants
using techniques of four-dimensional physics.

This is a companion paper to~\cite{Dasgupta:2016rhc}, covering all but section 3.3.
It presents a detailed mathematical derivation of the main results there,
as well as additional material. 
Among the new insights, those related to supersymmetry and the topological twist are highlighted.
This paper offers an alternative, complementary formulation of the contents in~\cite{Dasgupta:2016rhc}, but is
self-contained and can be read independently.}


\begin{document}

\maketitle 

\section{Introduction \label{introsect}}

Knot theory is the branch of topology that studies knots.
In this context, a knot is an embedding of a circle in three-dimensional Euclidean space
or its compact analogue: the three-sphere.
Two such knots are said to be equivalent iff there exists an ambient isotopy transforming one to the other.
This formal definition of equivalent knots is, unfortunately, insufficient in practice.
To such a great degree that one of the main unresolved problems in knot theory consists on distinguishing knots.
That is, determining when two knots are (or are not) equivalent.
This is known as the ``classification problem of knots''.
Very elaborate algorithms exist to this end, yet the problem persists.

Another approach to the knot differentiation puzzle involves knot invariants:
numbers, polynomials or homologies defined for each knot which remain unchanged for equivalent knots.
Interestingly, invariants such as Khovanov and Floer homologies are capable of telling apart the unknot
from any other non-equivalent knot.
Although this is a phenomenal achievement, there is still much to be accomplished.
So much so that, at present, it is not known whether a knot invariant exists which
is capable of distinguishing all inequivalent knots.

There are various ways to compute knot invariants.
Mathematicians use recursive relations, known as skein relations, to compute the Conway~\cite{Conway,Prytycki},
Alexander~\cite{Alexander} and Jones~\cite{Jones} polynomials, among others.
The first physics understanding of knot invariants appeared much later, in the groundbreaking work~\cite{Witten:1988hf}.
In it, knot polynomials are obtained as expectation values
of the holonomy of a Chern-Simons gauge field around a knot carrying a representation
of the underlying (compact) gauge group.
In particular, the Jones and HOMFLY-PT~\cite{Prytycki,HOMFLY} polynomials follow from considering
the defining representations of $SU(2)$ and $SU(N)$, respectively.

Starting roughly at the same time and up to now, there have been a number of works that address the study of knot invariants
from the point of view of four-dimensional physics:~\cite{Gopakumar:1998ki,Ooguri:1999bv,Gukov:2007ck,
Witten:2011zz,Gaiotto:2011nm,Aganagic:2012jb}, to mention a few.
It is within this context that the present work attempts to provide a unifying and neat scheme of the results
obtained so far and contribute new insights. Specifically, we will first establish a precise
connection between the models in~\cite{Witten:2011zz} and~\cite{Ooguri:1999bv}.
Then, we will reproduce the conclusions of~\cite{Witten:2011zz} in the low energy supergravity description
of a given M-theory model. As we shall see, our approach leads to a strikingly simple analysis
in the context of the usual classical Hamiltonian formalism.

This paper and~\cite{Dasgupta:2016rhc} constitute the first step in the path of computing knot invariants
from M-theory, compactified down to four dimensions.
We here lay the (fertile) ground for embedding knots in our setting.
The simplest knot invariant, the so called linking number, is computed in~\cite{Dasgupta:2016rhc}.
We leave the realization of more challenging invariants to the sequel.

\subsection{Organization of the paper \label{orgsect}}

The paper is arranged as follows. In part \ref{parta}, we construct two distinct M-theory configurations
that have all necessary features to harbor knots. We refer to these as (M, 1) and (M, 5). Specifically,
section \ref{ns5d3sect1} is devoted to the construction of (M, 1), starting from the well-known
D3-NS5 system in type IIB superstring theory considered in~\cite{Witten:2011zz}.
The very same D3-NS5 system is also the basis for the construction of (M, 5), presented in section \ref{ns5d3sect2}.
It is worth pointing out that (M, 5) is dual to the resolved conifold in the presence of fluxes considered
in~\cite{Ooguri:1999bv}.

Part \ref{partb} focuses on the study of the four-dimensional gauge theory that follows from appropriately compactifying
model (M, 1). In particular, section \ref{actionsec} deals with the derivation of its action.
The corresponding Hamiltonian is obtained in section \ref{hamilsec}, where we also minimize its
energy for static configurations of the fields. We thus find the BPS conditions for the gauge theory.
After the energy minimization process, the Hamiltonian reduces to an action in a three-dimensional subspace,
as proved in section \ref{bndsec}. Further, a careful analysis of the symmetries and physics of this
three-dimensional space shows that knots can be consistently embedded in its Euclidean version.

Due to the considerable length of the computational details and arguments presented, we have included
a graphical summary of the paper. It works in the following manner.
By looking at the twelve figures (and their captions) here shown,
the reader can quickly grasp the fundamental logic articulating each part and section.
Additionally, most of the figures refer to equations in the text: these constitute our main results.
Hence, the figures can be used to efficiently localize any particular information of interest within the text.

\subsection{Relation between the present work and~\cite{Dasgupta:2016rhc}}

This is a companion paper to~\cite{Dasgupta:2016rhc}. As such, it aims to clarify the main results stated there,
providing precise mathematical computations to endorse them. The complete list of equations in~\cite{Dasgupta:2016rhc}
that are here proven is shown in table \ref{table3}.
Broadly speaking,
the following are the key points we address exhaustively:

\begin{enumerate}

\item The details of the construction of the M-theory configurations (M, 1) and (M, 5).
In~\cite{Dasgupta:2016rhc}, these are called {\bf Model A} and {\bf Model B}, respectively
and are, to a large extent, simply stated rather than derived. Part \ref{parta} is devoted to rectifying this situation.
Specifically, a special effort is made to quantify all the intermediate geometries and fluxes
that one encounters in constructing (M, 1) and (M, 5) from the D3-NS5 system of~\cite{Witten:2011zz}.
Additionally, we emphasize how all considered configurations are exactly related to each other. 
It should be noted that the figures in part \ref{parta} are conceived to help in this respect.

\begin{table}[ht]
\begin{center}
\begin{tabular}{|c|c|l|c|}
\hline
\multicolumn{3}{|c|}{Present work} 
& Equations in~\cite{Dasgupta:2016rhc} \\
\hline\hline
\multirow{8}{*}{Part \ref{parta}} & 
& & 
(3.4), (3.5), (3.19)-(3.25), (3.33)-(3.39), \\ 
&  & \multirow{-2}{*}{Section \ref{ncsect}}& 
(3.41)-(3.44), (3.46), (3.47), (3.49), (3.53)-(3.55)\\ \cline{3-4}
&  & Section \ref{nonabsec}  &  (3.85), (3.86), (3.89), (3.90)\\ \cline{3-4}
& \multirow{-4}{*}{Section \ref{ns5d3sect1}}  & Section  \ref{rrdefsect} & 
(3.26), (3.29)-(3.32), (3.56)-(3.58)\\ \cline{2-4}
&  & Section \ref{torsionsec}  &  (4.1), (4.20), (4.23)\\ \cline{3-4}
&  & Section \ref{boostsec}   &  (4.2), (4.3)\\ \cline{3-4}
&  & \ &  (4.9), (4.10), (4.13), (4.16), \\
&  & \multirow{-2}{*}{Section \ref{choicesec}} &  (4.17), (4.24)-(4.26), (4.30)\\
\cline{3-4}
&  & Section \ref{oogurivafasec}  &  (4.8)\\
\cline{3-4}
& \multirow{-6}{*}{Section \ref{ns5d3sect2}} & Section \ref{M5sec}  &  (4.39), (4.40), (4.48), (4.51), (4.52)\\ \hline\hline
\multirow{13}{*}{Part \ref{partb}} & \multirow{5}{*}{Section \ref{actionsec}} & Section \ref{kintermsec} &
(3.52), (3.91)-(3.98)\\ \cline{3-4}
& & Section \ref{I11} & (3.76), (3.78)-(3.81)\\ \cline{3-4}
& & \multirow{2}{*}{Section \ref{I12}} & (3.101), (3.102), (3.105)-(3.111), \\
& & & (3.114)-(3.119), (3.121), (3.124)-(3.128)\\ \cline{3-4}
& & Section \ref{masstermsec} & (3.63), (3.67), (3.68)\\ \cline{3-4}
& & Section \ref{thirdtermsec} & (3.136)-(3.139), (3.142)-(3.148), (3.153)\\ \cline{2-4}
& \multirow{4}{*}{Section \ref{hamilsec}} & Section \ref{c20sec} & (3.158), (3.160)\\ \cline{3-4}
& & \multirow{2}{*}{Section \ref{minimsec}} & (3.161), (3.162), (3.167), (3.169),\\
& & & (3.171)-(3.173), (3.177)-(3.182)\\ \cline{3-4}
& & Section \ref{loceqsec} & (3.174), (3.207), (3.218), (3.220), (3.252)-(3.275)\\ \cline{3-4}
& & Section \ref{c2not0sec} & (3.183), (3.187), (3.225)\\ \cline{2-4}
& \multirow{4}{*}{Section \ref{bndsec}} & Section \ref{fsbndsec} & (3.233)\\ \cline{3-4}
& & Section \ref{bcsec} & (3.155), (3.222), (3.223), (3.243), (3.251)\\ \cline{3-4}
& & \multirow{3}{*}{Section \ref{twistsec}} & (3.156), (3.157), (3.184), (3.191), \\
& & & (3.224), (3.232), (3.236), (3.237), \\
& & &  (3.240)-(3.242), (3.246), (3.346)-(3.350)\\ \cline{3-4}
& & Section \ref{twistbulksec} & (3.282), (3.287)\\ \hline
\end{tabular}
\end{center}
\caption{List of equations in~\cite{Dasgupta:2016rhc} for which a detailed derivation can be found in the present work
and the section where this is done. The listed equations are the main results in~\cite{Dasgupta:2016rhc}
and they cover all but section 3.3 there.}
\label{table3}
\end{table}

\item A meticulous explanation (missing in~\cite{Dasgupta:2016rhc})
on what is the four-dimensional gauge theory action associated to (M, 1).
Ultimately, the action is given by (\ref{totaction}), or by (3.153) in the language of~\cite{Dasgupta:2016rhc}.
It depends on various coefficients, summarized in table \ref{table1}, that can be traced to the supergravity
parameters in (M, 1). An important side result is the derivation of these coefficients,
which were merely asserted in~\cite{Dasgupta:2016rhc}.

\item The ins and outs involved in rewriting the action (\ref{totaction}) as a Hamiltonian
that consists on a sum of squared terms, plus contributions from a three-dimensional boundary.
The Hamiltonian in question is first obtained as (\ref{3158}), for a particularly simple limiting case of the gauge theory.
This corresponds to (3.158) in~\cite{Dasgupta:2016rhc}. Right afterwards, 
it is generalized to (\ref{c2notzerohamtot}), a novel result from the perspective of~\cite{Dasgupta:2016rhc}.

\item The present work includes a comprehensive study of the supersymmetry of the gauge theory following from (M, 1).
In particular, it obtains the boundary conditions that the fields must obey so that the theory is $\mathcal{N}=2$
supersymmetric. Such discussion and results are not part of~\cite{Dasgupta:2016rhc}.

\item A basic review of the technique of topological twist and a careful investigation of its
compatibility with the desired amount of supersymmetry is another relevant addendum
to~\cite{Dasgupta:2016rhc} that we elaborate on. The main advantage of doing so results into
further insight into the origin and relevance of the all-important parameter $\hat{t}$
(or simply $t$ in~\cite{Witten:2011zz,Dasgupta:2016rhc})
defined in (\ref{twisvar}).
\end{enumerate}

In spite of its companion paper nature, the present work is self-contained and coherent by itself.
Consequently, it may be read independently of~\cite{Dasgupta:2016rhc}.
Nonetheless, an attempt is made to present all results in a different manner from~\cite{Dasgupta:2016rhc},
so that both works are mutually enriching. In this way, it should be fruitful to
check~\cite{Dasgupta:2016rhc} at times and so complement the present reading.

It is worth mentioning that the mathematical notation, albeit mostly coincident with the one used in~\cite{Dasgupta:2016rhc},
at times differs from it. The reason is simple: to avoid repetition of characters and thus prevent possible confusions that
may arise while reading through~\cite{Dasgupta:2016rhc}.
Nevertheless,
since a one to one mapping of equations is done, the reader should have no difficulty in going from one work to the other.

There is a part of~\cite{Dasgupta:2016rhc} which is not touched upon: it is section 3.3.
No complementary material to section 3.3 applies: it is detailed enough in its own right.
In it, knots are embedded in the aforementioned gauge theory.
This is achieved by introducing M2-branes along some particular directions in the M-theory configuration (M, 1).
From the four-dimensional point of view, such M2-branes are surface operators,
extensively studied co-dimension two objects (for example, see~\cite{Gukov:2008sn}).
Further, the M2-brane surface operators are used to obtain the linking numbers of any arbitrary knot.
The present paper is written so as to allow the interested reader to directly jump from the end of section \ref{bndsec}
to section 3.3 in~\cite{Dasgupta:2016rhc} without any hurdle.

\FloatBarrier

\part{Two M-theory constructions to study knot invariants: (M, 1) and (M, 5) \label{parta}}

As the title suggests, in this first part we will construct two different M-theory configurations
that provide an appropriate framework for the study of knots and their invariants.
We will refer to these configurations as (M, 1) and (M, 5).
Both of them will be directly obtained from the well-known type IIB system of a D3-brane ending on an NS5-brane
considered in~\cite{Witten:2011zz}.
Section \ref{ns5d3sect1} contains the construction of (M, 1) from the D3-NS5 system,
while section \ref{ns5d3sect2} derives (M, 5).
As will be argued towards the end of this first part, in section \ref{oogurivafasec}, (M, 5)
is intimately related to the model in~\cite{Ooguri:1999bv}.
Consequently, this part lays the ground for an explicit connection between the two seemingly different
approaches to study knot invariants of~\cite{Witten:2011zz} and~\cite{Ooguri:1999bv}.

\begin{figure}[hb]
\centering
\includegraphics[width=1\textwidth]{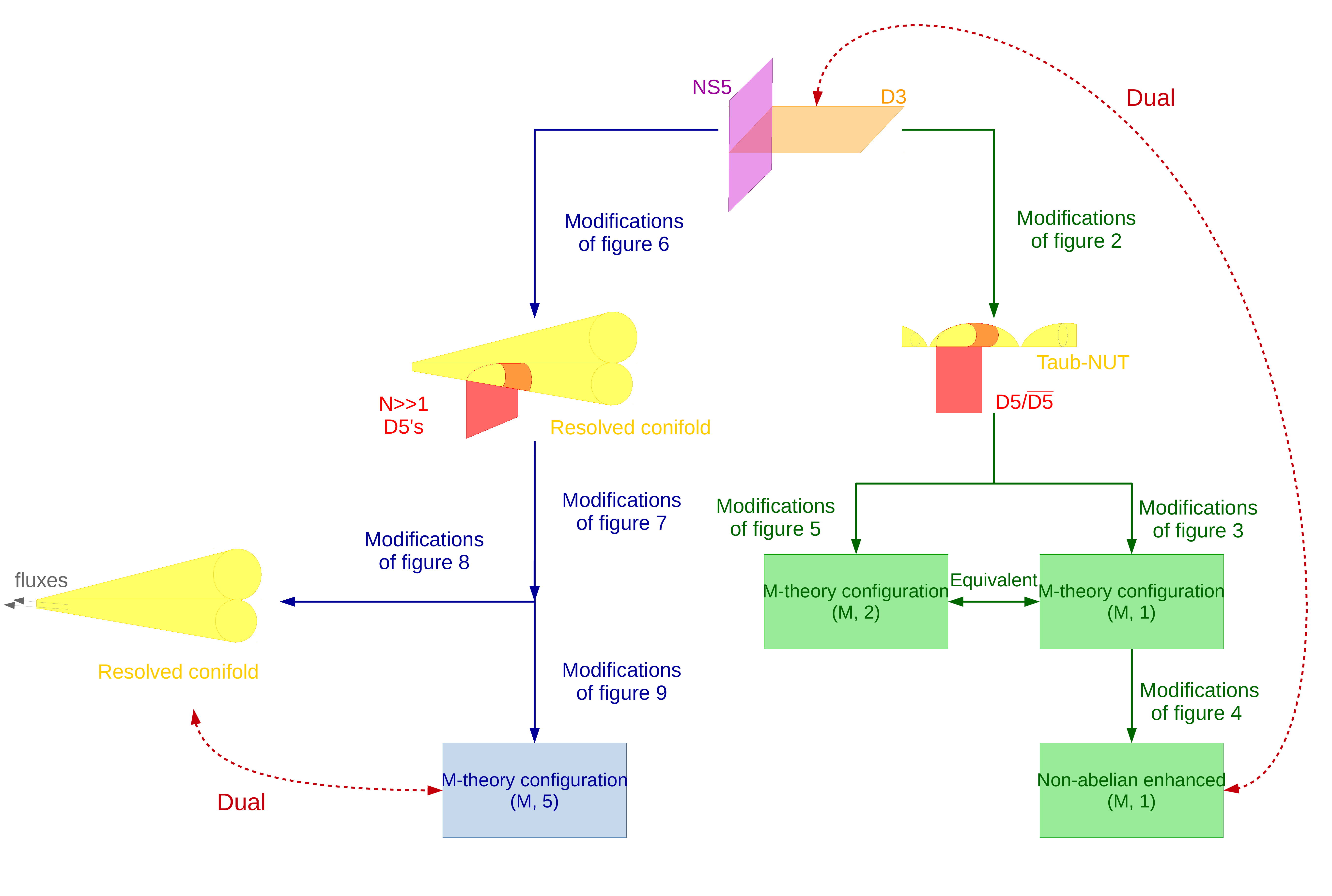}
\caption{Graphical summary of part \ref{parta}.
Starting from the type IIB D3-NS5 system of~\cite{Witten:2011zz},
we construct two different M-theory configurations where knots and their invariants can be studied.
We refer to these as (M, 1) (and its non-abelian enhancement) and (M, 5).
(The configuration (M, 2) is equivalent to (M, 1) for the purposes of our work, yet computationally tougher to handle.
We will thus focus our efforts in the study of (M, 1) only.) Note that (M, 1) is dual to~\cite{Witten:2011zz}.
Similarly, (M, 5) is dual to the resolved conifold in the presence of fluxes considered
in~\cite{Ooguri:1999bv}.
The right-hand side of the figure, colored green, schematizes the contents of section \ref{ns5d3sect1}.
The left-hand side, in blue, depicts the discussion in section \ref{ns5d3sect2}.}
\label{fig8}
\end{figure}

Before proceeding to the details, a word of warning: we will consider multiple type IIA, IIB and M-theory configurations.
Figure \ref{fig8} provides a visual sketch of the overall logic in this part.
Hence, the reader may find it clarifying to come back to this image while reading through
sections \ref{ns5d3sect1} and \ref{ns5d3sect2}.

\FloatBarrier

\section{The D3-NS5 system modified \label{ns5d3sect1}}

As we just mentioned, the starting point of our analysis is the well-known type IIB superstring theory configuration
of a D3-brane ending on an NS5-brane.
In more detail, we consider Minkowski spacetime $\mathbb{R}^{1,9}$, with mostly positive metric signature.
We denote the coordinates as $(t,\,x_1,\,x_2,\,x_3,\theta_1,\,\phi_1,\,\psi,\,r,\,x_8,\,x_9)$.
(The identifications $(x_4\equiv\theta_1,\,x_5\equiv\phi_1,\,x_6\equiv\psi,\,x_7\equiv r)$ will shortly become sensible.)
We take the D3-brane to stretch along $(t,\,x_1,\,x_2,\,\psi)$ and the NS5-brane along $(t,\,x_1,\,x_2,\,x_3,\,x_8,\,x_9)$.
The $U(1)$ gauge theory on the D3-brane has $\mathcal{N}=4$ supersymmetry and
the intersecting NS5-brane provides a half-BPS boundary condition.
The world-volume gauge theory thus has $\mathcal{N}=2$ supersymmetry.
This is, essentially, the starting point of~\cite{Witten:2011zz} as well.
(The only difference is that, in~\cite{Witten:2011zz}, an axionic background $\mathcal{C}_0$
is switched on. We will elaborate on this point in section \ref{rrdefsect}.)

\begin{figure}[t]
\centering
\includegraphics[width=0.9\textwidth]{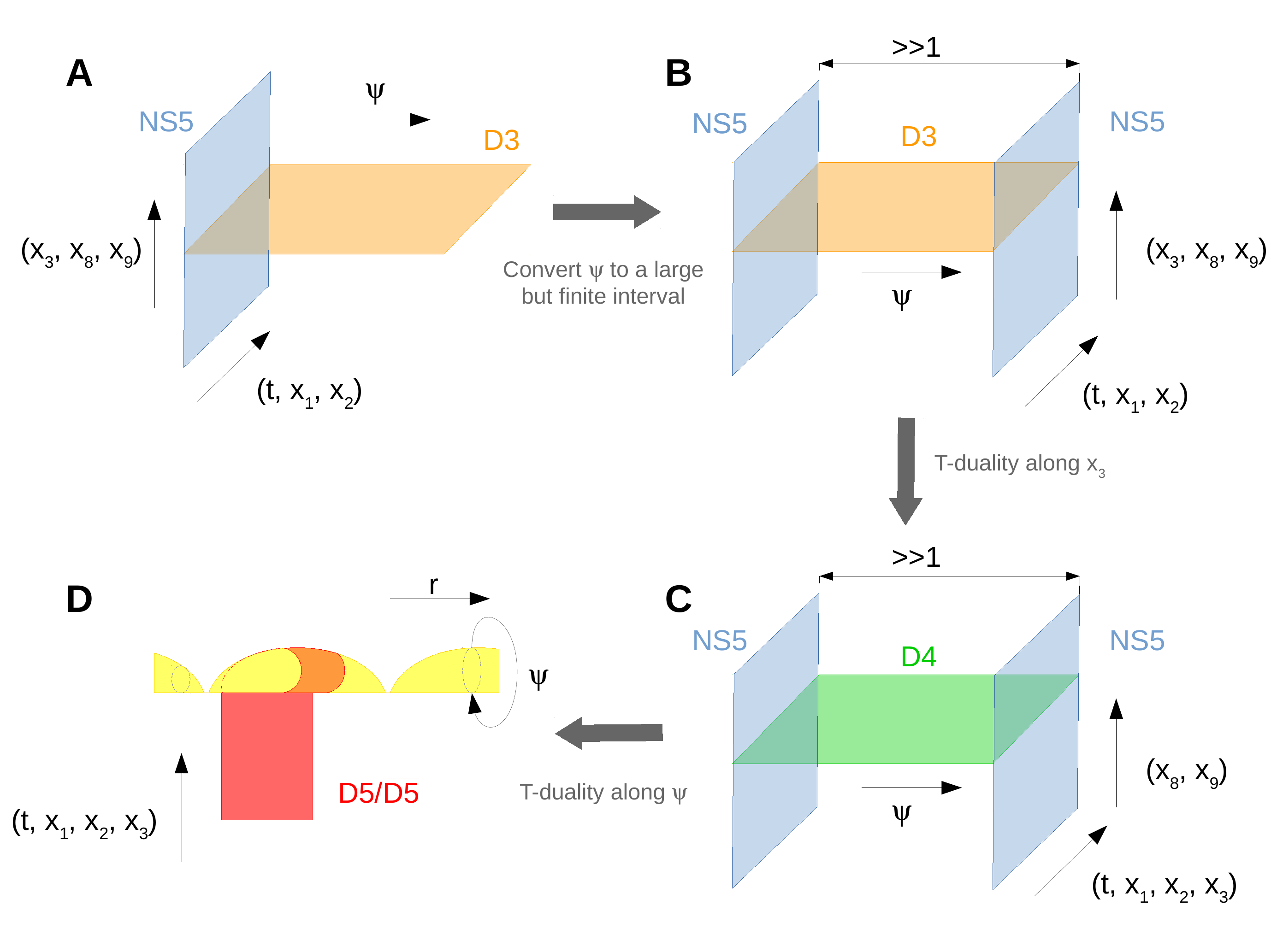}
\caption{Caricature of the modifications to the D3-NS5 system described in section \ref{ns5d3sect1}.
This chain of dualities is done so that the corresponding metric can be written: the geometry of {\bf D} is well-known.
{\bf A:} The well-known type IIB D3-NS5 system. The corresponding world-volume gauge theory has $\mathcal{N}=2$
supersymmetry.
The D3-brane spans the $(t,\,x_1,\,x_2,\,\psi)$ directions and the NS5-brane the
$(t,\,x_1,\,x_2,\,x_3,\,x_8,\,x_9)$ directions.
The $(\theta_1,\,\phi_1,\,r)$ directions are suppressed. 
{\bf B:} Introducing a second NS5-brane, parallel to the first one, converts the $\psi$ direction into
an interval. We take this interval to be large (but finite) in order to effectively retain the same amount of
supersymmetry.
{\bf C:} A T-duality along $x_3$ does not affect the parallel NS5-branes, but converts the D3-brane into a D4-brane.
{\bf D:} A T-duality along $\psi$ converts the parallel NS5-branes to a warped
Taub-NUT space along $(\theta_1,\,\phi_1,\,\psi,\,r)$.
The D4-brane converts to a D5/${\overline{\rm D5}}$ pair that wraps the $\psi$ direction and stretches along $r$.
The $(\theta_1,\,\phi_1,\,x_8,\,x_9)$ directions are suppressed.}
\label{fig1}
\end{figure}

Next, we do three modifications to the above set up.
These are depicted schematically in figure \ref{fig1} and discussed in the following.

\begin{itemize}

\item First, we introduce a second NS5-brane, parallel to the first one and which also intersects the D3-brane.
This means that the orthogonal direction to the NS5-branes of the D3-brane, namely $\psi$, is now a finite interval.
The inclusion of the second NS5-brane halves the amount of supersymmetry of the gauge theory on the D3-brane.
However, we consider the case when the $\psi$ interval is very large (that is, the two NS5-branes are far from each other).
Then, near the original NS5-brane, effectively no supersymmetry is lost in this step.

\item Second, we do a T-duality to type IIA superstring theory along $x_3$.
As a result, we now have a D4-brane (instead of a D3-brane) between the same two NS5-branes of before.

\item Third, we do a T-duality back to type IIB along $\psi$.
The NS5-branes thus disappear and give rise to a warped Taub-NUT space in the $(\theta_1,\,\phi_1,\,\psi,\,r)$ directions.
(This justifies the coordinate relabeling above.)
As argued in~\cite{Dasgupta:1999wx}, because $\psi$ is a finite interval and because
our construction leads to an $\mathcal{N}=2$ supersymmetric world-volume gauge theory, the D4-brane converts to a
D5/${\overline{\rm D5}}$
pair which wraps the $\psi$ direction and stretches along $r$.
\end{itemize}

The geometry corresponding to this last configuration is well-known
(in fact, the three modifications above were done only to be able to write the corresponding metric)
and is given by (3.4) and (3.5) in~\cite{Dasgupta:2016rhc}:
\begin{align}
ds_{(B,1)}^2=&e^{-\phi}(-dt^2+dx_1^2+dx_2^2+dx_3^2)+e^\phi F_4(dx_8^2+dx_9^2) \nonumber \\
&+e^\phi[F_1dr^2+F_2(d\psi+\cos\theta_1 d\phi_1)^2
+F_3(d\theta_1^2+\sin^2\theta_1d\phi_1^2)], \label{3.43.5}
\end{align}
where $e^{-\phi}$ is the usual type IIB dilaton.
(Since we will consider many metrics in the ongoing, we adopt the notation $ds_{(X,n)}^2$.
Here $X=A,\,B,\,M$ stands for type IIA, type IIB and M-theory, respectively and $n\in\mathbb{N}$ is an index to label
the different metrics that will occur.)
We consider, for simplicity, the following dependence of the warp factors and dilaton\footnote{As made more
precise in section \ref{torsionsec}, a definite choice of the warp factors and dilaton
will in general not preserve the $\mathcal{N}=2$ supersymmetry of the world-volume gauge theory.
Consequently, any concrete choice one may wish to consider must be checked to indeed
preserve the desired amount of supersymmetry.}:
\begin{align}
F_i=F_i(r), \quad\quad F_4=F_4(r,x_8,x_9), \quad\quad \phi=\phi(\theta_1,r,x_8,x_9), \quad\quad  i=1,2,3.
\label{easychoice}
\end{align}
The warped Taub-NUT space metric is, quite obviously, the second line in (\ref{3.43.5}).

Let us move the ${\overline{\rm D5}}$-brane far away along the $(x_8,\,x_9)$ directions 
(the Coulomb branch) and consider only the D5-brane. This will simplify the flux discussion
in the construction of the M-theory configurations (M, 1) (and its non-abelian enhanced version) and (M, 2)
that concern us in the present section \ref{ns5d3sect1} (see figure \ref{fig8}).
Nonetheless, in section \ref{masstermsec}, we will ``move back'' this ${\overline{\rm D5}}$-brane
and appropriately account for its effects. We will then see that the ${\overline{\rm D5}}$-brane plays 
an important, non-trivial role in our investigations.

It has been known for quite some time now that D-branes carry Ramond-Ramond (RR) charges~\cite{Polchinski:1995mt}.
In this case that concerns us, the D5-brane sources an RR three-form flux $\mathcal{F}_3^{(B,1)}$ 
that can be computed as\footnote{For a review on how fluxes can be determined, see~\cite{BLT}.}
\begin{align}
\mathcal{F}_3^{(B,1)}=e^{2\phi}\ast d\mathcal{J}_{(B,1)},
\end{align}
where $\mathcal{J}_{(B,1)}$ stands for the fundamental form of the metric $e^{-\phi}ds_{(B,1)}^2$
along the directions $(\theta_1,\,\phi_1,\,\psi,\,r,\,x_8,\,x_9)$, which we call $ds_{(1)}^2$:
\begin{align}
ds_{(1)}^2\equiv  F_1dr^2+F_2(d\psi+\cos\theta_1 d\phi_1)^2
+F_3(d\theta_1^2+\sin^2\theta_1d\phi_1^2)+F_4(dx_8^2+dx_9^2).
\label{ds1}
\end{align}
Let us calculate $\mathcal{F}_3^{(B,1)}$ in details next.

We take the vielbeins of (\ref{ds1}) to be
\begin{eqnarray}
\label{viB1}
\begin{array}{lllll}
&E_{\theta_1}^{(B,1)}=\sqrt{F_3}e_{\theta_1}^{(B,1)}=\sqrt{F_3}d\theta_1, \quad\quad
& E_{\phi_1}^{(B,1)}=\sqrt{F_3}e_{\phi_1}^{(B,1)}=\sqrt{F_3}\sin\theta_1 d\phi_1, \\ 
&E_\psi^{(B,1)}=\sqrt{F_2}e_\psi^{(B,1)}=\sqrt{F_2}(d\psi+\cos\theta_1 d\phi_1), \quad\quad
& E_r^{(B,1)}=\sqrt{F_1}e_r^{(B,1)}=\sqrt{F_1}dr, \\ 
&E_8^{(B,1)}=\sqrt{F_4}e_8^{(B,1)}=\sqrt{F_4}dx_8, \quad\quad
& E_9^{(B,1)}=\sqrt{F_4}e_9^{(B,1)}=\sqrt{F_4}dx_9.
\end{array}
\end{eqnarray}
These vielbeins can be used to compute the fundamental form $\mathcal{J}_{(B,1)}$.
The result is (3.19) in~\cite{Dasgupta:2016rhc}:
\begin{align}
\mathcal{J}_{(B,1)}=& E_{\theta_1}^{(B,1)}\wedge E_{\phi_1}^{(B,1)}+E_\psi^{(B,1)}\wedge E_r^{(B,1)}+E_8^{(B,1)}
\wedge E_9^{(B,1)} \nonumber \\
=&F_3\sin\theta_1 d\theta_1\wedge d\phi_1+\sqrt{F_1F_2}(d\psi+\cos\theta_1 d\phi_1)\wedge dr+F_4dx_8\wedge dx_9.
\end{align}
The exterior derivative of $\mathcal{J}_{(B,1)}$ is given by
\begin{align}
d\mathcal{J}_{(B,1)}=&F_{3,r}\sin\theta_1 dr\wedge d\theta_1\wedge d\phi_1-\sqrt{F_1F_2}\sin\theta_1 d\theta_1
\wedge d\phi_1\wedge dr
+F_{4,r}dr\wedge dx_8\wedge dx_9 \nonumber \\
=&(F_{3,r}-\sqrt{F_1F_2})\sin\theta_1 dr\wedge d\theta_1 \wedge d\phi_1+F_{4,r} dr\wedge dx_8\wedge dx_9,
\label{dJB1}
\end{align}
where $(F_{3,r},\, F_{4,r})$ stand for the derivatives of $(F_3,\,F_4)$ with respect to $r$.
In order to take the Hodge dual of $d\mathcal{J}_{(B,1)}$, we start by showing a few intermediate steps.
First, we write the metric (\ref{ds1}) in matrix form:
\begin{eqnarray}
M=\left(
\begin{array}{ccccccccc}
F_3 & 0 & 0 & 0 & 0 & 0 \\
0 & F_2\cos^2\theta_1+F_3\sin^2\theta_1 & F_2\cos\theta_1 & 0 & 0 & 0 \\
0 & F_2\cos\theta_1 & F_2 & 0 & 0 & 0 \\
0 & 0 & 0 & F_1 & 0 & 0 \\
0 & 0 & 0 & 0 & F_4 & 0 \\
0 & 0 & 0 & 0 & 0 & F_4
\end{array}
\right). \label{B1partmetric}
\end{eqnarray}
We denote as $\mathcal{M}$ the square root of the determinant of this matrix:
\begin{align}
\mathcal{M}\equiv \sqrt{\textrm{det} M}=\sqrt{F_1F_2}F_3F_4\sin\theta_1.
\end{align}
The inverse of $M$ is
\begin{eqnarray}
M^{-1}=\left(
\begin{array}{ccccccccc}
\frac{1}{F_3} & 0 & 0 & 0 & 0 & 0 \\
0 & \frac{\csc^2\theta_1}{F_3} & -\frac{\cot\theta_1\csc\theta_1}{F_3} & 0 & 0 & 0 \\
0 & -\frac{\cot\theta_1\csc\theta_1}{F_3} & \frac{1}{F_2}+\frac{\cot^2\theta_1}{F_3} & 0 & 0 & 0 \\
0 & 0 & 0 & \frac{1}{F_1} & 0 & 0 \\
0 & 0 & 0 & 0 & \frac{1}{F_4} & 0 \\
0 & 0 & 0 & 0 & 0 & \frac{1}{F_4}
\end{array}
\right).
\end{eqnarray}
The above three equations allow us to compute the Hodge dual of the wedge products
in (\ref{dJB1}). We obtain
\begin{align}
\ast(dr\wedge d\theta_1\wedge d\phi_1)=&\mathcal{M} M^{-1}_{rr} M^{-1}_{\theta_1\theta_1}(M^{-1}_{\phi_1\phi_1}
\epsilon_{r\theta_1\phi_1\psi x_8x_9}d\psi +M^{-1}_{\phi_1\psi}
\epsilon_{r\theta_1\psi\phi_1 x_8 x_9}d\phi_1)\wedge dx_8\wedge dx_9 \nonumber \\
=&-\sqrt{\frac{F_2}{F_1}}\frac{F_4}{F_3}\left(\csc\theta_1 d\psi
+\cot\theta_1 d\phi_1\right)\wedge dx_8\wedge dx_9, \\
\ast(dr\wedge dx_8\wedge dx_9)=&\mathcal{M}M^{-1}_{rr} M^{-1}_{x_8x_8} M^{-1}_{x_9x_9}\epsilon_{rx_8x_9\psi\theta_1\phi_1}
d\psi\wedge d\theta_1 \wedge d\phi_1\nonumber \\
=&\sqrt{\frac{F_2}{F_1}}\frac{F_3}{F_4}\sin\theta_1d\psi\wedge d\theta_1\wedge d\phi_1.
\end{align}
Consequently, the Hodge dual of $d\mathcal{J}_{(B,1)}$ is
\begin{align}
\ast d\mathcal{J}_{(B,1)}=e^{-2\phi}\left[
k_2(d\psi+\cos\theta_1d\phi_1)\wedge dx_8\wedge dx_9 
+k_1\sin\theta_1 d\psi\wedge d\theta_1 \wedge d\phi_1 \right],
\end{align}
where we have defined, following (3.21) in~\cite{Dasgupta:2016rhc},
\begin{align}
k_1\equiv e^{2\phi}\sqrt{\frac{F_2}{F_1}}\frac{F_3}{F_4}F_{4,r}, \quad\quad
k_2\equiv e^{2\phi}\sqrt{\frac{F_2}{F_1}}\frac{F_4}{F_3}(\sqrt{F_1F_2}-F_{3,r}).
\label{k12def}
\end{align}
Further using the vielbeins (\ref{viB1}),
we obtain the desired result, the RR three-form flux $\mathcal{F}_3^{(B,1)}$,
which precisely matches (3.20) in~\cite{Dasgupta:2016rhc}:
\begin{align}
\mathcal{F}_3^{(B,1)}=e_\psi^{(B,1)} \wedge \left(k_1 e_{\theta_1}^{(B,1)}\wedge e_{\phi_1}^{(B,1)}
+k_2  e_8^{(B,1)}\wedge e_9^{(B,1)}\right).
\label{320}
\end{align}
It is important to note that this three-form is not closed: $d\mathcal{F}_3^{(B,1)}\neq0$.
This reflects the presence of the D5-brane in this configuration.

Summing up, the type IIB configuration shown in figure \ref{fig1}{\bf D}
can be obtained directly from the well-known D3-NS5 system. It
has the metric (\ref{3.43.5}), dilaton $e^{-\phi}$ and an RR three-form flux (\ref{320}).

An essential ingredient that makes the study of knots using the D3-NS5 system possible is the presence of a
$\Theta$-term in the D3-brane gauge theory.
In the case of~\cite{Witten:2011zz}, this term is sourced by an axionic background $\mathcal{C}_0$.
In the following (section \ref{ncsect}), we will present an {\it alternative} 
(and computationally simpler) way to source the required $\Theta$-term:
by further modifying the above set up switching on a non-commutative deformation.
The fact that we do not need to (though, of course, we can) switch on $\mathcal{C}_0$ in order to have
an M-theory construction on which knot invariants can be studied will be the focus of section \ref{rrdefsect}.

\FloatBarrier

\subsection{Sourcing the $\Theta$-term: a non-commutative deformation \label{ncsect}}

\begin{figure}[t]
\centering
\includegraphics[width=0.9\textwidth]{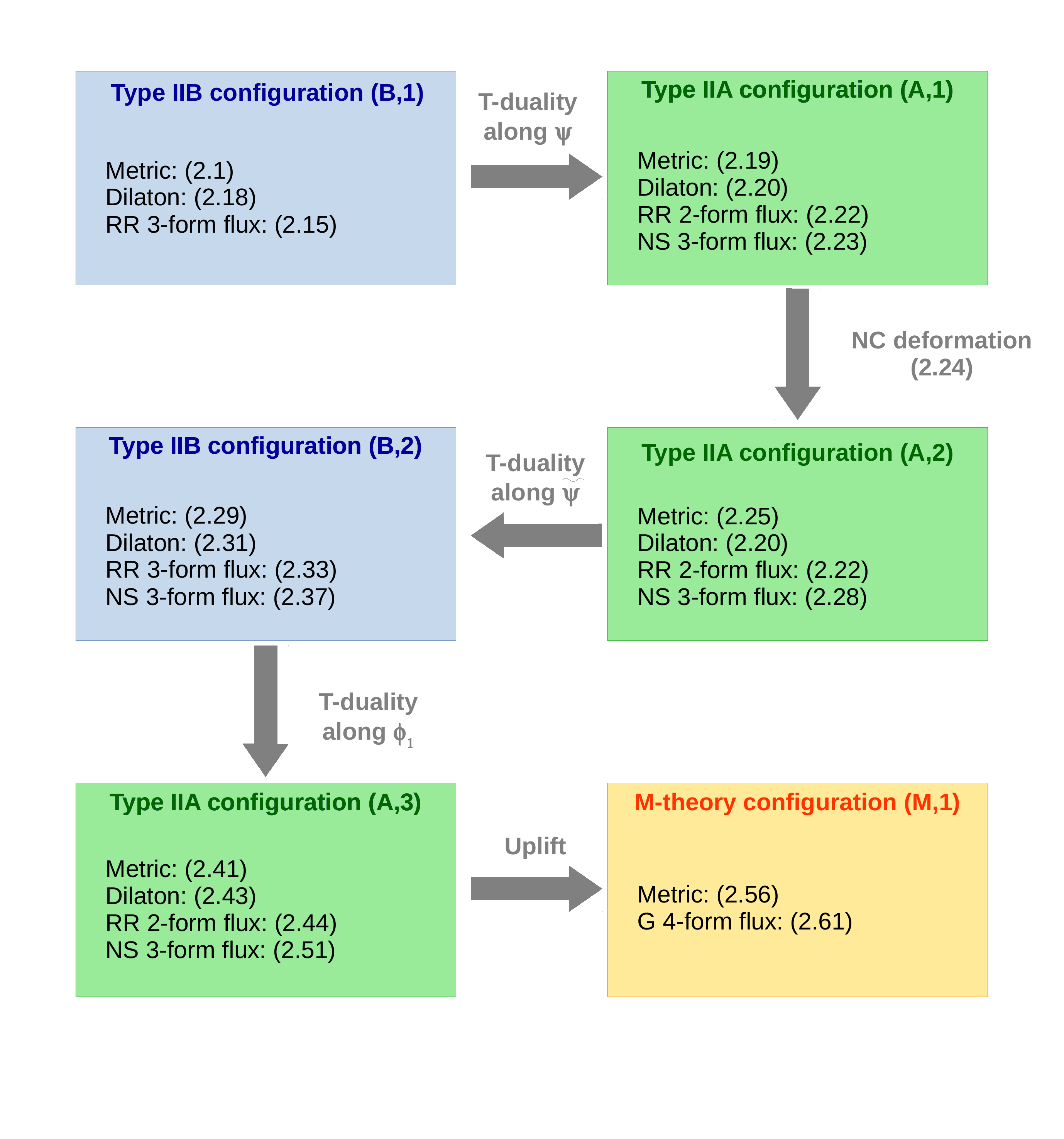}
\caption{Graphical summary of section \ref{ncsect}.
To the type IIB configuration of figure \ref{fig1}{\bf D} we do a series of modifications in order to source
a $\Theta$-term in the $U(1)$ world-volume gauge theory. This is achieved in going from the configuration (B,1) to (B,2).
The presence of a $\Theta$-term is essential to, later on, construct a three-dimensional space
with the required features to allow for the realization of knots.
The (B,2) configuration is then lifted to M-theory.
The configuration (M, 1) (and its non-abelian enhanced version, studied in section \ref{nonabsec})
is the first M-theory construction where knots can be studied.}
\label{fig3}
\end{figure}

The starting point in this section is, of course, the just discussed type IIB geometry in (\ref{3.43.5}).
We will first T-dualize this to type IIA along $\psi$.
(This means we will move from {\bf D} to {\bf C} in figure \ref{fig1}.)
Here, we will do the non-commutative deformation, which will only affect the $(x_3,\,\psi)$ directions:
$(x_3,\,\psi)\rightarrow(\tilde{x}_3,\,\tilde{\psi})$.
This will be followed by another T-duality along $\tilde{\psi}$.
At this point, we will have a type IIB configuration capable of sourcing the required $\Theta$-term in
the $U(1)$ world-volume gauge theory.
Then, we will T-dualize along $\phi_1$ to type IIA.
Finally, we will lift the resulting configuration to M-theory.
Along the way, we will also study the NS B-field, dilaton and fluxes associated to each geometry considered,
which will in turn shed some light into the connection between the non-commutative deformation and the $\Theta$-term.
(The precise connection between these two will be shown early in section \ref{c2not0sec},
see (\ref{gaugeparam}).)
Figure \ref{fig3} summarizes this chain of modifications and points out what the most relevant equations in this section are.

Let us go ahead and show in details the above outlined M-theory construction.
We start by rewriting the metric (\ref{3.43.5}) in a more convenient way for our present purposes\footnote{All through
this paper, we will use the formulae in section 6.5 of~\cite{Head} to perform T- and S-dualities
and to go from (to) type IIA to (from) M-theory.
Accordingly, we will always write the relevant metrics in the form that makes it straightforward to apply those formulae.}:
\begin{align}
ds_{(B,1)}^2=ds_{(2)}^2+e^{-\phi}dx_3^2 
+e^\phi F_2(d\psi+\cos\theta_1 d\phi_1)^2,
\end{align}
with $ds_{(2)}^2$ defined as
\begin{align}
ds_{(2)}^2\equiv e^{-\phi}(-dt^2+dx_1^2+dx_2^2)+e^\phi[F_1dr^2
+F_3(d\theta_1^2+\sin^2\theta_1d\phi_1^2)+F_4(dx_8^2+dx_9^2)].
\label{ds22}
\end{align}
We recall that the dilaton here is
\begin{align}
e^{\phi_{(B,1)}}=e^{-\phi}
\label{B1dilaton}
\end{align}
and the RR three-form flux was given in (\ref{320}).

T-dualizing along $\psi$, we get the metric
\begin{align}
ds_{(A,1)}^2=ds_{(2)}^2+e^{-\phi}dx_3^2 +\frac{e^{-\phi}}{F_2}d\psi^2,
\end{align}
with associated NS B-field and dilaton
\begin{align}
B_{(A,1)}=\cos\theta_1 d\psi\wedge d\phi_1, \quad\quad e^{\phi_{(A,1)}}=(e^{3\phi}F_2)^{-1/2}.
\end{align}
We take the relevant vielbeins associated to $ds_{(A,1)}^2$ to be
\begin{eqnarray}
\label{viA1}
\begin{array}{lllll}
&e_{\theta_1}^{(A,1)}=d\theta_1, \quad\quad
&e_{\phi_1}^{(A,1)}=\sin\theta_1 d\phi_1, \quad\quad
&e_\psi^{(A,1)}=d\psi+\cos\theta_1 d\phi_1, \\
&e_r^{(A,1)}=dr, \quad\quad
&e_8^{(A,1)}=dx_8, \quad\quad
&e_9^{(A,1)}=dx_9.
\end{array}
\end{eqnarray}
As for the fluxes, the RR three-form flux in (\ref{320}) now gives rise to the following RR two-form flux:
\begin{align}
\mathcal{F}_2^{(A,1)}=k_1 e_{\theta_1}^{(A,1)}\wedge e_{\phi_1}^{(A,1)}
+k_2  e_8^{(A,1)}\wedge e_9^{(A,1)}. \label{F2A1}
\end{align}
Note that, for an arbitrary value of the warp factors and $\phi$, the above flux is not closed: $d\mathcal{F}_2^{(A,1)}\neq0$.
This is consistent with having a D4-brane as a source (see figure \ref{fig1}{\bf C}).
The NS three-form flux is given by
\begin{align}
\mathcal{H}_3^{(A,1)}=dB_{(A,1)}=-\sin\theta_1 d\theta_1\wedge d\psi\wedge d\phi_1.
\label{H3A1}
\end{align}

We will now deform the above type IIA configuration.
The non-commutative deformation $(x_3,\,\psi)\rightarrow(\tilde{x}_3,\,\tilde{\psi})$ that we will consider
is
\begin{align}
\psi=\cos\theta_{nc}\tilde{\psi}, \quad\quad x_3=\sec\theta_{nc}\tilde{x}_3+\sin\theta_{nc}\tilde{\psi},
\label{ncdeformation}
\end{align}
where $\theta_{nc}\in[0,\,2\pi)$ is the deformation parameter.
Note that the $(x_3,\,\psi)$ directions in $ds_{(A,1)}^2$ form a square torus;
that is, a geometry which is isometric to a square with opposite sides identified.
Hence, the non-commutative deformation simply inclines the torus.
This same deformation was considered in~\cite{Chakravarty:2000qd}, albeit in a different context.
Under this deformation, the above type IIA metric changes to
\begin{align}
ds_{(A,2)}^2=&ds_{(2)}^2+e^{-\phi}(\sec\theta_{nc}d\tilde{x}_3+\sin\theta_{nc}d\tilde{\psi})^2
+\frac{e^{-\phi}}{F_2}\cos^2\theta_{nc} d\tilde{\psi}^2 \nonumber \\
=&ds_{(2)}^2+e^{-\phi}\left[\frac{\tilde{F}_2}{F_2}\sec^2\theta_{nc}d\tilde{x}_3^2+\frac{\cos^2\theta_{nc}}{\tilde{F}_2}
(d\tilde{\psi}+\tilde{F}_2\sec^2\theta_{nc}\tan\theta_{nc}d\tilde{x}_3)^2\right],
\label{deformetric}
\end{align}
where we have defined
\begin{align}
\label{tildeF2}
\tilde{F}_2\equiv&\frac{F_2}{1+F_2\tan^2\theta_{nc}},
\end{align}
as in (3.35) in~\cite{Dasgupta:2016rhc}
and the last rewriting of $ds_{(A,2)}^2$ was done in anticipation to the T-duality along $\tilde{\psi}$
that will soon follow.
The NS B-field is also affected by the deformation and now takes the form
\begin{align}
B_{(A,2)}=\cos\theta_{nc}\cos\theta_1 d\tilde{\psi}\wedge d\phi_1.
\end{align}
On the other hand, due to our simplifying choices in (\ref{easychoice}), the dilaton remains
unchanged: $e^{\phi_{(A,2)}}=e^{\phi_{(A,1)}}$.
The RR two-form flux (\ref{F2A1}) is also not affected by this deformation, $\mathcal{F}_2^{(A,2)}=\mathcal{F}_2^{(A,1)}$,
but the NS three-form flux in (\ref{H3A1}) changes to
\begin{align}
\mathcal{H}_3^{(A,2)}=dB_{(A,2)}=-\cos\theta_{nc}\sin\theta_1 d\theta_1\wedge d\tilde{\psi}\wedge d\phi_1.
\label{H3A2}
\end{align}

T-dualizing the metric (\ref{deformetric}) along $\tilde{\psi}$, one obtains the type IIB metric given in (3.22) in~\cite{Dasgupta:2016rhc}:
\begin{align}
ds_{(B,2)}^2=ds_{(2)}^2+e^{-\phi}\frac{\tilde{F}_2}{F_2}\sec^2\theta_{nc}d\tilde{x}_3^2
+e^\phi\tilde{F}_2(\frac{d\tilde{\psi}}{\cos\theta_{nc}}+\cos\theta_1d\phi_1)^2. \label{322}
\end{align}
The NS B-field associated to $ds_{(B,2)}^2$ is 
\begin{align}
B_{(B,2)}=\tilde{F}_2\sec^2\theta_{nc}\tan\theta_{nc}(d\tilde{\psi}+\cos\theta_{nc}\cos\theta_1d\phi_1)\wedge d\tilde{x}_3
\label{nssource}
\end{align}
and the dilaton is that suggested in (3.25) in~\cite{Dasgupta:2016rhc}:
\begin{align}
e^{\phi_{(B,2)}}=\sqrt{\frac{\tilde{F}_2}{F_2}}\sec\theta_{nc}e^{-\phi}.
\end{align}
To the $ds_{(B,2)}^2$ metric, we associate the following relevant vielbeins:
\begin{eqnarray}
\label{viB2}
\begin{array}{lllll}
&e_{\tilde{3}}^{(B,2)}=d\tilde{x}_3, \quad\quad
&e_{\theta_1}^{(B,2)}=d\theta_1, \quad\quad 
&e_{\phi_1}^{(B,2)}=\sin\theta_1 d\phi_1, \\ 
&e_{\tilde{\psi}}^{(B,2)}=d\tilde{\psi}+\cos\theta_{nc}\cos\theta_1 d\phi_1, \quad\quad
&e_8^{(B,2)}=dx_8, \quad\quad
&e_9^{(B,2)}=dx_9.
\end{array}
\end{eqnarray}
In terms of these, it is not hard to see that the RR three-form flux $\mathcal{F}_3^{(B,2)}$ dual to $\mathcal{F}_2^{(A,2)}$
can be written as in (3.23) in~\cite{Dasgupta:2016rhc}:
\begin{align}
\mathcal{F}_3^{(B,2)}=e_{\tilde{\psi}}^{(B,2)} \wedge \left(k_1 e_{\theta_1}^{(B,2)}\wedge e_{\phi_1}^{(B,2)}
+k_2  e_8^{(B,2)}\wedge e_9^{(B,2)}\right).
\label{323}
\end{align}
Once again, it is important to note that the flux $\mathcal{F}_3^{(B,2)}$ is not closed: $d\mathcal{F}_3^{(B,2)}\neq 0$.
This implies that indeed there is a D5-brane in this set up.
For completeness, we give the expression of $d\mathcal{F}_3^{(B,2)}$.
Rewriting (\ref{323}) as
\begin{align}
\mathcal{F}_3^{(B,2)}=k_1\sin\theta_1 d\tilde{\psi}\wedge d\theta_1\wedge d\phi_1
+k_2(d\tilde{\psi}+\cos\theta_{nc}\cos\theta_1 d\phi_1)\wedge dx_8\wedge dx_9,
\label{F3B2}
\end{align}
it is easy to see that its exterior derivative is that in (3.38) in~\cite{Dasgupta:2016rhc}:
\begin{align}
\nonumber
d\mathcal{F}_3^{(B,2)}=&k_{1,a}\sin\theta_1 da\wedge d\tilde{\psi}\wedge d\theta_1\wedge d\phi_1
+k_{2,a} da\wedge (d\tilde{\psi}+\cos\theta_{nc}\cos\theta_1 d\phi_1)\wedge dx_8\wedge dx_9 \\
&-k_2\cos\theta_{nc}\sin\theta_1 d\theta_1\wedge d\phi_1\wedge dx_8\wedge dx_9,
\end{align}
where we have defined $a\equiv (\theta_1,\,r,\, x_8,\, x_9)$ since, due to our choices (\ref{easychoice}),
$(k_1,\,k_2)$ only depend on these coordinates (and on the deformation parameter $\theta_{nc}$).
Determining $\mathcal{H}_3^{(B,2)}$ is also not hard.
Taking the exterior derivative of $B_{(B,2)}$, we obtain
\begin{align}
\mathcal{H}_3^{(B,2)}=\sec\theta_{nc}\tan\theta_{nc}\left[\tilde{F}_{2,r}\sec\theta_{nc} dr\wedge
(d\tilde{\psi}+\cos\theta_{nc}\cos\theta_1d\phi_1)
-\tilde{F}_2\sin\theta_1d\theta_1\wedge d\phi_1\right]\wedge d\tilde{x}_3,
\end{align}
which is a closed form by definition.
From (\ref{tildeF2}) it can be easily checked that $\tilde{F}_{2,r}=(\tilde{F}_2/F_2)^2F_{2,r}$.
Also using the vielbeins in (\ref{viB2}),
we can rewrite the NS flux as in (3.24) in~\cite{Dasgupta:2016rhc}:
\begin{align}
\mathcal{H}_3^{(B,2)}=\tilde{F}_2\sec\theta_{nc}\tan\theta_{nc}\left(\frac{\tilde{F}_2F_{2,r}}{F_2^2}\sec\theta_{nc}
e_r^{(B,2)}\wedge e_{\tilde{\psi}}^{(B,2)}-e_{\theta_1}^{(B,2)}\wedge e_{\phi_1}^{(B,2)}\right)\wedge e_{\tilde{3}}^{(B,2)}.
\end{align}

So far, all we have done in this section boils down to introducing an NS B-field to the type IIB configuration
that was our starting point (described in section \ref{ns5d3sect1} and depicted in figure \ref{fig1}{\bf D}).
This NS B-field, in turn, sources the NS three-form flux we just determined.
In section \ref{finalsect}, we will see how this NS flux sources the desired $\Theta$-term in the $U(1)$ world-volume gauge theory.
For the time being, however, let us focus on the construction of the M-theory configuration associated to this set up.

The following step in the duality chain outlined at the beginning of this section is to take
the T-dual along $\phi_1$ of (\ref{322}).
In order to make this step easy, we rewrite the aforementioned metric as
\begin{align}
ds_{(B,2)}^2=&ds_{(3)}^2
+e^\phi(\tilde{F}_2\cos^2\theta_1+F_3\sin^2\theta_1)
\Bigg(d\phi_1+\frac{\tilde{F}_2\cos\theta_1\sec\theta_{nc}}{\tilde{F}_2\cos^2\theta_1+F_3\sin^2\theta_1}d\tilde{\psi}\Bigg)^2,
\end{align}
where we have defined 
\begin{align}
\nonumber
ds_{(3)}^2\equiv& e^{-\phi}(-dt^2+dx_1^2+dx_2^2+\frac{\tilde{F}_2}{F_2}\sec^2\theta_{nc}d\tilde{x}_3^2)
+e^\phi\frac{\tilde{F}_2F_3\sec^2\theta_{nc}\sin^2\theta_1}{\tilde{F}_2\cos^2\theta_1+F_3\sin^2\theta_1}d\tilde{\psi}^2 \\
&+e^\phi[F_1dr^2+F_3d\theta_1^2+F_4(dx_8^2+dx_9^2)].
\end{align}
Similarly, a rewriting of its associated NS B-field will make the next duality straightforward:
\begin{align}
B_{(B,2)}=\tilde{F}_2\sec\theta_{nc}\tan\theta_{nc}\left[\sec\theta_{nc}
\frac{(F_3\sin^2\theta_1+\tilde{F}_2\cos^2\theta_1)}{\tilde{F}_2\cos^2\theta_1+F_3\sin^2\theta_1}
d\tilde{\psi}+\cos\theta_1 d\phi_1\right]\wedge d\tilde{x}_3.
\end{align}

T-dualizing along $\phi_1$, we obtain the type IIA geometry of (3.33) in~\cite{Dasgupta:2016rhc}:
\begin{align}
\label{333}
ds_{(A,3)}^2=&ds_{(3)}^2
+e^{-\phi}\frac{(d\phi_1
+\tilde{F}_2\sec\theta_{nc}\tan\theta_{nc}\cos\theta_1d\tilde{x}_3)^2}{\tilde{F}_2\cos^2\theta_1+F_3\sin^2\theta_1}.
\end{align}
The NS B-field associated to the $ds_{(A,3)}^2$ metric is that in (3.34) in~\cite{Dasgupta:2016rhc}:
\begin{align}
B_{(A,3)}=\frac{\tilde{F}_2\sec\theta_{nc}}{\tilde{F}_2\cos^2\theta_1+F_3\sin^2\theta_1}
(F_3\sec\theta_{nc}\tan\theta_{nc}\sin^2\theta_1 d\tilde{\psi}\wedge d\tilde{x}_3
+\cos\theta_1 d\phi_1\wedge d\tilde{\psi}).
\end{align}
The corresponding dilaton is (3.36) in~\cite{Dasgupta:2016rhc}:
\begin{align}
e^{\phi_{(A,3)}}=\sqrt{\frac{\tilde{F}_2}{F_2}}
\frac{\sec\theta_{nc}}{\sqrt{\tilde{F}_2\cos^2\theta_1+F_3\sin^2\theta_1}}e^{-3\phi/2}.
\label{336}
\end{align}
Coming to the fluxes, the type IIA two-form flux $\mathcal{F}_2^{(A,3)}$ dual to $\mathcal{F}_3^{(B,2)}$ in (\ref{F3B2})
can be easily seen to be
\begin{align}
\mathcal{F}_2^{(A,3)}=k_1\sin\theta_1 d\tilde{\psi}\wedge d\theta_1+k_2\cos\theta_{nc}\cos\theta_1 dx_8\wedge
dx_9.
\end{align}
It is again important to note that, of course, this two-form flux is not closed: $d\mathcal{F}_2^{(A,3)}\neq0$,
which reflects the presence of a D6-brane (dual to the D5-brane in the previous type IIB configuration).
Thus, if we denote as ${\bf A}_1$ the type IIA gauge field for this configuration, then it follows that $\mathcal{F}_2^{(A,3)}$
can be written as in (3.53) in~\cite{Dasgupta:2016rhc}:
\begin{align}
\mathcal{F}_2^{(A,3)}=d{\bf A}_1+\Delta, \quad\quad d\Delta=\textrm{sources}.
\label{353}
\end{align}
The explicit expression of the $d\Delta=d\mathcal{F}_2^{(A,3)}$ sources is that in (3.39) in~\cite{Dasgupta:2016rhc}:
\begin{align}
d\mathcal{F}_2^{(A,3)}=k_{1,a}\sin\theta_1 da\wedge d\tilde{\psi}\wedge d\theta_1
+(k_{2,a}\cos\theta_{nc}\cos\theta_1 da
-k_2\cos\theta_{nc}\sin\theta_1 d\theta_1)\wedge dx_8\wedge dx_9.
\end{align}
We define ${\bf A}_1$ as
\begin{align}
{\bf A}_1\equiv {{\bf A}_1}_{\theta_1}d\theta_1 +{{\bf A}_1}_{8}dx_8+{{\bf A}_1}_9dx_9,
\label{A1def}
\end{align}
with $({{\bf A}_1}_{\theta_1},\,{{\bf A}_1}_8,\,{{\bf A}_1}_9)$ depending only on the $(\theta_1,\,x_8,\,x_9)$ coordinates.
We further define
\begin{align}
\alpha_1\equiv\frac{\partial {{\bf A}_1}_9}{\partial x_8}-\frac{\partial {{\bf A}_1}_8}{\partial x_9}, \quad\quad
\alpha_2\equiv\frac{\partial {{\bf A}_1}_{\theta_1}}{\partial x_8}-\frac{\partial {{\bf A}_1}_8}{\partial \theta_1}, \quad\quad
\alpha_3\equiv \frac{\partial {{\bf A}_1}_{\theta_1}}{\partial x_9}-\frac{\partial {{\bf A}_1}_9}{\partial \theta_1}.
\end{align}
Using the above quantities, the exterior derivative of ${\bf A}_1$
is (3.42) in~\cite{Dasgupta:2016rhc}:
\begin{align}
d{\bf A}_1\equiv \alpha_1 dx_8\wedge dx_9+\alpha_2 dx_8\wedge d\theta_1+\alpha_3 dx_9\wedge d\theta_1.
\label{342}
\end{align}
Since $d(d{\bf A}_1)=0$, the $\alpha$'s just introduced are subject to the constraint
\begin{align}
\frac{\partial\alpha_1}{\partial\theta_1}-\frac{\partial\alpha_2}{\partial x_9}
+\frac{\partial\alpha_3}{\partial x_8}=0,
\label{343}
\end{align}
mentioned in (3.43) in~\cite{Dasgupta:2016rhc}. The definition (\ref{A1def}) will become sensible in the M-theory uplift that follows.
But first let us finish the flux discussion for this type IIA configuration.
We note that the corresponding NS three-form flux is given by the exterior derivative of
$B_{(A,3)}$. This is
\begin{align}
\mathcal{H}_3^{(A,3)}= db\wedge(\hat{k}_{1,b}d\tilde{\psi}\wedge d\tilde{x}_3+\hat{k}_{2,b}d\phi_1\wedge d\tilde{\psi}),
\label{H3A3}
\end{align}
where we have defined
\begin{align}
\hat{k}_1\equiv\frac{\tilde{F}_2F_3\sec^2\theta_{nc}\tan\theta_{nc}\sin^2\theta_1}{\tilde{F}_2\cos^2\theta_1+F_3\sin^2\theta_1},
\quad\quad
\hat{k}_2\equiv\frac{\tilde{F}_2F_3\sec\theta_{nc}\cos\theta_1}{\tilde{F}_2\cos^2\theta_1+F_3\sin^2\theta_1}
\end{align}
and $b\equiv (\theta_1,\,r)$ are the only coordinates on which the above two functions depend (recall our choices in
(\ref{easychoice})).

Finally, we will uplift the above type IIA configuration to M-theory.
To this aim, we rewrite the metric $ds_{(A,3)}^2$ in (\ref{333}) in a more convenient way.
We first introduce the quantities of (3.41) in~\cite{Dasgupta:2016rhc}:
\begin{eqnarray}
\label{Hs}
\begin{array}{llllll}
&H_1\equiv(H_2H_3)^{-1/3}, \quad\quad
&H_2\equiv (\cos^2\theta_{nc}+F_2\sin^2\theta_{nc})^{-1}, \\
&H_3\equiv(\tilde{F}_2\cos^2\theta_1+F_3\sin^2\theta_1)^{-1}, \quad\quad
&H_4\equiv H_3\tilde{F}_2F_3\sec^2\theta_{nc}\sin^2\theta_1, \\
&f_3\equiv\tilde{F}_2\sec\theta_{nc}\tan\theta_{nc}\cos\theta_1.
\end{array}
\end{eqnarray}
In terms of these, the metric $ds_{(A,3)}^2$ can be written as
\begin{align}
\nonumber
ds_{(A,3)}^2=&\frac{e^{-\phi}}{H_1}\left\{H_1[-dt^2+dx_1^2+dx_2^2+H_2d\tilde{x}_3^2+H_3(d\phi_1+f_3d\tilde{x}_3)^2]\right. \\
&+e^{2\phi}H_1\left.[F_1dr^2+F_3d\theta_1^2+F_4(dx_8^2+dx_9^2)+H_4d\tilde{\psi}^2]\right\}.
\label{dsA3}
\end{align}
It is essential to note that the M-theory uplift will only be able to capture the dynamics of the type IIA theory
in the strong coupling limit of the latter.
For us, that means that we can only rely on the M-theory description when
$e^{\phi_{(A,3)}}$ is of order one or bigger.
However, we will be interested in having a {\it finite} radius for the eleventh direction after we uplift.
Therefore, we will be careful to avoid the infinite coupling limit where
\begin{align}
e^{\phi_{(A,3)}}\rightarrow\infty.
\label{Mlimit}
\end{align}
From (\ref{336}) it follows that the above is true when $e^{-\phi}\rightarrow\infty$, for an arbitrary choice
of $(F_2,\,F_3)$. Additionally, the infinite coupling limit also applies at two isolated points $(p_1,\,p_2)$
given by $p_1=(\theta_1=0,\,r=r_1)$ and $p_2=(\theta_1=\pi/2,\,r=r_2)$ (for any value of the remaining
coordinates), where $(r_1,\,r_2)$ are the values of the radial coordinate for which $F_2(r_1)=0$ and
$F_3(r_2)=0$, respectively. (These are the same two points in (3.37) in~\cite{Dasgupta:2016rhc}.)

The M-theory metric corresponding to (\ref{dsA3}) is
\begin{align}
ds_{(M,1)}^2=&H_1[-dt^2+dx_1^2+dx_2^2+H_2d\tilde{x}_3^2+H_3(d\phi_1+f_3d\tilde{x}_3)^2
+e^{2\phi}(F_1dr^2+H_4d\tilde{\psi}^2)] \nonumber \\
&+e^{2\phi}H_1[F_3d\theta_1^2+F_4(dx_8^2+dx_9^2)]+e^{-2\phi}H_1^{-2}(dx_{11}+{\bf A}_1)^2,
\label{340}
\end{align}
where ${\bf A}_1$ is the type IIA gauge field defined in (\ref{A1def}).
We note that, due to (\ref{easychoice}) and (\ref{A1def}), for a fixed value of the radial coordinate, $r=r_0$,
the second line above describes a warped Taub-NUT space in the $(\theta_1,\,x_8,\,x_9,\,x_{11})$ directions.
(Indeed, this is what motivated the definition (\ref{A1def}).)
This is most easily seen by introducing the quantities in (3.45) in~\cite{Dasgupta:2016rhc},
\begin{align}
\label{defGs}
G_1\equiv e^{2\phi}H_1F_3\Big|_{r=r_0}, \quad\quad
G_2,G_3\equiv e^{2\phi}H_1F_4\Big|_{r=r_0}, \quad\quad
G_4\equiv e^{-2\phi}H_1^{-2}\Big|_{r=r_0}
\end{align}
and writing the warped Taub-NUT metric as in (3.44) in~\cite{Dasgupta:2016rhc}:
\begin{align}
ds_{{TN}_1}^2=G_1d\theta_1^2+G_2dx_8^2+G_3dx_9^2+G_4(dx_{11}+{\bf A}_1)^2.
\label{mettb}
\end{align}
Note that, as we just explained,
\begin{align}
G_i=G_i(\theta_1,x_8,x_9), \quad\quad i=1,2,3,4.
\end{align}
We take the vielbeins of (\ref{mettb}) as
\begin{align}
e_{\theta_1}^{(M,1)}=\sqrt{G_1}d\theta_1, \quad e_{8}^{(M,1)}=\sqrt{G_2}dx_8, \quad
e_9^{(M,1)}=\sqrt{G_3}dx_9, \quad e_{11}^{(M,1)}=\sqrt{G_4}(dx_{11}+{\bf A}_1).
\label{vielM1}
\end{align}
To better understand this Taub-NUT space, recall that, before the M-theory uplift,
we had a D6-brane in our type IIA configuration.
The M-theory uplift then converts this D6-brane to geometry.
In particular, we obtain the metric (\ref{340}), where (\ref{mettb}) is a {\it single-centered}
(warped) Taub-NUT space. In other words, in (\ref{mettb}), $G_4^{-1}=0$ occurs {\it once}
and the coordinate singularity at this point is the location of the D6-brane in the dual type IIA picture.
This is an important observation and essential to the G-flux computation that follows.

As we just hinted,
the remaining of this section will be devoted to the determination of the G-flux corresponding to this M-theory configuration.
As it is well-known,
there exists a unique, normalizable (anti-)self-dual harmonic two-form $\omega$ associated to 
a single-centered (warped) Taub-NUT space~\cite{Ruback}.
Using which, the G-flux\footnote{We remind the reader that the computation of fluxes
is nicely summarized in~\cite{BLT}.} for our M-theory configuration is given by (3.55) in~\cite{Dasgupta:2016rhc}:
\begin{align}
\mathcal{G}_4^{(M,1)}=\langle\mathcal{G}_4^{(M,1)}\rangle+\mathcal{F}\wedge \omega,
\label{355}
\end{align}
where $\langle\mathcal{G}_4^{(M,1)}\rangle= \mathcal{H}_3^{(A,3)}\wedge dx_{11}$ is the background
G-flux ($\mathcal{H}_3^{(A,3)}$ was determined in (\ref{H3A3})) and $\mathcal{F}=d\mathcal{A}$
is the field strength of the $U(1)$ world-volume gauge theory ($\mathcal{A}$ is the corresponding gauge field).
Thus, in order to obtain the explicit form of $\mathcal{G}_4^{(M,1)}$, we have one task left: $\omega$ must be computed.
We do so in the following.

We start by making the ansatz in (3.46) in~\cite{Dasgupta:2016rhc} for $\omega$
\begin{align}
\omega=d\zeta, \quad\quad \zeta=g(\theta_1,x_8,x_9)(dx_{11}+{\bf A}_1)
\label{zetadef}
\end{align}
and proceed to determine its precise value from the (anti-)self-duality requirement: $\omega=\pm\ast\omega$, where
the Hodge dual is taken with respect to the metric (\ref{mettb}).
Let us see this in details.
Using (\ref{342}) and (\ref{vielM1}), $\omega$ can be written as
\begin{align}
\label{expliom}
\omega=&\frac{1}{\sqrt{G_4}}\left(\frac{1}{\sqrt{G_1}}\frac{\partial g}{\partial\theta_1}e_{\theta_1}^{(M,1)}
+\frac{1}{\sqrt{G_2}}\frac{\partial g}{\partial x_8}e_{8}^{(M,1)}
+\frac{1}{\sqrt{G_3}}\frac{\partial g}{\partial x_9}e_{9}^{(M,1)}\right)\wedge e_{11}^{(M,1)} \\
&+g\left(\frac{\alpha_1}{\sqrt{G_2G_3}}e_8^{(M,1)}\wedge e_9^{(M,1)}
+\frac{\alpha_2}{\sqrt{G_1G_2}}e_8^{(M,1)}\wedge e_{\theta_1}^{(M,1)}
+\frac{\alpha_3}{\sqrt{G_1G_3}}e_9^{(M,1)}\wedge e_{\theta_1}^{(M,1)}\right).
\nonumber
\end{align}
Quite obviously,
\begin{eqnarray}
\begin{array}{lllll}
&\ast(e_{\theta_1}^{(M,1)}\wedge e_{11}^{(M,1)})=e_8^{(M,1)}\wedge e_9^{(M,1)}, \quad\quad
&\ast(e_{8}^{(M,1)}\wedge e_{11}^{(M,1)})=e_9^{(M,1)}\wedge e_{\theta_1}^{(M,1)}, \\
&\ast(e_{9}^{(M,1)}\wedge e_{11}^{(M,1)})=e_8^{(M,1)}\wedge e_{\theta_1}^{(M,1)}
\end{array}
\end{eqnarray}
and so, the Hodge dual of $\omega$ is
\begin{align}
&\ast\omega= 
+g\left(\frac{\alpha_1}{\sqrt{G_2G_3}}e_{\theta_1}^{(M,1)}
+\frac{\alpha_3}{\sqrt{G_1G_3}}e_8^{(M,1)}
-\frac{\alpha_2}{\sqrt{G_1G_2}}e_9^{(M,1)}\right)\wedge e_{11}^{(M,1)} \\
&+\frac{1}{\sqrt{G_4}}\left(\frac{1}{\sqrt{G_1}}\frac{\partial g}{\partial\theta_1}e_{8}^{(M,1)}\wedge e_{9}^{(M,1)}
+\frac{1}{\sqrt{G_3}}\frac{\partial g}{\partial x_9}e_{8}^{(M,1)}\wedge e_{\theta_1}^{(M,1)}
+\frac{1}{\sqrt{G_2}}\frac{\partial g}{\partial x_8}e_{9}^{(M,1)}\wedge e_{\theta_1}^{(M,1)}\right).
\nonumber
\end{align}
Imposing (anti-)self-duality of $\omega$ leads to three partial differential equations (PDEs):
\begin{align}
\frac{1}{g}\frac{\partial g}{\partial \theta_1}=\pm\alpha_1\sqrt{\frac{G_1G_4}{G_2G_3}}, \quad\quad
\frac{1}{g}\frac{\partial g}{\partial x_8}=\pm\alpha_3\sqrt{\frac{G_2G_4}{G_1G_3}}, \quad\quad
\frac{1}{g}\frac{\partial g}{\partial x_9}=\mp\alpha_2\sqrt{\frac{G_3G_4}{G_1G_2}}.
\end{align}
Using (\ref{Hs}) and (\ref{defGs}) in the above, we can rewrite these equations in terms of
the warp factors and $\phi$, as in (3.47) in~\cite{Dasgupta:2016rhc}:
\begin{align}
\label{PDEsalphas}
&\frac{1}{g}\frac{\partial g}{\partial \theta_1}=\pm \left.e^{-2\phi}\frac{\alpha_1}{F_4}
\sqrt{\frac{\tilde{F}_2F_3}{F_2}}\sec\theta_{nc}(\tilde{F}_2\cos^2\theta_1+F_3\sin^2\theta_1)^{-1/2}\right|_{r=r_0}, \\
\nonumber
&\frac{1}{g}\frac{\partial g}{\partial x_8}=\pm\left.e^{-2\phi}\alpha_3
\sqrt{\frac{\tilde{F}_2}{F_2F_3}}\sec\theta_{nc}(\tilde{F}_2\cos^2\theta_1+F_3\sin^2\theta_1)^{-1/2}\right|_{r=r_0}
=-\frac{\alpha_3}{\alpha_2}\frac{1}{g}\frac{\partial g}{\partial x_9}.
\end{align}
Solving the above set of PDEs generically is not easy.
Consequently, we will do some more simplifying assumptions. To begin with, let us
take, as in (3.49) in~\cite{Dasgupta:2016rhc},
\begin{align}
\alpha_1=0, \quad\quad \alpha_2=\beta_2(x_9)f(\theta_1,r,x_8,x_9)\Big|_{r=r_0}, \quad\quad
\alpha_3=\beta_3(x_8)f(\theta_1,r,x_8,x_9)\Big|_{r=r_0},
\label{alphasel}
\end{align}
where we have defined
\begin{align}
f=f(\theta_1,r,x_8,x_9)\equiv e^{2\phi}\sqrt{\tilde{F}_2\cos^2\theta_1+F_3\sin^2\theta_1}.
\end{align}
If we now choose $e^{2\phi}$ as in (3.54) in~\cite{Dasgupta:2016rhc},
\begin{align}
e^{2\phi}=\frac{e^{2\phi_0}Q(r,x_8,x_9)}{\sqrt{\tilde{F}_2\cos^2\theta_1+F_3\sin^2\theta_1}},
\label{354}
\end{align}
with $\phi_0$ some constant, then $(\alpha_2,\,\alpha_3)$ become independent of $\theta_1$ (that is,
functions of the coordinates $(x_8,\,x_9)$ only).
Recall that the $\alpha$'s were subject to the constraint (\ref{343}).
Hence, $Q=Q(r,x_8,x_9)$ above must satisfy
\begin{align}
\left.Q\left(\frac{d\beta_3}{dx_8}-\frac{d\beta_2}{dx_9}\right)+\beta_3\frac{\partial Q}{\partial x_8}-\beta_2
\frac{\partial Q}{\partial x_9}\right|_{r=r_0}=0.
\end{align}
Additionally, we define
\begin{align}
c_0\equiv\left.\sqrt{\frac{\tilde{F}_2}{F_2F_3}}\sec\theta_{nc}\right|_{r=r_0},
\end{align}
which is a constant that only depends on the deformation parameter $\theta_{nc}$.
Inserting all our choices and definitions in (\ref{PDEsalphas}), these PDEs reduce to
\begin{align}
\frac{1}{g}\frac{\partial g}{\partial x_8}=\pm c_0\beta_3(x_8), \quad\quad \frac{1}{g}\frac{\partial g}{\partial x_9}
=\mp c_0\beta_2(x_9),
\end{align}
where $g$ is now independent of $\theta_1$ and thus $g=g(x_8,x_9)$.
It is finally easy to use separation of variables to solve the above.
Assuming $g=\tilde{g}_1(x_8)\tilde{g}_2(x_9)$, we obtain two ordinary differential equations,
\begin{align}
\frac{d\tilde{g}_1}{\tilde{g}_1}=\pm c_0\beta_3(x_8)dx_8, \quad\quad \frac{d\tilde{g}_2}{\tilde{g}_2}=\mp c_0\beta_2(x_9)dx_9,
\end{align}
which can readily be solved to yield
\begin{align}
g=g_0\textrm{exp}\left[\pm c_0\left(\int_0^{x_8}\beta_3(x_8^\prime)dx_8^\prime-\int_0^{x_9}\beta_2(x_9^\prime)dx_9^\prime
\right)\right],
\label{gsol}
\end{align}
with $g_0$ some integration constant. This completes the computation of $\omega$ in (\ref{zetadef}),
which in turn gives us the explicit form of the G-flux in (\ref{355}). 

\FloatBarrier

\subsubsection{Enhancing the symmetry of the world-volume gauge theory: tensionless M2-branes \label{nonabsec}}

It is an intrinsically interesting question to ask whether our first M-theory construction above can be
generalized to account for non-abelian world-volume gauge theories (and not just the particularly simple
$U(1)$ case discussed so far). The answer is yes and the way to do so is discussed in~\cite{Sen:1997kz}.
Consequently, in this section we review and adapt the arguments in~\cite{Sen:1997kz} to our case.

But before we jump into the details of non-abelian enhancement in M-theory, it is instructive to
recall the well-known equivalent discussion in type IIA superstring theory~\cite{Witten:1995im}.
Consider $N$ parallel D6-branes ($N=2,3,4,\ldots$). Consider there are open strings stretched between these D6-branes.
In this case, the symmetry group of the corresponding world-volume gauge theory is
\begin{align}
\underbrace{U(1)\times U(1)\times\ldots\times U(1)}_{N\textrm{ times}}.
\end{align}
In the limit when the open strings become tensionless, the D6-branes come on top of each other (we thus have
$N$ coincident D6-branes). Then, the symmetry group of the corresponding world-volume gauge theory becomes
$SU(N)$.

If we lift the above type IIA configuration to M-theory, then the D6-branes convert to geometry and we obtain
the metric (\ref{340})\footnote{Since we never determined our warp factors and $Q$ function in (\ref{354}),
we can absorb the changes in the geometry due to the inclusion of the D6-branes and open strings
in these quantities.}, with (\ref{mettb}) a {\it multi-centered} (warped) Taub-NUT space.
Indeed, $G_4^{-1}=0$ now occurs $N$ times in (\ref{mettb}),
the coordinate singularities at these points denoting the location of the D6-branes in the dual type IIA picture.
As for the open strings, they convert to M2-branes wrapping the two-cycles in the Taub-NUT space (\ref{mettb}).
In the limit of tensionless M2-branes, the two-cycles vanish and the world-volume gauge theory
symmetry group becomes $SU(N)$.

Let us see how the above discussion applies to our set up in details.
The first step will be to construct the independent two-cycles in the space (\ref{mettb}).
In order to do so, let us start by rewriting the metric (\ref{mettb}) in a more convenient way.
Defining, as in (3.86) in~\cite{Dasgupta:2016rhc},
\begin{align}
U\equiv e^{2\phi}H_1^2\Big|_{r=r_0}, \quad\quad d\vec{{\bf x}}^2\equiv H_1^{-1}[F_3d\theta_1^2+F_4(dx_8^2+dx_9^2)]\Big|_{r=r_0},
\end{align}
we can rewrite (\ref{mettb}) as in (3.85) in~\cite{Dasgupta:2016rhc}:
\begin{align}
ds_{{TN}_1}^2=Ud\vec{{\bf x}}^2+U^{-1}(dx_{11}+{\bf A}_1)^2.
\end{align}
Recall that now this warped Taub-NUT space is a multi-centered one.
Using (\ref{Hs}) and (\ref{354}), $U$ above can be written in terms of the warp factors and $Q$ as
\begin{align}
U=e^{2\phi_0}Q(\cos^2\theta_{nc}+F_2\sin^2\theta_{nc})^{2/3}(\tilde{F}_2\cos^2\theta_1+F_3\sin^2\theta_1)^{1/6}\Big|_{r=r_0}.
\end{align}
For simplicity, we will do two assumptions next: we will take the deformation parameter to be sufficiently small
(that is, $\theta_{nc}<<1$) and we will consider
\begin{align}
F_2\Big|_{r=r_0}=F_3\Big|_{r=r_0}.
\label{f23r0}
\end{align}
Then, expanding to first order around $\theta_{nc}=0$ and using (\ref{f23r0}), $U$ becomes independent of $\theta_1$:
\begin{align}
\tilde{U}=\tilde{U}(x_8,x_9)\equiv\lim_{\theta_{nc}\rightarrow0}U=e^{2\phi_0}Q(r,x_8,x_9)F_3^{1/6}\Big|_{r=r_0}.
\end{align}
$\tilde{U}=0$ has $N$ solutions, which we denote as $\vec{l}_i=({x_8}_i,\, {x_9}_i)$ ($i=1,2,\ldots,N$).
Consider two such points $\vec{l}_i$ and $\vec{l}_j$ ($i\neq j$) and a geodesic $\mathcal{C}_g$ in the $(x_8,\,x_9)$ space
joining them. Attaching to each point in $\mathcal{C}_g$ a circle labeled by $x_{11}$, we obtain a minimal area two-cycle $X_{ij}$.
We take $X_{k,k+1}$ ($k=1,2,\ldots,N-1$) as the (minimal area) independent two-cycles.

It is well-known that
to each such two-cycle $X_{k,k+1}$, with $k$ fixed, we can associate a unique, normalizable, (anti-)self-dual two-form $\omega_k$.
Obtaining the explicit form of $\omega_k$ is straightforward, in view of our earlier results.
We only need to modify (\ref{zetadef}) to
\begin{align}
\omega_k=d\zeta_k, \quad\quad \zeta_k=g_k(x_8,x_9)(dx_{11}+{\bf A}_1)
\label{omegak}
\end{align}
and restrict the integrals in (\ref{gsol}) to the $X_{k,k+1}$ two-cycle:
\begin{align}
g_k=\tilde{g}_0\textrm{exp}\left[\pm c_0\int_{\vec{l}_k}^{\vec{l}_{k+1}}(\beta_3-\beta_2)|d\vec{l}_{\mathcal{C}_g}|\right],
\end{align}
where $\tilde{g}_0$ is some integration constant and $d\vec{l}_{\mathcal{C}_g}$ denotes line element along the geodesic $\mathcal{C}_g$
joining $\vec{l}_k$ and $\vec{l}_{k+1}$. 

Let us now compute the areas of the two-cycles $X_{k,k+1}$ and derive their intersection matrix. 
It will soon be clear why we do so.
As measured in the Taub-NUT metric, the area of $X_{k,k+1}$ is given by
\begin{align}
S_{k,k+1}=\int_{X_{k,k+1}}(\tilde{U}^{-1/2}dx_{11})(\tilde{U}^{1/2}\sqrt{F_4}\Big|_{r=r_0}|d\vec{l}_{\mathcal{C}_g}|)
=\tilde{\beta}R_{11}\int_{\vec{l}_k}^{\vec{l}_{k+1}}\sqrt{F_4}\Big|_{r=r_0}|d\vec{l}_{\mathcal{C}_g}|,
\end{align}
with $\tilde{\beta}$ a constant that avoids possible conical singularities along ${\mathcal{C}_g}$ and $R_{11}$ the
physical radius of the $x_{11}$ coordinate.
It is easy to see that the self-intersection number for each $S_{k,k+1}$ is two:
the $S_{k,k+1}$'s self-intersect at $\vec{l}_k$ and $\vec{l}_{k+1}$, with geodesics transversed in the same direction.
$S_{k,k+1}$ intersects $S_{k-1,k}$ only at $\vec{l}_k$, their geodesics being transversed in opposite directions.
No other two-cycles' areas intersect.
Thus, the $(N-1)\times(N-1)$ intersection matrix of the areas of the two-cycles $X_{k,k+1}$ is
\begin{eqnarray}
\left(
\begin{array}{ccccccc}
2 & -1 & 0 & 0 & \ldots & 0 & 0 \\
-1 & 2 & -1 & 0 & \ldots & 0 & 0 \\
0 & -1 & 2 & -1 & \ldots & 0 & 0 \\
\vdots & \vdots & \vdots & \vdots & \ddots & \vdots & \vdots \\
0 & 0 & 0 & 0 & \ldots & 2 & -1 \\
0 & 0 & 0 & 0 & \ldots & -1 & 2 
\end{array}
\right). \label{interma}
\end{eqnarray}
Or, written more compactly, as in (3.89) in~\cite{Dasgupta:2016rhc},
\begin{eqnarray}
[S_{k,k+1}]\circ [S_{l,l+1}]=
\begin{cases}
2\delta_{k,l} \\
-\delta_{l,k-1}
\end{cases}.
\end{eqnarray}
This is, of course, the Cartan matrix of the $A_{N-1}$ algebra.

\begin{figure}[t]
\centering
\includegraphics[width=0.9\textwidth]{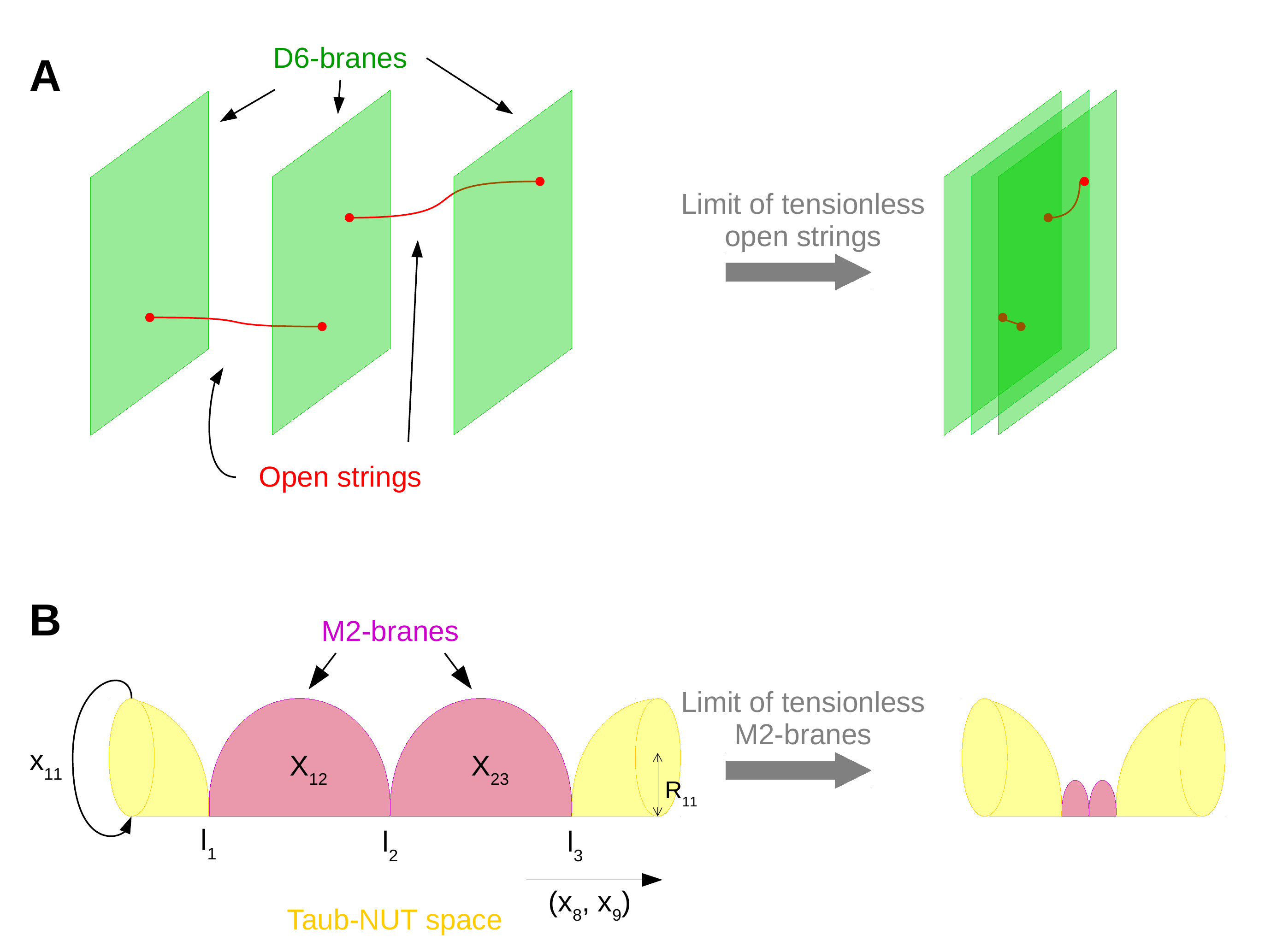}
\caption{Schematics of the non-abelian enhancement of the world-volume gauge symmetry
from $U(1)\times U(1)\times U(1)$ to $SU(3)$ in type IIA (top) and in M-theory (bottom).
{\bf A:} Three parallel D6-branes in type IIA, with open strings stretching between them.
The D6-branes span the $(t,\,x_1,\,x_2,\,\tilde{x}_3,\,\phi_1,\,\tilde{\psi},\,r)$ directions and the open
strings are in the $(x_8,\,x_9)$ plane.
The $\theta_1$ direction is suppressed.
When the open strings become tensionless, the D6-branes coincide.
This produces the non-abelian enhancement.
{\bf B:} Uplift to M-theory of the type IIA configurations in {\bf A}.
The D6-branes convert to geometry, giving rise to a multi-centered warped Taub-NUT space
along $(\theta_1,\,x_8,\,x_9,\,x_{11})$, for a fixed value of the radial coordinate: $r=r_0$.
$R_{11}$ is the physical radius of the coordinate $x_{11}$.
The $(t,\,x_1,\,x_2,\,\tilde{x}_3,\,\theta_1,\,\phi_1,\,\tilde{\psi},\,r)$ directions are suppressed in the figure.
The singularities in the Taub-NUT space lie at $(\vec{l}_1,\,\vec{l}_2,\,\vec{l}_3)$:
the position of the D6-branes in the dual type IIA configuration.
The open strings become M2-branes wrapping the minimal area, independent two-cycles $(X_{12},\,X_{23})$
between the singularities.
In the limit of tensionless M2-branes, these two-cycles vanish,
leading to the non-abelian enhancement.}
\label{fig5}
\end{figure}

Recall that there are M2-branes in this configuration.
They wrap the $X_{k,k+1}$ two-cycles and thus their intersection matrix is (\ref{interma}).
As previously explained, when the area of all these two-cycles tends to zero, the limit of tensionless M2-branes
sets in.
This corresponds to an $A_{N-1}$ singularity, which in turn is responsible for enhancing the world-volume
gauge symmetry to $SU(N)$, as shown in~\cite{Sen:1997js}.
Figure \ref{fig5} schematically depicts the above discussion for $N=3$, both in the type IIA and M-theory pictures.

To finish this section, we use all the above results to write the G-flux
of this non-abelian enhanced M-theory configuration as in (3.90) in~\cite{Dasgupta:2016rhc}:
\begin{eqnarray}
\mathcal{G}_4^{(M,1)}=\langle\mathcal{G}_4^{(M,1)}\rangle +\sum_{k=1}^{N-1}\mathcal{F}_k\wedge \omega_k. \label{multiflux}
\end{eqnarray}
Here, $\mathcal{F}_k$'s are the Cartan algebra values of the world-volume field strength $\mathcal{F}$,
the background G-flux $\langle\mathcal{G}_4^{(M,1)}\rangle $ is as earlier\footnote{Remember, however, that
the warp factors and $Q$ function introduced in (\ref{354}) are different from those in the abelian case,
due to the inclusion of the D6-branes and open strings
in the dual type IIA theory.} in (\ref{355}) and the two-forms $\omega_k$
were computed in (\ref{omegak}).

\FloatBarrier

\subsection{Accounting for an axionic background: an additional RR B-field \label{rrdefsect}}

Suppose we follow the prescription of~\cite{Witten:2011zz} to source the $\Theta$-term in the world-volume
gauge theory. That is, suppose we consider the type IIB D3-NS5 system {\it with} an axionic background $\mathcal{C}_0$.
How would that affect the results in the previous section (section \ref{ncsect}),
where $\mathcal{C}_0=0$?

Long story made short, we need to follow $\mathcal{C}_0$ along the modifications of section \ref{ns5d3sect1},
depicted in figure \ref{fig1}.
We note that $\mathcal{C}_0$ would not be affected while going from {\bf A} to {\bf B} in figure \ref{fig1}.
However, on going from {\bf B} to {\bf C}, $\mathcal{C}_0$ would dualize to a gauge field
in the $x_3$ direction. Finally, on going from {\bf C} to {\bf D},
the gauge field would lead to an RR B-field in the $(x_3,\,\psi)$ directions.
Schematically, 
\begin{align}
\mathcal{C}_0\,\,\xrightarrow[\textrm{but finite interval}]{\textrm{Convert }\psi\textrm{ to a large}} \,\,
\mathcal{C}_0 \,\,\xrightarrow[\textrm{along }x_3]{\textrm{T-duality}}\,\,
\mathcal{C}_1=(\mathcal{C}_1)_3dx_3\,\,\xrightarrow[\textrm{along }\psi]{\textrm{T-duality}}\,\,
\mathcal{C}_2=(\mathcal{C}_2)_{3\psi}dx_3\wedge d\psi.
\end{align}
Thus, in our construction, switching on an axionic background in the usual type IIB D3-NS5 system of~\cite{Witten:2011zz},
shown in figure \ref{fig1}{\bf A},
amounts to adding an RR B-field in the $(x_3,\,\psi)$ directions to the type IIB configuration
shown in figure \ref{fig1}{\bf D}.

In this section, however, we will see a {\it different} way in which we can obtain such an RR B-field in the
type IIB configuration before we uplift to M-theory. This will involve
another, distinct (although similar) chain of dualities and modifications to the type IIB configuration of
figure \ref{fig1}{\bf D} to that considered before, in section \ref{ncsect}.
In the following, we make precise this idea.

\begin{figure}[ht]
\centering
\includegraphics[width=0.9\textwidth]{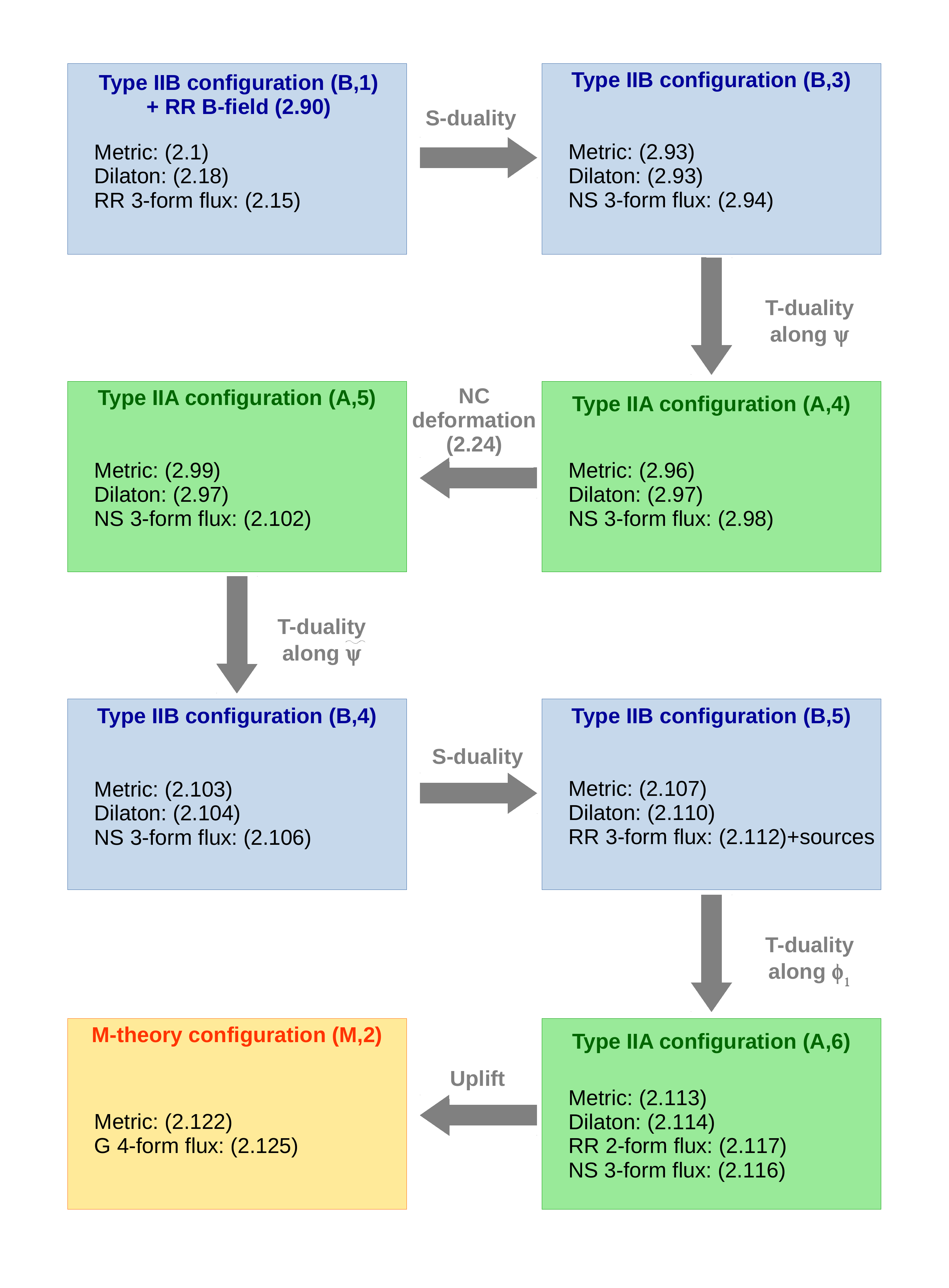}
\caption{Graphical summary of section \ref{rrdefsect}.
To the type IIB configuration of figure \ref{fig1}{\bf D} we associate an RR B-field
and then proceed to do a series of modifications in order to account for
the axionic background considered in~\cite{Witten:2011zz}.
This is achieved in going from the configuration (B,1), with the mentioned RR B-field added, to (B,5).
The (B,5) configuration is then lifted to M-theory.
However, as argued in the text, it will suffice to study the M-theory configuration (M,1) of figure \ref{fig3}.}
\label{fig4}
\newpage
\end{figure}

The starting point here is the starting point of section \ref{ncsect} as well:
the last configuration of section \ref{ns5d3sect1}, schematically depicted in figure \ref{fig1}{\bf D}.
To this configuration we will associate an RR B-field. 
We will then do an S-duality.
The next step will be a T-duality along $\psi$ to type IIA,
where we will do the same non-commutative deformation $(x_3,\,\psi)\rightarrow(\tilde{x}_3,\,\tilde{\psi})$
that was considered in section \ref{ncsect}.
Next, we will consider a T-duality along $\tilde{\psi}$ back to type IIB, followed by an S-duality.
At this point we will have a type IIB configuration with an RR B-field along $(\tilde{x}_3,\,\tilde{\psi})$.
Thus, effectively we will have accounted for the axionic background, as we wished to do.
The last T-duality will be along $\phi_1$ to type IIA.
The resulting configuration will be then lifted to M-theory.
As in section \ref{ncsect}, the NS and RR B-fields, dilaton and fluxes of all the above geometries will be determined.
Figure \ref{fig4} serves as a summary of the chain of modifications just described
and indicates the key equations in this section.

As just explained, we start by considering the type IIB geometry $ds_{(B,1)}^2$ in (\ref{3.43.5}), which has a
dilaton $e^{\phi_{(B,1)}}$ in (\ref{B1dilaton}) and an RR three-form flux $\mathcal{F}_3^{(B,1)}$ in (\ref{320}).
We will associate an RR B-field $\mathcal{C}_2^{(B,1)}$ to this set up as in (3.29) in~\cite{Dasgupta:2016rhc}:
\begin{align}
\mathcal{F}_3^{(B,1)}=d\mathcal{C}_2^{(B,1)}+\tilde{\Delta}, \quad\quad d\tilde{\Delta}=\textrm{sources}.
\label{329}
\end{align}
Note that the sources above are required to keep consistent with the fact that $\mathcal{F}_3^{(B,1)}$ is not closed.
These sources, of course, refer to the D5-brane present in this configuration.
For concreteness and as a particularly simple case, we will assume that $\mathcal{C}_2^{(B,1)}$ is
of the form in (3.26) in~\cite{Dasgupta:2016rhc}. That is, we consider 
\begin{align}
\mathcal{C}_2^{(B,1)}=b_{\theta_1\phi_1}d\theta_1\wedge d\phi_1+b_{89}dx_8\wedge dx_9,
\label{C2B1}
\end{align}
where $(b_{\theta_1\phi_1},\,b_{89})$ are functions of only $(\theta_1,\,r,\,x_8,\,x_9)$,
in order to respect all isometries in (\ref{3.43.5}).
It follows then that its exterior derivative is
\begin{align}
\nonumber
d\mathcal{C}_2^{(B,1)}=&d\theta_1\wedge d\phi_1\wedge\left(
\frac{\partial b_{\theta_1\phi_1}}{\partial r}dr
+\frac{\partial b_{\theta_1\phi_1}}{\partial x_8}dx_8
+\frac{\partial b_{\theta_1\phi_1}}{\partial x_9}dx_9\right)\\
&+\left(\frac{\partial b_{89}}{\partial \theta_1}d\theta_1
+\frac{\partial b_{89}}{\partial r}dr\right)\wedge dx_8\wedge dx_9.
\label{H3B3}
\end{align}
Using (\ref{viB1}), (\ref{320}) and the above, $\tilde{\Delta}$ in (\ref{329}) can be easily
checked to be
\begin{align}
\nonumber
\tilde{\Delta}=&d\theta_1\wedge d\phi_1\wedge\left( k_1\sin\theta_1 d\psi
-\frac{\partial b_{\theta_1\phi_1}}{\partial r}dr
-\frac{\partial b_{\theta_1\phi_1}}{\partial x_8}dx_8
-\frac{\partial b_{\theta_1\phi_1}}{\partial x_9}dx_9\right) \\
&+\left( k_2d\psi+k_2\cos\theta_1d\phi_1
-\frac{\partial b_{89}}{\partial \theta_1}d\theta_1
-\frac{\partial b_{89}}{\partial r}dr\right)\wedge dx_8\wedge dx_9.
\end{align}

S-dualizing the above, we obtain a type IIB configuration with metric, dipole and NS B-field given by
\begin{align}
ds_{(B,3)}^2=e^{\phi}ds_{(B,1)}^2, \quad\quad
e^{\phi_{(B,3)}}=e^{-\phi_{(B,1)}}, \quad\quad
B_{(B,3)}=\mathcal{C}_2^{(B,1)},
\end{align}
respectively. The corresponding NS three-form flux is the exterior derivative of $B_{(B,3)}$,
plus sources coming from the NS5-brane (dual to the D5-brane before).
Consequently, this is
\begin{align}
\mathcal{H}_3^{(B,3)}=d\mathcal{C}_2^{(B,1)}+\tilde{\Delta}=\mathcal{F}_3^{(B,1)},
\end{align}
not closed: $d\mathcal{H}_3^{(B,3)}\neq0$.
In other words, after the S-duality, the RR three-form flux becomes an NS one.
This is of course very convenient (and the reason to take the S-dual to begin with):
NS B-fields and fluxes are easier to deal with than RR ones.
In preparation to the T-duality along $\psi$ that will follow, we rewrite this metric as
\begin{align}
ds_{(B,3)}^2=e^{\phi}ds_{(2)}^2+dx_3^2 
+e^{2\phi} F_2(d\psi+\cos\theta_1 d\phi_1)^2,
\end{align}
where $ds_{(2)}^2$ was defined in (\ref{ds22}).

A T-duality along $\psi$ leads to the type IIA geometry
\begin{align}
ds_{(A,4)}^2=e^{\phi}ds_{(2)}^2+dx_3^2 
+\frac{e^{-2\phi}}{F_2}d\psi^2,
\end{align}
with associated dilaton and NS B-field
\begin{align}
e^{\phi_{(A,4)}}=(F_2)^{-1/2}, \quad\quad
B_{(A,4)}=\mathcal{C}_2^{(B,1)}+\cos\theta_1 d\psi\wedge d\phi_1.
\end{align}
The NS three-form flux is then given by
\begin{align}
\mathcal{H}_3^{(A,4)}=dB_{(A,4)}=d\mathcal{C}_2^{(B,1)}-\sin\theta_1 d\theta_1\wedge d\psi\wedge d\phi_1.
\end{align}
Note that this NS three-form flux is closed: $d\mathcal{H}_3^{(A,4)}=0$.
This is because, under the T-duality, the NS5-brane sources turn to geometry, as is well-known
(see, for example,~\cite{Tong:2002rq}).

Under the non-commutative deformation in (\ref{ncdeformation}), the type IIA metric changes to
\begin{align}
\nonumber
ds_{(A,5)}^2=&e^{\phi}ds_{(2)}^2+(\sec\theta_{nc}d\tilde{x}_3+\sin\theta_{nc}d\tilde{\psi})^2 
+\frac{e^{-2\phi}}{F_2}\cos^2\theta_{nc}d\tilde{\psi}^2 \\
=&e^{\phi}ds_{(2)}^2+\frac{\hat{F}_2}{F_2}\sec^2\theta_{nc}d\tilde{x}_3^2
+\frac{e^{-2\phi}}{\hat{F}_2}\cos^2\theta_{nc}\left(d\tilde{\psi}+e^{2\phi}\hat{F}_2\sec^2\theta_{nc}\tan\theta_{nc}d\tilde{x}_3
\right)^2,
\end{align}
where we have defined
\begin{align}
\hat{F}_2\equiv \frac{F_2}{1+e^{2\phi}F_2\tan^2\theta_{nc}}
\label{hatF2def}
\end{align}
and the last rewriting of the metric was done in anticipation to the T-duality along $\tilde{\psi}$ that we will soon
perform. Note the resemblance between $\hat{F}_2$ and $\tilde{F}_2$, defined in (\ref{tildeF2}).
Due to our choices in (\ref{easychoice}), the dilaton is not affected by the non-commutative deformation:
$e^{\phi_{(A,5)}}=e^{\phi_{(A,4)}}$. Similarly, our choice in (\ref{C2B1}) ensures that 
$\mathcal{C}_2^{(B,1)}$ remains unchanged too. The NS B-field, however, does change to
\begin{align}
B_{(A,5)}=\mathcal{C}_2^{(B,1)}+\cos\theta_{nc}\cos\theta_1 d\tilde{\psi}\wedge d\phi_1,
\end{align}
which in turn induces the NS three-form flux to change accordingly:
\begin{align}
\mathcal{H}_3^{(A,5)}=dB_{(A,5)}=d\mathcal{C}_2^{(B,1)}-\cos\theta_{nc}\sin\theta_1 d\theta_1\wedge d\tilde{\psi}\wedge d\phi_1.
\end{align}
Needless to say, this flux remains closed: $d\mathcal{H}_3^{(A,5)}=0$.

Upon a T-duality along $\tilde{\psi}$, we obtain the type IIB geometry
\begin{align}
ds_{(B,4)}^2=e^{\phi}ds_{(2)}^2+\frac{\hat{F}_2}{F_2}\sec^2\theta_{nc}d\tilde{x}_3^2
+e^{2\phi}\hat{F}_2\sec^2\theta_{nc}(d\tilde{\psi}+\cos\theta_{nc}\cos\theta_1d\phi_1)^2
\end{align}
with dilaton
\begin{align}
e^{\phi_{(B,4)}}=\sqrt{\frac{\hat{F}_2}{F_2}}\sec\theta_{nc}e^\phi.
\end{align}
The NS B-field $B_{(A,5)}$ dualizes to
\begin{align}
B_{(B,4)}=\mathcal{C}_2^{(B,1)}+e^{2\phi}\hat{F}_2\sec^2\theta_{nc}\tan\theta_{nc}(d\tilde{\psi}
+\cos\theta_{nc}\cos\theta_1d\phi_1)\wedge d\tilde{x}_3,
\end{align}
which contributes to the NS three-form flux
\begin{align}
\mathcal{H}_3^{(B,4)}=\frac{\tan\theta_{nc}}{\cos\theta_{nc}}\left[ k_{3,a} da\wedge 
(\frac{d\tilde{\psi}}{\cos\theta_{nc}}+\cos\theta_1 d\phi_1)-k_3\sin\theta_1 d\theta_1\wedge d\phi_1\right]\wedge d\tilde{x}_3
+\textrm{sources},
\end{align}
where we have defined $k_3\equiv e^{2\phi}\hat{F}_2$ and we recall that $a\equiv (\theta_1,\,r,\, x_8,\, x_9)$.
These are the only coordinates on which $k_3$ depends, as a consequence of our choices in (\ref{easychoice}).
The above flux is not closed, owing to the sources which denote the presence of an NS5-brane.
We do not determine the precise form of the sources here, for reasons that will soon become clear.

Next, we do an S-duality. This changes the metric to that in (3.30) in~\cite{Dasgupta:2016rhc}:
\begin{align}
ds_{(B,5)}^2=\frac{e^{-\phi}}{\sec\theta_{nc}}\sqrt{\frac{F_2}{\hat{F}_2}}\left[
e^{\phi}ds_{(2)}^2+\frac{\hat{F}_2}{F_2}\sec^2\theta_{nc}d\tilde{x}_3^2
+k_3\sec^2\theta_{nc}(d\tilde{\psi}+\cos\theta_{nc}\cos\theta_1d\phi_1)^2\right].
\end{align}
In preparation to the T-duality along $\phi_1$ that will follow, we rewrite $ds_{(B,5)}^2$ in a more convenient way:
\begin{align}
ds_{(B,5)}^2=ds_{(4)}^2+e^{\phi}\sqrt{\frac{F_2}{\hat{F}_2}}\cos\theta_{nc}
(F_3\sin^2\theta_1+\hat{F}_2\cos^2\theta_1)
\left(d\phi_1+\frac{\hat{F}_2\sec\theta_{nc}\cos\theta_1}{F_3\sin^2\theta_1+\hat{F}_2\cos^2\theta_1}d\tilde{\psi}\right)^2,
\end{align}
where we have defined
\begin{align}
ds_{(4)}^2\equiv& e^{-\phi}\sqrt{\frac{F_2}{\hat{F}_2}}\cos\theta_{nc}\left\{
-dt^2+dx_1^2+dx_2^2+\frac{\hat{F}_2}{F_2}\sec^2\theta_{nc}d\tilde{x}_3^2 \right.  \\
&\left.+e^{2\phi}\left[
F_1dr^2+F_3d\theta_1^2+F_4(dx_8^2+dx_9^2)
+\frac{\hat{F}_2F_3\sec^2\theta_{nc}\sin^2\theta_1}{F_3\sin^2\theta_1+\hat{F}_2\cos^2\theta_1}d\tilde{\psi}^2
\right]
\right\}.
\nonumber
\end{align}
The corresponding dilaton is that in (3.31) in~\cite{Dasgupta:2016rhc},
\begin{align}
e^{\phi_{(B,5)}}=\sqrt{\frac{F_2}{\hat{F}_2}}\cos\theta_{nc}e^{-\phi}.
\end{align}
The NS B-field now dualizes to an RR two-form flux given by (3.32) in~\cite{Dasgupta:2016rhc}:
\begin{align}
\mathcal{C}_2^{(B,5)}=-B_{(B,5)}=-\mathcal{C}_2^{(B,1)}
+k_3\sec^2\theta_{nc}\tan\theta_{nc}d\tilde{x}_3\wedge (d\tilde{\psi}
+\cos\theta_{nc}\cos\theta_1d\phi_1).
\end{align}
The above contributes to an RR three-form flux as $\mathcal{F}_3^{(B,5)}=d\mathcal{C}_2^{(B,5)}+\textrm{sources}$,
where
\begin{align}
d\mathcal{C}_2=-d\mathcal{C}_2^{(B,1)}
+\frac{\tan\theta_{nc}}{\cos\theta_{nc}}\left[
k_3\cos\theta_1 d\theta_1\wedge d\phi_1
-k_{3,a} da\wedge(\frac{d\tilde{\psi}}{\cos\theta_{nc}}+\cos\theta_1d\phi_1)
\right]\wedge d\tilde{x}_3
\label{F3B6}
\end{align}
and the sources reflect the presence of a D5-brane (S-dual to the previous NS5-brane), thus
leading to $d\mathcal{F}_3^{(B,5)}\neq0$.

All the modifications considered so far in this section have at this stage satisfied the desired goal:
to source an RR 2-form flux along $(\tilde{x}_3,\,\tilde{\psi})$ in our type IIB configuration before
the uplift to M-theory. As we explained in the beginning of the section, this is equivalent to
switching on an axionic background $\mathcal{C}_0$ in the usual D3-NS5 system. 
Having noted this important point, let us proceed with the remaining dualities to obtain the M-theory uplift
of the above configuration.

Upon a T-duality along $\phi_1$, the type IIB configuration above leads to a type IIA geometry given by
\begin{align}
ds_{(A,6)}^2=ds_{(4)}^2
+\sqrt{\frac{\hat{F}_2}{F_2}}\frac{e^{-\phi}\sec\theta_{nc}}{F_3\sin^2\theta_1+\hat{F}_2\cos^2\theta_1}d\phi_1^2.
\label{dsA6}
\end{align}
The type IIA dilaton in this case is
\begin{align}
e^{\phi_{(A,6)}}=\left(\frac{F_2}{\hat{F}_2}\right)^{1/4}
\left(\frac{e^{-3\phi}\sec\theta_{nc}}{F_3\sin^2\theta_1+\hat{F}_2\cos^2\theta_1}\right)^{1/2}.
\end{align}
There is an NS B-field associated to this metric,
\begin{align}
B_{(A,6)}=k_4d\phi_1\wedge d\tilde{\psi}, \quad\quad
k_4\equiv\frac{\hat{F}_2\sec\theta_{nc}\cos\theta_1}{F_3\sin^2\theta_1+\hat{F}_2\cos^2\theta_1},
\end{align}
which gives rise to an NS three-form flux of the form
\begin{align}
\mathcal{H}_3^{(A,6)}=dB_{(A,6)}=k_{4,a} da\wedge d\phi_1\wedge d\tilde{\psi}.
\label{H3A6}
\end{align}
Note that, as a consequence of our choices in (\ref{easychoice}) and
because $\hat{F}_2$ depends on $\phi$ (see (\ref{hatF2def})), $k_4=k_4(a)$ with $a\equiv (\theta_1,\,r,\, x_8,\, x_9)$.
The RR three-form flux $\mathcal{F}_3^{(B,5)}$ dualizes to an RR two-form flux. Using (\ref{H3B3}), this can be written as
\begin{align}
\nonumber
\mathcal{F}_2^{(A,6)}=&d\theta_1\wedge\left(\frac{\partial b_{\theta_1\phi_1}}{\partial r}dr
+\frac{\partial b_{\theta_1\phi_1}}{\partial x_8}dx_8
+\frac{\partial b_{\theta_1\phi_1}}{\partial x_9}dx_9\right)\\
&+\frac{\tan\theta_{nc}}{\cos\theta_{nc}}\left( k_{3,a}\cos\theta_1 da
-k_3\sin\theta_1d\theta_1\right)\wedge d\tilde{x}_3
+\textrm{sources}
\end{align}
and, of course, is not closed: $d\mathcal{F}_2^{(A,6)}\neq 0$, denoting a D6-brane source.
This is dual to the D5-brane sourcing $\mathcal{F}_3^{(B,5)}$ before.
Denoting as $\tilde{{\bf A}}_1$ the type IIA gauge field for this configuration, we can further
rewrite the above as
\begin{align}
\mathcal{F}_2^{(A,6)}=d\tilde{{\bf A}}_1+\Delta^\prime, \quad\quad d\Delta^\prime= \textrm{sources},
\label{F2A6gauge}
\end{align}
with $\tilde{{\bf A}}_1$ as in (3.58) in~\cite{Dasgupta:2016rhc}:
\begin{align}
\tilde{{\bf A}}_1=b_{\theta_1\phi_1}d\theta_1+k_3\frac{\tan\theta_{nc}}{\cos\theta_{nc}}\cos\theta_1d\tilde{x}_3.
\label{tildeA1def}
\end{align}

At last, we will uplift the above type IIA configuration to M-theory.
For this purpose, we start by rewriting $ds_{(A,6)}^2$ in a more convenient way.
Defining 
\begin{eqnarray}
\label{tildeHs}
\begin{array}{llllll}
&\tilde{H}_1\equiv(F_3\sin^2\theta_1+\hat{F}_2\cos^2\theta_1)^{1/3}\tilde{H}_2^{-1/3}, \quad\quad
&\tilde{H}_2\equiv \hat{F}_2F_2^{-1}\sec^2\theta_{nc}, \\
&\tilde{H}_3\equiv\tilde{H}_1^{-3}, \quad\quad
&\tilde{H}_4\equiv F_2F_3\sin^2\theta_1\tilde{H}_1^{-3}, 
\end{array}
\end{eqnarray}
as in (3.57) in~\cite{Dasgupta:2016rhc}, we can rewrite (\ref{dsA6}) as
\begin{align}
\nonumber
ds_{(A,6)}^2=&\frac{e^{-\phi}}{\tilde{H}_1\sqrt{\tilde{H}_2}}\left\{
\tilde{H}_1\left(-dt^2+dx_1^2+dx_2^2+\tilde{H}_2d\tilde{x}_3^2+\tilde{H}_3d\phi_1^2\right)\right. \\
&\left.+e^{2\phi}\tilde{H}_1\left[ F_1dr^2+F_3d\theta_1^2+F_4(dx_8^2+dx_9^2)+\tilde{H}_4d\tilde{\psi}^2\right]\right\}.
\end{align}
Again it should be borne in mind that the following M-theory only captures the dynamics of this type IIA theory in the
strong coupling limit where $e^{\phi_{(A,6)}}$ is, at least, of order one.
Being once more interested in having a finite radius for the eleventh direction, we shall be careful to avoid the
$e^{\phi_{(A,6)}}\rightarrow\infty$ limit.
This limit applies in the same cases as discussed in (\ref{Mlimit}) before.

The corresponding M-theory metric is that in (3.56) in~\cite{Dasgupta:2016rhc}:
\begin{align}
\nonumber
ds_{(M,2)}^2=&\tilde{H}_1\left[-dt^2+dx_1^2+dx_2^2+\tilde{H}_2d\tilde{x}_3^2+\tilde{H}_3d\phi_1^2
+e^{2\phi}(F_1dr^2+\tilde{H}_4d\tilde{\psi}^2)\right] \\
&+e^{2\phi}\tilde{H}_1[F_3d\theta_1^2+F_4(dx_8^2+dx_9^2)]
+\frac{e^{-2\phi}}{\tilde{H}_1^2\tilde{H}_2}(dx_{11}+\tilde{{\bf A}}_1)^2.
\label{M2metric}
\end{align}
In analogy to (\ref{defGs}) earlier, fixing $r=r_0$ and defining
\begin{align}
\label{deftildeGs}
\tilde{G}_1\equiv e^{2\phi}\tilde{H}_1F_3\Big|_{r=r_0}, \quad\quad
\tilde{G}_2,\tilde{G}_3\equiv e^{2\phi}\tilde{H}_1F_4\Big|_{r=r_0}, \quad\quad
\tilde{G}_4\equiv e^{-2\phi}\tilde{H}_1^{-2}\tilde{H}_2^{-1}\Big|_{r=r_0},
\end{align}
the last line above can be easily seen to be a warped Taub-NUT space with metric
\begin{align}
ds_{{TN}_2}^2=\tilde{G}_1d\theta_1^2+\tilde{G}_2dx_8^2+\tilde{G}_3dx_9^2+\tilde{G}_4
\left(dx_{11}+\tilde{{\bf A}}_1\Big|_{r=r_0}\right)^2.
\label{mettb2}
\end{align}
The G-flux corresponding to this second M-theory construction
is very similar to that in (\ref{355}):
\begin{align}
\mathcal{G}_4^{(M,2)}=\langle \mathcal{G}_4^{(M,2)}\rangle +\tilde{\mathcal{F}}\wedge \tilde{\omega},
\end{align}
where $\langle \mathcal{G}_4^{(M,2)}\rangle= \mathcal{H}_3^{(A,6)}\wedge dx_{11}$ is the background
G-flux ($\mathcal{H}_3^{(A,6)}$ is given by (\ref{H3A6})) and $\tilde{\omega}$ is the unique, normalizable
(anti-)self-dual harmonic two-form associated to 
the single-centered (warped) Taub-NUT space in (\ref{mettb2}).
Here, $\tilde{\mathcal{F}}$
stands for the field strength of the $U(1)$ world-volume gauge theory.

It would not be hard to adapt the computation of $\omega$ in section \ref{ncsect}
to the present case and obtain the explicit form of $\tilde{\omega}$.
In fact, we could adapt the discussion of section \ref{nonabsec} to the present case
and obtain a non-abelian enhancement of the world-volume gauge theory in this setup too.
However, before doing any more computations, let us compare the two M-theory metrics: (\ref{340}) and (\ref{M2metric}).
They are very similar. In fact, they just differ in the warp factors.
It is important to note that both of them break the Lorentz invariance
along the $(t,\,x_1,\,x_2)$ and the $\tilde{x}_3$ directions.
Moreover, both M-theories capture the dynamics of their dual type IIA configurations in the same
limit, as we noted a bit earlier.
Since the supergravity analysis that we will perform in part \ref{partb} will only depend
on the metric deformations, the above noted similarities are enough to consider that, for our purposes,
both M-theory configurations are equivalent.
Nonetheless, it is clear from our calculations so far that the first M-theory configuration is
computationally simpler to handle.
Indeed, as we already anticipated, the non-commutative deformation by itself sources the required $\Theta$-term
in the world-volume theory and that is all we will really need.
The present section explicitly has shown that (\ref{340}) captures all the information needed from
the type IIB configuration in~\cite{Witten:2011zz} to embed knots and study their invariants. 
Consequently, we will drop any further study of the M-theory configuration in (\ref{M2metric})
and instead carry all our investigations in the configuration with metric (\ref{340}).
That is, the first M-theory construction to study knot invariants is (M, 1) in figure \ref{fig3} and
its non-abelian enhancement in section \ref{nonabsec}.

It is important to bear in mind that the configuration (M, 1)
has been obtained from the D3-NS5 system of~\cite{Witten:2011zz}
using the well-defined chain of dualities depicted in figures \ref{fig1} and \ref{fig3}
(along with figure \ref{fig5}, for the non-abelian enhanced case).
Consequently, (M, 1) is {\it dual} to the model in~\cite{Witten:2011zz}, by construction.

Part \ref{partb} will be devoted to the study of the physics following from (M, 1).
A special emphasize will be made on what and why this is a suitable framework for
the realization of knots.
Before proceeding in this direction, however, we shall first construct yet another M-theory configuration,
which we will refer to as (M, 5). The configuration (M, 5) also follows from~\cite{Witten:2011zz}, but is {\it not} dual to
it, as we shall see. Instead, we will show that it is {\it dual} to the model in~\cite{Ooguri:1999bv}
and thus provides a second, independent natural framework for the realization of knots and the computation of knot invariants.

\FloatBarrier

\section{A different modification to the D3-NS5 system \label{ns5d3sect2}}

As was the case in section \ref{ns5d3sect1}
and as schematically shown in figure \ref{fig8}, the starting point of our analysis here too is the well-known type
IIB superstring theory configuration of a D3-brane ending on an NS5-brane considered in~\cite{Witten:2011zz}.
For the time being, we will not consider an axionic background: $\mathcal{C}_0=0$.
The notation and orientation of the branes are exactly as before, but with the further identifications
$(x_8\equiv\theta_2,\,x_9\equiv\phi_2)$, which will soon become sensible.

Next, we do five modifications to the above set up.
Figure \ref{fig2} schematically depicts them.
The modifications aim to ultimately make a precise connection between~\cite{Witten:2011zz} and~\cite{Ooguri:1999bv}.
We will discuss such connection later on.
For the time being, let us just discuss the modifications.

\begin{itemize}
\item First, we introduce a second NS5-brane, oriented along $(t,\,x_1\,x_2,\,x_3,\,\theta_1,\,\phi_1)$
and which intersects the D3-brane.

In analogy to the first modification in section \ref{ns5d3sect1},
this makes the direction orthogonal to both NS5-branes of the D3-brane, namely $\psi$, a finite interval.
The $\psi$ interval in this case is taken to be not too large.
Consequently, the $U(1)$ gauge theory on the D3-brane has only $\mathcal{N}=1$ supersymmetry now.

\begin{figure}[H]
\centering
\includegraphics[width=0.9\textwidth]{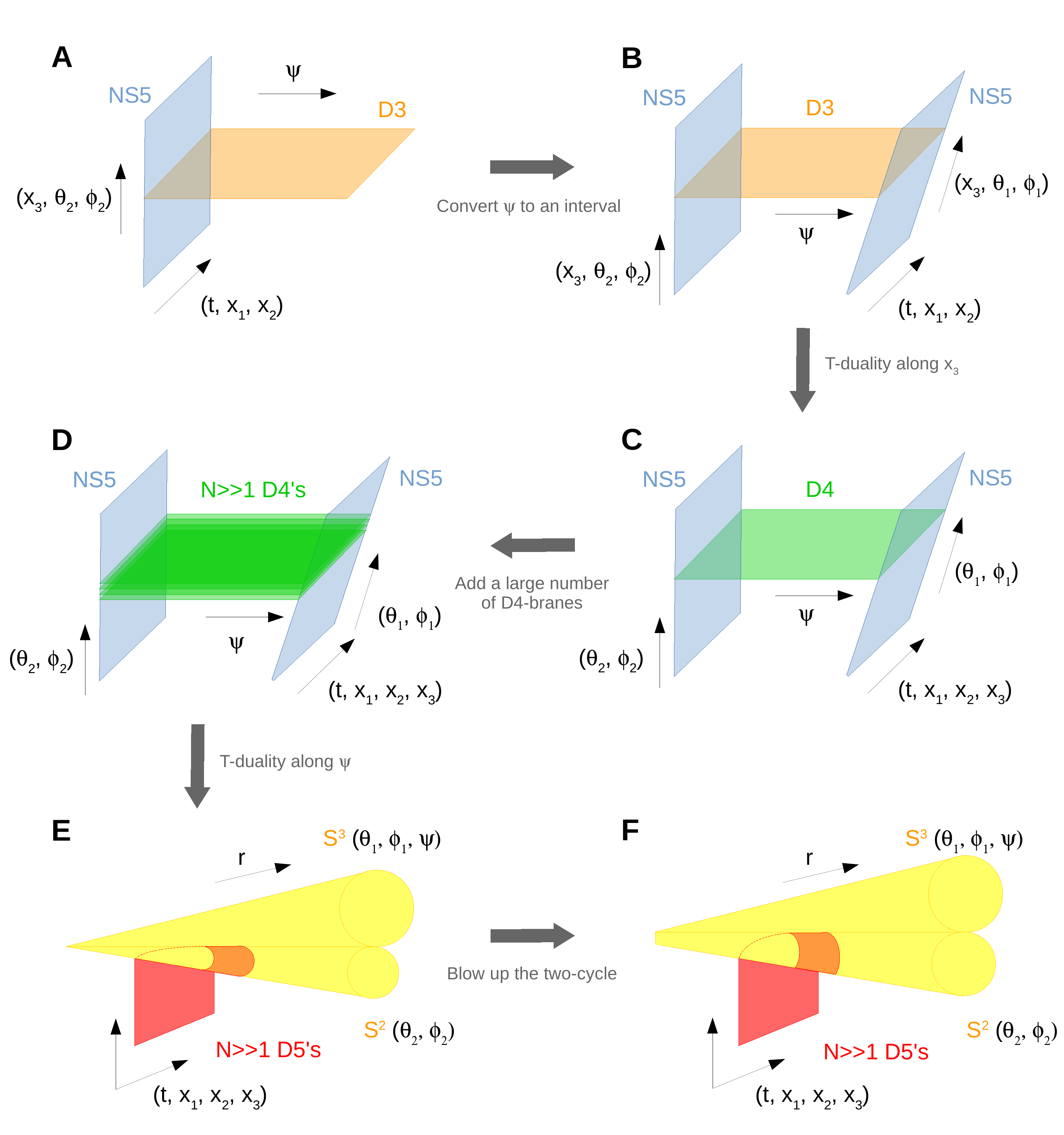}
\caption{Caricature of the modifications to the D3-NS5 system described in section \ref{ns5d3sect2}.
The reason to consider this chain of dualities is twofold: to be able to write the corresponding metric
(the geometry of {\bf F} is well-known) and to ultimately connect~\cite{Witten:2011zz} and~\cite{Ooguri:1999bv}.
{\bf A:} The well-known type IIB D3-NS5 system.
The D3-brane spans the $(t,\,x_1,\,x_2,\,\psi)$ directions and the NS5-brane the
$(t,\,x_1,\,x_2,\,x_3,\,\theta_2,\,\phi_2)$ directions.
The $(\theta_1,\,\phi_1,\,r)$ directions are suppressed. 
The gauge theory on the D3-brane has $\mathcal{N}=2$ supersymmetry.
{\bf B:} Introducing a second NS5-brane, oriented along $(t,\,x_1,\,x_2,\,x_3,\,\theta_1,\,\phi_1)$
converts the $\psi$ direction into an interval.
This reduces the amount of supersymmetry of the gauge theory on the D3-brane from $\mathcal{N}=2$ to $\mathcal{N}=1$.
The $r$ direction is suppressed.
{\bf C:} A T-duality along $x_3$ does not affect the NS5-branes, but converts the D3-brane into a D4-brane.
{\bf D:} We add a large amount of coincident D4-branes to the previous configuration.
The aim of this step is to later on establish a precise connection with the configuration studied in~\cite{Ooguri:1999bv}.
{\bf E:} A T-duality along $\psi$ converts the NS5-branes to a singular conifold along $(\theta_1,\,\phi_1,\,\psi,\,r,\,
\theta_2,\,\phi_2)$.
The D4-branes convert to as many D5-branes that wrap the vanishing two-cycle of the conifold.
{\bf F:} The blowing up of the two-cycle of the singular conifold leads to a resolved conifold.
The D5-branes are not affected.}
\label{fig2}
\end{figure}

\item Second, we do a T-duality to type IIA superstring theory along $x_3$,
which results in the D3-brane converting to a D4-brane. The NS5-branes are not affected by this T-duality.
This same duality was discussed at length in~\cite{Dasgupta:2001um,Dasgupta:2001fg}.

\item Third, we introduce a large number of coincident D4-branes, so that
we have a stuck of $N$ (where $N\in\mathbb{N}$ and $N>>1$) D4-branes between the two NS5-branes.

\item Fourth, we do a T-duality back to type IIB along $\psi$.
As a result, the NS5-branes disappear and give rise to a singular conifold in the $(\theta_1,\,\phi_1,\,\psi,\,r,\,
\theta_2,\,\phi_2)$ directions, which explains the coordinate relabeling above.
The $N$ D4-branes convert to $N$ D5-branes which wrap the vanishing two-cycle of the conifold.
This T-duality has been carefully discussed in~\cite{Dasgupta:1999wx, Maldacena:2009mw}.
Note that, unlike in section \ref{ns5d3sect1} (see figure \ref{fig1}{\bf D}), there are no ${\overline{\rm D5}}$-branes here.
This is because there is no Coulomb branch in this set up (the associated world-volume gauge theory is an
$\mathcal{N}=1$ supersymmetric one).

\item Finally, we blow up the two-cycle of the singular conifold and thus obtain a resolved conifold.
The metric on the resolved conifold is a {\it non-K\"{a}hler} one, as succinctly pointed out in~\cite{Maldacena:2009mw}
and as discussed in details in~\cite{Chen:2010bn}.
\end{itemize}

The geometry corresponding to this last configuration is known
(which also explains why the above modifications were done) and is given by (4.1) in~\cite{Dasgupta:2016rhc}:
\begin{align}
\label{metric6B}
ds_{(B,7)}^2=&e^{-\tilde{\phi}}(-dt^2+dx_1^2+dx_2^2+dx_3^2)\\
&+e^{\tilde{\phi}}
\left[\mathscr{F}_1dr^2+\mathscr{F}_2(d\psi+\sum_{i=1}^2\cos\theta_id\phi_i)^2
+\sum_{i=1}^2 \mathscr{F}_{2+i}(d\theta_i^2+\sin^2\theta_id\phi_i^2)\right]. 
\nonumber
\end{align}
Here, $e^{-\tilde{\phi}}$ is the usual type IIB dilaton:
\begin{align}
\label{dilB7}
e^{\tilde{\phi}_{(B,7)}}=e^{-\tilde{\phi}}.
\end{align}
For simplicity, we assume that the warp factors and
the dilaton only depend on the radial coordinate $r$:
\begin{align}
\mathscr{F}_i=\mathscr{F}_i(r), \quad\quad \tilde{\phi}=\tilde{\phi}(r), \quad\quad i=1,2,3,4.
\label{choicesB}
\end{align}
Under such assumption and for a fixed value of the radial coordinate, $r=r_0$,
the second line in (\ref{metric6B}) is the resolved conifold metric.
As was the case in section \ref{ns5d3sect1}, the D5-branes in this configuration source an RR three-form flux
$\mathcal{F}_3^{(B,7)}$ which can be computed as
\begin{align}
\label{sourceB}
\mathcal{F}_3^{(B,7)}=e^{2\tilde{\phi}}\ast d\mathcal{J}_{(B,7)},
\end{align}
where $\mathcal{J}_{(B,7)}$ is the fundamental two-form of the warped internal six-dimensional manifold
(note the dilaton is taken care of in (\ref{sourceB}) already)
with metric
\begin{align}
ds_{(7)}^2\equiv\mathscr{F}_1dr^2+\mathscr{F}_2(d\psi+\sum_{i=1}^2\cos\theta_id\phi_i)^2
+\sum_{i=1}^2 \mathscr{F}_{2+i}(d\theta_i^2+\sin^2\theta_id\phi_i^2).
\label{ds72}
\end{align}
We determine $\mathcal{F}_3^{(B,7)}$ in the following. (Note the coming calculation is
very similar to that presented earlier, between (\ref{viB1}) and (\ref{320}), so we will be succincter now.)

We start by defining the vielbeins associated to $ds_{(7)}^2$ as
\begin{eqnarray}
\label{vielb7}
\begin{array}{llll}
& E_{\theta_i}^{(B,7)}=\sqrt{\mathscr{F}_{2+i}}e_{\theta_i}^{(B,7)}=\sqrt{\mathscr{F}_{2+i}}d\theta_i,
\qquad
E_{\phi_i}^{(B,7)}=\sqrt{\mathscr{F}_{2+i}}e_{\phi_i}^{(B,7)}=\sqrt{\mathscr{F}_{2+i}}\sin\theta_id\phi_i, \\ 
& E_\psi^{(B,7)}=\sqrt{\mathscr{F}_2}e_\psi^{(B,7)}=\sqrt{\mathscr{F}_2}(d\psi+\sum\limits_{i=1}^2\cos\theta_i d\phi_i),
\qquad
E_r^{(B,7)}=\sqrt{\mathscr{F}_1} e_r^{(B,7)}=\sqrt{\mathscr{F}_1}dr,
\end{array}
\end{eqnarray}
where $i=1,\,2$. Using these vielbeins, it is easy to write down the fundamental two-form of our interest:
\begin{align}
\nonumber
\mathcal{J}_{(B,7)}=&\sum_{i=1}^2 E_{\theta_i}^{(B,7)}\wedge E_{\phi_i}^{(B,7)}+E_\psi^{(B,7)}\wedge E_r^{(B,7)} \\
=&\sum_{i=1}^2 \mathscr{F}_{2+i}\sin\theta_id\theta_i\wedge d\phi_i
+\sqrt{\mathscr{F}_1\mathscr{F}_2}(d\psi+\sum_{i=1}^2\cos\theta_i d\phi_i)\wedge dr. \label{mathj}
\end{align}
The exterior derivative of the above is
\begin{align}
d\mathcal{J}_{(B,7)}=\sum_{i=1}^2 (\mathscr{F}_{2+i,r}-\sqrt{\mathscr{F}_1\mathscr{F}_2})\sin\theta_i dr\wedge d\theta_i \wedge d\phi_i, 
\label{dJ2}
\end{align}
where, quite obviously, $\mathscr{F}_{2+i,r}$ stands for the derivative with respect to $r$ of $\mathscr{F}_{2+i}$ ($i=1,\,2$). 
Next, we wish to take the Hodge dual of the above. For this purpose, let us begin by writing (\ref{ds72}) in matrix form:
\begin{eqnarray}
\tilde{M}=\left(
\begin{array}{cccccccccc}
\mathscr{F}_3 & 0 & 0 & 0 & 0 & 0 \\
0 & \mathscr{F}_2\cos^2\theta_1+\mathscr{F}_3\sin^2\theta_1 & 0 & \mathscr{F}_2\cos\theta_1 & 0 & \mathscr{F}_2\cos\theta_1\cos\theta_2 \\
0 & 0 & \mathscr{F}_1 & 0 & 0 & 0 \\
0 & \mathscr{F}_2\cos\theta_1 & 0 & \mathscr{F}_2 & 0 & \mathscr{F}_2\cos\theta_2 \\
0 & 0 & 0 & 0 & \mathscr{F}_4 & 0 \\
0 & \mathscr{F}_2\cos\theta_1\cos\theta_2 & 0 & \mathscr{F}_2\cos\theta_2 & 0 & \mathscr{F}_2\cos^2\theta_2+\mathscr{F}_4\sin^2\theta_2
\end{array}
\right).
\end{eqnarray}
The inverse of the above metric is
\begin{eqnarray}
\tilde{M}^{-1}=\left(
\begin{array}{cccccccc}
\frac{1}{\mathscr{F}_3} & 0 & 0 & 0 & 0 & 0 \\
0 & \frac{\csc^2\theta_1}{\mathscr{F}_3} & 0 & -\frac{\cot\theta_1\csc\theta_1}{\mathscr{F}_3} & 0 & 0 \\
0 & 0 & \frac{1}{\mathscr{F}_1} & 0 & 0 & 0 \\
0 & -\frac{\cot\theta_1\csc\theta_1}{\mathscr{F}_3} & 0 & \frac{1}{\mathscr{F}_2}+\frac{\cot^2\theta_1}{\mathscr{F}_3}+\frac{\cot^2\theta_2}{\mathscr{F}_4}  & 0 & 
-\frac{\cot\theta_2\csc\theta_2}{\mathscr{F}_4}\\
0 & 0 & 0 & 0 & \frac{1}{\mathscr{F}_4} & 0 \\
0 & 0 & 0 & -\frac{\cot\theta_2\csc\theta_2}{\mathscr{F}_4} & 0 & \frac{\csc^2\theta_2}{\mathscr{F}_4} 
\end{array}
\right)
\end{eqnarray}
and the square root of its determinant is
\begin{align}
\tilde{\mathcal{M}}\equiv \sqrt{\det \tilde{M}}=\sqrt{\mathscr{F}_1\mathscr{F}_2}\mathscr{F}_3\mathscr{F}_4\sin\theta_1\sin\theta_2.
\end{align}
All this information can now be used to compute the Hodge dual of the wedge products in (\ref{dJ2}). For a fixed value of $i$ ($i=1$
or $i=2$),

\begin{align}
\nonumber
\ast(dr\wedge d\theta_i\wedge d\phi_i)=&\tilde{\mathcal{M}} \tilde{M}^{-1}_{rr}\tilde{M}^{-1}_{\theta_i\theta_i}( 
\tilde{M}^{-1}_{\phi_i\phi_i}
\epsilon_{r\theta_i\phi_i \psi \theta_j\phi_j}d\psi
+\tilde{M}^{-1}_{\phi_i\psi}\epsilon_{r\theta_i\psi\phi_i\theta_j\phi_j}
 d\phi_i)\wedge d\theta_j\wedge d\phi_j \\
=&\sqrt{\frac{\mathscr{F}_2}{\mathscr{F}_1}}\frac{\mathscr{F}_{2+j}}{\mathscr{F}_{2+i}}\csc\theta_i\sin\theta_j(d\psi
+\cos\theta_i d\phi_i)\wedge d\theta_j\wedge d\phi_j,
\end{align}
with $j$ fixed and not equal to $i$. That is, either $(i,\,j)=(1,\,2)$ or $(i,\,j)=(2,\,1)$.
Putting everything together, the three-form flux in (\ref{sourceB}) can be easily seen to be
\begin{align}
\mathcal{F}_3^{(B,7)}=e^{2\tilde{\phi}} \sqrt{\frac{\mathscr{F}_2}{\mathscr{F}_1}}\sum_{\substack{i,j=1 \\ i\neq j}}^2
\frac{\mathscr{F}_{2+j}}{\mathscr{F}_{2+i}} (\mathscr{F}_{2+i,r}-\sqrt{\mathscr{F}_1\mathscr{F}_2})\sin\theta_j(d\psi
+\cos\theta_i d\phi_i)\wedge d\theta_j\wedge d\phi_j.
\label{F3B7}
\end{align}
Note that, in good agreement with the previously pointed out presence of D5-branes in this configuration, the above flux
is not closed: $d\mathcal{F}_3^{(B,7)}\neq0$.

Later on, in section \ref{choicesec}, we will be interested in making a fully precise
choice of the warp factors and dilaton in (\ref{choicesB}). Accordingly, we note that
not any such choice will
eventually lead to a world-volume gauge theory with $\mathcal{N}=1$ supersymmetry. The story is in fact a bit
more involved: the warp factors and dilaton must satisfy a particular constraint equation so that we indeed have
$\mathcal{N}=1$ supersymmetry. In the following section, we derive this constraint equation.

\FloatBarrier

\subsection{Demanding $\mathcal{N}=1$ supersymmetry: torsion classes \label{torsionsec}}

The aforementioned constraint equation relating the warp factors and dilaton in (\ref{choicesB}) that ensures
$\mathcal{N}=1$ supersymmetry in the associated world-volume gauge theory is most easily derived using the
technique of torsion classes. A detailed yet concise review of the technique and its applications to
string theory can be found in~\cite{LopesCardoso:2002vpf}. A more mathematical approach to the same
material is~\cite{Chiossi}. In this section, we review and adapt the results in these references to the present case
and thus obtain the desired constraint equation. (This is, essentially, the content of section 3.1
in~\cite{Dasgupta:2014txa} too.)

We start by noting that the type IIB configuration determined in the previous section has an internal six-dimensional
manifold, whose (Riemannian) metric was given in (\ref{ds72}). This manifold is equipped with a fundamental two-form,
given in (\ref{mathj}). In a more mathematical language, we say that this is a six-dimensional manifold with a $U(3)$
structure $J$. An $SU(3)$ structure is then determined by a real three-form $\Omega_+$, which we will soon compute.
There is an {\it intrinsic torsion} associated to each of these structures. For our purposes, only
the intrinsic torsion $\tau_1$ of the $SU(3)$ structure will be relevant. $\tau_1$ belongs to a space
which can be decomposed into five classes:
\begin{align}
\tau_1\in\mathcal{W}_1\oplus \mathcal{W}_2\oplus \mathcal{W}_3\oplus \mathcal{W}_4\oplus \mathcal{W}_5,
\label{torsions}
\end{align}
according to its decomposition into the irreps of $SU(3)$
\begin{align}
(\bf{1}+\bf{1})+(\bf{8}+\bf{8})+(\bf{6}+\bar{\bf{6}})+(\bf{3}+\bar{\bf{3}})+(\bar{\bf{3}}+\bf{3}).
\end{align}
We denote the component of $\tau_1$ in $\mathcal{W}_i$ as $W_i$ ($i=1,\,2,\,3,\,4,\,5$).

Before proceeding further, let us introduce the so called contraction operator $\lrcorner$, which will immediately
become useful to us. 
Let $(e_1,\,e_2,\,\ldots,\,e_i)$ be an orthonormal basis of the cotangent space $T^\ast M$ of any $i$-dimensional 
manifold $M$.
Given a $j$-form $\omega_1$ and a $k$-form $\omega_2$ in $T^\ast M$ (with $i\geq j\geq k\geq 0$),
\begin{align}
\omega_1\equiv (\omega_1)_{12\ldots j}\prod_{l=1}^j e_l, \quad\quad \omega_2\equiv (\omega_2)_{12\dots k}\prod_{l=1}^ke_l,
\end{align}
the contraction operator
$\lrcorner$ is a map from the pair $(\omega_1,\,\omega_2)$
to a $(j-k)$-form given by
\begin{eqnarray}
\label{lrcorner}
\omega_2\,\,\lrcorner\,\, \omega_1\equiv  \frac{1}{j!}\left(
\begin{array}{cc} j \\ k \end{array}
\right) (\omega_1)^{12\ldots j}(\omega_2)_{12\dots k}\prod_{l=k+1}^j e_l,
\end{eqnarray}
with the convention that $e_1\wedge e_2\,\, \lrcorner\,\, e_1\wedge e_2\wedge e_3=e_3$, etc.
Having introduced the contraction operator, we now have all the ingredients required to derive the desired constraint
equation.

The necessary and sufficient conditions to ensure $\mathcal{N}=1$ supersymmetry in the world-volume gauge theory
corresponding to the geometry (\ref{metric6B}) have long been known~\cite{Strominger}\footnote{
The conditions in~\cite{Strominger} are actually a bit too stringent.
Later on, examples
of $\mathcal{N}=1$ supersymmetric theories which did not satisfy all these conditions were found
(see, for example~\cite{Chakravarty:2000qd}). For our case, however, the list in~\cite{Strominger}
will suffice.}. These conditions were then reformulated in~\cite{LopesCardoso:2002vpf} in
terms of the torsion classes we just introduced in (\ref{torsions}). For the present case,
they amount to demanding that (4.23) in~\cite{Dasgupta:2016rhc} should hold true:
\begin{align}
2W_4+W_5=0, \label{susycons}
\end{align}
with $(W_4,\,W_5)$ defined as
\begin{align}
W_4\equiv \frac{1}{2}J\,\,\lrcorner\,\, dJ, \quad\quad W_5\equiv \frac{1}{2}\Omega_+\,\,\lrcorner\,\, d\Omega_+.
\label{w45def}
\end{align}
The remaining of this section is devoted to the calculation of (\ref{susycons}) in terms of the warp
factors and dilaton in (\ref{choicesB}).

In order to match the conventions in~\cite{Dasgupta:2014txa}, where the interested reader can find
an elaboration of the present discussion, we take the complex vielbeins of the internal six-manifold of (\ref{metric6B})
as in there:
\begin{align}
\label{complexvielb}
\mathcal{E}_1^{(B,7)}= e^{\tilde{\phi}}(\sqrt{\mathscr{F}_1}e_r^{(B,7)}+i\sqrt{\mathscr{F}_2}e_\psi^{(B,7)}), \quad\quad
\mathcal{E}_{1+i}^{(B,7)}=e^{\tilde{\phi}+i\psi/2}\sqrt{\mathscr{F}_{2+i}}(e_{\theta_i}^{(B,7)}+ie_{\phi_i}^{(B,7)}), 
\end{align}
where the vielbeins $e^{(B,7)}$ where defined in (\ref{vielb7}) and $i=1,\,2$.
In terms of these vielbeins, the $U(3)$ structure $J$ of the internal space is given by
\newpage
\begin{align}
\nonumber
J=&\overline{\mathcal{E}}_1^{(B,7)}\wedge \mathcal{E}_1^{(B,7)}+
\sum_{i=1}^2\mathcal{E}_{1+i}^{(B,7)}\wedge \overline{\mathcal{E}}_{1+i}^{(B,7)} \\
=&2ie^{2\tilde{\phi}}\Big(\sqrt{\mathscr{F}_1\mathscr{F}_2}e_r^{(B,7)}\wedge e_\psi^{(B,7)}+\sum_{i=1}^2
\mathscr{F}_{2+i} e_{\phi_i}^{(B,7)}\wedge e_{\theta_i}^{(B,7)}\Big),
\label{JBfinal}
\end{align}
where the bar denotes complex conjugation.
We also define the three-form $\Omega$ as
\begin{align}
\Omega \equiv \mathcal{E}_1\wedge \mathcal{E}_2\wedge \mathcal{E}_3
=e^{3\tilde{\phi}+i\psi}\sqrt{\mathscr{F}_3\mathscr{F}_4}\left(\sqrt{\mathscr{F}_1}e_r^{(B,7)}+i\sqrt{\mathscr{F}_2}e_\psi^{(B,7)}\right)\wedge
\prod_{i=1}^2\left(e_{\theta_i}^{(B,7)}+ie_{\phi_i}^{(B,7)}\right).
\label{omegaB}
\end{align}
The $SU(3)$ structure $\Omega_+$ of the internal space is just the real part of the above three-form: $\Omega_+\equiv\textrm{Re}(\Omega)$.
Using Euler's formula, it is not hard to show that
\begin{align}
\nonumber
\Omega_+=&e^{3\tilde{\phi}}\sqrt{\mathscr{F}_3\mathscr{F}_4}\left[
\left(\sqrt{\mathscr{F}_1}\cos\psi e_r^{(B,7)}-\sqrt{\mathscr{F}_2}\sin\psi e_\psi^{(B,7)}\right)\wedge
\left(e_{\theta_1}^{(B,7)}\wedge e_{\theta_2}^{(B,7)}-e_{\phi_1}^{(B,7)}\wedge e_{\phi_2}^{(B,7)}\right)\right. \\
&\left.-\left(\sqrt{\mathscr{F}_1}\sin\psi e_r^{(B,7)}+\sqrt{\mathscr{F}_2}\cos\psi e_\psi^{(B,7)}\right)\wedge
\left(e_{\theta_1}^{(B,7)}\wedge e_{\phi_2}^{(B,7)}+e_{\phi_1}^{(B,7)}\wedge e_{\theta_2}^{(B,7)}\right)
\right].
\label{omegaplusB}
\end{align}
In order to obtain the exterior derivative of the two structures of our interest, $(J,\, \Omega_+)$,
it is necessary to use the explicit form of the vielbeins in (\ref{vielb7}). Rather tedious algebra yields
\begin{align}
dJ=&2ie^{2\tilde{\phi}}\sum_{i=1}^2\left(\sqrt{\mathscr{F}_1\mathscr{F}_2}-\mathscr{F}_{2+i,r}-2\tilde{\phi}_r\mathscr{F}_{2+i}\right)
e_r^{(B,7)}\wedge e_{\theta_i}^{(B,7)}\wedge e_{\phi_i}^{(B,7)},  \label{dJB}\\
d\Omega_+=& k_1^\prime
 e_r^{(B,7)}\wedge e_{\phi_1}^{(B,7)}\wedge e_{\phi_2}^{(B,7)}\wedge \sum_{i=1}^2\cot\theta_i e_{\theta_i}^{(B,7)} \nonumber \\
&+ k_1^\prime d\psi\wedge e_r^{(B,7)}\wedge \left(e_{\theta_1}^{(B,7)}\wedge e_{\phi_2}^{(B,7)}
+e_{\phi_1}^{(B,7)}\wedge e_{\theta_2}^{(B,7)}\right) \nonumber \\
& +k_2^\prime
 e_r^{(B,7)}\wedge e_{\theta_1}^{(B,7)}\wedge e_{\theta_2}^{(B,7)}\wedge \sum_{i=1}^2\cot\theta_i e_{\phi_i}^{(B,7)} \nonumber \\
&+ k_2^\prime d\psi\wedge e_r^{(B,7)}\wedge \left(e_{\theta_1}^{(B,7)}\wedge e_{\theta_2}^{(B,7)}
-e_{\phi_1}^{(B,7)}\wedge e_{\phi_2}^{(B,7)}\right), \label{dompl}
\end{align}
where the subscript $r$, as before, denotes derivation with respect to the radial coordinate and we have defined
\begin{align}
k_1^\prime\equiv e^{3\tilde{\phi}}\sqrt{\mathscr{F}_2\mathscr{F}_3\mathscr{F}_4}\cos\psi\left(
3\tilde{\phi}_r-\sqrt{\frac{\mathscr{F}_1}{\mathscr{F}_2}}+\sum_{i=2}^4\frac{\mathscr{F}_{i,r}}{2\mathscr{F}_i}
\right), \quad\quad 
k_2^\prime\equiv-\tan\psi k_1^\prime.
\end{align}
Using (\ref{lrcorner}) and all the above in (\ref{w45def}), it is a matter of care and patience to obtain
the relevant components of the intrinsic torsion of $\Omega_+$ as in 
(4.20) in~\cite{Dasgupta:2016rhc}:
\begin{align}
W_4=\left(\tilde{\phi}_r+\sum_{i=3}^4\frac{\mathscr{F}_{i,r}-\sqrt{\mathscr{F}_1\mathscr{F}_2}}{4\mathscr{F}_i}\right) e_r^{(B,7)}, \quad\quad
W_5=\frac{1}{2}\left(\tilde{\phi}_r-\frac{1}{3}\sqrt{\frac{\mathscr{F}_1}{\mathscr{F}_2}}+\sum_{i=2}^4\frac{\mathscr{F}_{i,r}}{6\mathscr{F}_i}\right) e_r^{(B,7)}.
\end{align}
Finally, inserting these values of $(W_4,\,W_5)$ in (\ref{susycons}), the desired constraint ensuring $\mathcal{N}=1$ supersymmetry is
\begin{align}
\label{n1cons}
30\tilde{\phi}_r-2\sqrt{\frac{\mathscr{F}_1}{\mathscr{F}_2}}+\frac{\mathscr{F}_{2,r}}{\mathscr{F}_2}+
\sum_{i=3}^4\left(7\frac{\mathscr{F}_{i,r}}{\mathscr{F}_i}-6\frac{\sqrt{\mathscr{F}_1\mathscr{F}_2}}{\mathscr{F}_i}\right)=0.
\end{align}

At this point one may wonder if similar constraints should not have been worked out for our configuration (M, 1)
with metric (\ref{3.43.5}) in section \ref{ns5d3sect1} as well. Surely if $\mathcal{N}=1$ supersymmetry
constrains the choice of warp factors and dilaton in (\ref{choicesB}), $\mathcal{N}=2$ supersymmetry will also
constrain the choice in (\ref{easychoice}). The resolution to this issue is, unfortunately, beyond the scope
of this work, as the powerful technique of torsion classes has not yet been generalized to the case of
$\mathcal{N}=2$ supersymmetry. Consequently, any specific choice for the warp factors in (\ref{easychoice})
and $Q$ in (\ref{354}) that one may want to
consider will require an explicit verification that it indeed preserves the desired amount of
supersymmetry\footnote{We will discuss how this is achieved in the gauge theory following from (M, 1)
in section \ref{bcsec} later on.}.

To sum things up, so far we have obtained from
the well-known D3-NS5 system (with no axion) of~\cite{Witten:2011zz} 
the type IIB configuration with metric (\ref{metric6B}), dilaton $e^{-\tilde{\phi}}$ and an RR three-form flux (\ref{F3B7}).
In order for this configuration to lead to a $\mathcal{N}=1$ supersymmetric world-volume gauge theory,
the constraint (\ref{n1cons}) should be satisfied.
However,
we would like to consider a type IIB configuration which, besides having an RR three-form flux, also has
an NS three-form flux.
This is, in principle, not an easy task. However, the series of dualities first
presented in~\cite{Maldacena:2009mw}
and later on further studied in~\cite{Chen:2010bn} and~\cite{Dasgupta:2014txa}, when applied to our above configuration,
precisely serves this purpose. In the following section,
we explain these dualities in details and obtain a type IIB configuration with both RR and NS fluxes.
Such a generalization will then, in section \ref{oogurivafasec}, allow us to establish a direct connection with
the model to study knots presented in~\cite{Ooguri:1999bv}.

\subsection{Obtaining a type IIB configuration with RR and NS fluxes: a boost in M-theory \label{boostsec}}

\begin{figure}[t]
\centering
\includegraphics[width=0.9\textwidth]{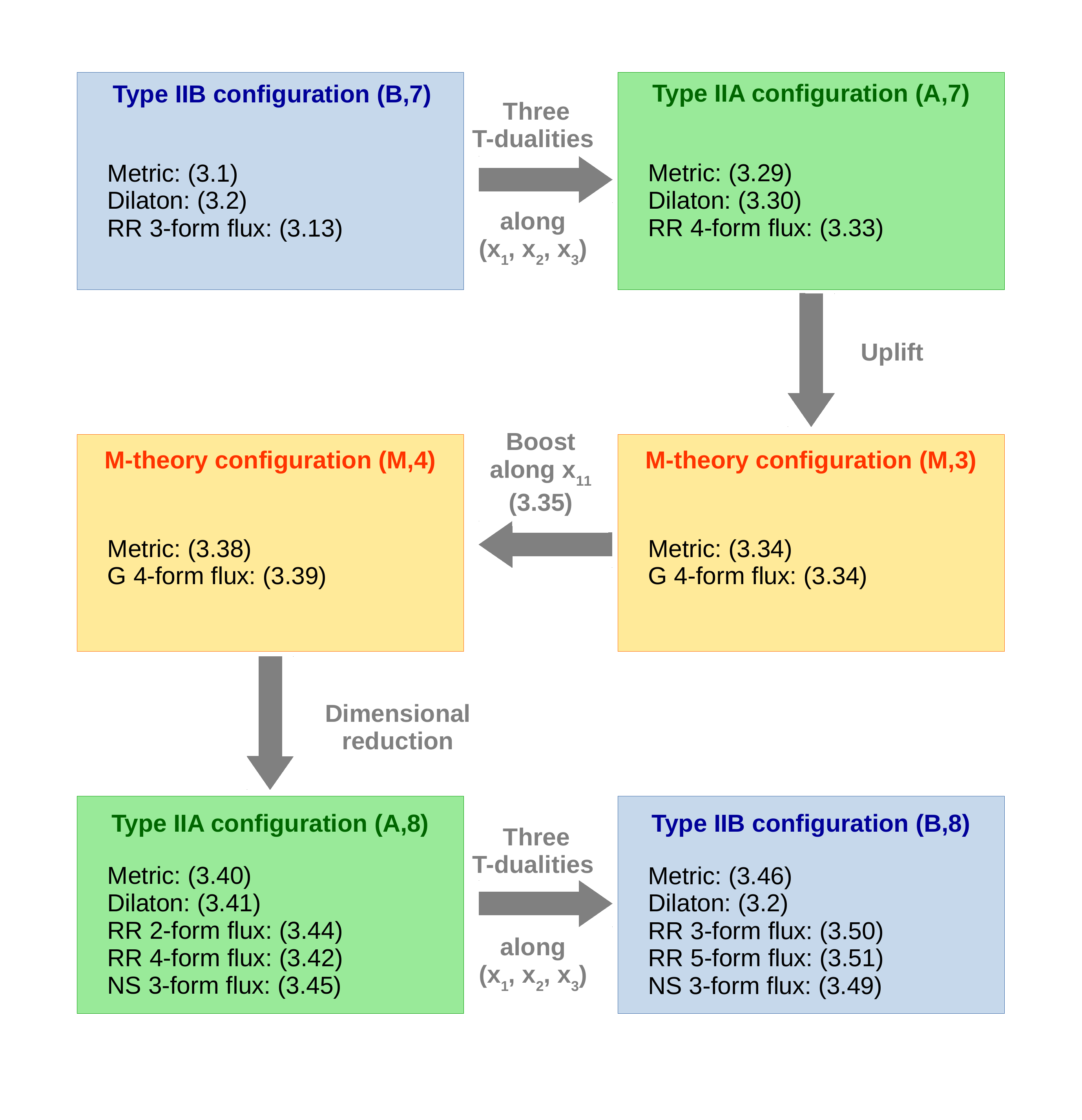}
\caption{Graphical summary of section \ref{boostsec}.
To the type IIB configuration of figure \ref{fig2}{\bf F} we do a series of modifications.
In this manner, we obtain a type IIB configuration that, besides RR fluxes, has NS fluxes as well.}
\label{fig6}
\end{figure}

We start this section considering the type IIB configuration described in section \ref{ns5d3sect2} and
depicted in figure \ref{fig2}{\bf F}.
We will first perform three T-dualities, along $(x_1,\,x_2,\,x_3)$, to type IIA.
The resulting configuration will then be lifted to M-theory, where we will perform a boost
along the $(t,\,x_{11})$ directions: $(t,\,x_{11})\rightarrow (\tilde{t},\,\tilde{x}_{11})$.
This will be followed by a dimensional reduction to type IIA. The last step will be to 
T-dualize along $(x_1,\,x_2,\,x_3)$ back to type IIB. Of course, we will work out the NS B-field, dilaton and RR and NS fluxes
associated to each geometry considered along this chain of modifications. As we already pointed out,
starting from a type IIB configuration which only has RR fluxes,
we will thus obtain a type IIB configuration with RR and NS fluxes.
As already said and as we shall show, the additional NS fluxes are required in order to precisely reproduce the model
in~\cite{Ooguri:1999bv}.
Figure \ref{fig6} outlines the just described
chain of modifications and serves as a summary of the key results in the present section.

As just mentioned, to the type IIB configuration shown in figure \ref{fig2}{\bf F} we do three T-dualities, along $(x_1,\,x_2,\,x_3)$.
It is rather straightforward to see that the metric 
then becomes
\begin{align}
ds_{(A,7)}^2=-e^{-\tilde{\phi}}dt^2+e^{\tilde{\phi}} (dx_1^2+dx_2^2+dx_3^2+ ds_{(7)}^2),
\end{align}
where $ds_{(7)}^2$ was defined in (\ref{ds72}). 
Coming to the dilaton, its changes can be summarized as follows:
\begin{align}
e^{\tilde{\phi}_{(B,7)}}=e^{-\tilde{\phi}}\,\xrightarrow[\textrm{along } x_1]{\textrm{T-duality}}\,
e^{-\tilde{\phi}/2}\,\xrightarrow[\textrm{along } x_2]{\textrm{T-duality}}\, 1\,
\xrightarrow[\textrm{along } x_3]{\textrm{T-duality}} \,e^{\tilde{\phi}/2}=e^{\tilde{\phi}_{(A,7)}}.
\end{align}
This can be used to rewrite our type IIA metric in a form that will soon make it straightforward to uplift it to M-theory:
\begin{align}
ds_{(A,7)}^2=e^{\tilde{\phi}/3}\left[-e^{-4\tilde{\phi}/3}dt^2+e^{2\tilde{\phi}/3} (dx_1^2+dx_2^2+dx_3^2+ ds_{(7)}^2)\right].
\label{dsA7}
\end{align}
Regarding the $\mathcal{F}_3^{(B,7)}$ flux, we note that each T-duality will 
add a leg to it along its corresponding Minkowskian direction $(x_1,\,x_2,\,x_3)$. That is,
\begin{align}
\nonumber
&\mathcal{F}_3^{(B,7)}\,\xrightarrow[\textrm{along } x_1]{\textrm{T-duality}}\,
dx_1\wedge \mathcal{F}_3^{(B,7)}\, \xrightarrow[\textrm{along } x_2]{\textrm{T-duality}} \,dx_2\wedge dx_1\wedge \mathcal{F}_3^{(B,7)} \\
&\qquad\quad \xrightarrow[\textrm{along } x_3]{\textrm{T-duality}} \,dx_3\wedge dx_2\wedge dx_1\wedge \mathcal{F}_3^{(B,7)}= \mathcal{F}_6^{(A,7)}.
\label{F33T}
\end{align}
We thus obtain an RR six-form flux.
This flux is not closed ($d\mathcal{F}_6^{(A,7)}\neq 0$),
which is to be expected, since the three T-dualities convert the $N$ coincident D5-branes of the previous type IIB configuration to $N$ coincident D2-branes that source
$\mathcal{F}_6^{(A,7)}$.
The Hodge dual of this six-form flux then gives us the more convenient (for the coming uplift)
RR four-form flux of this type IIA configuration:
\begin{align}
\nonumber
\mathcal{F}_4^{(A,7)}=&\ast \mathcal{F}_6^{(A,7)}= \ast \mathcal{F}_3^{(B,7)}\wedge dt= d\mathcal{J}_{(B,7)}\wedge dt\\
=&\sum_{i=1}^2 (\mathscr{F}_{2+i,r}-\sqrt{\mathscr{F}_1\mathscr{F}_2})\sin\theta_i dr\wedge d\theta_i \wedge d\phi_i\wedge dt,
\end{align}
where the first Hodge dual is with respect to the full ten-dimensional metric (\ref{dsA7}), whereas the second one is with respect to (\ref{ds72}).
The above result makes use of (\ref{sourceB}), (\ref{dJ2}) and (\ref{F33T}). 

We wrote our type IIA configuration so that the uplift to M-theory would be effortless. We get the following metric and G-flux:
\begin{align}
ds_{(M,3)}^2=-e^{-4\tilde{\phi}/3}dt^2+e^{2\tilde{\phi}/3} (dx_1^2+dx_2^2+dx_3^2+ ds_{(7)}^2+dx_{11}^2), \quad\quad
\mathcal{G}_{4}^{(M,3)}=\mathcal{F}_4^{(A,7)}.
\end{align}
Note that the D2-branes now convert to $N$ coincident M2-branes.

The key step in this chain of dualities comes next: we perform a boost in the eleventh direction. Explicitly,
\begin{align}
x_{11}=\cosh\beta \tilde{x}_{11}-\sinh\beta \tilde{t}, \quad\quad t=-\sinh\beta \tilde{x}_{11} +\cosh\beta \tilde{t},
\label{boost}
\end{align}
with $\beta$ the boost parameter.
Following equation (4.3) in~\cite{Dasgupta:2016rhc}, we define the quantity
\begin{eqnarray}
\Upsilon\equiv \sinh^2\beta(e^{2\tilde{\phi}/3}-e^{-4\tilde{\phi}/3}).
\label{deltadef}
\end{eqnarray}
Using the above two equations in $ds_{(M,3)}^2$, it is a matter of simple algebra to check that the boosted M-theory metric
is given by
\begin{align}
\nonumber
ds_{(M,4)}^2=&e^{2\tilde{\phi}/3} (dx_1^2+dx_2^2+dx_3^2+ ds_{(7)}^2)+(\Upsilon-e^{-4\tilde{\phi}/3})d\tilde{t}^2 \\
&+(\Upsilon+e^{2\tilde{\phi}/3})d\tilde{x}_{11}^2
-2\Upsilon \coth\beta d\tilde{x}_{11} d\tilde{t}.
\end{align}
Note that the boost has now generated a gauge field in the M-theory. This is most clearly seen upon rewriting the
above metric as
\newpage
\begin{align}
\nonumber
ds_{(M,4)}^2=&e^{2\tilde{\phi}/3} (dx_1^2+dx_2^2+dx_3^2+ ds_{(7)}^2)-\frac{e^{-2\tilde{\phi}/3}}{\Upsilon+e^{2\tilde{\phi}/3}}d\tilde{t}^2 \\
&+(\Upsilon+e^{2\tilde{\phi}/3})(d\tilde{x}_{11}-\frac{\Upsilon\coth\beta}{\Upsilon+e^{2\tilde{\phi}/3}}d\tilde{t})^2.
\end{align}
This rewriting is convenient for the coming dimensional reduction too.
Similarly, the boosted G-flux can be easily seen to be
\begin{eqnarray}
\mathcal{G}_{4}^{(M,4)}= d\mathcal{J}_{(B,7)}\wedge (\cosh\beta d\tilde{t}-\sinh\beta d\tilde{x}_{11}),
\end{eqnarray}
with $d\mathcal{J}_{(B,7)}$ as in (\ref{dJ2}).

The next step in the chain of dualities outlined in the beginning of the section is
to dimensionally reduce the above to type IIA. The metric corresponding to this configuration
is
\begin{align}
ds_{(A,8)}^2=-\frac{e^{-2\tilde{\phi}/3}}{\sqrt{\Upsilon+e^{2\tilde{\phi}/3}}}d\tilde{t}^2+
e^{2\tilde{\phi}/3}\sqrt{\Upsilon+e^{2\tilde{\phi}/3}}(dx_1^2+dx_2^2+dx_3^2+ ds_{(7)}^2)
\label{dsA8}
\end{align}
and the corresponding dilaton is
\begin{align}
e^{\tilde{\phi}_{(A,8)}}=\left(\Upsilon+e^{2\tilde{\phi}/3}\right)^{3/4}.
\end{align}
Coming now to the fluxes, we note that
the M2-branes of the previous M-theory setup now convert to D2-branes, which source an RR four-form flux
given by
\begin{align}
\mathcal{F}_4^{(A,8)}=\cosh\beta d\mathcal{J}_{(B,7)}\wedge  d\tilde{t}.
\end{align}
The Hodge dual of the above will soon be useful. This is an RR six-form flux of the form
\begin{align}
\mathcal{F}_6^{(A,8)}=\ast \mathcal{F}_4^{(A,8)}=\cosh\beta dx_1\wedge dx_2\wedge dx_3\wedge \mathcal{F}_3^{(B,7)},
\end{align}
which is clearly not closed, $d\mathcal{F}_6^{(A,8)}\neq 0$, as expected. (Recall $\mathcal{F}_3^{(B,7)}$ was given in (\ref{F3B7}).)
Additionally, the M-theory gauge field generated by the boost (\ref{boost}), effectively converts to a ``D0-charge''. This D0-charge sources
a closed RR two-form flux: the exterior derivative of the just mentioned gauge field. Explicitly,
\begin{align}
\mathcal{F}_2^{(A,8)}=-d\left(\frac{\Upsilon\coth\beta}{\Upsilon+e^{2\tilde{\phi}/3}}d\tilde{t}\right)=
\coth\beta\frac{d}{dr}\left(\frac{\Upsilon}{\Upsilon+e^{2\tilde{\phi}/3}}\right)d\tilde{t}\wedge dr,
\end{align}
where we have used the fact that, as a consequence of our choices in (\ref{choicesB}), the gauge field only depends on the
radial coordinate $r$ (and the boost parameter $\beta$). To finish this flux discussion, we note that the boost generates a closed
NS three-form flux, just as we wanted:
\begin{align}
\mathcal{H}_3^{(A,8)}=-\sinh\beta d\mathcal{J}_{(B,7)}.
\end{align}

To finish this section, the only remaining task is to perform three T-dualities, along $(x_1,\,x_2,\,x_3)$, back to type IIB.
From (\ref{dsA8}), it follows that the geometry corresponding to our final configuration is
\begin{align}
ds_{(B,8)}^2=\frac{e^{-2\tilde{\phi}/3}}{\sqrt{\Upsilon+e^{2\tilde{\phi}/3}}}(-d\tilde{t}^2+dx_1^2+dx_2^2+dx_3^2)
+e^{2\tilde{\phi}/3}\sqrt{\Upsilon+e^{2\tilde{\phi}/3}}ds_{(7)}^2.
\label{newmetricB}
\end{align}
The changes in the dilaton can be summarized as follows:
\begin{align}
e^{\tilde{\phi}_{(A,8)}}\,\xrightarrow[\textrm{along } x_1]{\textrm{T-duality}}\,
e^{-\tilde{\phi}/3}\left(\Upsilon+e^{2\tilde{\phi}/3}\right)^{1/2}\,\xrightarrow[\textrm{along } x_2]{\textrm{T-duality}}\, 
e^{-2\tilde{\phi}/3}\left(\Upsilon+e^{2\tilde{\phi}/3}\right)^{1/4}\,
\xrightarrow[\textrm{along } x_3]{\textrm{T-duality}} \,e^{-\tilde{\phi}}.
\end{align}
Hence, the dilaton remains as in the beginning:
\begin{align}
e^{\tilde{\phi}_{(B,8)}}=e^{\tilde{\phi}_{(B,7)}}=e^{-\tilde{\phi}}.
\label{dilB8}
\end{align}
It is rather obvious that, since the dualities are along diagonal directions of the metric,
the NS three-form flux will not be affected in this case:
\begin{align}
\mathcal{H}_3^{(B,8)}=\mathcal{H}_3^{(A,8)}=-\sinh\beta d\mathcal{J}_{(B,7)}.
\label{H3B8}
\end{align}
Regarding the $\mathcal{F}_6^{(A,8)}$ flux, we note that each T-duality will 
remove a leg to it along its corresponding Minkowskian direction $(x_1,\,x_2,\,x_3)$. That is,
we have the reverse process to that earlier in (\ref{F33T}):
\begin{align}
\nonumber
&\mathcal{F}_6^{(A,8)}=\cosh\beta dx_1\wedge dx_2\wedge dx_3\wedge \mathcal{F}_3^{(B,7)}
\,\xrightarrow[\textrm{along } x_1]{\textrm{T-duality}}\,
\cosh\beta dx_2\wedge dx_3\wedge \mathcal{F}_3^{(B,7)} \\
&\qquad \xrightarrow[\textrm{along } x_2]{\textrm{T-duality}} \,
\cosh\beta dx_3\wedge \mathcal{F}_3^{(B,7)}
\, \xrightarrow[\textrm{along } x_3]{\textrm{T-duality}} \,
\cosh\beta \mathcal{F}_3^{(B,7)}=\mathcal{F}_3^{(B,8)}.
\label{F3B8}
\end{align}
We thus obtain a non-closed RR three-form flux, an indication of the $N$ coincident D5-branes present in this configuration.
Finally, the D0-charge previously sourcing $\mathcal{F}_2^{(A,8)}$ now converts to  a D3-charge. The D3-charge then sources
an RR five-form flux which, in analogy to (\ref{F33T}), is given by $\mathcal{F}_2^{(A,8)}\wedge dx_1\wedge dx_2\wedge dx_3$, plus
its Hodge dual (since the D3-charge is self-dual, the corresponding RR flux must be self-dual too). We thus obtain
\begin{align}
\mathcal{F}_5^{(B,8)}=\coth\beta(1+\ast)\frac{d}{dr}\left(\frac{\Upsilon}{\Upsilon+e^{2\tilde{\phi}/3}}\right)d\tilde{t}
\wedge dr\wedge dx_1\wedge dx_2\wedge dx_3,
\label{F5B8}
\end{align}
where the Hodge dual is, of course, with respect to the metric (\ref{newmetricB}).
The geometry and fluxes of this final type IIB configuration are precisely those in (4.2) in~\cite{Dasgupta:2016rhc}.
As a consistency check, one may verify that setting $\beta=0$ (no boost), we recover the initial type IIB configuration with only dilaton and
RR three-form flux:
\begin{align}
\textrm{configuration }(B,8) \xrightarrow[\beta=0]{}\textrm{configuration }(B,7).
\end{align}

It is important to note that none of the modifications performed in this section affects the supersymmetry of the starting
configuration (configuration (B, 7)).
In other words, the previously derived constraint equation (\ref{n1cons}) is enough to ensure that the end configuration
(configuration (B, 8))
is associated to an $\mathcal{N}=1$ supersymmetric world-volume gauge theory too.
We refer the interested reader to section 3.2 in~\cite{Dasgupta:2014txa} for an enlightening discussion
on the difficulties to derive this constraint equation in the context of the configuration (B, 8),
where the internal 6-dimensional manifold is not complex, unlike in the configuration (B, 7).

\FloatBarrier

\subsubsection{Exact results: a specific choice of the warp factors \label{choicesec}}

At this stage, we would like to make our discussion fully precise.
Thus, following (4.9) in~\cite{Dasgupta:2016rhc}, we choose our warp factors as
\begin{eqnarray}
\mathscr{F}_1=\frac{e^{-\tilde{\phi}}}{2F}, \quad\quad \mathscr{F}_2=\frac{r^2e^{-\tilde{\phi}}F}{2}, \quad\quad \mathscr{F}_3
=\frac{r^2e^{-\tilde{\phi}}}{4}+a^2,
\quad\quad \mathscr{F}_4=\frac{r^2e^{-\tilde{\phi}}}{4}, \label{warpchoice}
\end{eqnarray}
where, in good agreement with our previous choices in (\ref{choicesB}), 
\begin{eqnarray}
F=F(r), \quad\quad a^2\equiv a_0^2+\tilde{a}(r).
\end{eqnarray}
The constant $a_0^2$ is to be interpreted as the resolution parameter of the blown up two-cycle in the resolved conifold.
(This choice was already studied in~\cite{Dasgupta:2014txa} and~\cite{Chen:2013nma}.)
In this section, we work out three constraint equations that ultimately allow us to compute
$(F,\, e^{\tilde{\phi}},\, a)$ above and thereby fully determine our type IIB configuration in this case.
We will do so for a particularly simple case, as the most general scenario is computationally hard to handle.

The first constraint equation follows from demanding that the choice (\ref{warpchoice})
leads to a world-volume gauge theory with $\mathcal{N}=1$ supersymmetry.
As we argued in section \ref{torsionsec}, this amounts to requiring that (\ref{n1cons}) holds true.
Using (\ref{warpchoice}) in (\ref{n1cons}), it is quite straightforward to show that the first constraint
can be written as in (4.25) in~\cite{Dasgupta:2016rhc}:
\begin{align}
\label{421}
\left(15+88\frac{a^2e^{\tilde{\phi}}}{r^2}\right)\tilde{\phi}_r+{56e^{\tilde{\phi}}}\frac{a}{r^2}a_r+\frac{2}{r}
+\left(\frac{4}{r}+\frac{1}{F}F_r-\frac{2}{rF}\right)\left(1+\frac{4a^2e^{\tilde{\phi}}}{r^2}\right)=0,
\end{align}
where $(\tilde{\phi}_r,\,a_r,\,F_r)$ stand for the derivatives with respect to the radial coordinate $r$ of
$(\tilde{\phi},\,a,\,F)$.

For the second constraint equation, we will demand quantization of the magnetic charge
of the D5-branes in our configuration.
Recall that, in spite of the duality chain of figure \ref{fig6}, our D5-branes
remain as in figure \ref{fig2}{\bf F}: oriented along $(t,\,x_1,\,x_2,\,x_3)$ and wrapping the two-cycle
parametrized by $(\theta_2,\,\phi_2)$.
As it is well-known\footnote{A succinct and clear review on charge quantization
of D-branes can be found in~\cite{BLT}.}, the D5-branes' charge stems from the RR three-form flux $\mathcal{F}_3^{(B,8)}$.
Accordingly, let us begin by giving the explicit form of this flux
when the warp factors are chosen as just mentioned. This amounts to inserting (\ref{warpchoice}) in (\ref{F3B8})
and further using (\ref{vielb7}) and (\ref{F3B7}). Rather easy and quick algebra then gives
\begin{align}
\mathcal{F}_3^{(B,8)}=-\frac{e^{\tilde{\phi}} r^3 F}{4}\cosh\beta \left(\tilde{k}_1
e_{\theta_1}^{(B,7)}\wedge e_{\phi_1}^{(B,7)}
+\tilde{k}_2e_{\theta_2}^{(B,7)}\wedge e_{\phi_2}^{(B,7)}\right)
\wedge  e_\psi^{(B,7)},
\label{F3B8choices}
\end{align}
where we have defined
\begin{align}
\tilde{k}_1\equiv \tilde{\phi}_r\left(1+\frac{4a^2e^{\tilde{\phi}}}{r^2}\right), \quad\quad
\tilde{k}_2\equiv \frac{r^2\tilde{\phi}_r-8aa_re^{\tilde{\phi}}}{r^2+4a^2e^{\tilde{\phi}}}.
\end{align}
This is (4.10) in~\cite{Dasgupta:2016rhc}. Now, the magnetic charge of the D5-branes in our setup can be calculated
as the integral of their RR three-form flux over the three cycle orthogonal to them:
\begin{align} 
q_m=\int_{S^3} \mathcal{F}_3^{(B,8)},
\end{align}
with $S^3$ the three cycle labeled by $(\theta_1,\,\phi_1,\psi)$ and depicted in figure \ref{fig2}{\bf F}.
It is easy to see that only the first term in (\ref{F3B8choices}) will contribute to the magnetic charge.
Normalizing the three cycle volume as
\begin{align}
V_{S^3}\equiv \int_{S^3} e_{\theta_1}^{(B,7)}\wedge e_{\phi_1}^{(B,7)}\wedge  e_\psi^{(B,7)}=1
\end{align}
and demanding $q_m\in\mathbb{Z}$, we obtain the second constraint equation:
\begin{align}
\tilde{c}_0\equiv \frac{e^{\tilde{\phi}} r^3 F}{4} \tilde{k}_1\cosh\beta\in\mathbb{Z}.
\label{constraint1}
\end{align}

The third and last constraint follows from $d^2\mathcal{F}_3^{(B,8)}=0$.
For simplicity, we will consider the limit when $(a,\,a_r)$ are of the same order and sufficiently small,
$a\sim a_r <<1$. Under this assumption, we can expand $\tilde{k}_2$ around $a^2=0$ and obtain
\begin{align}
\tilde{k}_2= \tilde{\phi}_r\left(1-\frac{4a^2e^{\tilde{\phi}}}{r^2}\right)-\frac{8aa_re^{\tilde{\phi}}}{r^2}+\mathcal{O}(a^3).
\end{align}
Further introducing the quantities in (4.13) and (4.17) in~\cite{Dasgupta:2016rhc},
\begin{align}
\eta_3\equiv \left(e_{\theta_1}^{(B,7)}\wedge e_{\phi_1}^{(B,7)}
-e_{\theta_2}^{(B,7)}\wedge e_{\phi_2}^{(B,7)}\right)\wedge  e_\psi^{(B,7)}, \quad
G\equiv e^{2\tilde{\phi}}rF\cosh\beta\left(2aa_r-\frac{e^{-\tilde{\phi}}r^2\tilde{\phi}_r}{2}\right),
\end{align}
it is not hard to convince oneself that $\mathcal{F}_3^{(B,8)}$ can be written in the very suggestive way
\begin{eqnarray}
\mathcal{F}_3^{(B,8)}=-\tilde{c}_0\eta_3
+G  e_{\theta_2}^{(B,7)}\wedge e_{\phi_2}^{(B,7)}\wedge e_\psi^{(B,7)}, \label{F3final}
\end{eqnarray}
where we have used our first constraint (\ref{constraint1}).
Note that $\eta_3$ is a closed form ($d\eta_3=0$). Consequently, the exterior derivative of the above comes solely
from the second term.
Denoting as $G_r$ the derivative of $G$ with respect to $r$, we obtain $d\mathcal{F}_3^{(B,8)}$ as in
(4.16) in~\cite{Dasgupta:2016rhc}:
\begin{align}
d\mathcal{F}_3^{(B,8)}=&G_r e_r^{(B,7)} \wedge e_{\theta_2}^{(B,7)}\wedge e_{\phi_2}^{(B,7)}\wedge e_\psi^{(B,7)}
+G d\left( e_{\theta_2}^{(B,7)}\wedge e_{\phi_2}^{(B,7)}\wedge e_\psi^{(B,7)}\right) \nonumber \\
=&G_r e_r^{(B,7)} \wedge e_\psi^{(B,7)} \wedge e_{\theta_2}^{(B,7)}\wedge e_{\phi_2}^{(B,7)}
-G e_{\theta_1}^{(B,7)} \wedge e_{\phi_1}^{(B,7)} \wedge e_{\theta_2}^{(B,7)} \wedge e_{\phi_2}^{(B,7)},
\end{align}
where in the last step we have made use of (\ref{vielb7}). Of course, the exterior derivative of the above must vanish
and this leads to our third constraint equation:
\begin{align}
0=d^2\mathcal{F}_3^{(B,8)}=-G_r e_r^{(B,7)} \wedge e_{\theta_1}^{(B,7)} \wedge e_{\phi_1}^{(B,7)} \wedge e_{\theta_2}^{(B,7)}
\wedge e_{\phi_2}^{(B,7)}
\implies G_r=0.
\label{constraint2}
\end{align}

Having derived the three constraints of our interest, (\ref{421}), (\ref{constraint1}) and (\ref{constraint2}),
we will now solve them under the assumption $a\sim a_r <<1$, keeping only terms up to order $\mathcal{O}(a)$.
(Other solutions to these equations are of course possible, but we will not attempt them here.)
In this case, (\ref{421}) reduces to (4.24) in~\cite{Dasgupta:2016rhc},
\begin{align}
r\tilde{\phi}_r+\frac{r}{15F}F_r-\frac{2}{15F}+\frac{2}{5}+\mathcal{O}(a^2)=0 \label{420}
\end{align}
and (\ref{constraint1}) becomes 
\begin{align}
\tilde{c}_0= \frac{e^{\tilde{\phi}} r^3 F}{4} \tilde{\phi}_r\cosh\beta+\mathcal{O}(a^2), 
\end{align}
which immediately ensures that (\ref{constraint2}) is satisfied in the limit here considered.
Defining $Z\equiv e^{\tilde{\phi}}$ and $\hat{c}_0\equiv\tilde{c}_0/\cosh\beta$, we can solve for $F$ in the above
\begin{align}
F=\frac{4\hat{c}_0}{r^3Z_r}+\mathcal{O}(a^2).
\end{align}
Substitution in (\ref{420}) then yields (4.26) in~\cite{Dasgupta:2016rhc}:
\begin{align}
rZ_{rr}-3Z_r+r\left(\frac{r^2}{2c_0}-\frac{15}{Z}\right)Z_r^2+\mathcal{O}(a^2)=0,
\label{422}
\end{align}
with $Z_{rr}\equiv d^2Z/dr^2$. One may easily verify that a solution to (\ref{422}) is given by
$Z=24\hat{c}_0r^{-2}$. It follows then that (4.30) in~\cite{Dasgupta:2016rhc},
\begin{align}
e^{\tilde{\phi}}=\frac{24\hat{c}_0}{r^2}+\mathcal{O}(a^2), \quad\quad F=-\frac{1}{12}+\mathcal{O}(a^2),
\label{expphi}
\end{align}
fully determines our choices in (\ref{warpchoice}), up to order $\mathcal{O}(a^2)$.
The explicit form of the type IIB configuration (B, 8) in figure \ref{fig6}
can then be obtained by simply using (\ref{warpchoice}) and (\ref{expphi}) in (\ref{newmetricB}) and in
(\ref{dilB8})-(\ref{F5B8}).

\FloatBarrier

\subsubsection{Connection to the model in~\cite{Ooguri:1999bv} \label{oogurivafasec}}

\begin{figure}[t]
\centering
\includegraphics[width=0.9\textwidth]{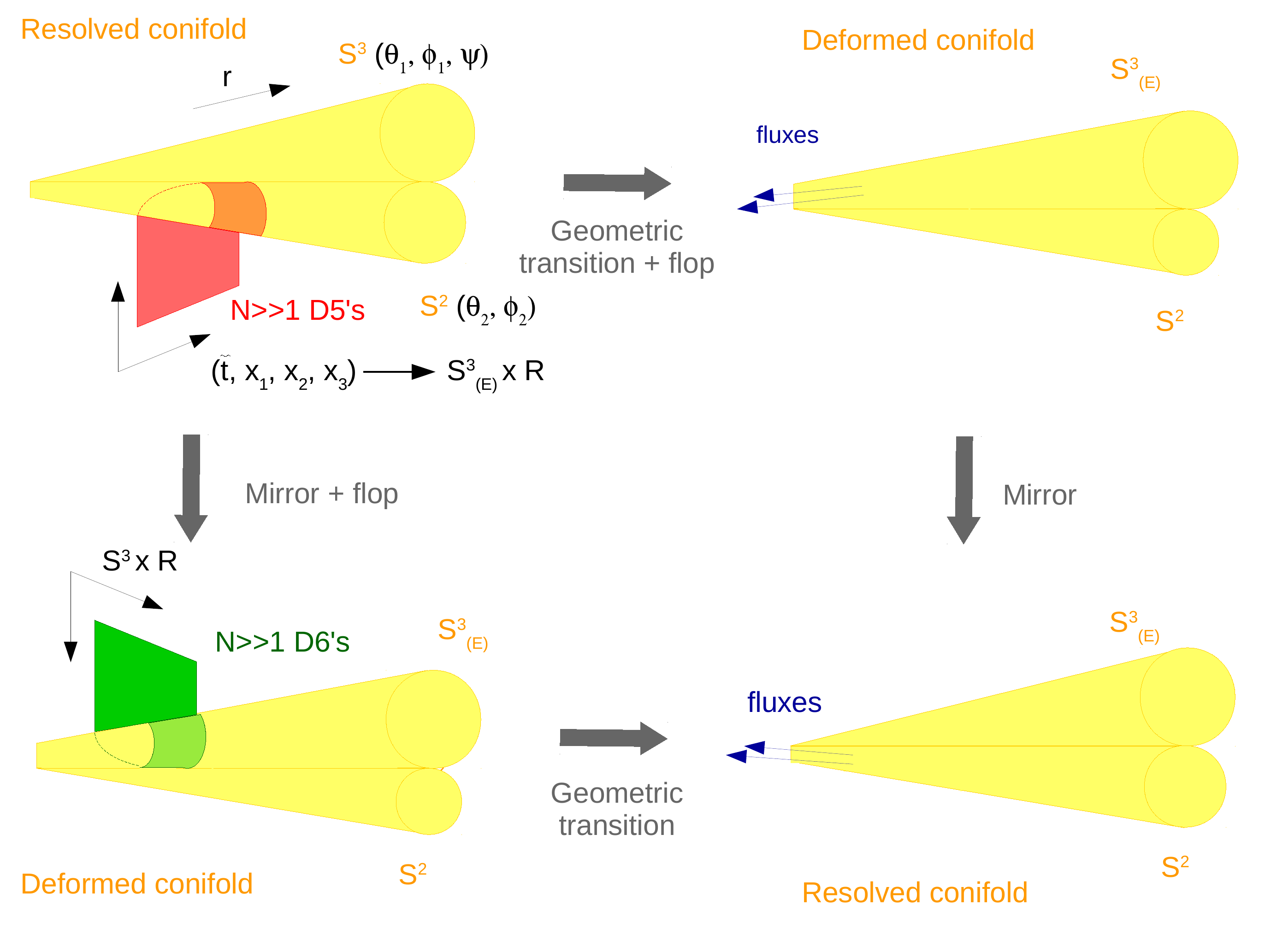}
\caption{Depiction of the discussions in section \ref{oogurivafasec}.
To the configuration (B, 8) of figure \ref{fig6} we do the following modifications modifications:
Euclideanize and compactify the $(\tilde{t},\,x_1,\,x_2)$ directions, go to the mirror picture,
perform a flop operation and take the gravity dual.
The resulting configuration is that of a resolved conifold in the presence of fluxes
studied in~\cite{Ooguri:1999bv}. Our configuration (B, 8) is that on the top, left corner, whereas
the most well-known realization of 
the model in~\cite{Ooguri:1999bv} is drawn on the bottom, right corner.
It should be noted that, as explained in the text, the mirror operations
here shown are only valid in a certain energy range.}
\label{fig11}
\end{figure}

The present section is devoted to sketching how the configuration (B, 8) of figure \ref{fig6}
is related to the resolved conifold in the presence of fluxes considered by Ooguri and Vafa in~\cite{Ooguri:1999bv}.
Here, we will clearly point out the modifications needed to obtain the model in~\cite{Ooguri:1999bv} from (B, 8).
These are depicted in figure \ref{fig11}, which serves as a graphical summary of the present section too.
Nonetheless, unlike in previous sections, we will not present a thorough derivation of the geometries and fluxes for each
intermediate configuration considered in the process. Such exhaustive study is beyond the scope of this work and
is deferred to the sequel. In the sequel, following~\cite{Ooguri:1999bv}, we also intend to explore
knot invariants in the configuration (M, 5), which follows from (B, 8) and which is constructed in details in section
\ref{M5sec}. For the time being,
we refer the interested reader to section 4.4 in~\cite{Dasgupta:2016rhc} for a preliminary discussion of the physics stemming
from (M, 5) and the realization of knots in this set up.

As we just mentioned, our starting point in this section is the configuration (B, 8) summarized in figure \ref{fig6}.
Essentially, this is the same configuration as that drawn in figure \ref{fig2}{\bf F},
but in the presence of both RR and NS fluxes. In figure \ref{fig11},
this is shown in the top, left corner.
As can be seen, (B, 8) consists on a large number $N$ of D5-branes wrapping the two-cycle $S^{2}$ of a non-K\"{a}hler
resolved conifold.
Let us start by making an observation that will soon be relevant to us.
From the orientation of the D5-branes shown in figure \ref{fig2}{\bf F}
it is clear that, upon a dimensional reduction, we expect to obtain an
$SU(N)$ world-volume gauge theory along
$(\tilde{t},\,x_1,\,x_2,\,x_3)$. Loosely speaking, the physics following from (B, 8)
are encoded in the directions $(\tilde{t},\,x_1,\,x_2,\,x_3)$.

Next, recall that the metric corresponding to (B, 8) was given in (\ref{newmetricB}).
Note in particular that the spacetime directions $(\tilde{t},\,x_1,\,x_2)$
in this geometry parametrize a three-dimensional Minkowski subspace.
The first modification to (B, 8) that one needs to consider in order to obtain the model in~\cite{Ooguri:1999bv}
consists on Euclideanizing and compactifying these directions, so that they parametrize a sphere:
$(\tilde{t},\,x_1,\,x_2)\rightarrow S^{3}_{(E)}$. Then, the corresponding physical theory will
lie in $S^{3}_{(E)}\times\mathbb{R}$, where $\mathbb{R}$ stands for the line labeled by the coordinate $x_3$.

Secondly, we must perform a series of T- and SYZ-dualities to the resulting configuration, which will take us to the so-called
mirror picture. The required dualities are far from trivial, involving many subtleties.
Nevertheless, the works~\cite{Strominger:1996it,Becker:2004qh,Alexander:2004eq,Becker:2005ef}
deal with all difficulties exhaustively and show that the mirror picture consists on $N$ D6-branes 
wrapping the three-cycle $S^{3}$ of a non-K\"{a}hler deformed conifold.
This is true only for energies higher than the inverse size
of the two-cycle $S^{2}$ of the dual resolved conifold. As a consequence, we will restrict ourselves in the ongoing
to this energy regime\footnote{As argued around (2.5) in~\cite{Dasgupta:2016rhc}, for energies lower than the size
of $S^{2}$, the mirror picture will lead to D4-branes instead of D6-branes. Although such scenario may be interesting
as well, it does not relate to the model in~\cite{Ooguri:1999bv} and thus we are presently not concerned with it.}.

In the described mirror picture of our interest, the $N$ D6-branes are oriented along the seven-dimensional
subspace $S^{3}_{(E)}\times S^{3}\times\mathbb{R}$. The third and last modification required to obtain the model
in~\cite{Ooguri:1999bv} is given by a flop operation, that exchanges $S^{3}_{(E)}$ and $S^{3}$
as described in (4.8) in~\cite{Dasgupta:2016rhc}: $S^{3}_{(E)}\leftrightarrow S^{3}$.
Clearly, this does not affect the orientation of the D6-branes, yet it {\it transfers} the physics from 
$S^{3}_{(E)}\times\mathbb{R}$ to $S^{3}\times\mathbb{R}$, thus yielding the D6-brane realization of the model
in~\cite{Ooguri:1999bv} depicted on the bottom, left corner of figure \ref{fig11}.

A more well-known realization of the set up in~\cite{Ooguri:1999bv} is obtained by simply taking the large $N$ dual
(in other words, performing a geometric transition)
of the above configuration. In this case, the deformed conifold becomes a resolved one. The D6-branes
disappear in the dual picture, giving rise to fluxes. This configuration is precisely that shown on the bottom, right corner
of figure \ref{fig11}.

Alternatively, one may take the large $N$ dual of (B, 8) first and consider the mirror picture afterwards.
The result is the same: we obtain the deformed conifold with fluxes of~\cite{Ooguri:1999bv}.
This equivalent procedure is depicted on the top, right corner of figure \ref{fig11}.

At this stage, we have argued that our configuration (B, 8) is related to the model
in~\cite{Ooguri:1999bv} by a simple chain of dualities. That is, (B, 8) is {\it dual} to~\cite{Ooguri:1999bv}.
In the next section, we will build an M-theory configuration (M, 5) from (B, 8).
As we shall see, (B, 8) is dual to (M, 5) and so this will allow us to conclude that
(M, 5) is dual to~\cite{Ooguri:1999bv} too.

\subsection{Non-commutative deformation and M-theory uplift \label{M5sec}}

\begin{figure}[t]
\centering
\includegraphics[width=0.9\textwidth]{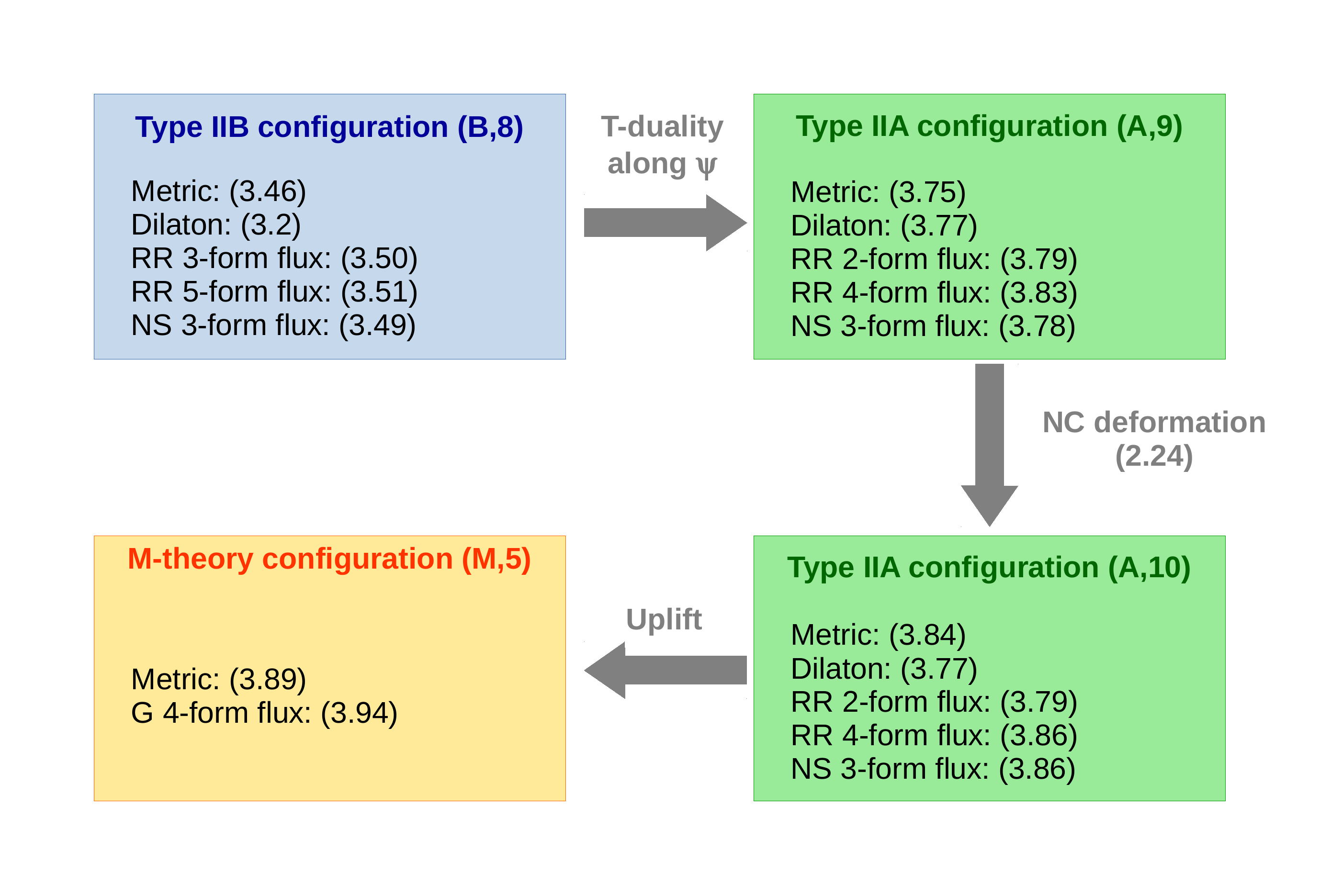}
\caption{Graphical summary of section \ref{M5sec}.
To the configuration (B, 8) of figure \ref{fig6} we do a series of modifications, so as to source
a $\Theta$-term in the corresponding world-volume gauge theory. The resulting configuration is then lifted
to M-theory. 
The configuration (M, 5)
is the second M-theory construction where knots can be studied.}
\label{fig7}
\end{figure}

In this section we will obtain the second M-theory construction where knot invariants can be studied: (M, 5).
Clearly, the starting point will be the configuration (B, 8) in figure \ref{fig6}.
We will first do a T-duality along $\psi$ to type IIA, where we will perform the same
non-commutative deformation we considered in section \ref{ncsect}: $(x_3,\,\psi)\rightarrow(\tilde{x}_3,\,\tilde{\psi})$.
As we argued in both sections \ref{ncsect} and \ref{rrdefsect}, this deformation sources
the $\Theta$-term in the associated world-volume gauge theory, which is crucial for allowing
the embedding of knots in our model.
Finally, we will uplift the resulting configuration to M-theory. 
As has been the case so far, the dilaton and fluxes for each geometry considered will be worked out here too.
Figure \ref{fig7} provides a graphical summary of this chain of modifications and indicates what the main results
in this section are.

In order to obtain the T-dual of the (B, 8) configuration, we first rewrite its geometry in (\ref{newmetricB})
in a convenient form for our present purposes:
\begin{align}
ds_{(B,8)}^2=\frac{1}{\sqrt{h}}\left(ds_{t12}^2+dx_3^2\right)+\sqrt{h}\left[\mathscr{F}_2\left(d\psi+\sum_{i=1}^2\cos\theta_i
d\phi_i\right)^2+ds_{(8)}^2\right],
\label{dsb8}
\end{align}
where we have defined
\begin{align}
ds_{t12}^2\equiv-d\tilde{t}^2+dx_1^2+dx_2^2, \quad\quad ds_{(8)}^2\equiv \mathscr{F}_1dr^2
+\sum_{i=1}^2 \mathscr{F}_{2+i}\left(d\theta_i^2+\sin^2\theta_i d\phi_i^2\right)
\label{dst12}
\end{align}
and, following (4.40) in~\cite{Dasgupta:2016rhc}, we have also introduced
\begin{align}
h\equiv e^{4\tilde{\phi}/3}\left(\Upsilon+e^{2\tilde{\phi}/3}\right). \label{4352}
\end{align}
(We remind the reader that $\Upsilon$ was defined in (\ref{deltadef}).)
As can be easily inferred from (\ref{H3B8}), the above geometry is associated to an NS B-field
\begin{align}
B_{(B,8)}=\sinh\beta\sum_{i=1}^2\left(\sqrt{\mathscr{F}_1\mathscr{F}_2}\cos\theta_idr
-\mathscr{F}_{2+i}\sin\theta_id\theta_i\right)\wedge d\phi_i.
\end{align}

It is now straightforward to T-dualize along $\psi$ the metric (\ref{dsb8}). We thus obtain the type IIA
geometry in (4.39) in~\cite{Dasgupta:2016rhc}:
\begin{align}
ds_{(A,9)}^2=\frac{1}{\sqrt{h}}\left(ds_{t12}^2+dx_3^2+\frac{1}{\mathscr{F}_2}d\psi^2\right)+\sqrt{h}ds_{(8)}^2,
\label{dsa9}
\end{align}
with associated NS B-field
\begin{align}
B_{(A,9)}=B_{(B,8)}+\sum_{i=1}^2\cos\theta_i d\psi\wedge d\phi_i.
\end{align}
The dilaton for this type IIA configuration is, quite obviously, that in (4.40) in~\cite{Dasgupta:2016rhc}:
\begin{align}
e^{\tilde{\phi}_{(A,9)}}=h^{-1/4}\mathscr{F}_2^{-1/2}e^{-\tilde{\phi}}.
\end{align}
The NS three-form flux can be easily derived to be
\begin{align}
\mathcal{H}_3^{(A,9)}=dB_{(A,9)}=\mathcal{H}_3^{(B,8)}+\sum_{i=1}^2\sin\theta_i  d\theta_i\wedge d\phi_i\wedge d\psi,
\end{align}
with $\mathcal{H}_3^{(B,8)}$ as in (\ref{dJ2}) and (\ref{H3B8}).
Coming to the RR fluxes now, we note that the T-duality converts the D5-branes which wrap the two-cycle of the
resolved conifold in the configuration (B, 8) to $N$ coincident D6-branes that wrap the two-sphere
parametrized by $(\theta_1,\,\phi_1)$ in the dual type IIA picture\footnote{Actually,
this T-duality is more subtle and can also lead to D4-branes. We discuss this important point in section
\ref{oogurivafasec}.}. Consequently, the RR three-form flux (\ref{F3B8}) 
(where $\mathcal{F}_3^{(B,7)}$ was given in (\ref{F3B7})) that was sourced by the D5-branes
now gives rise to the RR two-form flux
\begin{align}
\mathcal{F}_2^{(A,9)}=e^{2\tilde{\phi}} \cosh\beta \sqrt{\frac{\mathscr{F}_2}{\mathscr{F}_1}}
\sum_{\substack{i,j=1 \\ i\neq j}}^2
\frac{\mathscr{F}_{2+j}}{\mathscr{F}_{2+i}} (\mathscr{F}_{2+i,r}-\sqrt{\mathscr{F}_1\mathscr{F}_2})\sin\theta_j
d\theta_j\wedge d\phi_j,
\end{align}
as well as to the RR four-form flux
\begin{align}
\mathcal{F}_4^{(1)}=e^{2\tilde{\phi}}\cosh\beta  \sqrt{\frac{\mathscr{F}_2}{\mathscr{F}_1}}\sum_{\substack{i,j=1 \\ i\neq j}}^2
\frac{\mathscr{F}_{2+j}}{\mathscr{F}_{2+i}} (\mathscr{F}_{2+i,r}-\sqrt{\mathscr{F}_1\mathscr{F}_2})\sin\theta_j
\cos\theta_i d\psi\wedge d\phi_i\wedge d\theta_j\wedge d\phi_j.
\label{F41}
\end{align}
Both are sourced by the dual D6-branes (and hence, $d\mathcal{F}_2^{(A,9)}\neq0\neq d\mathcal{F}_4^{(1)}$).
On the other hand, the D3-charge that sourced the self-dual RR five-form flux in (\ref{F5B8})
converts to a D4-charge after the T-duality. They now source RR four- and six-form fluxes, which are Hodge dual to
each other (with respect to the metric (\ref{dsa9})). Starting from (\ref{F5B8}) and using (\ref{4352}), it is clear that the RR six-form flux is
\begin{align}
\mathcal{F}_6^{(A,9)}=\coth\beta\frac{d}{dr}\left(\frac{e^{2\tilde{\phi}}}{h}\right) d\tilde{t}\wedge dx_1\wedge
dx_2\wedge dx_3\wedge d\psi\wedge dr.
\end{align}
However, its Hodge-dual four-form will become more convenient once we perform the uplift to M-theory,
with views to computing the G-flux there. Since the metric (\ref{dsa9}) is diagonal,
it is not hard to show that the flux of our interest is given by
\begin{align}
\mathcal{F}_4^{(2)}=\ast\mathcal{F}_6^{(A,9)}=-\coth\beta\frac{d}{dr}\left(\frac{e^{2\tilde{\phi}}}{h}\right)h^2\sqrt{\frac{\mathscr{F}_2}{\mathscr{F}_1}}
\prod_{i=1}^2 \mathscr{F}_{2+i}\sin\theta_i d\theta_i\wedge d\phi_i.
\label{F42}
\end{align}
The total RR four-form flux for this configuration is thus
\begin{align}
\mathcal{F}_4^{(A,9)}=\mathcal{F}_4^{(1)}+\mathcal{F}_4^{(2)}.
\end{align}

We will now apply the non-commutative deformation $(x_3,\,\psi)\rightarrow(\tilde{x}_3,\,\tilde{\psi})$
in (\ref{ncdeformation}) to the above type IIA configuration.
The metric (\ref{dsa9}) then changes to
\begin{align}
ds_{(A,10)}^2=&\frac{1}{\sqrt{h}}ds_{t12}^2+\frac{1}{\sqrt{h}}(\sec\theta_{nc}d\tilde{x}_3+\sin\theta_{nc}d\tilde{\psi})^2
+\frac{\cos^2\theta_{nc}}{\sqrt{h}\mathscr{F}_2}d\tilde{\psi}^2 +\sqrt{h}ds_{(8)}^2\nonumber \\
=&(e^{\tilde{\phi}_{(A,9)}})^{2/3}
\left(\frac{\mathscr{F}_2e^{2\tilde{\phi}}}{h}\right)^{1/3}\left[ds_{t12}^2
+(\frac{d\tilde{x}_3}{\cos\theta_{nc}}+\sin\theta_{nc}d\tilde{\psi})^2
+\frac{\cos^2\theta_{nc}}{\sqrt{h}\mathscr{F}_2}d\tilde{\psi}^2
+hds_{(8)}^2\right],\label{dtildes2IIA}
\end{align}
where the last rewriting was done in preparation to the M-theory uplift that will follow.
The dilaton and RR two--form flux can be readily seen not to be affected by the deformation:
\begin{align}
e^{\tilde{\phi}_{(A,10)}}=e^{\tilde{\phi}_{(A,9)}}, \quad\quad
\mathcal{F}_2^{(A,10)}=\mathcal{F}_2^{(A,9)}. 
\end{align}
However, the RR four-form flux and the NS three-form flux do change to
\begin{align}
\mathcal{F}_4^{(A,10)}=&e^{2\tilde{\phi}}\cosh\beta\cos\theta_{nc}  \sqrt{\frac{\mathscr{F}_2}{\mathscr{F}_1}}
 d\tilde{\psi}\wedge \left(\hat{\hat{k}}_1d\phi_1\wedge d\theta_2\wedge d\phi_2+
 \hat{\hat{k}}_2 d\theta_1\wedge d\phi_1\wedge d\phi_2\right)
+\mathcal{F}_4^{(2)}, \nonumber \\
\vspace*{-0.3cm}
\mathcal{H}_3^{(A,10)}=&\mathcal{H}_3^{(B,8)}+\cos\theta_{nc}\sum_{i=1}^2\sin\theta_i d\theta_i \wedge
d\phi_i\wedge d\tilde{\psi},
\end{align}
where we have defined
\begin{align}
\hat{\hat{k}}_1\equiv
\frac{\mathscr{F}_{4}}{\mathscr{F}_{3}} (\mathscr{F}_{3,r}-\sqrt{\mathscr{F}_1\mathscr{F}_2})\sin\theta_2
\cos\theta_1, \quad\quad
\hat{\hat{k}}_2\equiv
\frac{\mathscr{F}_{3}}{\mathscr{F}_{4}} (\mathscr{F}_{4,r}-\sqrt{\mathscr{F}_1\mathscr{F}_2})\sin\theta_1
\cos\theta_2.
\end{align}
Once more, the RR two-form flux not being closed, we can rewrite it in a similar fashion to what we did earlier in
(\ref{353}) and (\ref{F2A6gauge}):
\begin{align}
\mathcal{F}_2^{(A,10)}=d\hat{{\bf A}}_1+\hat{\Delta}, \quad\quad
\hat{{\bf A}}_1\equiv\cosh\beta\sum_{i=1}^2\cos\theta_i d\phi_i,
\quad\quad d\hat{\Delta}=\textrm{sources},
\label{gaugefieldBdef}
\end{align}
with $\hat{{\bf A}}_1$ the type IIA gauge field for this configuration (A, 10).
We will soon see that it is opportune to define $\hat{{\bf A}}_1$ as we just did, which is
(4.51) in~\cite{Dasgupta:2016rhc}. Before we proceed, let us make one last observation:
the subsequent M-theory uplift will only capture the dynamics of this type IIA theory
when $e^{\tilde{\phi}_{(A,10)}}$ is of order one, or bigger.

The M-theory metric corresponding to (\ref{dtildes2IIA}) is (4.48) in~\cite{Dasgupta:2016rhc}:
\begin{align}
ds_{(M,5)}^2=(e^{\tilde{\phi}_{(A,9)}})^{-2/3}
ds_{(A,10)}^2+(h\mathscr{F}_2^2e^{4\tilde{\phi}})^{-1/3}(dx_{11}+\hat{{\bf A}}_1)^2.
\end{align}
We note that, due to (\ref{choicesB}) and (\ref{gaugefieldBdef}), for a fixed value of the $\phi_1$ coordinate,
$\phi_1=\phi_1^\ast$, the metric along the directions $(r,\,\theta_2,\,\phi_2,\,x_{11})$ describes a warped
Taub-NUT space. Introducing the quantities
\begin{align}
\hat{G}_1\equiv\mathscr{F}_1(h^2\mathscr{F}_2 e^{2\tilde{\phi}})^{1/3}, \quad\quad
\hat{G}_2\equiv\frac{\mathscr{F}_4}{\mathscr{F}_1}\hat{G}_1, \quad\quad
\hat{G}_3\equiv\sin^2\theta_2\hat{G}_2, \quad\quad
\hat{G}_4\equiv(h\mathscr{F}_2^2e^{4\tilde{\phi}})^{-1/3}, \label{hatGs}
\end{align}
which are only functions of the coordinates $(r,\,\theta_2)$ (and the boost parameter $\beta$),
we can write the metric for the Taub-NUT space as
\begin{align}
ds_{{TN}_3}^2=\hat{G}_1dr^2+\hat{G}_2d\theta_2^2+\hat{G}_3d\phi_2^2+\hat{G}_4(dx_{11}+\hat{{\bf A}}_1^\ast)^2,
\label{tn3}
\end{align}
where we have defined
\begin{align}
 \hat{{\bf A}}_1^\ast\equiv \hat{{\bf A}}_1\Big|_{\phi_1=\phi_1^\ast}=
\cosh\beta\cos\theta_2 d\phi_2.
\end{align}
To the metric (\ref{tn3}), we associate the following vielbeins:
\begin{align}
e_{r}^{(M,5)}=\sqrt{\hat{G}_1}dr, \quad
e_{\theta_2}^{(M,5)}=\sqrt{\hat{G}_2}d\theta_2, \quad
e_{\phi_2}^{(M,5)}=\sqrt{\hat{G}_3}d\phi_2, \quad
e_{11}^{(M,5)}=\sqrt{\hat{G}_4}(dx_{11}+\hat{{\bf A}}_1^\ast).
\end{align}
As was the case in section \ref{nonabsec}, this is a {\it multi-centered} (warped) Taub-NUT space.
Recall that we had $N$ D6-branes in the configuration (A, 10) prior to the uplift.
Hence, $\hat{G}_4^{-1}=0$ happens $N$ times, leading to coordinate singularities that denote
the location of the D6-branes in the dual type IIA picture. Further, the D6-branes in (A, 10)
were coincident and consequently we are, by construction, at the non-abelian enhanced scenario
discussed in \ref{nonabsec}: the symmetry group of the associated world-volume gauge theory is $SU(N)$.
It follows then that the G-flux for this M-theory configuration is of the same form as that in (\ref{multiflux}):
\begin{align}
\mathcal{G}_4^{(M,5)}=\langle\mathcal{G}_4^{(M,5)}\rangle+\sum_{k=1}^{N-1}\hat{\mathcal{F}}_k\wedge \hat{\omega}_k,
\label{G4M5}
\end{align}
where $\hat{\mathcal{F}}_k$'s are the Cartan algebra values of the world-volume field strength $\hat{\mathcal{F}}$, the
$\hat{\omega}_k$'s are the  unique, normalizable, (anti-)self-dual two-forms associated to the minimal area
independent two-cycles in the space (\ref{tn3}) and the background G-flux is given by
\begin{align}
\langle\mathcal{G}_4^{(M,5)}\rangle=\mathcal{F}_4^{(A,10)}+\mathcal{H}_3^{(A,10)}\wedge dx_{11}.
\end{align}
Writing it explicitly, we obtain (4.52) in~\cite{Dasgupta:2016rhc}\footnote{Note that the contribution
to the G-flux stemming from the RR five-form flux $\mathcal{F}_5^{(B,8)}$
(this is the second line in (\ref{backGM5})) is written
in a different yet equivalent manner in~\cite{Dasgupta:2016rhc}.
In this reference, the relationship $d\mathcal{F}_5^{(B,8)}\propto\mathcal{H}_3^{(B,8)}\wedge
\mathcal{F}_3^{(B,8)}\propto d\theta_1\wedge d\phi_1\wedge d\psi\wedge dr\wedge d\theta_2\wedge d\phi_2$ is used.
Then, $\mathcal{F}_5^{(B,8)}$ is expressed as a sum of two contributions,
obtained by integration over $\theta_1$ and $\theta_2$, respectively.
In this language, our approach consists of integrating over $r$ instead.}:
\begin{align}
\langle\mathcal{G}_4^{(M,5)}\rangle=&e^{2\tilde{\phi}}\cosh\beta\cos\theta_{nc}\sqrt{\frac{\mathscr{F}_2}{\mathscr{F}_1}}
d\tilde{\psi}\wedge\left(\hat{\hat{k}}_1d\phi_1\wedge d\theta_2\wedge d\phi_2+\hat{\hat{k}}_2d\theta_1\wedge d\phi_1
\wedge d\phi_2\right) \nonumber \\
&-\coth\beta\frac{d}{dr}\left(\frac{e^{2\tilde{\phi}}}{h}\right)h^2\sqrt{\frac{\mathscr{F}_2}{\mathscr{F}_1}}
\prod_{i=1}^2 \mathscr{F}_{2+i}\sin\theta_i d\theta_i\wedge d\phi_i \label{backGM5} \\ \nonumber
&+\sum_{i=1}^2 \sin\theta_i  d\theta_i \wedge d\phi_i \wedge dx_{11} \wedge
\left[\sinh\beta (\mathscr{F}_{2+i,r}-\sqrt{\mathscr{F}_1\mathscr{F}_2}) dr-\cos\theta_{nc}d\tilde{\psi}\right].
\end{align}
It can be readily seen that the only quantities left to be computed are the $\hat{\omega}_k$'s. We do so in the following.
The discussion is analogous to that in section \ref{nonabsec}, so we will be brief.

We begin the computation of the $\hat{\omega}_k$'s by constructing the minimal area independent two-cycles of (\ref{tn3})
to which they are associated. Note that $\hat{G}_4=\hat{G}_4(r)$. Thus, we can call the $N$ solutions to $\hat{G}_4^{-1}=0$
as $r_{(i)}$, where $i=1,\,2,\,\ldots,\, N$. Consider two such solutions, $r_{(i)}$ and $r_{(j)}$ (where $i\neq j$) and
the straight line in the $r$ direction connecting them, $\mathcal{C}_r$. Attaching to each point in $\mathcal{C}_r$ a circle
labeled by $x_{11}$, we obtain the corresponding minimal area two-cycle $X_{ij}$. We take $X_{k,k+1}$
(with $k=1,\,2,\,\ldots,\,N-1$)
as the {\it independent} minimal area two-cycles where the $\hat{\omega}_k$'s are defined and consider the following ansatze for them:
\begin{align}
\hat{\omega}_k=d\hat{\zeta}_k, \quad\quad \hat{\zeta}_k=\hat{g}_k(dx_{11}+\hat{{\bf A}}_1^\ast).
\end{align}
Easy algebra then yields
\begin{align}
\hat{\omega}_k=&\frac{\hat{g}_{k,r}}{\sqrt{\hat{G}_1\hat{G}_4}}e_r^{(M,5)}\wedge e_{11}^{(M,5)}
-\frac{\hat{g}_k}{\sqrt{\hat{G}_2\hat{G}_3}}\cosh\beta\sin\theta_2 e_{\theta_2}^{(M,5)}\wedge e_{\phi_2}^{(M,5)}, \nonumber \\
\ast \hat{\omega}_k=&\frac{\hat{g}_{k,r}}{\sqrt{\hat{G}_1\hat{G}_4}}e_{\theta_2}^{(M,5)}\wedge e_{\phi_2}^{(M,5)}
-\frac{\hat{g}_k}{\sqrt{\hat{G}_2\hat{G}_3}}\cosh\beta\sin\theta_2 e_r^{(M,5)}\wedge e_{11}^{(M,5)},
\end{align}
where, obviously, the Hodge dual is with respect to the metric (\ref{tn3}) and $\hat{g}_{k,r}$
stands for the derivative of $\hat{g}_{k}$ with respect to the radial coordinate $r$. Using (\ref{hatGs}) and demanding (anti-)self-duality of $\hat{\omega}_k$
we obtain the ordinary differential equation
\begin{align}
\frac{1}{\hat{g}_k}\frac{d\hat{g}_k}{dr}=\mp\cosh\beta\frac{e^{-\tilde{\phi}}}{\mathscr{F}_4}\sqrt{\frac{\mathscr{F}_1}{h\mathscr{F}_2}},
\end{align}
which can be readily solved to give
\begin{align}
\hat{g}_k=\hat{g}_0\textrm{exp}\left(\mp\int_{r_{(k)}}^{r_{(k+1)}} \frac{e^{-\tilde{\phi}}}{\mathscr{F}_4}\sqrt{\frac{\mathscr{F}_1}{h\mathscr{F}_2}}dr\right),
\end{align}
with $\hat{g}_0$ some integration constant where we have absorbed the contribution of $\cosh\beta$. The above fully determines the G-flux in (\ref{G4M5}).

We remind the reader that all the discussion in this section (so far) is subject to the constraint
(\ref{n1cons}) so as to ensure $\mathcal{N}=1$ supersymmetry in the corresponding world-volume gauge theory.

The configuration (M, 5) is the second and last theory we construct for the study of knots
and their invariants. (The first one is (M, 1) and its non-abelian enhancement, discussed earlier in
sections \ref{ncsect} and \ref{nonabsec}, respectively.) In the remaining of this work, we will
only study the configuration (M, 1). Indeed, in part \ref{partb}, we will understand in details the
four-dimensional gauge theory
stemming from (M, 1). In doing so, we will argue how and why (M, 1) provides a natural framework to realize knots.
All investigation of the embedding of knots in (M, 5)
is deferred to the sequel.

Before proceeding further, it is important to emphasize that, in constructing (M, 1) and (M, 5), we have already
achieved a very major result in this work.
Note that, as depicted in figure \ref{fig8}, the configuration (M, 1) is dual to the D3-NS5 system
of~\cite{Witten:2011zz}. On the other hand, the configuration (M, 5) follows from the very same D3-NS5 system
and is dual to the resolved conifold in the presence of fluxes considered in~\cite{Ooguri:1999bv}.
Hence, we have made explicit the modifications that directly connect the  seemingly  very
distinct models in~\cite{Witten:2011zz} and~\cite{Ooguri:1999bv}.
In plain English, we have provided a {\it unifying picture}
between the two existing approaches to computing knot invariants in string theory.

\FloatBarrier

\part{Study of the four-dimensional gauge theory following from the configuration (M, 1)\label{partb}}

As hinted by the title itself, this second part focuses on the (non-abelian enhanced)
M-theory configuration (M, 1) constructed in section \ref{ns5d3sect1}.
The fundamental purpose here will be to show that indeed (M, 1) provides a suitable framework
for the realization of knots. 
To this aim, we shall derive and investigate the four-dimensional, $\mathcal{N}=2$
supersymmetric, $SU(N)$ gauge theory associated to (M, 1).
Such study is presented in three main steps. In section \ref{actionsec},
we obtain the action of the aforementioned gauge theory. 
Section \ref{hamilsec} is devoted to the associated Hamiltonian and the minimization of its energy,
which yields the BPS conditions for the theory.
This analysis naturally leads to a three-dimensional subspace, which we denote as $X_3$ and which is the main object
of interest in section
\ref{bndsec}. As we shall see, the physics in $X_3$ are governed by a Chern-Simons action.
Consequently, $X_3$ (or, more precisely, its Euclideanization) constitutes a suitable space where knots can be embedded.

\begin{figure}[hb]
\centering
\includegraphics[width=0.8\textwidth]{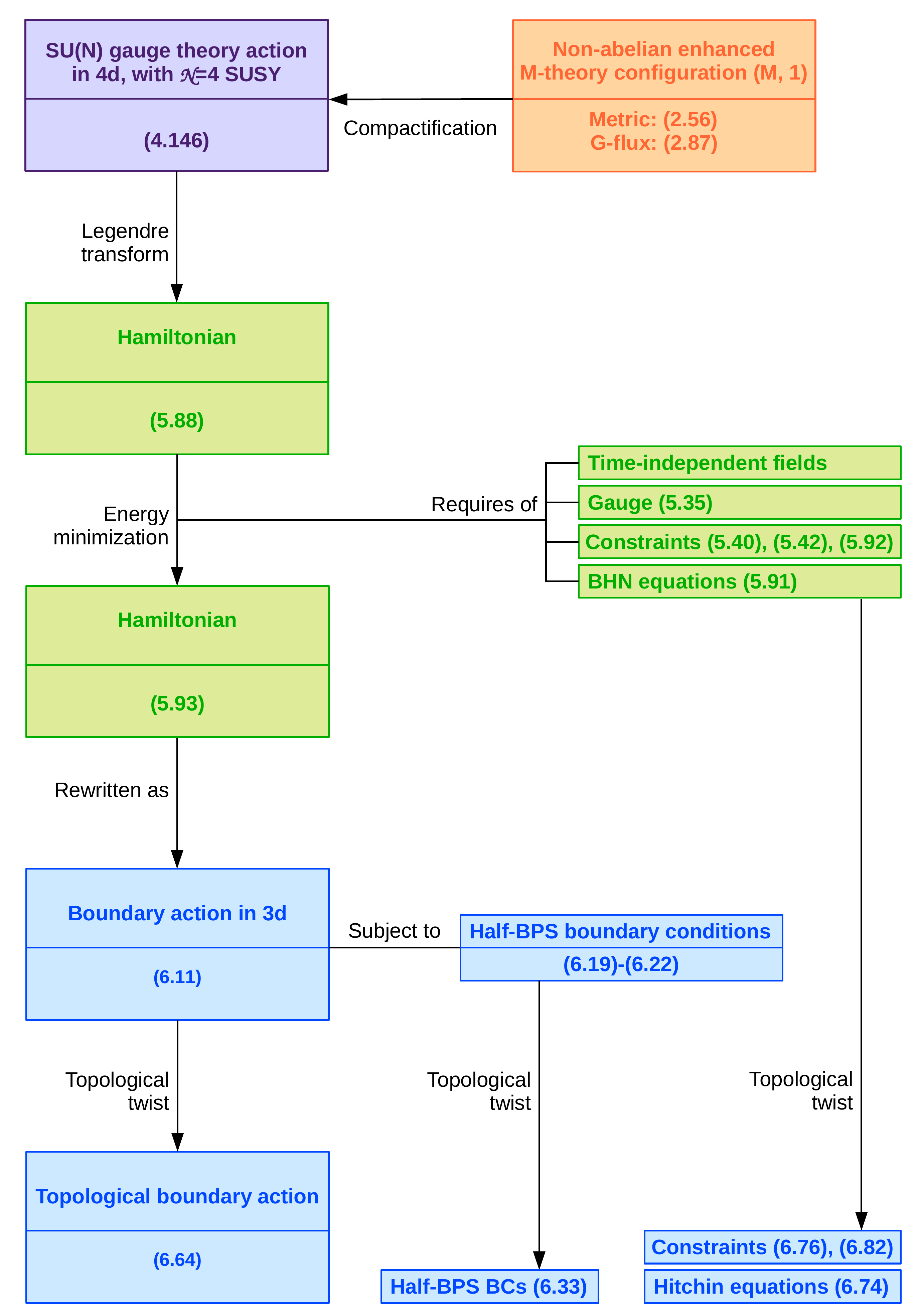}
\caption{Graphical summary of part \ref{partb}.
In orange, the starting point: the non-abelian enhanced
M-theory configuration (M, 1) of section \ref{ns5d3sect1}.
In purple, the contents of section \ref{actionsec}: the derivation of the four-dimensional gauge theory
stemming from (M, 1). Colored green, the obtention and minimization of the corresponding Hamiltonian, presented in section
\ref{hamilsec}. Blue is associated to section \ref{bndsec}, which focuses
on the study of the three-dimensional subspace where knots can be embedded.
}
\label{fig12}
\end{figure}

Figure \ref{fig12} provides a visual sketch of the overall logic and key results in this part.
Given the considerable length of the calculations involved,
the reader may find it useful to keep an eye in this image while reading through
the following three sections. In this way, the underlying principal flow of ideas shall hopefully not be lost
during the presentation of the corresponding computational details.

\FloatBarrier

\section{Bosonic action for the four-dimensional $SU(N)$ gauge theory \label{actionsec}}

\begin{figure}[hb]
\centering
\includegraphics[width=1\textwidth]{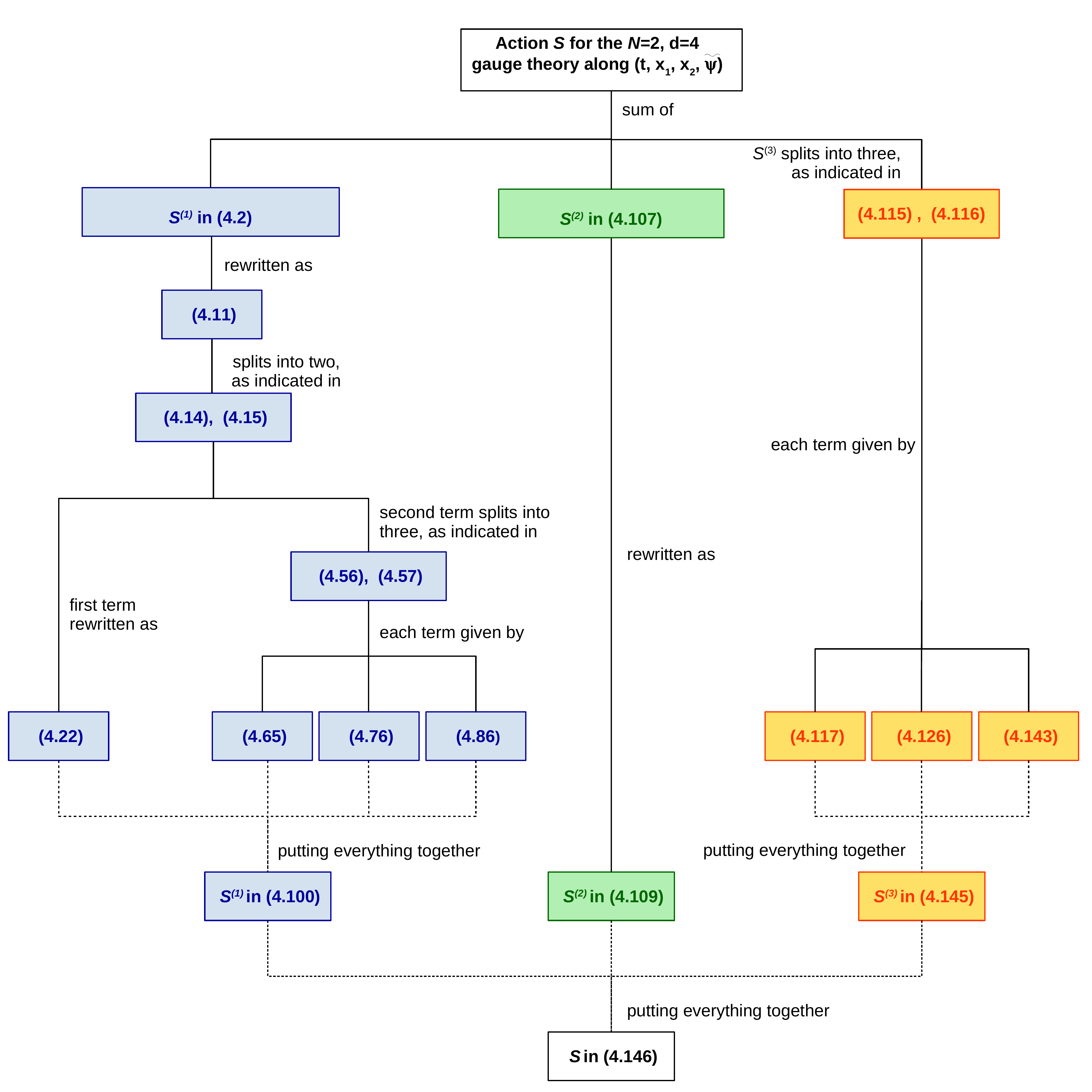}
\caption{Graphical summary of section \ref{actionsec}, where we obtain the bosonic action for the four-dimensional $SU(N)$ gauge theory
following from the non-abelian M-theory configuration (M, 1) of part \ref{parta}.
This figure sketches the connection between the very many terms whose addition gives the aforementioned action.
The colors correspond to the subsections where the mentioned equations can be found:
in blue results derived in section \ref{kintermsec}, in green those explained in section \ref{masstermsec}
and in yellow the terms worked out in section \ref{thirdtermsec}.}
\label{fig9}
\end{figure}

In accordance to the plan above outlined,
in this section we argue what the bosonic action is for the $SU(N)$ world-volume gauge theory along
$(t,\,x_1,\,x_2,\,\tilde{\psi})$ that
follows from the non-abelian enhanced M-theory configuration (M, 1).
This gauge theory has $\mathcal{N}=2$ supersymmetry by construction.
(We will not be interested in doing so here, but supersymmetry 
could be used to obtain the fermionic sector of the theory.)
In principle, one could explicitly write the eleven-dimensional M-theory action
and then work out the desired four-dimensional reduction\footnote{
Compactification is done via the G-flux (\ref{multiflux}) and metric (\ref{340}) reduced over the normalizable
internal harmonic forms. The Taub-NUT subspace has
normalizable harmonic two-forms (\ref{omegak}). For our case, compactification can thus be defined.}.
However, this is more easily said than done.
We will thus follow a different approach here:
we will obtain the total action as the sum of three distinct contributions,
providing ample motivation for each term.

The first two of these three terms directly stem from our construction of (M, 1)
in section \ref{ns5d3sect1} and are indeed initially written in terms of only quantities there defined.
Writing these terms as functions of the vector multiplet
of the $\mathcal{N}=4$ supersymmetric (with half-BPS boundary conditions) $SU(N)$ world-volume gauge theory is, however,
far from trivial. In achieving this task, we further split the two terms in many parts.

The third and last term is, unluckily, hard to present in such a manner.
Consequently, we start by directly writing it in terms of the aforementioned vector multiplet.
Nonetheless, the length and complexity of the term lead us to further divide it into smaller pieces too.

To help the reader make sense of the very many terms that follow, we include
figure \ref{fig9}. This figure provides a graphical summary of this section \ref{actionsec}, pointing out
all the different contributions to the total action and their origin. 

A last important remark before jumping into computation.
To avoid as much as possible dragging long prefactors, we set the Planck length to one
right from the onset: $l_p\equiv 1$.

\FloatBarrier

\subsection{Kinetic term of the G-flux \label{kintermsec}}

The first contribution to the aforementioned bosonic action we will consider is 
the kinetic term of the G-flux (\ref{multiflux}).
Our approach will be to work out in details this term for the abelian configuration (M, 1) of section \ref{ncsect}
and then generalize the result to the non-abelian scenario of section \ref{nonabsec}.
With this aim in mind, let us first recall the main features of both the abelian
and non-abelian configurations (M, 1).

The geometry of the configuration (M, 1) was given in (\ref{340}), be it for the abelian or non-abelian case.
By simple inspection, it can be readily seen that the eleven-dimensional
manifold $X_{11}$ on which this metric is defined naturally decomposes into three subspaces:
\begin{align}
X_{11}=X_4\otimes\Sigma_3\otimes TN, \quad\quad X_4=X_3\otimes \mathbb{R}^+.
\label{decom}
\end{align}
Here, $X_4$ is the four-dimensional subspace where we will define our gauge theory. This further decomposes into
$X_3$ (the Minkowski-type three-dimensional subspace along $(t,\,x_1,\,x_2)$) and $\mathbb{R}^+$ (the half real line
labeled by $\tilde{\psi}$). This second decomposition clearly denotes that
there is no Lorentz invariance along $\tilde{\psi}$. On the other hand, $\Sigma_3$ is the three-cycle parametrized by
$(\tilde{x}_3,\,\phi_1,\,r)$ and $TN$ stands for the warped Taub NUT space spanning $(\theta_1,\,x_8,\,x_9,\,
x_{11})$. For the abelian (M, 1), this is a single-centered Taub NUT, whereas for the non-abelian (M, 1)
it is an $N$-centered one.

After the non-abelian enhancement, there are $N$ coincident M2-branes oriented along $(x_8,\,x_9,\,x_{11})$
in the configuration (M, 1), as depicted in figure \ref{fig5}{\bf B}.
Following the notation of section \ref{nonabsec}, we denote as $\vec{l}_1$ the location of
these M2-branes in the $(x_8,\,x_9)$ plane.
It is around this point $\vec{l}_1$ that we shall determine the action of the non-abelian world-volume gauge theory.

Coming to the fluxes, the G-flux for the non-abelian enhanced (M, 1) was given in (\ref{multiflux}).
This G-flux consists of two pieces: the {\it delocalized} background flux $\langle\mathcal{G}_4^{(M,1)}\rangle$
and the {\it localized} contribution of $\sum\limits_{k=1}^{N-1}\mathcal{F}_k\wedge \omega_k$, sharply peaked
around $\vec{l}_1$. As it is common practice in the literature,
we will assume the delocalized piece is such that its contribution around $\vec{l}_1$ is negligible.

In the abelian case, the situation is essentially the same.
The only difference being that the G-flux is now given by (\ref{355}).
The Taub-NUT space has a unique singularity,
whose location we can denote as $\vec{l}_1$ as well. The G-flux again splits into delocalized and
localized parts. We assume the delocalized part's contribution is inconsequential around $\vec{l}_1$.

We will now use all the above remarks to obtain
the first term for the $U(1)$ world-volume gauge theory action:
\begin{align}
S^{(1)}\equiv \int_{X_{11}} \textrm{Tr}\left(\mathcal{G}_4^{(M,1)}\wedge \ast \mathcal{G}_4^{(M,1)}\right),
\label{initialact1}
\end{align}
where the Hodge dual is with respect to the eleven-dimensional metric (\ref{340}).
Using (\ref{355}) and
because we are interested in the gauge theory around $\vec{l}_1$, where $\langle\mathcal{G}_4^{(M,1)}\rangle$
is negligible, the above reduces to
\begin{align}
S^{(1)}=
\int_{X_{11}}\textrm{Tr}
\left(\mathcal{F} \wedge \omega\right)\wedge\ast
\left(\mathcal{F}\wedge \omega\right),
\end{align}
with $\mathcal{F}$ the seven-dimensional abelian field strength.
By definition, $\omega$ is (anti-)self-dual
and is restricted to the subspace $TN$. For concreteness, we take it to be self-dual
in the ongoing. On the other hand, $\mathcal{F}$ spans $X_4\otimes \Sigma_3$.
Then, we can rewrite $S^{(1)}$ as
\begin{align}
S^{(1)}=\int_{TN}\omega\wedge\omega\int_{X_4\otimes\Sigma_3}\mathcal{F}\wedge\ast\mathcal{F},
\label{Sg1}
\end{align}
where the Hodge duals are taken with respect to the subspaces of (\ref{340}) indicated by the corresponding integrals.
This drastic simplification where the Taub-NUT completely decouples is not as trivial as we just made it sound.
Hence, before proceeding further, let us carefully show how this can be made to happen consistently.

Naively, the decoupling happens if the following two conditions are satisfied:
\begin{itemize}
\item The integral over $TN$ above only depends on the $(\theta_1,\,x_8,\,x_9,\,x_{11})$ coordinates.
\item The integral over $X_4\otimes\Sigma_3$ is independent of $(\theta_1,\,x_8,\,x_9,\,x_{11})$.
\end{itemize}
The first condition can easily be seen to hold true. The two-form $\omega$
was defined in (\ref{zetadef}), with the gauge field ${\bf A}_1$ given by (\ref{A1def}).
It is clear from these expressions that the integrand $\omega\wedge\omega$ only depends
on the Taub-NUT coordinates, as desired.
The metric for the space $TN$ was given in (\ref{mettb})
and, as pointed out there, only depends on $(\theta_1,\,x_8,\,x_9,\,x_{11})$.
This implies the measure for the integral over $TN$ will have the same coordinate dependence.
The second condition, however does not hold true.
An inspection of the metric (\ref{340}) along the directions of $X_4$ and $\Sigma_3$
leads us to conclude that the measure of the second integral in (\ref{Sg1})
will depend on $(\theta_1, \, x_8,\, x_9)$.
(Recall our choices for the warp factors in (\ref{easychoice}) and for the dilaton in (\ref{354})
to understand this last statement.)
Nevertheless, this desired decoupling can be {\it effectively} made to happen.
Let us see how.

A careful inspection of (\ref{340}) restricted to $X_4\otimes\Sigma_3$ shows that the dependence
of the second integral in (\ref{Sg1}) on $(x_8,\,x_9)$
comes solely from the dilaton (\ref{354}). We can therefore remove this $(x_8,\,x_9)$ dependence by
assuming that the dilaton is given, to leading order, by its constant piece:
\begin{align}
e^{2\phi}\approx e^{2\phi_0}.
\label{consdil}
\end{align}
(Note that the above assumption is in excellent agreement to the strong coupling limit discussed around
(\ref{Mlimit}), required for our M-theory configuration to be valid, if we consider
$e^{2\phi_0}$ to be of order one.)
The $\theta_1$ dependence of the second integral in (\ref{Sg1}) is, however, not ``removable''.
Let us thus turn to the $\theta_1$ dependence of the first integral in (\ref{Sg1}).

To match the notation in~\cite{Dasgupta:2016rhc}, we will
call the first integral in (\ref{Sg1}) as
\begin{align}
\frac{c_1}{v_3}\equiv \int_{TN}\omega\wedge\omega.
\label{c1overv3}
\end{align}
Using (\ref{A1def}), (\ref{vielM1}), the first equation in (\ref{alphasel}) and (\ref{gsol}) in (\ref{expliom}),
it is a matter of easy algebra to obtain the two-form $\omega$ as
\begin{align}
\omega=&\sum_{i=8}^9\frac{\partial g}{\partial x_i}dx_i\wedge(dx_{11}+{{\bf A}_1}_{\theta_1}d\theta_1)
+\left(\frac{\partial g}{\partial x_8}{{\bf A}_1}_{9}-\frac{\partial g}{\partial x_9}{{\bf A}_1}_{8}\right)dx_8\wedge dx_9
\nonumber \\
&+g(\alpha_2 dx_8+\alpha_3 dx_9)\wedge d\theta_1.
\end{align}
Then, $(g, \, \alpha_2,\,\alpha_3)$ being all functions of only $(x_8,\,x_9)$,
it follows that (\ref{c1overv3}) is actually independent of $\theta_1$:
\begin{align}
\omega\wedge \omega=2g\left(\alpha_3\frac{\partial g}{\partial x_8}-\alpha_2\frac{\partial g}{\partial x_9}\right)
d\theta_1\wedge dx_8\wedge dx_9\wedge dx_{11}.
\end{align}
(The above is (3.52) in~\cite{Dasgupta:2016rhc}.)
Consequently, choosing (\ref{consdil}) and {\it transferring} the $\theta_1$ integral to the second
integral in (\ref{Sg1}) as an average, we can consistently decouple the contribution to this term of the action
of the Taub-NUT space:
\begin{align}
S^{(1)}=\frac{c_1}{v_3}\int_0^\pi \frac{d\theta_1}{2\pi}\int_{X_4\otimes\Sigma_3}\mathcal{F}\wedge\ast\mathcal{F},
\label{fins1}
\end{align}
where this prefactor should be understood, in this abelian case, as
\begin{align}
\frac{c_1}{v_3}=\int_0^{R_8} dx_8\int_0^{R_9} dx_9\int_0^{R_{11}}dx_{11}\,\,
2g\left(\alpha_3\frac{\partial g}{\partial x_8}-\alpha_2\frac{\partial g}{\partial x_9}\right),
\label{c1v3pre}
\end{align}
with $R_i$ denoting the radius of the $x_i$ direction (for $i=8,\,9,\,11$).
Note that $(x_8,\,x_9)$ are non-compact directions, while $x_{11}$ is compact.

At this point, it is easy to infer what the generalization of (\ref{fins1}) is to the non-abelian case:
\begin{align}
S^{(1)}=\frac{C_1}{V_3}I^{(1)}, \quad\quad
I^{(1)}\equiv \int_0^\pi \frac{d\theta_1}{2\pi}\int_{X_4\otimes\Sigma_3}\textrm{Tr}(\mathcal{F}\wedge\ast\mathcal{F}),
\label{nonabs1}
\end{align}
where $\mathcal{F}$ is now the {\it non-abelian} seven-dimensional field strength and the trace is
taken in the adjoint representation of $SU(N)$. There are just two subtleties
in going from (\ref{fins1}) to (\ref{nonabs1}) that we better discuss.

The first one is regarding the prefactor $({C_1}/{V_3})$.
This prefactor is, of course, no longer given by (\ref{c1v3pre}).
Instead, it depends on the two-forms $\omega_k$ in (\ref{omegak}).
Its explicit form is rather tedious to work out and we will not attempt to compute it here.
For our purposes, it suffices to note that,
by construction (see the details in section \ref{nonabsec}),
we are guaranteed  its independence on the
$\theta_1$ coordinate. So we can transfer the $\theta_1$ integral to the subspace orthogonal to $TN$ as an average
and indeed obtain (\ref{nonabs1}).

The second subtlety is regarding the appearance of the trace. (Note that the non-abelian
G-flux in (\ref{multiflux}) only involves the Cartan algebra values of $\mathcal{F}$.)
Let us try to shed some light to this point by first
recalling how the non-abelian enhancement was achieved in section \ref{nonabsec} (perhaps it suffices to
take a second look at figure \ref{fig5}{\bf B}).
There, we wrapped
M2-branes around the (minimal area, independent) two-cycles of the $N$-centered Taub-NUT space (\ref{mettb}).
The two-cycles were then shrunk to zero size, making the M2-branes tensionless.
From this point of view, internal fluctuations of the Taub-NUT space are supposed to provide the Cartan values
of the field strength. Fluctuations of the M2-branes along the Taub-NUT directions
would then contribute the remaining roots and weights, thus leading to the full trace in (\ref{nonabs1}).
A more detailed version of this argument may be found in~\cite{Sen:1997kz,Witten:1995im,Sen:1997js}
and references therein. However, no rigorous proof of this conjecture
exists. The argument between (3.91) and (3.98) in~\cite{Dasgupta:2016rhc}
in terms of a sigma model may well be the most solid evidence for this claim.

The fact that the trace should be in the adjoint representation has a simple enough heuristic explanation.
Additionally, this very argument settles what the bosonic matter content is in our non-abelian world-volume gauge theory.
Recall figure \ref{fig1}{\bf B}. There, to the usual type IIB D3-NS5 system we added a second, parallel NS5-brane.
The distance between the two NS5-branes being large enough then allows for effectively retaining $\mathcal{N}=2$
supersymmetry in the whole of the system. By the same logic, deep in the bulk of the D3-brane, far away from both the
NS5-branes, we expect $\mathcal{N}=4$ supersymmetry effectively.
As is well-known, any $\mathcal{N}=4$ supersymmetric gauge theory has a vector multiplet consisting on
four gauge fields and six real scalars, all of them in the adjoint representation. Certainly, this is the matter
content we expect in the bosonic sector for our D3-brane gauge theory too, far from the NS5-branes.
On the other hand, the bosonic matter content of any $\mathcal{N}=2$ supersymmetric gauge theory
is arranged in a vector multiplet of four gauge fields and two real scalars in the adjoint representation
and a chiral multiplet containing four real scalars in any representation.
Needless to say, this is the matter content we expect in the bosonic sector of our gauge theory nearby the NS5-branes.
It then stands to reason that, if we are to reconcile these two limits in our set up, we require
the four scalars of the $\mathcal{N}=2$ chiral multiplet to be in the adjoint representation.
Therefore, the bosonic matter content of our $SU(N)$ gauge theory is settled to that of the $\mathcal{N}=4$ vector multiplet:
four gauge fields and six real scalars, all of them in the adjoint representation. 

Subtleties aside, we take (\ref{nonabs1}) as our starting point and
devote the remaining of this section to writing $I^{(1)}$ in terms of
the just discussed $\mathcal{N}=4$ vector multiplet, which spans the directions $(t,\,x_1,\,x_2,\,\tilde{\psi})$.
To begin with, we assume that the seven-dimensional non-abelian field strength $\mathcal{F}$ only depends
on these coordinates:
\begin{align}
\mathcal{F}=\mathcal{F}(t,\,x_1,\,x_2,\,\tilde{\psi}).
\label{depenF}
\end{align}
Secondly, and owing to the decomposition (\ref{decom}),
we make a distinction between the seven-dimensional field strengths along $X_4$ and $\Sigma_3$:
\begin{align}
\mathcal{F}=\mathcal{F}^{(X_4)}+\mathcal{F}^{(\Sigma_3)}.
\label{F34decom}
\end{align}
Using such distinction in (\ref{nonabs1}), we naturally split the first contribution to the non-abelian action
into two pieces:
\begin{align}
S^{(1)}=\frac{C_1}{V_3}\left(I^{(1,1)}+I^{(1,2)}\right),
\label{split1}
\end{align}
with
\begin{align}
I^{(1,1)}\equiv \int_0^\pi \frac{d\theta_1}{2\pi}\int_{X_4\otimes\Sigma_3}\textrm{Tr}(\mathcal{F}^{(X_4)}
\wedge\ast\mathcal{F}^{(X_4)}), \quad
I^{(1,2)}\equiv \int_0^\pi \frac{d\theta_1}{2\pi}\int_{X_4\otimes\Sigma_3}\textrm{Tr}(\mathcal{F}^{(\Sigma_3)}
\wedge\ast\mathcal{F}^{(\Sigma_3)}).
\label{I11I12}
\end{align}
Rather obviously, the Hodge dual in both $I^{(1,1)}$ and $I^{(1,2)}$ is (still) with respect to
the seven-dimensional metric of $X_4\otimes\Sigma_3$.

Note that the crossed terms  $(\mathcal{F}^{(X_4)}
\wedge\ast\mathcal{F}^{(\Sigma_3)})$ and $(\mathcal{F}^{(\Sigma_3)}
\wedge\ast\mathcal{F}^{(X_4)})$ are zero and thus have not been included in (\ref{split1}).
The argument for the vanishing of the first such term is as follows. Each component of $\mathcal{F}^{(\Sigma_3)}$ spans
two directions of $\Sigma_3$. Consequently, the corresponding term of $\ast\mathcal{F}^{(\Sigma_3)}$ is oriented along
all four directions of $\Sigma_4$ and the remaining direction of $\Sigma_3$. As the components of $\mathcal{F}^{(X_4)}$
span two directions of $\Sigma_4$, the term $(\mathcal{F}^{(X_4)}
\wedge\ast\mathcal{F}^{(\Sigma_3)})$ necessarily contains the wedge product of two same $X_4$ directions and
thus yields zero. The argument for the vanishing of the second crossed term is similar.

At this stage, the only quantities left to be determined to explicitly write $S^{(1)}$ are $I^{(1,1)}$ and $I^{(1,2)}$,
defined in (\ref{I11I12}).
Their computation is quite long and involved. Consequently, we will do so in separate sections.
In the end, we will
put together in (\ref{split1}) the $I^{(1,1)}$ and $I^{(1,2)}$ we shall obtain,
thereby expressing the first term for the gauge theory action
in terms of the $\mathcal{N}=4$ vector multiplet's matter content.

\subsubsection{Determining $I^{(1,1)}$: the contribution of gauge field strengths \label{I11}}

As the title suggests, this section is devoted to the computation of $I^{(1,1)}$ in (\ref{I11I12})
in terms of the field strengths associated to the $\mathcal{N}=4$ vector multiplet's gauge fields.
But before jumping into the details of the calculation, let us introduce some quantities that will soon be useful.

We begin by taking a closer look at the seven-dimensional space $X_4\otimes\Sigma_3$, where $I^{(1,1)}$ is defined.
Its metric can be directly read from (\ref{340}) to be
\begin{align}
ds_{X_4\otimes\Sigma_3}^2=H_1[-dt^2+dx_1^2+dx_2^2+H_2d\tilde{x}_3^2+H_3(d\phi_1+f_3d\tilde{x}_3)^2
+e^{2\phi_0}(F_1dr^2+H_4d\tilde{\psi}^2)],
\label{7dmetric}
\end{align}
where we have made use of our assumption (\ref{consdil}).
Following the spirit of the language in~\cite{Dasgupta:2016rhc}, we denote as $g_7$ the determinant of the above metric:
\begin{align}
g_7\equiv \textrm{det}(ds_{X_4\otimes\Sigma_3}^2)=e^{4\phi_0}F_1H_1^7H_2H_3H_4=e^{4\phi_0}F_1H_1^4H_4,
\label{g7}
\end{align}
where in the last step we have used the fact that $H_1^3H_2H_3=1$, which follows from (\ref{Hs}).
It will also come in handy to write the metric along the subspace $X_4$, albeit in matrix form:
\begin{align}
g_{ab}=H_1\textrm{diag}(-1,1,1), \quad\quad g_{\tilde{\psi}\tilde{\psi}}=e^{2\phi_0}H_1H_4.
\end{align}
Here, the subscripts $(a,\,b)$ take values $(0,\,1,\,2)$ and stand for the Lorentz-invariant directions $(t,\,x_1,\,x_2)$.
Being diagonal, it is straightforward to see that the inverse of the $X_4$ metric, in matrix form, is given by
\begin{align}
g^{ab}=\frac{1}{H_1}\textrm{diag}(-1,1,1), \quad\quad g^{\tilde{\psi}\tilde{\psi}}=\frac{e^{-2\phi_0}}{H_1H_4}.
\label{matrentr}
\end{align}
Calling $g_4$ the (absolute value of the) determinant of the $X_4$ metric, this is
\begin{align}
g_4\equiv\left| \textrm{det}(ds_{X_4}^2)\right|=e^{2\phi_0}H_1^4H_4.
\label{g4}
\end{align}

Having introduced our notation, we may now proceed to the determination of $I^{(1,1)}$.
First of all, we explicitly write the wedge product of its integrand as
\begin{align}
\mathcal{F}^{(X_4)}\wedge\ast\mathcal{F}^{(X_4)}=&\sqrt{g_7}\sum_{a,b,c,d=0}^2 g^{ab}\left(g^{cd}\mathcal{F}_{ac}\mathcal{F}_{bd}
+g^{\tilde{\psi}\tilde{\psi}}\mathcal{F}_{a\tilde{\psi}}\mathcal{F}_{b\tilde{\psi}}\right) \nonumber \\
=&\sqrt{\frac{F_1}{H_4}}\Big(e^{2\phi_0}H_4\sum_{\substack{a,b=0 \\ a<b}}^2\mathcal{F}_{ab}^2+\sum_{a=0}^2
\mathcal{F}_{a\tilde{\psi}}^2\Big).
\end{align}
Using the above in (\ref{I11I12}), we have that
\begin{align}
\label{genform}
I^{(1,1)}=c_{11}\int d^4x \sum_{\substack{a,b=0 \\ a<b}}^2\textrm{Tr}(\mathcal{F}_{ab}^2)+c_{12}\int d^4x
\sum_{a=0}^2 \textrm{Tr}(\mathcal{F}_{a\tilde{\psi}}^2), 
\end{align}
where the integration is with respect to the world-volume coordinates
$(t,\,x_1,\,x_2,\,\tilde{\psi})$ and where
we have defined the coefficients $c_{11}$ and $c_{12}$ as
\begin{align}
\label{intc11}
c_{11}\equiv\displaystyle e^{2\phi_0}\int d^4\tilde{\zeta}
\sqrt{F_1H_4}, \quad\quad
c_{12}\equiv \displaystyle \int d^4\tilde{\zeta}
\sqrt{\frac{F_1}{H_4}}.
\end{align}
As a short-hand notation that will keep appearing, we have introduced
\begin{align}
\int d^4\tilde{\zeta}\equiv \int_0^{R_3}d\tilde{x}_3\int_0^{2\pi}d\phi_1\int_0^\infty dr\int_0^\pi \frac{d\theta_1}{2\pi}
\label{dtildezeta}
\end{align}
above, with $R_3$ the radius of the non-compact direction $\tilde{x}_3$.
Note that these coefficients have been taken out of the integral over the world-volume coordinates in (\ref{genform})
because $F_1$ and $H_4$ are only functions of the radial coordinate and $\theta_{nc}$
(recall our choice in (\ref{easychoice}) and the definitions in (\ref{tildeF2}) and (\ref{Hs})).
For this same reason, we can right away perform the $(\tilde{x}_3,\,\phi_1)$ integrals above.
Further using (\ref{Hs}), we can express $c_{11}$ and $c_{12}$ as
\begin{align}
c_{11}=2 R_3e^{2\phi_0}\sec\theta_{nc} \int_0^\infty dr\, \sqrt{F_1\tilde{F}_2F_3} \mathcal{I}^{(1)}, \quad\quad
c_{12}=2R_3\cos\theta_{nc}\int_0^\infty dr\,\sqrt{\frac{F_1}{\tilde{F}_2F_3}}\mathcal{I}^{(2)},
\label{c11c12al}
\end{align}
where we have defined
\begin{align}
\mathcal{I}^{(1)}\equiv \int_0^{\pi/2}
\frac{\sin\theta_1d\theta_1}{\sqrt{\tilde{F}_2\cos^2\theta_1+F_3\sin^2\theta_1}}, \quad\quad
\mathcal{I}^{(2)}\equiv \int_0^{\pi/2}
d\theta_1\,\csc\theta_1\sqrt{\tilde{F}_2\cos^2\theta_1+F_3\sin^2\theta_1}.
\label{mathcalI12}
\end{align}

Since they will keep showing up, it is useful to introduce the functions
\begin{align}
\chi(\theta_1)\equiv \sqrt{\tilde{F}_2+F_3+(\tilde{F}_2-F_3)\cos2\theta_1}, \quad\quad
\tilde{\chi}(\theta_1)\equiv \sqrt{2(\tilde{F}_2-F_3)}\cos\theta_1.
\label{chitheta1}
\end{align}
Using these, the first of these integrals can be readily performed to yield
\begin{align}
\mathcal{I}^{(1)}=-\frac{1}{\sqrt{\tilde{F}_2-F_3}}\ln\left|
\chi(\theta_1)+\tilde{\chi}(\theta_1)\right|\Bigg|_{\theta_1=0}^{\theta_1=\pi/2} 
=\frac{\mathcal{J}_3}{2\sqrt{\tilde{F}_2-F_3}},
\label{mathI1sol}
\end{align}
where we have defined
\begin{align}
\mathcal{J}_3\equiv \ln\left|\frac{\sqrt{\tilde{F}_2}+\sqrt{\tilde{F}_2-F_3}}{\sqrt{\tilde{F}_2}-\sqrt{\tilde{F}_2-F_3}}\right|,
\label{mathj3}
\end{align}
a quantity which will appear in the present analysis very often.
It is clear that the above will be real if and only if we require that $\tilde{F}_2\geq F_3$,
for all values of $(r,\,\theta_{nc})$. Thus, we will demand this holds true in the ongoing.
Using the above in (\ref{c11c12al}), we obtain $c_{11}$ as in (3.76) in~\cite{Dasgupta:2016rhc}:
\begin{align}
\label{c11final}
c_{11}= R_3e^{2\phi_0}\sec\theta_{nc} \int_0^\infty dr\,\,\mathcal{J}_3 \sqrt{\frac{F_1\tilde{F}_2F_3}{\tilde{F}_2-F_3}}.
\end{align}
It is important to note that the above coefficient is just a number. The numerical value of $c_{11}$
depends only on the choice of warp factors one would like to consider in (\ref{easychoice}). This choice is subject to
the constraint $\tilde{F}_2\geq F_3$ and should be checked to preserve the desired $\mathcal{N}=4$
supersymmetry in the world-volume (later on reduced to $\mathcal{N}=2$ supersymmetry via half-BPS boundary conditions).

Coming now to $\mathcal{I}^{(2)}$, we start by defining the soon
to be useful three quantities in (3.79) in~\cite{Dasgupta:2016rhc}:
\begin{align}
b_1\equiv \sqrt{\frac{F_3}{\tilde{F}_2-F_3}}, \quad
b_2\equiv \frac{1}{2}\sqrt{\frac{F_1(\tilde{F}_2-F_3)}{\tilde{F}_2F_3}}, \quad
b_3\equiv \frac{2}{b}\sqrt{\frac{F_3+b^2(\tilde{F}_2-F_3)}{\tilde{F}_2-F_3}}, \quad b\in(\mathbb{R}^+-\{1\}).
\label{379}
\end{align}
We can use $b_1$ to rewrite the integral of our interest in the more convenient form
\begin{align}
\mathcal{I}^{(2)}=\sqrt{\tilde{F}_2-F_3}\int_0^{\pi/2}d\theta_1\,\sqrt{\frac{b_1^2+\cos^2\theta_1}{1-\cos^2\theta_1}}.
\label{429}
\end{align}
Under the change of variables
\begin{align}
\cos\theta_1= z, \quad\quad d\theta_1=-\frac{dz}{\sqrt{1-z^2}},
\label{theta1toz}
\end{align}
the above can be further rewritten as
\begin{align}
\mathcal{I}^{(2)}=\frac{\sqrt{\tilde{F}_2-F_3}}{2}\int_{-1}^1dz\,\frac{\sqrt{b_1^2+z^2}}{b^2-z^2},
\label{I2inter}
\end{align}
where $b$ as defined in (\ref{379}) is a regularization factor that we have introduced by hand
in order to avoid the singularities
of $\mathcal{I}^{(2)}$ at $z=\pm1$. In the same spirit of $(\chi(\theta_1),\,\tilde{\chi}(\theta_1))$
before, let us introduce two more functions that will come in handy repeatedly:
\begin{align}
\eta(z)\equiv \arctanh\left(\frac{z}{b}\sqrt{\frac{b_1^2+b^2}{b_1^2+z^2}}\right), \quad\quad
\tilde{\eta}(z)\equiv \ln\left|z+\sqrt{b_1^2+z^2}\right|. \label{etasdef}
\end{align}
Finally, all the above can be used to integrate over $z$ in (\ref{I2inter}) and obtain
\begin{align}
\frac{2\mathcal{I}^{(2)}}{\sqrt{\tilde{F}_2-F_3}}=
\frac{\sqrt{b_1^2+b^2}}{b}\eta(z)
-\tilde{\eta}(z) \Bigg|_{z=-1}^{z=1}
=b_3\mathcal{J}_4
+\mathcal{J}_3^{-1},
\label{432}
\end{align}
where we have defined the many times to occur quantity $\mathcal{J}_4$ as
\begin{align}
\mathcal{J}_4\equiv \arctanh\left(\frac{1}{b}\sqrt{\frac{F_3+b^2(\tilde{F}_2-F_3)}{\tilde{F}_2}}\right).
\label{mathj4}
\end{align}
Plugging our result in (\ref{intc11}), the coefficient $c_{12}$ may be expressed as in (3.78) in~\cite{Dasgupta:2016rhc}:
\begin{align}
\label{c12final}
c_{12}=2R_3\cos\theta_{nc}\int_0^\infty\hspace*{-0.2cm} dr\, b_2\left(b_3
\mathcal{J}_4
+\mathcal{J}_3^{-1}\right).
\end{align}
As was the case for $c_{11}$ before, we want $c_{12}$ to be a well defined number for all choices
of warp factors in (\ref{easychoice}) satisfying the constraint $\tilde{F}_2\geq F_3$ (and preserving
$\mathcal{N}=2$ supersymmetry). It is not clear from our above result that this should be the case in the following two cases:
\begin{itemize}
\item $F_3\rightarrow 0$. This limit also includes the case $(\tilde{F}_2,\,F_3)\rightarrow0$ since, in order to
be consistent with the constraint $\tilde{F}_2\geq F_3$, we must demand that $F_3$ approaches zero {\it faster}
than $\tilde{F}_2$. Hence, the case $(\tilde{F}_2,\,F_3)\rightarrow0$ should be studied by first demanding
$F_3\rightarrow0$ and afterwards considering the $\tilde{F}_2\rightarrow 0$ limit of the resulting expression.
\item $\tilde{F}_2\rightarrow F_3\nrightarrow0$.
\end{itemize}
Let us thus study such subtle scenarios in details and show that $c_{12}$ in (\ref{c12final}) is indeed
a finite number even then.

To consider the first case, namely $F_3\rightarrow0$, we start by rewriting the argument of the
inverse hyperbolic tangent in (\ref{mathj4}) as
\begin{align}
\frac{1}{b}\sqrt{\frac{F_3+b^2(\tilde{F}_2-F_3)}{\tilde{F}_2}}=\sqrt{1+\left(\frac{1-b^2}{b^2}\right)\frac{F_3}{\tilde{F}_2}}.
\label{retan}
\end{align}
Next, we note that in the logarithmic term of (\ref{c12final}),
namely $\mathcal{J}_3$ in (\ref{mathj3}), only the numerator diverges as $F_3\rightarrow0$,
while the denominator is well defined in this limit. Hence, retaining only the divergent terms in the integrand of
(\ref{c12final}) and using (\ref{retan}), we focus on the study of
\begin{align}
\lim_{F_3\rightarrow0}c_{12}\sim\lim_{F_3\rightarrow0}
\left[b_2b_3\arctanh\left(\sqrt{1+\left(\frac{1-b^2}{b^2}\right)\frac{F_3}{\tilde{F}_2}}\right)
+b_2\ln\left|\sqrt{\tilde{F}_2}-\sqrt{\tilde{F}_2-F_3}\right|\right].
\label{lim1}
\end{align}
From our definitions in (\ref{379}) it follows that
\begin{align}
\lim_{F_3\rightarrow 0} b_2=\lim_{F_3\rightarrow 0}\sqrt{\frac{F_1}{F_3}}=
\lim_{F_3\rightarrow 0}b_2b_3
\label{b2b3lim}
\end{align}
which, used in (\ref{lim1}), gives
\begin{align}
\lim_{F_3\rightarrow0}c_{12}\sim\lim_{F_3\rightarrow 0}\sqrt{\frac{F_1}{F_3}}
\left[\arctanh\left(\sqrt{1+\left(\frac{1-b^2}{b^2}\right)\frac{F_3}{\tilde{F}_2}}\right)
+\ln\left|\sqrt{\tilde{F}_2}-\sqrt{\tilde{F}_2-F_3}\right|\right].
\end{align}
Applying L'H\^{o}pital's rule to the two terms above, it is easy to see that
\begin{align}
\lim_{F_3\rightarrow 0}\sqrt{\frac{F_1}{F_3}}
\arctanh\left(\sqrt{1+\left(\frac{1-b^2}{b^2}\right)\frac{F_3}{\tilde{F}_2}}\right)
=-\lim_{F_3\rightarrow 0}\sqrt{\frac{F_1}{F_3}}\ln\left|\sqrt{\tilde{F}_2}-\sqrt{\tilde{F}_2-F_3}\right|.
\end{align}
That is, the divergent contribution to $\big(\displaystyle\lim_{F_3\rightarrow0}c_{12}\big)$
is zero, as pointed out in (3.80) in~\cite{Dasgupta:2016rhc} too.
This implies $c_{12}$ takes some finite numerical value when $F_3\rightarrow0$.

If we now turn our attention to the $(\tilde{F}_2,\,F_3)\rightarrow0$ case, the above still holds true.
However, the denominator the of logarithmic term of (\ref{c12final}) is no longer well defined
and consequently, we must study it. As already argued, we first should consider the $F_3\rightarrow 0$
limit of this term and then impose $\tilde{F}_2\rightarrow0$ there.
Using (\ref{b2b3lim}) and applying L'H\^{o}pital's rule, this additional divergent term can also be
seen to vanish:
\begin{align}
\lim_{\tilde{F}_2,F_3\rightarrow0}c_{12}\sim\lim_{\tilde{F}_2,F_3\rightarrow0}
\frac{1}{\sqrt{F_3}}\ln\left|2\sqrt{\tilde{F}_2}\right|=
\lim_{\tilde{F}_2,F_3\rightarrow0}-\frac{F_3^{3/2}}{\tilde{F}_2}=0.
\label{limfin}
\end{align}
Thus, $c_{12}=0$ when $(\tilde{F}_2,\,F_3)\rightarrow0$.

Finally, we study the limit $\tilde{F}_2\rightarrow F_3\nrightarrow0$.
From (\ref{379}), it is not hard to work out the following two limits:
\begin{align}
\lim_{\tilde{F}_2\rightarrow F_3} b_2=0, \quad\quad 
\lim_{\tilde{F}_2\rightarrow F_3} b_2 b_3=\frac{1}{b}\sqrt{\frac{F_1}{F_3}}.
\end{align}
Inserting the above in (\ref{c12final}), we obtain (3.81) in~\cite{Dasgupta:2016rhc}:
\begin{align}
\lim_{\tilde{F}_2\rightarrow F_3} c_{12}=
2R_3\cos\theta_{nc}\int_0^\infty dr \frac{1}{b}\sqrt{\frac{F_1}{F_3}}\arctanh\left(\frac{1}{b}\right)\sim
\arctanh\left(\frac{1}{b}\right),
\end{align}
which can be very large, yet is finite (as the regularization factor satisfies $b\neq1$ by definition).
This proofs that $c_{12}$ is just some number as $\tilde{F}_2\rightarrow F_3$.

Summing up, $I^{(1,1)}$ is given by (\ref{genform}), with $c_{11}$ given by (\ref{c11final})
and $c_{12}$ by (\ref{c12final}). Both of the coefficients are well defined numbers for any choice of
the warp factors one may want to consider, as long as the constraint $\tilde{F}_2\geq F_3$ is respected.

\subsubsection{Determining $I^{(1,2)}$: the contribution of three scalar fields \label{I12}}

In this section we compute $I^{(1,2)}$ in (\ref{I11I12}) in terms
of the $\mathcal{N}=4$ vector multiplet's matter content.
As in the previous section \ref{I11}, it is convenient to first introduce
certain quantities, which will be necessary in the subsequent calculation.

Let us begin by looking at the three-cycle $\Sigma_3$, parametrized by $(\tilde{x}_3,\,\phi_1,\,r)$.
Its metric can be easily inferred from (\ref{7dmetric}) to be
\begin{align}
ds_{\Sigma_3}^2=H_1H_2 d\tilde{x}_3^2+H_1H_3(d\phi_1+f_3d\tilde{x}_3)^2+e^{2\phi_0}H_1F_1dr^2.
\label{metricsigma3}
\end{align}
We take the vielbeins associated to the above metric as in (3.102) in~\cite{Dasgupta:2016rhc}:
\begin{align}
e_{\tilde{3}}^{(\Sigma_3)}= \sqrt{H_1H_2}d\tilde{x}_3, \quad\quad
e_r^{(\Sigma_3)}= e^{\phi_0} \sqrt{H_1F_1}dr, \quad\quad
e_{\phi_1}^{(\Sigma_3)}=\sqrt{H_1H_3}(d\phi_1+f_3dx_3).
\label{vielsigma3}
\end{align}
It is not hard to see that these vielbeins satisfy
\begin{align}
\ast e_{\tilde{3}}^{(\Sigma_3)}=e_r^{(\Sigma_3)}\wedge e_{\phi_1}^{(\Sigma_3)}, \quad\quad
\ast e_r^{(\Sigma_3)}=e_{\phi_1}^{(\Sigma_3)}\wedge e_{\tilde{3}}^{(\Sigma_3)}, \quad\quad
\ast e_{\phi_1}^{(\Sigma_3)}=e_{\tilde{3}}^{(\Sigma_3)}\wedge  e_r^{(\Sigma_3)}, \label{ast3viel}
\end{align}
where the Hodge duals are with respect to the metric (\ref{metricsigma3}).

Let us now focus on $\mathcal{F}^{(\Sigma_3)}$ in (\ref{I11I12}).
This field strength is related to the corresponding three-dimensional non-abelian gauge field $\mathcal{A}^{(\Sigma_3)}$
in the usual manner
\begin{align}
\mathcal{F}^{(\Sigma_3)}=\mathcal{D}\mathcal{A}^{(\Sigma_3)}+\mathcal{A}^{(\Sigma_3)}\wedge\mathcal{A}^{(\Sigma_3)}, 
\label{F3A3}
\end{align}
where the covariant derivative is given by (3.116) in~\cite{Dasgupta:2016rhc}:
\begin{align}
\mathcal{D}_a\equiv \partial_a+i[\mathcal{A}_a,\quad], \quad\quad \mathcal{D}_{\tilde{\psi}}\equiv \partial_{\tilde{\psi}}
+i[\mathcal{A}_{\tilde{\psi}},\quad], 
\label{covder}
\end{align}
with $a=(0,\,1,\,2)$ standing for the Lorentz-invariant directions $(t,\,x_1,\, x_2)$ and 
$(\mathcal{A}_a,\,\mathcal{A}_{\tilde{\psi}})$ the world-volume gauge fields
associated to the field strengths in (\ref{genform}).
Following (3.101) in~\cite{Dasgupta:2016rhc}, we define $\mathcal{A}^{(\Sigma_3)}$ as
\begin{align}
\mathcal{A}^{(\Sigma_3)}\equiv\mathcal{A}_{\tilde{3}}d\tilde{x}_3+\mathcal{A}_{\phi_1}d\phi_1+\mathcal{A}_rdr
=\hat{\alpha}_1e_{\tilde{3}}^{(\Sigma_3)}+\hat{\alpha}_2e_r^{(\Sigma_3)}+\hat{\alpha}_3e_{\phi_1}^{(\Sigma_3)}.
\label{3101}
\end{align}
In the last step above we have used (\ref{vielsigma3}) and the one-forms
\begin{align}
\hat{\alpha}_1\equiv \frac{\mathcal{A}_{\tilde{3}}-f_3\mathcal{A}_{\phi_1}}{\sqrt{H_1H_2}}, \quad\quad
\hat{\alpha}_2\equiv \frac{e^{-\phi_0}\mathcal{A}_r}{\sqrt{H_1F_1}}, \quad\quad
\hat{\alpha}_3\equiv \frac{\mathcal{A}_{\phi_1}}{\sqrt{H_1H_3}}.
\label{alphasdef}
\end{align}
Because of (\ref{depenF}), $(\mathcal{A}_{\tilde{3}},\,\mathcal{A}_{\phi_1},\, \mathcal{A}_r)$
are functions of only $(t,\,x_1,\,x_2,\,\tilde{\psi})$. (Note that this also explains our definitions in (\ref{covder}).)
On the other hand, from (\ref{easychoice}), (\ref{tildeF2}) and (\ref{Hs}), it is clear that the
$\hat{\alpha}_i$'s (with $i=1,\,2,\,3$) additionally depend on $(\theta_1,\,r)$.
A vital remark follows:
from the point of view of the four-dimensional gauge theory,
$(\mathcal{A}_{\tilde{3}},\,\mathcal{A}_{\phi_1},\, \mathcal{A}_r)$ should be understood as three real scalar fields
in the adjoint representation. 

Our above discussion settles the ground to determine $I^{(1,2)}$ in (\ref{I11I12}) in terms of the real scalar fields
$(\mathcal{A}_{\tilde{3}},\,\mathcal{A}_{\phi_1},\, \mathcal{A}_r)$.
The integrand there is of the form 
\begin{align}
\mathcal{F}^{(\Sigma_3)}\wedge\ast\mathcal{F}^{(\Sigma_3)}=&
\mathcal{D}\mathcal{A}^{(\Sigma_3)}\wedge\ast\left(\mathcal{D}\mathcal{A}^{(\Sigma_3)}\right)
+\mathcal{A}^{(\Sigma_3)}\wedge\mathcal{A}^{(\Sigma_3)}\wedge\ast \left(\mathcal{A}^{(\Sigma_3)}
\wedge\mathcal{A}^{(\Sigma_3)}\right)\nonumber \\
&+\mathcal{A}^{(\Sigma_3)}\wedge\mathcal{A}^{(\Sigma_3)}\wedge \ast \left(\mathcal{D}\mathcal{A}^{(\Sigma_3)}\right)
+\mathcal{D}\mathcal{A}^{(\Sigma_3)}\wedge\ast \left(\mathcal{A}^{(\Sigma_3)}\wedge\mathcal{A}^{(\Sigma_3)}\right),
\end{align}
where all the Hodge duals are with respect to the seven-dimensional
metric (\ref{7dmetric}) and we have made use of (\ref{F3A3}).
Owing to the decomposition (\ref{decom}), it is easy to see that the last line above vanishes.
(The reason is analogous to that given around (\ref{I11I12}) for the vanishing of the there-called ``crossed terms''.)
Consider the first such term.
The two-form $\mathcal{D}\mathcal{A}^{(\Sigma_3)}$ spans one direction in $X_4$ and
another one in $\Sigma_3$. Consequently, its corresponding Hodge dual five-form
is defined along the remaining three directions of $X_4$ and two directions of $\Sigma_3$.
But, since $\mathcal{A}^{(\Sigma_3)}\wedge\mathcal{A}^{(\Sigma_3)}$ stretches along two directions of $\Sigma_3$,
the wedge product of these two last forms will necessarily contain the wedge product of one of the directions of $\Sigma_3$
with itself. Anti-symmetry of the wedge product then implies zero value for this first term.
A similar argument applies to the second term too.
The decomposition (\ref{decom}) also allows for a drastic simplification of the two terms in the first line above.
Indeed, we can decouple $X_4$ and $\Sigma_3$ completely and write
\begin{align}
\mathcal{F}^{(\Sigma_3)}\wedge\ast\mathcal{F}^{(\Sigma_3)}=&\sqrt{g_4}\,\,d^4x\left[
\sum_{a=0}^2\mathcal{D}_a\mathcal{A}^{(\Sigma_3)}\wedge\ast\left(\mathcal{D}_a\mathcal{A}^{(\Sigma_3)}\right)
+\mathcal{D}_{\tilde{\psi}}\mathcal{A}^{(\Sigma_3)}\wedge\ast\left(\mathcal{D}_{\tilde{\psi}}\mathcal{A}^{(\Sigma_3)}\right)
\right.\nonumber \\
&\left.+\mathcal{A}^{(\Sigma_3)}\wedge\mathcal{A}^{(\Sigma_3)}\wedge\ast \left(\mathcal{A}^{(\Sigma_3)}
\wedge\mathcal{A}^{(\Sigma_3)}\right)
\right],
\end{align}
where the Hodge dual on the left-hand side is with respect to the seven-dimensional metric (\ref{7dmetric}),
whereas the Hodge duals on the right-hand side are with respect to the three-dimensional metric (\ref{metricsigma3}).
We remind the reader that $g_4$ was defined in (\ref{g4}) and that $(d^4x\equiv dt\,dx_1\,dx_2\,d\tilde{\psi})$,
as in (\ref{genform}). Inserting the above in (\ref{I11I12}), we can split the computation of $I^{(1,2)}$
into three as
\begin{align}
I^{(1,2)}=\int d^4x\,\textrm{Tr}\left(I^{(1,2,1)}+I^{(1,2,2)}+I^{(1,2,3)}\right),
\label{I12split}
\end{align}
where we have defined
\newpage
\begin{align}
I^{(1,2,1)}\equiv&\int_0^\pi\frac{d\theta_1}{2\pi}\int_{\Sigma_3}\sqrt{g_4}\,\,
\mathcal{A}^{(\Sigma_3)}\wedge\mathcal{A}^{(\Sigma_3)}\wedge\ast \left(\mathcal{A}^{(\Sigma_3)}
\wedge\mathcal{A}^{(\Sigma_3)}\right), \nonumber \\
\vspace*{0.2cm}
I^{(1,2,2)}\equiv&\int_0^\pi\frac{d\theta_1}{2\pi}\int_{\Sigma_3}\sqrt{g_4}\,\,
\sum_{a=0}^2\mathcal{D}_a\mathcal{A}^{(\Sigma_3)}\wedge\ast\left(\mathcal{D}_a\mathcal{A}^{(\Sigma_3)}\right), \label{3int} \\
\vspace*{0.2cm}
I^{(1,2,3)}\equiv&\int_0^\pi\frac{d\theta_1}{2\pi}\int_{\Sigma_3}\sqrt{g_4}\,\,
\mathcal{D}_{\tilde{\psi}}\mathcal{A}^{(\Sigma_3)}\wedge\ast\left(\mathcal{D}_{\tilde{\psi}}\mathcal{A}^{(\Sigma_3)}\right),
\nonumber
\end{align}
Clearly, the Hodge duals here are with respect to (\ref{metricsigma3}).
In the following, we determine all these three terms separately.

\subsubsection*{Computation of $I^{(1,2,1)}$ in (\ref{3int})}

To begin with, we focus on $I^{(1,2,1)}$ in (\ref{3int}). Using (\ref{ast3viel}) and (\ref{3101}),
it is a matter or quick and easy algebra to obtain
\begin{align}
\mathcal{A}^{(\Sigma_3)}\wedge\mathcal{A}^{(\Sigma_3)}=&[\hat{\alpha}_1,\hat{\alpha}_2]e_{\tilde{3}}^{(\Sigma_3)}
\wedge e_r^{(\Sigma_3)}+[\hat{\alpha}_1,\hat{\alpha}_3]e_{\tilde{3}}^{(\Sigma_3)}\wedge e_{\phi_1}^{(\Sigma_3)}
+[\hat{\alpha}_2,\hat{\alpha}_3]e_r^{(\Sigma_3)}\wedge e_{\phi_1}^{(\Sigma_3)},\nonumber \\ 
\ast \left(\mathcal{A}^{(\Sigma_3)}\wedge\mathcal{A}^{(\Sigma_3)}\right)=&
[\hat{\alpha}_1,\hat{\alpha}_2]e_{\phi_1}^{(\Sigma_3)}
-[\hat{\alpha}_1,\hat{\alpha}_3]e_r^{(\Sigma_3)}
+[\hat{\alpha}_2,\hat{\alpha}_3]e_{\tilde{3}}^{(\Sigma_3)}. 
\end{align}
The wedge product of the above two quantities is then
\begin{align}
\mathcal{A}^{(\Sigma_3)}\wedge\mathcal{A}^{(\Sigma_3)}\wedge
\ast \left(\mathcal{A}^{(\Sigma_3)}\wedge\mathcal{A}^{(\Sigma_3)}\right)=
\left([\hat{\alpha}_1,\hat{\alpha}_2]^2+[\hat{\alpha}_1,\hat{\alpha}_3]^2+[\hat{\alpha}_2,\hat{\alpha}_3]^2\right)
e_{\tilde{3}}^{(\Sigma_3)}\wedge e_r^{(\Sigma_3)}\wedge e_{\phi_1}^{(\Sigma_3)}. 
\end{align}
From the above, as well as our definitions in (\ref{g4}), (\ref{vielsigma3}) and (\ref{alphasdef}),
it follows (without much algebraic effort)
that $I^{(1,2,1)}$ in (\ref{3int})
can be rewritten as in (3.105) in~\cite{Dasgupta:2016rhc}:
\begin{align}
 \label{Sint1}
I^{(1,2,1)}=a_1[\mathcal{A}_r,\mathcal{A}_{\phi_1}-\frac{a_3}{2a_1}\mathcal{A}_{\tilde{3}}]^2
+\frac{4a_1a_2-a_3^2}{4a_1}
[\mathcal{A}_{\tilde{3}},\mathcal{A}_r]^2+a_4[\mathcal{A}_{\tilde{3}},\mathcal{A}_{\phi_1}]^2,
\end{align}
where we have defined, using (\ref{dtildezeta}),
\begin{align}
\label{asdef}
\begin{array}{llll}
&a_1\equiv \displaystyle\int d^4\tilde{\zeta}\sqrt{\frac{H_4}{F_1}}\left(\frac{1}{H_3}+\frac{f_3^2}{H_2}\right), 
&\quad\quad a_2\equiv \displaystyle\int d^4\tilde{\zeta}\sqrt{\frac{H_4}{F_1}}\frac{1}{H_2}, \\ \\
&a_3\equiv 2\displaystyle\int d^4\tilde{\zeta}\sqrt{\frac{H_4}{F_1}}\frac{f_3}{H_2},
&\quad\quad a_4\equiv e^{2\phi_0} \displaystyle\int d^4\tilde{\zeta}\frac{\sqrt{H_4F_1}}{H_2H_3}.
\end{array}
\end{align}
These coefficients can be easily written in terms of the warp factors using
(\ref{Hs}). Further, remember our warp factor choices in (\ref{easychoice}),
the definition of $\tilde{F}_2$ in (\ref{tildeF2}) and our assumption of constant dilaton in (\ref{consdil}).
Then, it is clear that the $a_i$'s (with $i=1,\,2,\,3,\,4$) only depend on the $(r,\,\theta_1)$ coordinates
and so the $(\tilde{x}_3,\,\phi_1)$ integrals in (\ref{dtildezeta}) are trivial and can be carried out right away.
Altogether, we have that
\newpage
\begin{align}
\nonumber
&a_1= R_3\sec\theta_{nc}\int_0^\infty dr \sqrt{\frac{\tilde{F}_2F_3}{F_1}}\left(\mathcal{I}^{(3)}
+\tilde{F}_2^2\tan^2\theta_{nc}(1+F_2\tan^2\theta_{nc})\mathcal{I}^{(4)}\right), \\ 
\label{asfinal}
& a_2= 2R_3\sec\theta_{nc}\int_0^\infty dr \sqrt{\frac{\tilde{F}_2F_3}{F_1}}\mathcal{I}^{(1)}, \qquad\qquad
a_3\propto \mathcal{I}^{(5)},\\ 
\nonumber
&a_4= e^{2\phi_0}R_3\sec\theta_{nc} 
\int_0^\infty dr  \sqrt{F_1\tilde{F}_2F_3}(\cos^2\theta_{nc}+F_2\sin^2\theta_{nc})\mathcal{I}^{(3)},
\end{align}
where $\mathcal{I}^{(1)}$ was defined in (\ref{mathcalI12}) and where we have further defined
\begin{align}
&\mathcal{I}^{(3)}\equiv\int_0^\pi d\theta_1 \,\,\sin\theta_1\sqrt{\tilde{F}_2\cos^2\theta_1+F_3\sin^2\theta_1}, \quad\quad
\mathcal{I}^{(4)}\equiv\int_0^\pi 
\frac{\sin\theta_1\cos^2\theta_1d\theta_1}{\sqrt{\tilde{F}_2\cos^2\theta_1+F_3\sin^2\theta_1}}, \nonumber \\
&\mathcal{I}^{(5)}\equiv \int_0^\pi d\theta_1
\frac{\sin\theta_1\cos\theta_1}{\sqrt{\tilde{F}_2\cos^2\theta_1+F_3\sin^2\theta_1}}.\label{I34int}
\end{align}

It is most interesting to note that $a_3$ vanishes, since
\begin{align}
\mathcal{I}^{(5)}\propto 
1-\sqrt{\frac{\tilde{F}_2+F_3+(\tilde{F}_2-F_3)\cos2\theta_1}{\tilde{F}_2+F_3}}\Bigg|
_{\theta_1=0}^{\theta_1=\pi}=0, \label{zeroint1}
\end{align}
as noted in (3.108) in~\cite{Dasgupta:2016rhc} too.
This greatly simplifies $I^{(1,2,1)}$ in (\ref{Sint1}).
Specifically, (\ref{zeroint1}) implies that there are no crossed terms
for the interactions among the real scalars $(\mathcal{A}_{\tilde{3}},\,\mathcal{A}_{\phi_1},\,\mathcal{A}_r)$:
\begin{align}
\label{3114}
I^{(1,2,1)}=a_1[\mathcal{A}_r,\mathcal{A}_{\phi_1}]^2
+a_2[\mathcal{A}_{\tilde{3}},\mathcal{A}_r]^2+a_4[\mathcal{A}_{\tilde{3}},\mathcal{A}_{\phi_1}]^2,
\end{align}
in good agreement with (3.114) in~\cite{Dasgupta:2016rhc}.
In the ongoing, we shall focus in the determination of the remaining coefficients in (\ref{asfinal}) and show that they are
well defined numbers for any choice of the warp factors one may wish to consider.

With this aim in mind, we start by performing the integrals in (\ref{I34int}).
Using our definitions in (\ref{chitheta1}), we obtain for $\mathcal{I}^{(3)}$
\begin{align}
\mathcal{I}^{(3)}=-\frac{1}{4}\left(\sqrt{2}\cos\theta_1\chi(\theta_1) 
+\frac{2F_3}{\sqrt{\tilde{F}_2-F_3}}\ln\left|
\chi(\theta_1)+\tilde{\chi}(\theta_1)\right|\right)\Bigg|_{\theta_1=0}^{\theta_1=\pi} 
=\sqrt{\tilde{F}_2}+\frac{F_3\mathcal{J}_3}{2\sqrt{\tilde{F}_2-F_3}}, \label{I3res}
\end{align}
where $\mathcal{J}_3$ was defined in (\ref{mathj3}).
Similarly, $\mathcal{I}^{(4)}\equiv (\tilde{F}_2-F_3)\mathcal{I}^{(4)}$ gives
\begin{align}
\tilde{\mathcal{I}}^{(4)}=\frac{1}{4}\left(-\sqrt{2}\cos\theta_1\chi(\theta_1)
+\frac{2F_3}{\sqrt{\tilde{F}_2-F_3}}\ln\left|\chi(\theta_1)+\tilde{\chi}(\theta_1)
\right|\right)\Bigg|_{\theta_1=0}^{\theta_1=\pi} 
=\sqrt{\tilde{F}_2}
-\frac{F_3\mathcal{J}_3}{2\sqrt{\tilde{F}_2-F_3}}. \label{I4res}
\end{align}
We remind the reader that $\mathcal{I}^{(1)}$ was determined in (\ref{mathI1sol}) already.
Then, substitution of these results in (\ref{asfinal}) immediately gives us the coefficients
$(a_1,\,a_2,\,a_4)$ in the desired form:
\newpage
\begin{align}
&a_1=R_3\sec\theta_{nc}\int_0^\infty dr \sqrt{\frac{\tilde{F}_2F_3}{F_1}}
\left(\tilde{a}_+\sqrt{\tilde{F}_2}+
\frac{\tilde{a}_-F_3\mathcal{J}_3}{2\sqrt{\tilde{F}_2-F_3}}\right)  \label{a1a2a4} \\
&a_2=R_3\sec\theta_{nc}\int_0^\infty dr\,\, \tilde{a}_2\mathcal{J}_3, \quad\quad \nonumber
a_4=R_3\sec\theta_{nc}\int_0^\infty dr  \,\, \tilde{a}_4\left(\sqrt{\tilde{F}_2}+\frac{F_3\mathcal{J}_3}{2\sqrt{\tilde{F}_2-F_3}}
\right), 
\end{align}
which are (3.106), (3.109) and (3.110) in~\cite{Dasgupta:2016rhc}, respectively.
Following (3.107) and (3.111) in~\cite{Dasgupta:2016rhc}, the $(\tilde{a}_\pm,\,\tilde{a}_2,\,\tilde{a}_4)$ coefficients
appearing above are defined as
\begin{align}
\tilde{a}_{\pm}\equiv& 1\pm \frac{(\tilde{F}_2\tan\theta_{nc})^2}{\tilde{F}_2-F_3}(1+F_2\tan^2\theta_{nc}), \nonumber \\
\tilde{a}_2\equiv& (\cos^2\theta_{nc}+F_2\sin^2\theta_{nc})
\sqrt{\frac{\tilde{F}_2F_3}{F_1(\tilde{F}_2-F_3)}}, \label{tildea1a2} \\
\tilde{a}_4\equiv& e^{2\phi_0}(\cos^2\theta_{nc}+F_2\sin^2\theta_{nc})\sqrt{F_1\tilde{F}_2F_3}. \nonumber
\end{align}
Upon a careful inspection of the coefficients in (\ref{a1a2a4}), it is not hard to convince oneself that
these all are just numbers for any choice of the warp factors in (\ref{easychoice}).
The only constraint is that $\tilde{F}_2\geq F_3$ should hold true,
as was the case for the other coefficients as well.

In short, $I^{(1,2,1)}$ is given by (\ref{3114}), with $(a_1,\,a_2,\,a_3)$ in (\ref{a1a2a4}) well defined numbers
for any choice of warp factors satisfying $\tilde{F}_2\geq F_3$.

\subsubsection*{Computation of $I^{(1,2,2)}$ in (\ref{3int})}

We now turn our attention to $I^{(1,2,2)}$ in (\ref{3int}). From (\ref{3101}),
it is easy to obtain
\begin{align}
\mathcal{D}_a\mathcal{A}^{(\Sigma_3)}=
(\mathcal{D}_a\hat{\alpha}_1)e_{\tilde{3}}^{(\Sigma_3)}
+(\mathcal{D}_a\hat{\alpha}_2)e_r^{(\Sigma_3)}
+(\mathcal{D}_a\hat{\alpha}_3)e_{\phi_1}^{(\Sigma_3)}.
\label{DaA}
\end{align}
The Hodge dual of the above with respect to the metric (\ref{metricsigma3}) is straightforward, in view of (\ref{ast3viel})
and is given by
\begin{align}
\ast\mathcal{D}_a\mathcal{A}^{(\Sigma_3)}=(\mathcal{D}_a\hat{\alpha}_1)
e_r^{(\Sigma_3)}\wedge e_{\phi_1}^{(\Sigma_3)}
-(\mathcal{D}_a\hat{\alpha}_2)e_{\tilde{3}}^{(\Sigma_3)}\wedge e_{\phi_1}^{(\Sigma_3)}
+(\mathcal{D}_a\hat{\alpha}_3)e_{\tilde{3}}^{(\Sigma_3)}\wedge e_{r}^{(\Sigma_3)}.
\end{align}
The wedge product of the above two quantities is
\begin{align}
(\mathcal{D}_a\mathcal{A}^{(\Sigma_3)})\wedge\ast (\mathcal{D}_a\mathcal{A}^{(\Sigma_3)}) 
=\left[(\mathcal{D}_a\hat{\alpha}_1)^2+(\mathcal{D}_a\hat{\alpha}_2)^2+(\mathcal{D}_a\hat{\alpha}_3)^2\right]
e_{\tilde{3}}^{(\Sigma_3)}\wedge e_r^{(\Sigma_3)}\wedge e_{\phi_1}^{(\Sigma_3)}.
\label{tracekin}
\end{align}
Feeding the above to (\ref{3int}) and further using (\ref{g4}), (\ref{vielsigma3}) and (\ref{alphasdef}),
$I^{(1,2,2)}$ can be written as 
\begin{align}
\label{fullI122}
I^{(1,2,2)}=\sum_{a=0}^2\left[
c_{a\tilde{3}}(\mathcal{D}_a\mathcal{A}_{\tilde{3}}-\frac{\mu}{c_{a\tilde{3}}}\mathcal{D}_a\mathcal{A}_{\phi_1})^2
+c_{ar}(\mathcal{D}_a\mathcal{A}_r)^2
+c_{a\phi_1}(\mathcal{D}_a\mathcal{A}_{\phi_1})^2\right],
\end{align}
where, making use of (\ref{dtildezeta}),
we have further defined the coefficients
\begin{align}
\label{car}
\begin{array}{lll}
&c_{a\tilde{3}}\equiv  \displaystyle e^{2\phi_0}\int d^4\tilde{\zeta} \,\,\frac{\sqrt{H_4F_1}}{H_2}, &\quad\quad
\mu\equiv \displaystyle e^{2\phi_0} \int d^4\tilde{\zeta} \,\, \sqrt{H_4F_1}\frac{f_3}{H_2}, \\ \\
&c_{ar}\equiv \displaystyle \int d^4\tilde{\zeta} \,\,\sqrt{\frac{H_4}{F_1}}, &\quad\quad
c_{a\phi_1}\equiv \displaystyle e^{2\phi_0}\int d^4\tilde{\zeta} \,\, \frac{\sqrt{H_4F_1}}{H_3}.
\end{array}
\end{align}
These coefficients can be written in terms of the warp factors using
(\ref{Hs}). Exactly as was the case before with the coefficients in (\ref{asdef}),
the $(\tilde{x}_3,\,\phi_1)$ integrals are trivial here too.
Thus, we have that
\begin{align}
\label{cswarp}
\begin{array}{llll}
&c_{a\tilde{3}}=\displaystyle2R_3\sec\theta_{nc}\int_0^\infty dr\,\,\tilde{a}_4\mathcal{I}^{(1)}, &\quad\quad
\mu\displaystyle\propto\mathcal{I}^{(5)}, \\ \\
&c_{ar}=\displaystyle2R_3\sec\theta_{nc}\int_0^\infty dr\sqrt{\frac{\tilde{F}_2F_3}{F_1}}\mathcal{I}^{(1)}, &\quad\quad
c_{a\phi_1}=\displaystyle e^{2\phi_0}R_3\sec\theta_{nc}\int_0^\infty dr\,\,\sqrt{F_1\tilde{F}_2F_3}\mathcal{I}^{(3)},
\end{array}
\end{align}
where $(\mathcal{I}^{(1)},\,\mathcal{I}^{(3)},\,\mathcal{I}^{(5)},\,\tilde{a}_4)$ were defined in (\ref{mathcalI12}),
(\ref{I34int}) and (\ref{tildea1a2}), respectively.

In a similar fashion to what happened in the determination of $I^{(1,2,1)}$,
the result in (\ref{zeroint1}) makes $\mu$ vanish. This implies that there are no crossed terms
for the kinetic terms of $(\mathcal{A}_{\tilde{3}},\,\mathcal{A}_{\phi_1},\,\mathcal{A}_r)$
we presently study. In other words, (\ref{fullI122}) reduces to the second line of (3.115) in~\cite{Dasgupta:2016rhc}:
\begin{align}
\label{3115b}
I^{(1,2,2)}=\sum_{a=0}^2\left[
c_{a\tilde{3}}(\mathcal{D}_a\mathcal{A}_{\tilde{3}})^2
+c_{ar}(\mathcal{D}_a\mathcal{A}_r)^2
+\tilde{c}_{a\phi_1}(\mathcal{D}_a\mathcal{A}_{\phi_1})^2\right],
\end{align}
with $\tilde{c}_{a\phi_1}$ defined as
\begin{align}
\tilde{c}_{a\phi_1}\equiv c_{a\phi_1}+\frac{\mu^2}{c_{a\tilde{3}}}, \quad\quad
\frac{\mu^2}{c_{a\tilde{3}}}=R_3\sec^3\theta_{nc}\tan^2\theta_{nc}\int_0^\infty dr\,\,\tilde{a}_4\tilde{F}_2\mathcal{I}^{(4)}.
\end{align}
In writing the second equality above, we have made use of all (\ref{Hs}), (\ref{dtildezeta}), (\ref{I34int}),
(\ref{tildea1a2}) and (\ref{car}). At this point, we are left with only the task of computing 
$(c_{a\tilde{3}},\,c_{ar},\,\tilde{c}_{a\phi_1})$ and showing they are all some real number.

The computation part is straightforward, in view of our earlier results in (\ref{mathI1sol}), (\ref{I3res}) and
(\ref{I4res}). We thus obtain (3.117)-(3.119) in~\cite{Dasgupta:2016rhc}:
\begin{align}
&c_{a\tilde{3}}=R_3\sec\theta_{nc}\int_0^\infty dr\frac{\tilde{a}_4\mathcal{J}_3}{\sqrt{\tilde{F}_2-F_3}}, \quad\quad
c_{ar}=R_3\sec\theta_{nc}\int_0^\infty dr\,\,\mathcal{J}_3 \sqrt{\frac{\tilde{F}_2F_3}{F_1(\tilde{F}_2-F_3)}},
\nonumber \\
&\tilde{c}_{a\phi_1}=e^{2\phi_0}R_3\sec\theta_{nc}\int_0^\infty dr \sqrt{F_1\tilde{F}_2F_3}
\left(\tilde{a}_+\sqrt{\tilde{F}_2}+
\frac{\tilde{a}_-F_3\mathcal{J}_3}{2\sqrt{\tilde{F}_2-F_3}}\right), \label{csfinal}
\end{align}
where $(\mathcal{J}_3,\,\tilde{a}_\pm,\,\tilde{a}_4)$ were defined in (\ref{mathj3}) and (\ref{tildea1a2}), respectively.
On the other hand, the issue of proving that all three coefficients above are numbers is also simple enough.
Once again, one must demand that $\tilde{F}_2\geq F_3$ to prevent the ``blowing up'' of these
quantities. However, any value of the warp factors in (\ref{easychoice}) satisfying this constraint
can be readily seen to yield a finite, real result when used in (\ref{csfinal}).

Consequently, we conclude that $I^{(1,2,2)}$ is given by (\ref{3115b}), with $(c_{a\tilde{3}},\,c_{ar},\,\tilde{c}_{a\phi_1})$
there appearing given by (\ref{csfinal}). These are well defined numbers as long as the warp factors
are chosen such that $\tilde{F}_2\geq F_3$.

\subsubsection*{Computation of $I^{(1,2,3)}$ in (\ref{3int})}

At last, we consider  $I^{(1,2,3)}$ in (\ref{3int}). Its computation is very similar to that of $I^{(1,2,2)}$,
albeit algebraically more involved. In the following, we show all the relevant details.
With the aid of (\ref{ast3viel}) and (\ref{3101}), it is easy to see that
\begin{align}
\mathcal{D}_{\tilde{\psi}}\mathcal{A}^{(\Sigma_3)}\wedge
\ast \left(\mathcal{D}_{\tilde{\psi}}\mathcal{A}^{(\Sigma_3)}\right)=
\left[\left(\mathcal{D}_{\tilde{\psi}}\hat{\alpha}_1\right)^2
+\left(\mathcal{D}_{\tilde{\psi}}\hat{\alpha}_2\right)^2+\left(\mathcal{D}_{\tilde{\psi}}\hat{\alpha}_3\right)^2\right]
e_{\tilde{3}}^{(\Sigma_3)}\wedge e_r^{(\Sigma_3)}\wedge e_{\phi_1}^{(\Sigma_3)}. 
\end{align}
Using the above and the definitions in (\ref{g4}), (\ref{vielsigma3}) and (\ref{alphasdef}) in (\ref{3int}),
one can rewrite $I^{(1,2,3)}$ as
\begin{align}
I^{(1,2,3)}=
c_{\tilde{\psi} \tilde{3}}\left(\mathcal{D}_{\tilde{\psi}}\mathcal{A}_{\tilde{3}}-\frac{\nu}{c_{\tilde{\psi}\tilde{3}}}
\mathcal{D}_{\tilde{\psi}}\mathcal{A}_{\phi_1}\right)^2
+c_{\tilde{\psi} r}\left(\mathcal{D}_{\tilde{\psi}}\mathcal{A}_r\right)^2
+c_{\tilde{\psi}\phi_1}\left(\mathcal{D}_{\tilde{\psi}}\mathcal{A}_{\phi_1}\right)^2,
\label{I123mid}
\end{align}
where, making use of (\ref{dtildezeta}), we have defined
\begin{align}
\begin{array}{llll}
&\displaystyle c_{\tilde{\psi} \tilde{3}}\equiv \int \frac{d^4\tilde{\zeta}}{H_2}\sqrt{\frac{F_1}{H_4}}, &\quad\quad
\displaystyle\nu\equiv \int d^4\tilde{\zeta}\frac{f_3}{H_2}\sqrt{\frac{F_1}{H_4}}, \\ \\
&\displaystyle c_{\tilde{\psi}  r}\equiv e^{-2\phi_0}\int \frac{d^4\tilde{\zeta}}{\sqrt{H_4F_1}}, &\quad\quad
\displaystyle c_{\tilde{\psi} \phi_1}\equiv \int \frac{d^4\tilde{\zeta}}{H_3}\sqrt{\frac{F_1}{H_4}}.
\end{array}
\label{cpsir}
\end{align}
These coefficients can be expressed in terms of the warp factors in (\ref{easychoice}) by inserting (\ref{Hs})
in the above. It is again the case that the $(\tilde{x}_3,\,\phi_1)$ integrals are trivial
and so we obtain
\begin{align}
\begin{array}{llll}
&\displaystyle c_{\tilde{\psi} \tilde{3}}=2R_3\cos\theta_{nc}\int_0^\infty dr
\frac{\tilde{b}_2\mathcal{I}^{(2)}}{\sqrt{\tilde{F}_2-F_3}}, &\quad\quad
\displaystyle\nu\propto\mathcal{I}^{(6)}, \\ \\
&\displaystyle c_{\tilde{\psi}  r}=4e^{-2\phi_0}R_3\cos\theta_{nc}\int_0^\infty dr
\frac{b_2\mathcal{I}^{(2)}}{F_1\sqrt{\tilde{F}_2-F_3}},
&\quad\quad \displaystyle c_{\tilde{\psi} \phi_1}=R_3\cos\theta_{nc}\int_0^\infty dr\sqrt{\frac{F_1}{\tilde{F}_2F_3}}
\mathcal{I}^{(7)},
\end{array}
\end{align}
Here, we have defined $\tilde{b}_2$ as a slight variant of $b_2$ in (\ref{379}):
\begin{align}
\tilde{b}_2\equiv (\cos^2\theta_{nc}+F_2\sin^2\theta_{nc})\sqrt{\frac{F_1(\tilde{F}_2-F_3)}{\tilde{F}_2F_3}},
\label{tildeb2}
\end{align}
$\mathcal{I}^{(2)}$ is as in (\ref{mathcalI12}) and the remaining integrals there appearing are defined as
\begin{align}
\begin{array}{llll}
&\displaystyle \mathcal{I}^{(6)}\equiv \int_0^\pi d\theta_1\,\,
\cot\theta_1 (\tilde{F}_2\cos^2\theta_1+F_3\sin^2\theta_1)^{1/2}, \\ 
&\displaystyle \mathcal{I}^{(7)}\equiv \int_0^\pi d\theta_1\,\, \csc\theta_1(\tilde{F}_2\cos^2\theta_1+F_3\sin^2\theta_1)^{3/2}.
\end{array}
\end{align}

In view of our earlier results for $(a_3,\,\mu)$ in (\ref{asfinal}) and (\ref{cswarp}) respectively,
it will come as no surprise that $\nu$ above vanishes. To see this,
we simply need to use $b_1$ in (\ref{379}) and the change of variables in (\ref{theta1toz}).
Then, after regularization, $\mathcal{I}^{(6)}$ vanishes by symmetry:
\begin{align}
\mathcal{I}^{(6)}\propto
\int_{-1}^1 dz\frac{z(b_1^2+z^2)^{1/2}}{b^2-z^2}=0, \quad\quad b\in(\mathbb{R}^+-\{1\}).
\end{align}
Therefore, (\ref{I123mid}) simplifies considerably, leading to no crossed terms
between the kinetic terms of $(\mathcal{A}_{\tilde{3}},\,\mathcal{A}_{\phi_1},\,\mathcal{A}_r)$ here considered:
\begin{align}
I^{(1,2,3)}=
c_{\tilde{\psi} \tilde{3}}\left(\mathcal{D}_{\tilde{\psi}}\mathcal{A}_{\tilde{3}}\right)^2
+c_{\tilde{\psi} r}\left(\mathcal{D}_{\tilde{\psi}}\mathcal{A}_r\right)^2
+\tilde{c}_{\tilde{\psi}\phi_1}\left(\mathcal{D}_{\tilde{\psi}}\mathcal{A}_{\phi_1}\right)^2.
\label{3115a}
\end{align}
This is the first line of (3.115) in~\cite{Dasgupta:2016rhc}, with $\tilde{c}_{\tilde{\psi}\phi_1}$ defined as
\begin{align}
\tilde{c}_{\tilde{\psi}\phi_1}\equiv c_{\tilde{\psi}\phi_1}+\frac{\nu^2}{c_{\tilde{\psi} \tilde{3}}}, \quad\quad
\frac{\nu^2}{c_{\tilde{\psi}}}=R_3\sec\theta_{nc}\tan^2\theta_{nc}\int_0^\infty dr\,\,\tilde{a}_2\sqrt{\tilde{F}_2-F_3}
\frac{F_1\tilde{F}_2}{F_3}\mathcal{I}^{(8)}.
\end{align}
In order to obtain the second equality above, the definitions in (\ref{Hs}), (\ref{dtildezeta}),
(\ref{tildea1a2}) and (\ref{cpsir}) have been used and we have further introduced
\begin{align}
\mathcal{I}^{(8)}\equiv \int_0^\pi d\theta_1 \frac{\cos^2\theta_1}{\sin\theta_1}\sqrt{\tilde{F}_2\cos^2\theta_1
+F_3\sin^2\theta_1}.
\end{align}
Hence, we are only left with the task of computing
$(c_{\tilde{\psi} \tilde{3}},\,c_{\tilde{\psi} r},\,\tilde{c}_{\tilde{\psi}\phi_1})$.

To do so, we first recall $\mathcal{I}^{(2)}$ was already determined in (\ref{432}) and so we still need to perform
the integrals $(\mathcal{I}^{(7)},\,\mathcal{I}^{(8)})$.
For $\mathcal{I}^{(7)}$, it is convenient to do the same set of transformations that
we considered for $\mathcal{I}^{(2)}$ between (\ref{429}) and (\ref{432}) earlier on.
Namely, 
\begin{align}
\frac{\mathcal{I}^{(7)}}{(\tilde{F}_2-F_3)^{3/2}}=&
\int_{-1}^1 dz\frac{(b_1^2+z^2)^{3/2}}{b^2-z^2} 
=\frac{(b_1^2+b^2)^{3/2}}{b}\eta(z)
-\frac{3b_1^2+2b^2}{2}\tilde{\eta}(z)-\frac{z}{2}\sqrt{b_1^2+z^2}\Bigg|_{z=-1}^{z=1}
\nonumber \\
=&\frac{b^2}{4}b_3^3\mathcal{J}_4
-\frac{3b_1^2+2b^2}{2}\mathcal{J}_3-\sqrt{\frac{\tilde{F}_2}{\tilde{F}_2-F_3}},
\end{align}
where $b\in(\mathbb{R}^+-\{1\})$ is a regularization factor, $(\eta(z),\,\tilde{\eta}(z))$ were defined in (\ref{etasdef})
and in the last step we have used (\ref{mathj3}), (\ref{379})
and (\ref{mathj4}).
In fact, we can do essentially the same for $\mathcal{I}^{(8)}$ and obtain
\begin{align}
\frac{\mathcal{I}^{(8)}}{\sqrt{\tilde{F}_2-F_3}}=&\int_{-1}^1dz\,\,z^2\frac{\sqrt{b_1^2+z^2} }{b^2-z^2}
= b\sqrt{b_1^2+b^2}\eta(z)
-\frac{b_1^2+2b^2}{2}
\tilde{\eta}(z)-\frac{z}{2}\sqrt{b_1^2+z^2}\Bigg|_{z=-1}^{z=1} \nonumber \\
=&b^2b_3\mathcal{J}_4
-\frac{b_1^2+2b^2}{2}\mathcal{J}_3
-\sqrt{\frac{\tilde{F}_2}{\tilde{F}_2-F_3}}.
\end{align}
With all these results at hand,
it is now a matter of substitution and easy algebra to obtain the desired
coefficients as in (3.121) and (3.124) in~\cite{Dasgupta:2016rhc}:
\begin{align}
&c_{\tilde{\psi} \tilde{3}}=R_3\cos\theta_{nc}\int_0^\infty dr\,\,
\tilde{b}_2\left(
b_3\mathcal{J}_4
+\mathcal{J}_3^{-1}\right),
\quad\quad
\tilde{c}_{\tilde{\psi}\phi_1}=\int_0^\infty dr\,\,
\left(a_{01}\mathcal{J}_4+b_{01}\mathcal{J}_3^{-1}-c_{01}
\right), \nonumber \\
&c_{\tilde{\psi} r}=2e^{-2\phi_0}R_3\cos\theta_{nc}\int_0^\infty dr\,\, \frac{b_2}{F_1}\left(b_3\mathcal{J}_4
+\mathcal{J}_3^{-1}\right).
\label{cpsis}
\end{align}
Recall that $(\tilde{F}_2,\,b_2,\,b_3,\,\tilde{b}_2)$ were defined in (\ref{tildeF2}),
(\ref{379})
and (\ref{tildeb2}),
respectively. Following (3.125)-(3.128) in~\cite{Dasgupta:2016rhc}, the other factors in $\tilde{c}_{\tilde{\psi}\phi_1}$
are defined as
\newpage
\begin{align}
a_{01}\equiv & R_3b^2b_3(\tilde{F}_2-F_3)\left(\cos\theta_{nc}b_3^2\frac{\sqrt{F_1(\tilde{F}_2-F_3)}}{4\tilde{F}_2F_3}
+\tilde{a}_2\frac{\tan^2\theta_{nc}}{\cos\theta_{nc}}\frac{F_1\tilde{F}_2}{F_3}\right), \nonumber \\
b_{01}\equiv & \frac{R_3}{2}\sqrt{\frac{F_1}{F_3}}\left(\cos\theta_{nc}f^{(1)}
\sqrt{\frac{\tilde{F}_2-F_3}{\tilde{F}_2}}+\tilde{a}_2\tilde{F}_2f^{(2)}\frac{\tan^2\theta_{nc}}{\cos\theta_{nc}}
\sqrt{\frac{F_1}{F_3}}\right),
\label{abc01} \\
c_{01} \equiv &R_3(\tilde{F}_2-F_3)\sqrt{\frac{F_1}{F_3}}\left(\cos\theta_{nc}
+\tilde{a}_2\tilde{F}_2^2\frac{\tan^2\theta_{nc}}{\cos\theta_{nc}}\sqrt{\frac{F_1(\tilde{F}_2-F_3)}{\tilde{F}_2F_3}}\right),
\nonumber
\end{align}
with $(f^{(1)},\,f^{(2)})$ given by
\begin{align}
f^{(1)}\equiv 3F_3+2b^2(\tilde{F}_2-F_3), \quad\quad f^{(2)}\equiv f^{(1)}-2F_3.
\label{f12def}
\end{align}

In exactly the same way shown in the end of section \ref{I11} for $c_{12}$,
it follows that $(c_{\tilde{\psi} \tilde{3}},\,c_{\tilde{\psi} r})$ are just numbers for any choice of the warp factors
satisfying $\tilde{F}_2\geq F_3$. The scenario is more subtle in the case of $\tilde{c}_{\tilde{\psi}\phi_1}$.
It is not clear at all that this coefficient is finite when
\begin{itemize}
\item $F_3\rightarrow 0$. (As discussed after (\ref{c12final}), 
this limit also includes the case $(\tilde{F}_2,\,F_3)\rightarrow0$.)
\item $\tilde{F}_2\rightarrow F_3\nrightarrow0$.
\end{itemize}
However, it turns out that
\begin{align}
\lim_{F_3\rightarrow 0}\tilde{c}_{\tilde{\psi}\phi_1}=0,
\end{align}
the mathematical details precisely as in between (\ref{lim1}) and (\ref{limfin}) for $c_{12}$ before.
Consequently, we will just show that $\tilde{c}_{\tilde{\psi}\phi_1}$ is well defined when $\tilde{F}_2\rightarrow F_3$.
To do this, we call $\epsilon^2\equiv \tilde{F}_2-F_3$ and take the $\epsilon\rightarrow0$ limit.
Used in $(b_3,\,\tilde{a}_4)$ in (\ref{379}) and (\ref{tildea1a2}), we get
\begin{align}
\lim_{\epsilon\rightarrow0}b_3\sim\lim_{\epsilon\rightarrow0}\frac{1}{\epsilon}\sim\lim_{\epsilon\rightarrow0}\tilde{a}_2.
\end{align}
Then, feeding the above to (\ref{abc01}), we obtain
\begin{align}
\lim_{\epsilon\rightarrow 0}a_{01}\sim1, \quad\quad
\lim_{\epsilon\rightarrow 0}b_{01}\sim\lim_{\epsilon\rightarrow0}\frac{1}{\epsilon},
\quad\quad \lim_{\epsilon\rightarrow 0}c_{01}=0.
\end{align}
We consider this very same limit for $(\mathcal{J}_3,\,\mathcal{J}_4)$
in (\ref{mathj3}) and (\ref{mathj4}):
\begin{align}
\lim_{\epsilon\rightarrow 0}\mathcal{J}_3=\lim_{\epsilon\rightarrow 0}\ln\left|\frac{1+\epsilon}{1-\epsilon}\right|,
\quad\quad \lim_{\epsilon\rightarrow 0}\mathcal{J}_4=\arctanh\frac{1}{b},
\end{align}
which is finite, as $b\neq1$ by definition. All the above can be used in $\tilde{c}_{\tilde{\psi}\phi_1}$ in (\ref{cpsis}).
Retaining only the divergent part, we have that
\begin{align}
\lim_{\epsilon\rightarrow0}\tilde{c}_{\tilde{\psi}\phi_1}\sim \lim_{\epsilon\rightarrow0}\frac{1}{\epsilon}
\ln\left|\frac{1+\epsilon}{1-\epsilon}\right|=\lim_{\epsilon\rightarrow0}
\left(\frac{1}{1+\epsilon}+\frac{1}{1-\epsilon}\right)=2,
\end{align}
where in the last step we have applied L'H\^{o}pital's rule. In other words, the seemingly divergent
part of $\tilde{c}_{\tilde{\psi}\phi_1}$ is actually finite. Thus, $\tilde{c}_{\tilde{\psi}\phi_1}$
is a well defined number for any warp factors one may wish to consider, as long as $\tilde{F}_2\geq F_3$. 

Quickly summing up, $I^{(1,2,3)}$ si given by (\ref{3115a}) and the coefficients
$(c_{\tilde{\psi} \tilde{3}},\,c_{\tilde{\psi} r},\,\tilde{c}_{\tilde{\psi}\phi_1})$
there appearing are all well defined numbers if $\tilde{F}_2\geq F_3$. Their explicit form is that in (\ref{cpsis}).

We can finally collect all our results so far into a quite simple form.
First, we use (\ref{3114}), (\ref{3115b}) and (\ref{3115a}) in (\ref{I12split}) and write $I^{(1,2)}$ as
\begin{align}
I^{(1,2)}=&\int d^4x\,\textrm{Tr}\Bigg\{a_1[\mathcal{A}_r,\mathcal{A}_{\phi_1}]^2
+a_2[\mathcal{A}_{\tilde{3}},\mathcal{A}_r]^2+a_4[\mathcal{A}_{\tilde{3}},\mathcal{A}_{\phi_1}]^2 \nonumber \\
&+\sum_{a=0}^2\left[
c_{a\tilde{3}}(\mathcal{D}_a\mathcal{A}_{\tilde{3}})^2
+c_{ar}(\mathcal{D}_a\mathcal{A}_r)^2
+\tilde{c}_{a\phi_1}(\mathcal{D}_a\mathcal{A}_{\phi_1})^2\right] \\
&+c_{\tilde{\psi} \tilde{3}}\left(\mathcal{D}_{\tilde{\psi}}\mathcal{A}_{\tilde{3}}\right)^2
+c_{\tilde{\psi} r}\left(\mathcal{D}_{\tilde{\psi}}\mathcal{A}_r\right)^2
+\tilde{c}_{\tilde{\psi}\phi_1}\left(\mathcal{D}_{\tilde{\psi}}\mathcal{A}_{\phi_1}\right)^2
\Bigg\}. \nonumber
\end{align}
Now, inserting (\ref{genform}) and the above in (\ref{split1}), the first term of
the bosonic action for the $SU(N)$ world-volume gauge theory along
$(t,\,x_1,\,x_2,\,\tilde{\psi})$ can be readily seen to be
\begin{align}
\nonumber
S^{(1)}=&\frac{C_1c_{11}}{V_3}\int d^4x \sum_{\substack{a,b=0 \\ a<b}}^2\textrm{Tr}(\mathcal{F}_{ab}^2)+
\frac{C_1c_{12}}{V_3}\int d^4x
\sum_{a=0}^2 \textrm{Tr}(\mathcal{F}_{a\tilde{\psi}}^2) \\
&+\frac{C_1}{V_3}\int d^4x\,\textrm{Tr}\Bigg\{a_1[\mathcal{A}_r,\mathcal{A}_{\phi_1}]^2
+a_2[\mathcal{A}_{\tilde{3}},\mathcal{A}_r]^2+a_4[\mathcal{A}_{\tilde{3}},\mathcal{A}_{\phi_1}]^2 \nonumber \\
&+\sum_{a=0}^2\left[
c_{a\tilde{3}}(\mathcal{D}_a\mathcal{A}_{\tilde{3}})^2
+c_{ar}(\mathcal{D}_a\mathcal{A}_r)^2
+\tilde{c}_{a\phi_1}(\mathcal{D}_a\mathcal{A}_{\phi_1})^2\right] \nonumber \\ \label{term1action}
&+c_{\tilde{\psi} \tilde{3}}\left(\mathcal{D}_{\tilde{\psi}}\mathcal{A}_{\tilde{3}}\right)^2
+c_{\tilde{\psi} r}\left(\mathcal{D}_{\tilde{\psi}}\mathcal{A}_r\right)^2
+\tilde{c}_{\tilde{\psi}\phi_1}\left(\mathcal{D}_{\tilde{\psi}}\mathcal{A}_{\phi_1}\right)^2\Bigg\}.
\end{align}
It is important to bear in mind that
all the coefficients appearing in this first term of the action have been shown to be real numbers
for any choice of the warp factors satisfying $\tilde{F}_2\geq F_3$.
We remind the reader that any specific choice of warp factors must additionally ensure $\mathcal{N}=2$
supersymmetry. We will dwell into such considerations in section \ref{bcsec}. Presently and
without further delay, let us turn to the second term of this bosonic action.

\subsection{Mass term of the G-flux \label{masstermsec}}

In order to obtain the second term for the bosonic action of the $\mathcal{N}=2$ supersymmetric
gauge theory along $(t,\,x_1,\,x_2,\,\tilde{\psi})$, we first need to brush up a bit the construction
of the abelian M-theory configuration (M, 1) of section \ref{ns5d3sect1}. In particular, we need to
recall how we moved far away along the Coulomb branch the ${\overline{\rm D5}}$-brane of figure \ref{fig1}{\bf D}.
(Bear in mind that, as depicted, these branes stretch along the diretions $(t,\,x_1,\,x_2,\,x_3,\,\psi,\,r)$.)
In this manner, we managed to effectively ignore the presence of this ${\overline{\rm D5}}$-brane
in the configuration (B, 1) of figure \ref{fig3}, thereby simplifying the starting point of our quantitative 
derivation of (M, 1). It is now time to study the essential effects that the presence of this ${\overline{\rm D5}}$-brane
has for the gauge theory.

Let us begin by bringing back to its original position the ${\overline{\rm D5}}$-brane. In other words, 
let us consider that the D5-brane in the configuration (B, 1) has right next to it a parallel ${\overline{\rm D5}}$-brane.
To prevent the D5/${\overline{\rm D5}}$ pair from collapsing (thus giving rise to tachyons),
we switch on a small NS B-field $\tilde{B}_2^{(B,1)}$ along the directions $(x_3,\,r)$ in both the D5- and ${\overline{\rm D5}}$-branes.
As carefully explained in~\cite{Dasgupta:1999wx}, the D5/${\overline{\rm D5}}$ pair with such
an NS B-field on it can alternatively be interpreted as two fractional D3-branes spanning $(t,\,x_1,\,x_2,\,\psi)$.
(Note that our choice of orientation of the NS B-field leads to the stretching of
the fractional D3-branes along precisely the directions of the gauge theory.)
From this point of view, it is easy to infer that we must also switch on a small RR B-field $\tilde{\mathcal{C}}_2^{(B,1)}$
along the same directions $(x_3,\,r)$, so as to ensure the tadpole cancellation condition is satisfied\footnote{
The tadpole condition is, essentially, the statement that the charge of the fractional D3-branes should be conserved.
It follows directly from the Bianchi identity and equations of motion of the corresponding fluxes.
A neat derivation of the tadpole condition can be found in section 4.2 of~\cite{Grana:2005jc}.}.
As a particularly simple and consistent choice, we will consider both these fields to only depend
on the $(\theta_1,\,r)$ coordinates:
\begin{align}
\tilde{B}_2^{(B,1)}\equiv F^{(1)}dx_3\wedge dr, \quad\quad
\tilde{\mathcal{C}}_2^{(B,1)}\equiv F^{(2)}dx_3\wedge dr, \quad\quad F^{(i)}=F^{(i)}(\theta_1,r),
\quad\quad i=1,2.
\end{align}
With the goal of understanding how these new B-fields will affect the configuration (M, 1), in the following 
we will subject them to the chain of modifications in figure \ref{fig3}.

For our present purposes, it turns out we need not do the whole analysis in details, as in part \ref{parta} before.
Further, we need not worry about the NS B-field either. Rather, it suffices to note that, in going from (B, 1) to (B, 2),
the above RR B-field will be affected by the non-commutative deformation in (\ref{ncdeformation}) and will
also receive additional contributions along other directions. We shall not be interested in such additional terms,
so we will consider simply that
\begin{align}
\tilde{\mathcal{C}}_2^{(B,2)}=\sec\theta_{nc}  F^{(2)}d\tilde{x}_3\wedge dr+\textrm{other terms}.
\end{align}
(The reader should not be worried at the drastic simplification in the analysis at this point, since it will
shortly become clear why one can consistently do so.) Then, in T-dualizing along $\phi_1$ to the configuration
(A, 3), we obtain an RR three-form potential of the form
\begin{align}
\tilde{\mathcal{C}}_3^{(A,3)}=\sec\theta_{nc}  F^{(2)}d\phi_1\wedge d\tilde{x}_3\wedge dr+\textrm{other terms}.
\end{align}
Without loss of generality, the relevant part of $\tilde{\mathcal{C}}_3^{(A,3)}$ will be assumed to be of the form suggested
in (3.67) in~\cite{Dasgupta:2016rhc}:
\begin{align}
\tilde{\mathcal{C}}_3^{(A,3)}=\frac{N_r\sin2\theta_{nc}\cos\theta_{nc} p(\theta_1)q(\theta_{nc})}
{2(\cos^2\theta_{nc}+N\sin^2\theta_{nc})^2}
dr\wedge d\tilde{x}_3\wedge d\phi_1, \label{367}
\end{align}
with $(p,\,q)$ periodic functions of $(\theta_1,\,\theta_{nc})$ with period $(\pi,\,2\pi)$, respectively  and
$N=N(r,\,\theta_{nc})$ sufficiently small for all values of the radial coordinate and such that
\begin{align}
\lim_{r\rightarrow0}N=0, \quad\quad \lim_{r\rightarrow\infty}N=1. \label{limitsN}
\end{align}
Quite obviously, $N_r$ stands for the derivative of $N$ with respect to $r$.
Finally, in the uplift from (A, 3) to (M, 1), (\ref{367}) will lead to the background G-flux of (\ref{355})
receiving the additional contribution given by
\begin{align}
\delta \langle\mathcal{G}_4^{(M,1)}\rangle=d \tilde{\mathcal{C}}_3^{(A,3)}.
\end{align}
(For completeness, let us just mention that the NS B-field $\tilde{B}_2^{(B,1)}$ will also add to the background G-flux
of (M, 1), as roughly $d\tilde{B}_2^{(B,1)}\wedge dx_{11}$.)

Summing up, the inclusion of the ${\overline{\rm D5}}$-brane in such a way that tachyons are avoided
affects only the background G-flux of the abelian configuration (M, 1). As already argued in section \ref{kintermsec},
the background G-flux does not contribute at all to the first term of the action (\ref{term1action}).
Consequently, the ${\overline{\rm D5}}$-brane does {\it not} affect our results so far
(and hence there is no need to make more precise the above analysis).

However, the particular contribution (\ref{367}) to the RR three-form potential of the configuration (A, 3) does play a key role.
It sources a new term\footnote{
Actually, this second term for our bosonic action
is well-known and usually referred to as ``anomalous interaction term'' in the literature.
The interested reader can find a lucid review of its main features in section 4 of~\cite{Sen:1997pr} (and references
therein).} for the gauge theory action, which we can interpret as a mass term for the G-flux of (M, 1):
\begin{align}
S^{(2)}\equiv \int_{X_{11}}\,\,\tilde{\mathcal{C}}_3^{(A,3)}\wedge \mathcal{G}_4^{(M,1)}\wedge
\mathcal{G}_4^{(M,1)},
\label{Sboson2}
\end{align}
with $\mathcal{G}_4^{(M,1)}$ given by (\ref{355}) in this abelian scenario
and the eleven-dimensional manifold $X_{11}$ as described around (\ref{decom}).

Moving on to the non-abelian enhanced case (constructed in section \ref{nonabsec}), our entire discussion hitherto
straightforwardly goes through.
The only two differences are that we have $N$ number of D5/${\overline{\rm D5}}$ pairs instead of just one and
that $\mathcal{G}_4^{(M,1)}$ in (\ref{Sboson2}) is now the
non-abelian G-flux in (\ref{multiflux}). Since the background G-flux in (\ref{multiflux}) is negligible
and using the non-abelian generalization of (\ref{c1overv3}), the second term of the action reduces to
\begin{align}
S^{(2)}=\frac{C_1}{V_3}\int_0^\pi\frac{d\theta_1}{2\pi}
\int_{X_4\otimes \Sigma_3}\textrm{Tr}\left(\tilde{\mathcal{C}}_3^{(A,3)}\wedge
\mathcal{F}\wedge \mathcal{F}\right),
\label{S2int}
\end{align}
with $\mathcal{F}$ the non-abelian seven-dimensional field strength.
As was the case with the first term $S^{(1)}$ of the bosonic action, the trace is
taken in the adjoint representation of the gauge group, in this case $SU(N)$.
Also, note that we have transferred the $\theta_1$ integral (as an average) to the $X_4\otimes\Sigma_3$ subspace of $X_{11}$,
to consistently decouple the contribution of the Taub-NUT space to $S^{(2)}$.
Relevant comments regarding the appearance of
this trace and the decoupling of the Taub-NUT subspace are as discussed before, between equations (\ref{Sg1}) and
(\ref{depenF}).

The $S^{(2)}$ term in (\ref{S2int}) is actually very simple. Note that $\tilde{\mathcal{C}}_3^{(A,3)}$ spans
all three directions of the three-cycle $\Sigma_3$. Recall also the decomposition of $\mathcal{F}$ in (\ref{F34decom}).
It is clear that $\mathcal{F}^{(\Sigma_3)}$ cannot contribute to $S^{(2)}$, as it would then lead
to a (vanishing) wedge product between two same directions of $\Sigma_3$.
On the other hand, $\mathcal{F}^{(X_4)}$ does contribute, but is restricted to $X_4$ and does not depend on the
$\theta_1$ coordinate, both properties following by definition.
Thus, the integral over $X_4\otimes \Sigma_3$ naturally decomposes into independent integrals in $X_4$ and $\Sigma_3$
and (\ref{S2int}) is in reality just given by
\begin{align}
S^{(2)}=c_2I^{(2)}\quad\quad I^{(2)}\equiv \int_{X_4}\textrm{Tr}\left(\mathcal{F}^{(X_4)}\wedge \mathcal{F}^{(X_4)}\right),
\quad\quad c_2\equiv \frac{C_1}{V_3}\int_0^\pi\frac{d\theta_1}{2\pi}\int_{\Sigma_3}\tilde{\mathcal{C}}_3^{(A,3)}.
\label{bosonact2}
\end{align}
For the moment, the above form of $I^{(2)}$ will suffice.
We will work on further rewritings of this integral in due time, when the need arises.
Consequently, let us focus on the only task left: the determination of the coefficient $c_2$.

This too turns out to be quite easy.
Using (\ref{dtildezeta}) and (\ref{367}), we can rewrite $c_2$ as
\begin{align}
c_2=\frac{C_1}{V_3} \int d^4\tilde{\zeta}\,\, \frac{N_r\sin2\theta_{nc}\cos\theta_{nc} p(\theta_1)q(\theta_{nc})}
{2(\cos^2\theta_{nc}+N\sin^2\theta_{nc})^2}. \label{c2mid}
\end{align}
Once more, the integrals over $(\tilde{x}_3,\,\phi_1)$ here are trivial. To simplify the notation a bit,
we absorb the contribution of the $\theta_1$ integral in the radius of the $\tilde{x}_3$ non-compact direction as
\begin{align}
\tilde{R}_3\equiv  \frac{R_3}{2}\int_0^\pi d\theta_1\,\,p(\theta_1). \label{tildeR3def}
\end{align}
Then, $c_2$ can be seen to be exactly as suggested in (3.63) and (3.68) in~\cite{Dasgupta:2016rhc}:
\begin{align}
c_2=\frac{C_1}{V_3}\tilde{R}_3\sin2\theta_{nc}\cos\theta_{nc}q(\theta_{nc})\int_0^\infty 
\frac{N_rdr}{(\cos^2\theta_{nc}+N\sin^2\theta_{nc})^2}=2\frac{C_1\tilde{R}_3}{V_3}\sin\theta_{nc}q(\theta_{nc}),
\end{align}
where in the last step we have used the boundary values in (\ref{limitsN}).
Our final expression for $c_2$ leaves no room for doubt: this coefficient is just some well defined number.
To match the notation in~\cite{Dasgupta:2016rhc} and without loss of generality, one may set $2\tilde{R}_3=V_3$ and
thus simply consider $c_2$ as
\begin{align}
c_2=C_1\sin\theta_{nc}q(\theta_{nc}). \label{c2final}
\end{align}
Written in this manner, $C_1$ accounts for the dependence of the $c_2$ coefficient on the non-abelian version of the M-theory
configuration (M, 1) of section \ref{ns5d3sect1}. The factor $\sin\theta_{nc}$ ensures that $\theta_{nc}=0$ implies $c_2=0$
(recall that $\theta_{nc}$ was introduced to this aim precisely). Finally, $q(\theta_{nc})$ allows us to have
as complex a dependence on $\theta_{nc}$ of $c_2$ as one may wish. 

\subsection{Completing the four-dimensional vector multiplet: third term for the action \label{thirdtermsec}}

In this section, we compute the third and last term $S^{(3)}$ that contributes to the bosonic action
of the $\mathcal{N}=2$ four-dimensional gauge theory. As we already pointed out in the beginning of
section \ref{actionsec}, this third term is not easily derivable from the non-abelian M-theory configuration (M, 1).
(In fact, there is no rigorous derivation of this type of term in the literature.)
Nonetheless, all the knowledge we have gathered while deriving the first two terms, $S^{(1)}$ and $S^{(2)}$,
will now pay off and allow us to obtain the remaining third term.

Let us begin by recalling that in the end of section \ref{kintermsec} we argued that the bosonic matter content
in the gauge theory must be exactly that in the $\mathcal{N}=4$ vector multiplet. That is, in our action
we must have four gauge fields and six real scalars, all of them in the adjoint representation of $SU(N)$.
However, upon inspection of the already derived first two
terms in the gauge theory action (given by (\ref{term1action}) and (\ref{bosonact2})), we note that
so far only the gauge fields $(\mathcal{A}_t,\,\mathcal{A}_1,\,\mathcal{A}_2,\,\mathcal{A}_{\tilde{\psi}})$
and three real scalars $(\mathcal{A}_{\tilde{3}},\,\mathcal{A}_{\phi_1},\, \mathcal{A}_r)$
have appeared in our analysis. Hence, we are missing the contribution of the other three real scalars.
Following the notation of~\cite{Dasgupta:2016rhc}, we will refer to these as $(\varphi_1,\,\varphi_2,\,\varphi_3)$.
Accordingly, $S^{(3)}$ will capture the dynamics of these scalar fields.

Let us next note that the terms $S^{(1)}$ and $S^{(2)}$ originate from 
the G-flux of the non-abelian configuration (M, 1), which is given by (\ref{multiflux}).
Further, these two terms exhaust all possible contributions of the G-flux to the action.
(This is most clearly seen by looking at the initial form of $S^{(1)}$ and $S^{(2)}$
in (\ref{initialact1}) and (\ref{Sboson2}), respectively.)
In consequence, $S^{(3)}$ must emerge purely from the geometry of (M, 1).
In other words, we expect the scalar fields $\varphi_k$ (with $k=1,\,2,\,3$) to stem from fluctuations of
the eleven-dimensional supergravity Einstein term of (M, 1).
In terms of our non-abelian scenario of figure \ref{fig5}{\bf B},
this means that the Taub-NUT space $TN$ and the M2-branes wrapping its two-cycles
fluctuate along $X_4\otimes\Sigma_3$\footnote{To the forgetful reader, the subspaces $(TN,\,X_4,\,\Sigma_3)$ of the full
eleven-dimensional manifold $X_{11}$ were introduced
and described around (\ref{decom}).}. We will right away simplify the scenario and assume 
the fluctuations are restricted to $X_4$ only, so that
\begin{align}
\varphi_k=\varphi_k(t,x_1,x_2,\tilde{\psi}) \quad\quad \forall k=1,2,3.
\label{varphidepen}
\end{align}
We will further suppose that, in fluctuating along orthogonal directions of $X_{11}$, $TN$
itself does not get back-reacted. Or, more accurately, that the back-reaction of $TN$ is negligible
compared to the change that the metric of $X_4\otimes \Sigma_3$ experiences.
This last key assumption allows us to write $S^{(3)}$ as an integral over $X_4\otimes\Sigma_3$ only.
In the same vein as for the previous two terms of the action, we will also average over the contribution
of the $\theta_1$ coordinate.

Having shed sufficient qualitative light into the nature and content of $S^{(3)}$, we are now ready to
make this term in the action fully precise. Naturally, $S^{(3)}$ must contain the kinetic terms and the
self-interaction terms of $(\varphi_1,\,\varphi_2,\,\varphi_3)$, as well as their
interaction terms with $(\mathcal{A}_{\tilde{3}},\,\mathcal{A}_{\phi_1},\, \mathcal{A}_r)$:
\begin{align}
S^{(3)}=S_{kin}^{(\varphi)}+S_{int}^{(\varphi\varphi)}+S_{int}^{(\mathcal{A}\varphi)}.
\label{S3split}
\end{align}
This just mimics the well-known $\mathcal{N}=4$ vector multiplet's action for the $\varphi$ scalar fields.
In the same spirit of (\ref{3int}), we can write the above as
\newpage
\begin{align}
S_{kin}^{(\varphi)}=&\int_0^\pi\frac{d\theta_1}{2\pi}
\int_{X_4\otimes\Sigma_3}\textrm{Tr}\sum_{k=1}^3\left[\sum_{a=0}^2 g^{aa}(\mathcal{D}_a\varphi_k)^2
+g^{\tilde{\psi}\tilde{\psi}}(\mathcal{D}_{\tilde{\psi}}\varphi_k)^2\right], \nonumber \\
S_{int}^{(\varphi\varphi)}=&\int_0^\pi\frac{d\theta_1}{2\pi}
\int_{X_4\otimes\Sigma_3}\textrm{Tr}\sum_{k=1}^3[\varphi_k,\varphi_l]^2,  \label{3139} \\
S_{int}^{(\mathcal{A}\varphi)}=&\int_0^\pi\frac{d\theta_1}{2\pi}\int_{X_4\otimes\Sigma_3}\textrm{Tr}\sum_{k=1}^3\left(
[\mathcal{A}^{(\Sigma_3)},\varphi_k]\wedge\ast[\mathcal{A}^{(\Sigma_3)},\varphi_k]\right), \nonumber
\end{align}
where $(g^{aa},\,g^{\tilde{\psi}\tilde{\psi}})$ are given by (\ref{matrentr}), the covariant derivatives
were defined in (\ref{covder}), $\mathcal{A}^{(\Sigma_3)}$ stands for (\ref{3101})
and the Hodge dual is with respect to the three-dimensional metric of $\Sigma_3$ in 
(\ref{metricsigma3}). In the following, we work out these terms separately.

\subsection*{Computation of $S_{kin}^{(\varphi)}$ in (\ref{3139})}

This kinetic piece is rather unchallenging to work out. Simply writing out explicitly the integral over
$X_4\otimes\Sigma_3$ there appearing and using (\ref{g7}), (\ref{matrentr}) and (\ref{dtildezeta}), $S_{kin}^{(\varphi)}$
can be written as in (3.139) in~\cite{Dasgupta:2016rhc}:
\begin{align}
S_{kin}^{(\varphi)}=\int d^4x\,\,\textrm{Tr}\sum_{k=1}^3\left[\sum_{a=0}^2 b_{ak}(\mathcal{D}_a\varphi_k)^2
+b_{\tilde{\psi}k}(\mathcal{D}_{\tilde{\psi}}\varphi_k)^2\right] \label{Skinvar}
\end{align}
where, once more, $d^4x\equiv dtdx_1dx_2d\tilde{\psi}$ and the coefficients $(b_{ak},\,b_{\tilde{\psi}k})$ are defined as
\begin{align}
b_{ak}\equiv e^{2\phi_0}\int d^4\tilde{\zeta} H_1\sqrt{F_1H_4}, \quad\quad
b_{\tilde{\psi}k}\equiv \int d^4\tilde{\zeta} H_1\sqrt{\frac{F_1}{H_4}}. \label{bssdeff}
\end{align}
Further introducing (\ref{Hs}) in the above and noting that the integrands are independent of
$(\tilde{x}_3,\,\phi_1)$, these coefficients considerably simplify:
\begin{align}
\begin{array}{llll}
b_{ak}=&\displaystyle e^{2\phi_0}R_3\sec\theta_{nc}\int_0^\infty dr\,\,
(\cos^2\theta_{nc}+F_2\sin^2\theta_{nc})^{1/3}F_3^{1/3}\sqrt{F_1\tilde{F}_2}\,\mathcal{I}^{(9)}, \\
b_{\tilde{\psi}k}=& \displaystyle R_3\sec\theta_{nc}\int_0^\infty dr\,\,
(\cos^2\theta_{nc}+F_2\sin^2\theta_{nc})^{1/3}F_3^{1/3}\sqrt{\frac{F_1}{\tilde{F}_2}}\,\mathcal{I}^{(10)},
\end{array}
\end{align}
with the integrals there appearing defined as
\begin{align}
\mathcal{I}^{(9)}\equiv \int_0^\pi d\theta_1\,\,\frac{\sin\theta_1}{\hat{\chi}^{1/6}}, \quad\quad
\mathcal{I}^{(10)}\equiv \int_0^\pi d\theta_1\,\, \frac{\hat{\chi}^{5/6}}{\sin\theta_1}, \quad\quad
\hat{\chi}=\hat{\chi}(\theta_1)\equiv 1+\frac{\tilde{F}_2-F_3}{F_3}\cos^2\theta_1. \label{I910def}
\end{align}
These integrals are most easily performed after doing the by now familiar change of variables in (\ref{theta1toz}).
For $\mathcal{I}^{(9)}$ we obtain
\begin{align}
\mathcal{I}^{(9)}=\int_{-1}^1 dz\left(1+\frac{\tilde{F}_2-F_3}{F_3}z^2\right)^{-1/6} 
= z\,\, {}_2F_1\left(\frac{1}{6},\frac{1}{2},\frac{3}{2};\frac{F_3-\tilde{F}_2}{F_3}z^2\right)\Bigg|_{z=-1}^{z=1}
= 2\Theta_{12}. \label{I9sol}
\end{align}
Similarly, using (\ref{theta1toz}), introducing the regularization factor $b\in(\mathbb{R}^+-\{1\})$ in the same way as in
(\ref{I2inter}) previously and further changing variables as
\begin{align}
z^2=\hat{z}, \quad\quad dz=\frac{d\hat{z}}{2\sqrt{\hat{z}}},
\end{align}
the integral $\mathcal{I}^{(10)}$ yields
\begin{align}
\mathcal{I}^{(10)}=&2\int_{0}^1 \frac{dz}{b^2-z^2}\left(1+\frac{\tilde{F}_2-F_3}{F_3}z^2\right)^{5/6}
=\int_0^1\frac{d\hat{z}}{\sqrt{\hat{z}}}\frac{1}{b^2-\hat{z}}\left(1+\frac{\tilde{F}_2-F_3}{F_3}\hat{z}\right)^{5/6}
\nonumber \\
=&\frac{2\sqrt{\hat{z}}}{b^2}F_1\left(\frac{1}{2},-\frac{5}{6},1,\frac{3}{2};\frac{F_3-\tilde{F}_2}{F_3}\hat{z};
\frac{\hat{z}}{b^2}\right)\Bigg|_{\hat{z}=0}^{\hat{z}=1}=\frac{2}{b^2}\Theta_{34}.
\end{align}
Following the notation in (3.136) and (3.138) in~\cite{Dasgupta:2016rhc}, $(\Theta_{12},\,\Theta_{34})$ above
stand for the following hypergeometric functions:
\begin{align}
\Theta_{12}\equiv {}_2F_1\left(\frac{1}{6},\frac{1}{2},\frac{3}{2};\frac{F_3-\tilde{F}_2}{F_3}\right), \quad\quad 
\Theta_{34}\equiv F_1\left(\frac{1}{2},-\frac{5}{6},1,\frac{3}{2};\frac{F_3-\tilde{F}_2}{F_3};
\frac{1}{b^2}\right). \label{1234def}
\end{align}
Putting everything together, we obtain the coefficients $(b_{ak},\,b_{\tilde{\psi}k})$ exactly as in (3.135) and (3.137)
in~\cite{Dasgupta:2016rhc}:
\begin{align}
\label{bakfinal}
\begin{array}{llll}
b_{ak}=&\displaystyle 2e^{2\phi_0}R_3\sec\theta_{nc}\int_0^\infty dr (\cos^2\theta_{nc}+F_2\sin^2\theta_{nc})^{1/3}F_3^{1/3}
\sqrt{F_1\tilde{F}_2}\Theta_{12}, \\
b_{\tilde{\psi} k}=&\displaystyle 2\frac{R_3}{b^2}\cos\theta_{nc}\int_0^\infty dr(\cos^2\theta_{nc}+F_2\sin^2\theta_{nc})^{1/3}
F_3^{1/3}
\sqrt{\frac{F_1}{\tilde{F}_2}}\Theta_{34}.
\end{array}
\end{align}
Recalling the constraint $\tilde{F}_2\geq F_3$ of section \ref{kintermsec}, the reader will not have a hard time of
convincing himself that the above two coefficients are well defined numbers for any choice of warp factors
in (\ref{easychoice}).

\subsection*{Computation of $S_{int}^{(\varphi\varphi)}$ in (\ref{3139})}

The determination of this self-interaction term is a simplified version of the computation
we just presented for the kinetic term. As in there, all boils down to explicitly
writing the integral over $X_4\otimes\Sigma_3$ in (\ref{3139}) with the aid of (\ref{g7}) and (\ref{dtildezeta}):
\begin{align}
S_{int}^{(\varphi\varphi)}= \int d^4x\,\,\textrm{Tr}\sum_{k,l=1}^3 d_{kl}[\varphi_k,\varphi_l]^2, \quad\quad
d_{kl}\equiv e^{2\phi_0}\int d^4\tilde{\zeta} H_1^2\sqrt{F_1H_4} \qquad \forall k,l=1,2,3,
\label{Svarvar}
\end{align}
with  $d^4x\equiv dtdx_1dx_2d\tilde{\psi}$. For the determination of the $d_{kl}$ coefficients, the first step is to
use (\ref{Hs}) and carry out the trivial $(\tilde{x}_3,\,\phi_1)$ integrals. We thus find that
\begin{align}
d_{kl}=e^{2\phi_0}R_3\sec\theta_{nc}\int_0^\infty dr (\cos^2\theta_{nc}+F_2\sin^2\theta_{nc})^{2/3}
\sqrt{F_1\tilde{F}_2F_3}\mathcal{I}^{(11)} \qquad \forall k,l=1,2,3,
\end{align}
where we have defined, using $\hat{\chi}$ in (\ref{I910def}),
\begin{align}
\mathcal{I}^{(11)}\equiv F_3^{1/6}\int_0^\pi d\theta_1\,\,\sin\theta_1\hat{\chi}^{1/6}. 
\end{align}
Given the similarity between the above and $(\mathcal{I}^{(9)},\,\mathcal{I}^{(10)})$ before,
the attentive reader will already have guessed that the easiest way to perform the above integral
is by doing the change of variables in (\ref{theta1toz}):
\begin{align}
F_3^{-1/6}\mathcal{I}^{(11)}=& \int_{-1}^1 dz\Big(1+\frac{\tilde{F}_2-F_3}{F_3}z^2\Big)^{1/6} \\
=&\frac{3z}{4}\Big(1+\frac{\tilde{F}_2-F_3}{F_3}z^2\Big)^{1/6}+\frac{z}{4}{}_2F_1\big(\frac{1}{2},\frac{5}{6},
\frac{3}{2};\frac{F_3-\tilde{F}_2}{F_3}z^2\Big)\Bigg|_{z=-1}^{z=1}
=\frac{\Theta_{56}}{2F_3^{1/6}}, \nonumber
\end{align}
where $\Theta_{56}$ is as in (3.143) in~\cite{Dasgupta:2016rhc}:
\begin{align}
\Theta_{56}\equiv3\tilde{F}_2^{1/6}+F_3^{1/6}\,\,{}_2F_1\left(\frac{1}{2},\frac{5}{6},
\frac{3}{2};\frac{F_3-\tilde{F}_2}{F_3}\right). \label{56def}
\end{align}
As a result, we can write the $d_{kl}$ coefficients as suggested by (3.142) in~\cite{Dasgupta:2016rhc}:
\begin{align}
\label{dklfinal}
d_{kl}=\frac{e^{2\phi_0}}{2}R_3\sec\theta_{nc}\int_0^\infty dr \sqrt{F_1\tilde{F}_2F_3}
(\cos^2\theta_{nc}+F_2\sin^2\theta_{nc})^{2/3}\Theta_{56} \qquad \forall k,l=1,2,3,
\end{align}
which are just some number whatever choice of warp factors one may wish to consider in (\ref{easychoice}).

\subsection*{Computation of $S_{int}^{(\mathcal{A}\varphi)}$ in (\ref{3139})}

The final term to be computed, namely the interaction term between the two sets of three real scalars 
$\mathcal{A}^{(\Sigma_3)}$ and $\varphi_k$ ($k=1,\,2,\,3$), is mathematically more involved than its previous
two counterparts. Hence, let us first take a few preparatory baby steps. From (\ref{ast3viel}) and (\ref{3101})
it follows that
\begin{align}
[\mathcal{A}^{(\Sigma_3)},\varphi_k]=&
[\hat{\alpha}_1,\varphi_k]\,e_{\tilde{3}}^{(\Sigma_3)}+[\hat{\alpha}_2,\varphi_k]\,
e_r^{(\Sigma_3)}+[\hat{\alpha}_3,\varphi_k]\,e_{\phi_1}^{(\Sigma_3)}, \\ \nonumber
\ast [\mathcal{A}^{(\Sigma_3)},\varphi_k]=&
[\hat{\alpha}_1,\varphi_k]\,e_r^{(\Sigma_3)}\wedge e_{\phi_1}^{(\Sigma_3)}
-[\hat{\alpha}_2,\varphi_k]\,e_{\tilde{3}}^{(\Sigma_3)}\wedge e_{\phi_1}^{(\Sigma_3)}
+[\hat{\alpha}_3,\varphi_k]\,e_{\tilde{3}}^{(\Sigma_3)}\wedge e_r^{(\Sigma_3)},
\end{align}
the Hodge dual having been taken with respect to (\ref{metricsigma3}).
The wedge product between the above two quantities is then
\begin{align}
[\mathcal{A}^{(\Sigma_3)},\varphi_k]\wedge \ast [\mathcal{A}^{(\Sigma_3)},\varphi_k]=
([\hat{\alpha}_1,\varphi_k]^2+[\hat{\alpha}_2,\varphi_k]^2+[\hat{\alpha}_3,\varphi_k]^2)\,
e_{\tilde{3}}^{(\Sigma_3)}\wedge e_r^{(\Sigma_3)} e_{\phi_1}^{(\Sigma_3)}.
\end{align}
Since $H_1^3H_2H_3=1$, as a direct consequence of our definitions in (\ref{Hs}), and reversing
(\ref{vielsigma3}) and (\ref{3101}),
the above can be rewritten in the more convenient form
\newpage
\begin{align}
[\mathcal{A}^{(\Sigma_3)},\varphi_k]\wedge &\ast [\mathcal{A}^{(\Sigma_3)},\varphi_k]=e^{\phi_0}\frac{\sqrt{F_1}}{H_1}\Bigg\{
\frac{e^{-2\phi_0}}{F_1}[\mathcal{A}_r,\varphi_k]^2
+\frac{1}{H_2}[\mathcal{A}_{\tilde{3}},\varphi_k]^2 \nonumber \\ 
&+\left(\frac{f_3^2}{H_2}+\frac{1}{H_3}\right)[\mathcal{A}_{\phi_1},\varphi_k]^2 
-\frac{2f_3}{H_2}[\mathcal{A}_{\tilde{3}},\varphi_k][\mathcal{A}_{\phi_1},\varphi_k]\Bigg\}d\tilde{x}_3\wedge dr\wedge d\phi_1.
\end{align}
This is nothing but the integrand of $S_{int}^{(\mathcal{A}\varphi)}$ in (\ref{3139}). There, after
expanding the integral over $X_4\otimes\Sigma_3$ and using (\ref{g4}) and (\ref{dtildezeta}), we get the
interaction term as
\begin{align}
S_{int}^{(\mathcal{A}\varphi)}=\int d^4x\,\,\textrm{Tr}\sum_{k=1}^3\left(
c_{rk}[\mathcal{A}_r,\varphi_k]^2
+c_{\tilde{3}k}[\mathcal{A}_{\tilde{3}},\varphi_k]^2
+c_{\phi_1 k}[\mathcal{A}_{\phi_1},\varphi_k]^2 
-c_{kk}[\mathcal{A}_{\tilde{3}},\varphi_k][\mathcal{A}_{\phi_1},\varphi_k]\right).
\label{intAvarphi}
\end{align}
The four coefficients above (and these are the very last ones) are defined as
\begin{align}
\label{crk}
\begin{array}{lll}
&\displaystyle c_{rk}\equiv \int d^4\tilde{\zeta}\,\, H_1
\sqrt{\frac{H_4}{F_1}}, &\quad\quad
\displaystyle c_{\tilde{3}k}\equiv e^{2\phi_0}\int d^4\tilde{\zeta}\,\, 
\frac{H_1}{H_2}\sqrt{F_1H_4}, \\
&\displaystyle c_{\phi_1k}\equiv e^{2\phi_0}\int d^4\tilde{\zeta}\,\,
H_1\sqrt{F_1H_4}\left(\frac{f_3^2}{H_2}+\frac{1}{H_3}\right), &\quad\quad
\displaystyle c_{kk}\equiv 2e^{2\phi_0} \int d^4\tilde{\zeta}\,\,
\frac{f_3}{H_2}\sqrt{F_1H_4}.
\end{array}
\end{align}
Introducing (\ref{Hs}) and carrying out the trivial $(\tilde{x}_3,\,\phi_1)$ integrals,
these coefficients simplify to (3.144)-(3.146) in~\cite{Dasgupta:2016rhc}:
\begin{align}
c_{rk}=& 2R_3\sec\theta_{nc}\int_0^\infty dr\,\,F_3^{1/3}\sqrt{\frac{\tilde{F}_2}{F_1}}
(\cos^2\theta_{nc}+F_2\sin^2\theta_{nc})^{1/3}\Theta_{12}, \nonumber \\
c_{\tilde{3}k}=& 2e^{2\phi_0}R_3\sec\theta_{nc}\int_0^\infty dr\,\,F_3^{1/3} \sqrt{F_1\tilde{F}_2}
(\cos^2\theta_{nc}+F_2\sin^2\theta_{nc})^{4/3} \Theta_{12}, \label{csfinfin}\\
c_{\phi_1k}=& e^{2\phi_0}R_3\sec\theta_{nc}\int_0^\infty dr \,\,\sqrt{F_1\tilde{F}_2F_3}
(\cos^2\theta_{nc}+F_2\sin^2\theta_{nc})^{1/3}\Pi_{78} \nonumber
\end{align}
and $c_{kk}\propto\mathcal{I}^{(5)}$, with $\mathcal{I}^{(5)}$ defined in (\ref{I34int}).  Note that in the case of $(c_{rk},\,c_{\tilde{3}k})$
we have also integrated over $\theta_1$, using to this aim (\ref{I910def}), (\ref{I9sol}) and (\ref{1234def}). Also,
we have defined $\Pi_{78}$ as in (3.147) in~\cite{Dasgupta:2016rhc}:
\begin{align}
\Pi_{78}\equiv \hat{\Pi}_{78}+3\sec^2\theta_{nc}\tan^2\theta_{nc}\tilde{F}_2^2
(\cos^2\theta_{nc}+F_2\sin^2\theta_{nc})\tilde{\Pi}_{78}, \label{78def}
\end{align}
with $(\hat{\Pi}_{78},\,\tilde{\Pi}_{78})$ depending on the $\hat{\chi}$ function in (\ref{I910def}) as
\begin{align}
\label{hatpi78}
\hat{\Pi}_{78}\equiv F_3^{5/6}\int_0^\pi d\theta_1\,\,\sin\theta_1\hat{\chi}^{5/6}, \quad\quad 
\tilde{\Pi}_{78}\equiv \frac{1}{3F_3^{1/6}}\int_0^\pi d\theta_1\,\,\frac{\sin\theta_1\cos^2\theta_1}{\hat{\chi}^{1/6}}.
\end{align}
Once more, these integrals are most easily carried out
after doing the change of variables in (\ref{theta1toz}). For $\hat{\Pi}_{78}$ we get
\newpage
\begin{align}
\nonumber
&F_3^{-5/6}\hat{\Pi}_{78}=\int_{-1}^1 dz\left(1+\frac{\tilde{F}_2-F_3}{F_3}z^2\right)^{5/6} \nonumber \\
&=\frac{3z}{8}\left(1+\frac{\tilde{F}_2-F_3}{F_3}z^2\right)^{5/6}
+\frac{5z}{8}\, {}_2F_1\left(\frac{1}{6},\frac{1}{2},\frac{3}{2};\frac{F_3-\tilde{F}_2}{F_3}z^2\right)\Bigg|_{z=-1}^{z=1} 
=\frac{3}{4}\left(\frac{\tilde{F}_2}{F_3}\right)^{5/6}+\frac{5}{4}\Theta_{12},
\end{align}
where in the last step we have made use of (\ref{1234def}). Similarly, $\tilde{\Pi}_{78}$ gives
\begin{align}
\nonumber
F_3^{-5/6}\tilde{\Pi}_{78}=&
\frac{1}{3F_3}\int_{-1}^1 dz\,\, z^2
\left(1+\frac{\tilde{F}_2-F_3}{F_3}z^2\right)^{-1/6}
=\frac{z }{8(\tilde{F}_2-F_3)}\left[
\left(1+\frac{\tilde{F}_2-F_3}{F_3}z^2\right)^{5/6}\right. \\
&\left.-{}_2 F_1\left(\frac{1}{6},\frac{1}{2},\frac{3}{2};\frac{F_3-\tilde{F}_2}{F_3}z^2\right)
\right]_{z=-1}^{z=1} 
=\frac{ \left(\tilde{F}_2/F_3\right)^{5/6}-\Theta_{12}}{4(\tilde{F}_2-F_3)}.
\end{align}
The above two results recover (3.148) in~\cite{Dasgupta:2016rhc} and, used in (\ref{78def}), allow us to 
write $\Pi_{78}$ as
\begin{align}
\Pi_{78}= \frac{3}{4}\tilde{F}_2^{5/6}+\frac{5}{4}F_3^{5/6}\Theta_{12}+\frac{3}{4}\left(\frac{\tan\theta_{nc}}{\cos\theta_{nc}}\right)^2
(\cos^2\theta_{nc}+F_2\sin^2\theta_{nc})(\tilde{F}_2^{5/6}-F_3^{5/6}\Theta_{12})\frac{\tilde{F}_2^2}{\tilde{F}_2-F_3}. \label{pi78}
\end{align}
As we saw in (\ref{zeroint1}), $\mathcal{I}^{(5)}=0$ and so the coefficient $c_{kk}$ vanishes.
This reduces our interaction term in (\ref{intAvarphi}) to its final form:
\begin{align}
S_{int}^{(\mathcal{A}\varphi)}=\int d^4x\,\,\textrm{Tr}\sum_{k=1}^3\left(
c_{rk}[\mathcal{A}_r,\varphi_k]^2
+c_{\tilde{3}k}[\mathcal{A}_{\tilde{3}},\varphi_k]^2
+c_{\phi_1 k}[\mathcal{A}_{\phi_1},\varphi_k]^2 \right).
\label{SAvar}
\end{align}
For the very last time, we observe that the coefficients appearing above are, as a simple inspection of their form in
(\ref{csfinfin}) suggests, well defined numbers for any choice of the
warp factors one may wish to consider in (\ref{easychoice}). Just to make the entire analysis transparent,
we show that the only seemingly divergent term is actually finite. Defining $\epsilon\equiv (\tilde{F}_2-F_3)$,
we have that
\begin{align}
\lim_{\tilde{F}_2\rightarrow F_3}\tilde{\Pi}_{78}
=\lim_{\epsilon\rightarrow0}\frac{(F_3+\epsilon)^{5/6}-F_3^{5/6}}{4\epsilon}
\approx \frac{5}{24F_3^{1/6}},
\end{align}
a finite result as predicted. (Recall that $F_3\rightarrow0$ cannot be considered in this case, as we explained after
(\ref{c12final}) earlier on.)

It is now the time to collect all our results in this section.
First, we introduce all (\ref{Skinvar}), (\ref{Svarvar}) and (\ref{SAvar}) in (\ref{S3split}).
We then have that the third and last term for our gauge theory action is
\newpage
\begin{align}
S^{(3)}=&\int d^4x\,\,\textrm{Tr}\left\{\sum_{k=1}^3\left[\sum_{a=0}^2 b_{ak}(\mathcal{D}_a\varphi_k)^2
+b_{\tilde{\psi}k}(\mathcal{D}_{\tilde{\psi}}\varphi_k)^2\right]
+\sum_{k,l=1}^3 d_{kl}[\varphi_k,\varphi_l]^2\right.\nonumber \\
&+\left.\sum_{k=1}^3\left(
c_{rk}[\mathcal{A}_r,\varphi_k]^2
+c_{\tilde{3}k}[\mathcal{A}_{\tilde{3}},\varphi_k]^2
+c_{\phi_1 k}[\mathcal{A}_{\phi_1},\varphi_k]^2\right) \right\}.
\label{S3last}
\end{align}
At last, adding all three contributions $S^{(1)}$ in (\ref{term1action}), $S^{(2)}$ in (\ref{bosonact2}) and
$S^{(3)}$ right above, we obtain the total bosonic action for the four-dimensional gauge theory to be
that in (3.153) in~\cite{Dasgupta:2016rhc}:
\begin{align}
\nonumber
S=&\frac{C_1c_{11}}{V_3}\int d^4x \sum_{\substack{a,b=0 \\ a<b}}^2\textrm{Tr}(\mathcal{F}_{ab}^2)+
\frac{C_1c_{12}}{V_3}\int d^4x
\sum_{a=0}^2 \textrm{Tr}(\mathcal{F}_{a\tilde{\psi}}^2)
+c_2\int_{X_4}\textrm{Tr}\left(\mathcal{F}^{(X_4)}\wedge \mathcal{F}^{(X_4)}\right) \\
&+\frac{C_1}{V_3}\int d^4x\,\textrm{Tr}\Bigg\{a_1[\mathcal{A}_r,\mathcal{A}_{\phi_1}]^2
+a_2[\mathcal{A}_{\tilde{3}},\mathcal{A}_r]^2+a_4[\mathcal{A}_{\tilde{3}},\mathcal{A}_{\phi_1}]^2
+\sum_{a=0}^2\left[
c_{a\tilde{3}}(\mathcal{D}_a\mathcal{A}_{\tilde{3}})^2\right. \nonumber \\
&\left.+c_{ar}(\mathcal{D}_a\mathcal{A}_r)^2
+\tilde{c}_{a\phi_1}(\mathcal{D}_a\mathcal{A}_{\phi_1})^2\right]
+c_{\tilde{\psi} \tilde{3}}\left(\mathcal{D}_{\tilde{\psi}}\mathcal{A}_{\tilde{3}}\right)^2
+c_{\tilde{\psi} r}\left(\mathcal{D}_{\tilde{\psi}}\mathcal{A}_r\right)^2
+\tilde{c}_{\tilde{\psi}\phi_1}\left(\mathcal{D}_{\tilde{\psi}}\mathcal{A}_{\phi_1}\right)^2\Bigg\} \nonumber \\
&+\int d^4x\,\,\textrm{Tr}\left\{\sum_{k=1}^3\left[\sum_{a=0}^2 b_{ak}(\mathcal{D}_a\varphi_k)^2
+b_{\tilde{\psi}k}(\mathcal{D}_{\tilde{\psi}}\varphi_k)^2\right]
+\sum_{k,l=1}^3 d_{kl}[\varphi_k,\varphi_l]^2\right.\nonumber \\
&+\left.\sum_{k=1}^3\left(
c_{rk}[\mathcal{A}_r,\varphi_k]^2
+c_{\tilde{3}k}[\mathcal{A}_{\tilde{3}},\varphi_k]^2
+c_{\phi_1 k}[\mathcal{A}_{\phi_1},\varphi_k]^2\right) \right\}.
\label{totaction}
\end{align}
To finish this section, we include table \ref{table1}.
This is  a quick guide to finding the explicit form (in terms of the warp factors in
(\ref{easychoice}),  the deformation parameter $\theta_{nc}$ in (\ref{ncdeformation})
and the constant dilaton in (\ref{consdil})) of the abundant coefficients on which our above action depends.
These will keep appearing all through the remaining of part \ref{partb}. Recall that
we have explicitly shown that all these coefficients are well defined numbers for any choice of the warp factors,
as long as the constraint $\tilde{F}_2\geq F_3$ is satisfied, with $\tilde{F}_2$ as in (\ref{tildeF2}).

\begin{table}[t]
\begin{tabular}{| c||c |}
\hline
\smash{{\rlap{\hspace{5.4cm} (All the coefficients in blue depend on $\mathcal{J}_3$ in (\ref{mathj3}).)}}} 
Coefficient & Given in  \\  \hline  \hline
\rowcolor{lightblue}
$c_{11}$ & (\ref{c11final}) \\ \hline
\rowcolor{lightblue}
\smash{{\rlap{\hspace{4cm} $\longrightarrow$  \hspace*{-0.15cm}
Depends on $(b,\,b_2,\,b_3)$ in (\ref{379}) and $\mathcal{J}_4$ in (\ref{mathj4}).}}} 
$c_{12}$ & (\ref{c12final}) \\ \hline
\rowcolor{lightblue}
\smash{\raisebox{-8pt}{\rlap{\hspace{4.3cm} $\Bigg\}  \hspace*{-0.2cm} \longrightarrow$  
Depend on $(\tilde{a}_\pm,\,\tilde{a}_2,\,\tilde{a}_4)$ in (\ref{tildea1a2}).}}} 
$a_{1},\,a_2,\,a_4$ & (\ref{a1a2a4}) \\ \hline
\rowcolor{lightblue}
$c_{a\tilde{3}},\,c_{ar},\,\tilde{c}_{a\phi_1}$ & (\ref{csfinal}) \\ \hline
\rowcolor{lightblue}
\smash{\raisebox{-10pt}{\rlap{\hspace{3.7cm} $\longrightarrow$  
Depend on all the above via $(a_{01},\,b_{01},\,c_{01})$ in (\ref{abc01}),}}} 
\smash{\raisebox{-23pt}{\rlap{\hspace{4.3cm}  as well as $\tilde{b}_2$ in (\ref{tildeb2})
and $(f^{(1)},\,f^{(2)})$ in (\ref{f12def}).}}} 
& \\ \rowcolor{lightblue}
$c_{\tilde{\psi} \tilde{3}},\,c_{\tilde{\psi} r},\,\tilde{c}_{\tilde{\psi}\phi_1}$ & (\ref{cpsis}) \\ [0.5cm]\hline \hline
\rowcolor{celadon}
$c_2$ & (\ref{c2final}) \\ \hline \hline
\rowcolor{corn}
\smash{{\rlap{\hspace{4.4cm} $\longrightarrow$  \hspace*{-0.15cm}
Depend on $b$ in (\ref{379}) and $(\Theta_{12},\,\Theta_{34})$ in (\ref{1234def}).}}} 
$b_{ak},\,b_{\tilde{\psi}k}$ & (\ref{bakfinal}) \\ \hline
\rowcolor{corn}
\smash{{\rlap{\hspace{4cm} $\longrightarrow$  \hspace*{-0.15cm}
Depends on $\Theta_{56}$ in (\ref{56def}).}}} 
$d_{kl}$ & (\ref{dklfinal}) \\ \hline
\rowcolor{corn}
\smash{{\rlap{\hspace{4.82cm} $\longrightarrow$  \hspace*{-0.15cm}
Depend on $(\Theta_{12},\,\Pi_{78})$ in (\ref{1234def}) and (\ref{pi78}).}}} 
$c_{\tilde{3}k},\,c_{rk},\,c_{\phi_1k}$ & (\ref{csfinfin}) \\ \hline
\end{tabular}
\caption{List of coefficients appearing in the bosonic action (\ref{totaction}) and the equation numbers where
they are expressed in terms of only the warp factors in (\ref{easychoice}) and (\ref{tildeF2}), the deformation parameter in
(\ref{ncdeformation}) and the leading constant term of the dilaton in (\ref{consdil}).
Note that we don't compute $(C_1/V_3)$ explicitly. However, its abelian version $(c_1/v_3)$ is given by
(\ref{c1v3pre}). Note also that all the coefficients in blue require $\tilde{F}_2\geq F_3$ to be finite.
The colors in the table point to the origin of the coefficients: in blue those stemming from $S^{(1)}$
discussed in section \ref{kintermsec}, in green that related to $S^{(2)}$ in section \ref{masstermsec}
and in yellow the coefficients of $S^{(3)}$ in section \ref{thirdtermsec}.}
\label{table1}
\end{table}

Before proceeding ahead in our analysis, it is worth noting that in the present work we do not study the
four-dimensional bosonic action stemming from the configuration (M, 2) of section \ref{rrdefsect}.
This is because (M, 2) was shown to be equivalent
to the configuration (M, 1) of sections \ref{ncsect} and \ref{nonabsec} (see figure \ref{fig8}), the latter being
computationally easier to handle. However, this action is discussed in~\cite{Dasgupta:2016rhc}
and argued to be of the form (\ref{totaction}), the only difference being that the coefficients
of table \ref{table1} would in that case change. We refer the interested reader to~\cite{Dasgupta:2016rhc}
for the pertinent details.

\FloatBarrier

\section{The bulk theory: the Hamiltonian and its minimization \label{hamilsec}}

This section is devoted to the derivation of the BPS conditions for the $\mathcal{N}=2$ four-dimensional
gauge theory along $(t,\,x_1,\,x_2,\,\tilde{\psi})$, whose action we just obtained in (\ref{totaction}).
It goes without a saying that the BPS conditions follow from minimizing the energy of the system with
action (\ref{totaction}), considering {\it static} configurations of the fields there. Hence, it is quite clear
that the first step towards achieving our aim in this section will be to obtain the Hamiltonian
associated to (\ref{totaction}). The second and last step will be to minimize this Hamiltonian, under the assumption
that the gauge and scalar fields are time-independent.

\begin{figure}[ht]
\centering
\includegraphics[width=0.9\textwidth]{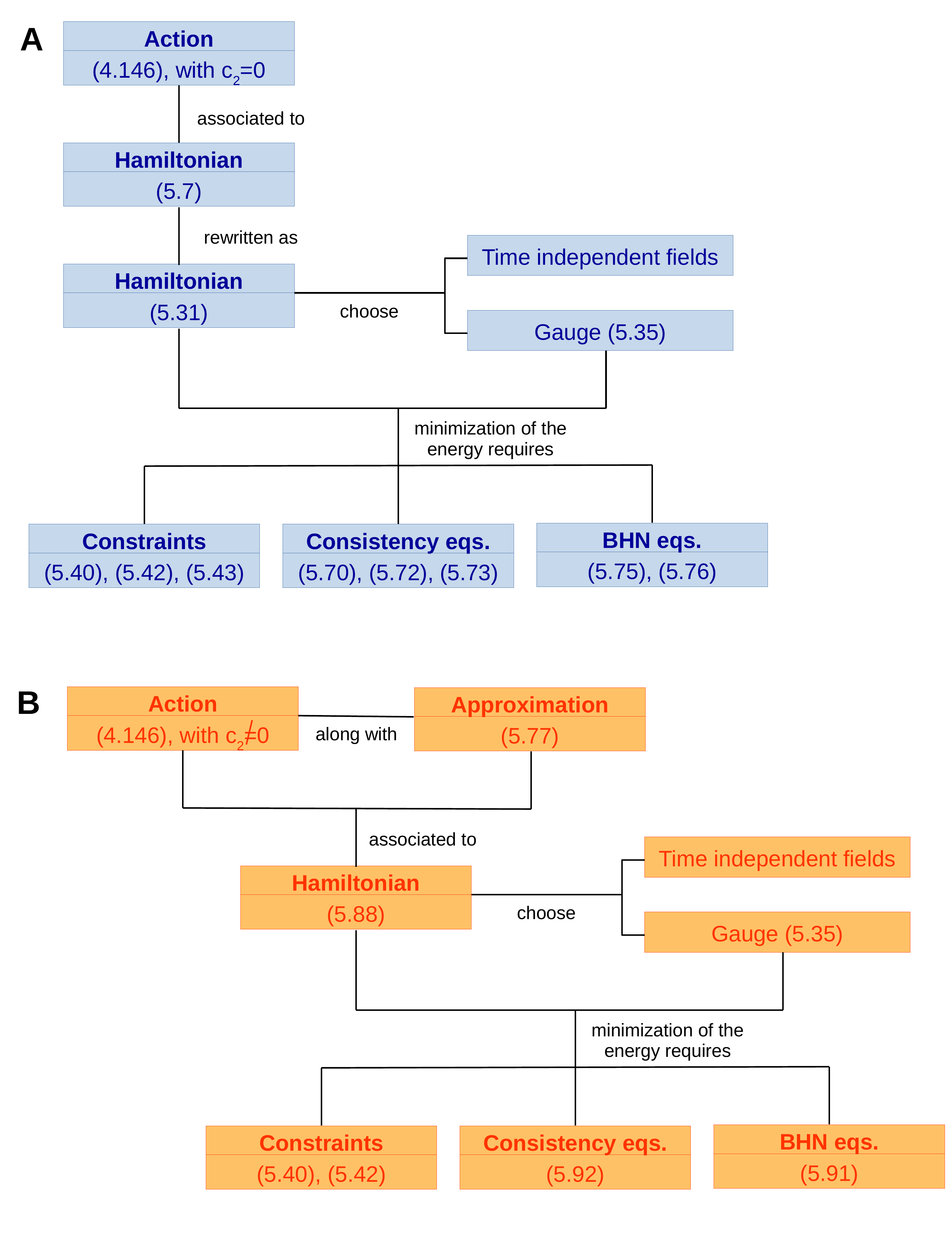}
\caption{Sketch of the main results in section \ref{hamilsec}, where we obtain the Hamiltonian
following from the gauge theory action (\ref{totaction}) and minimize its energy.
As a result, we obtain a set of equations the gauge and scalar fields in the theory must obey.
The so called BHN equations are particularly important, as they are related to knot invariants.
{\bf A:} Since the computation is a bit involved, in section \ref{c20sec} this is done
in a particularly simple limit: setting $c_2=0$ in (\ref{totaction}).
{\bf B:} The generalization to the case of interest, $c_2\neq0$ in (\ref{totaction}), is done in section \ref{c2not0sec} and follows
without much effort from the previous analysis.}
\label{fig10}
\end{figure}

Yet once more, this is more easily said than done. Consequently, we will do the following. First, we shall
determine and minimize the Hamiltonian following from (\ref{totaction}) in a particularly simple limit:
we will set $c_2=0$ there.
That is to say, we will begin by performing the analysis when there is no topological term in the action.
Then, we will use the insights thus gathered to generalize
the results to the $c_2\neq 0$ case we are really interested in.

This procedure is depicted in figure \ref{fig10}, where we also make reference to the main
results in the present section. As such, the reader may find it useful to look at figure \ref{fig10} as a guiding map
for section \ref{hamilsec}: it captures the main logic behind the computational details shown in the following.
 
\FloatBarrier

\subsection{Analysis for the case $c_2=0$ in (\ref{totaction}) \label{c20sec}}

Obtaining the Hamiltonian associated to a given action is a well defined problem in classical mechanics, which
our readers surely know by heart. As such, after setting $c_2=0$ in (\ref{totaction}), one could go ahead
with the standard procedure: infer the conjugate momenta and write the Hamiltonian as the Legendre transformation of
the Lagrangian. However, in view of the length and complexity of the action (\ref{totaction}), this procedure would be quite
a long and tiresome mathematical exercise for us. Therefore, we will use a different approach to obtain the Hamiltonian:
we will map our action to that in (2.1) in~\cite{Lee:2006gqa} and directly read off our Hamiltonian
from (2.4) in the same reference.

The Lagrangian density $\mathcal{L}$ of our theory can be directly inferred from the action (\ref{totaction}),
since
\begin{align}
S=\int d^4x\,\,\mathcal{L}.
\end{align}
With $c_2=0$, $\mathcal{L}$ in (\ref{totaction}) is precisely of the form of the Lagrangian (2.1) in~\cite{Lee:2006gqa},
up to relative factors, under the following identifications\footnote{In all the
identifications of our present work to other references we
will show the quantities of the cited source (our theory) on the left-hand (right-hand) side.}:
\begin{align}
x_M \rightarrow (t,x_1,x_2,\tilde{\psi}), \quad\quad \phi_A\rightarrow
(\mathcal{A}_{\tilde{3}},\mathcal{A}_{\phi_1},\mathcal{A}_r,
\varphi_1,\varphi_2,\varphi_3), \quad\quad \phi_5\rightarrow \mathcal{A}_{\tilde{3}}.
\label{idenkor}
\end{align}
Note that our definitions for the covariant derivatives in (\ref{covder}) differ from the covariant derivatives
in~\cite{Lee:2006gqa}. This mismatch is accounted for by replacing factors of $(i)$ there
by $(-i)$ in our case. Properly accounting for the additional prefactors in our theory as well, it is rather
simple to see that the different terms that compose the Hamiltonian (2.4) in~\cite{Lee:2006gqa}
are, in the language of the present paper, given by
\begin{align}
\begin{array}{llll}
\displaystyle\sum_a(F_{a0}-D_a\phi_5)^2 \rightarrow  \mathcal{T}_1, \quad\quad 
\sum_a(D_0\phi_a+i[\phi_5,\phi_a])^2  \rightarrow  \mathcal{T}_2, \quad\quad
(D_0\phi_5)^2 \rightarrow  \mathcal{T}_3, \\
\qquad \displaystyle  \frac{1}{2}\sum_{a\neq b}(F_{ab}-\epsilon_{abcd}D_c\phi_d+i[\phi_a,\phi_b])^2 \rightarrow 
\mathcal{T}_4, \quad\quad 
\Big(\sum_a D_a\phi_a\Big)^2 \rightarrow  0,
\end{array}
\end{align}
where we have defined $(\mathcal{T}_1,\,\mathcal{T}_2,\,\mathcal{T}_3)$ as
\begin{align}
\label{mathcalt1}
\begin{array}{rllll}
\mathcal{T}_1&\equiv&\displaystyle
\frac{C_1}{V_3}\sum_{\alpha=1}^2(\sqrt{c_{11}}\mathcal{F}_{\alpha0}-\sqrt{c_{\alpha \tilde{3}}}
\mathcal{D}_\alpha\mathcal{A}_{\tilde{3}})^2+\frac{C_1}{V_3}(\sqrt{c_{12}}\mathcal{F}_{\tilde{\psi}0}-
\sqrt{c_{\tilde{\psi}\tilde{3}}}\mathcal{D}_{\tilde{\psi}}\mathcal{A}_{\tilde{3}})^2, \\
\mathcal{T}_2&\equiv& \displaystyle\frac{C_1}{V_3}(\sqrt{c_{0r}}\mathcal{D}_0\mathcal{A}_r-i\sqrt{a_2}
[\mathcal{A}_{\tilde{3}},\mathcal{A}_r])^2+\frac{C_1}{V_3}
(\sqrt{\tilde{c}_{0\phi_1}}\mathcal{D}_0\mathcal{A}_{\phi_1}-i\sqrt{a_4}[\mathcal{A}_{\tilde{3}},\mathcal{A}_{\phi_1}])^2\\
&&\displaystyle+\sum_{k=1}^3(\sqrt{b_{0k}}\mathcal{D}_0\varphi_k-i\sqrt{c_{\tilde{3}k}}[\mathcal{A}_{\tilde{3}},\varphi_k])^2,
\qquad\qquad \mathcal{T}_3\equiv \frac{C_1}{V_3}c_{0\tilde{3}}(\mathcal{D}_0\mathcal{A}_{\tilde{3}})^2
\end{array}
\end{align}
and where $\mathcal{T}_4$ naturally splits into two, $\mathcal{T}_4=\mathcal{T}_4^{(1)}+\mathcal{T}_4^{(2)}$,
due to the decomposition of the subspace $X_4$ explained in (\ref{decom}):
\begin{align}
\mathcal{T}_4^{(1)}=\frac{1}{2}\sum_{\alpha,\beta=1}^2\Big(\sqrt{\frac{C_1}{V_3}}\tau^{(1)} +\tau^{(2)}\Big)^2, \quad\quad
\mathcal{T}_4^{(2)}=\frac{1}{2}\sum_{\alpha=1}^2\Big(\sqrt{\frac{C_1}{V_3}}\tau^{(3)}+\tau^{(2)}\Big)^2,
\label{mathcalt4}
\end{align}
with $(\tau^{(1)},\,\tau^{(2)},\,\tau^{(3)})$ standing for
\begin{align}
\tau^{(1)}\equiv& \sqrt{c_{11}}\mathcal{F}_{\alpha\beta}
-\sqrt{c_{\tilde{\psi} r}}\epsilon_{\alpha\beta\tilde{\psi} r}\mathcal{D}_{\tilde{\psi}}\mathcal{A}_r
-\sqrt{\tilde{c}_{\tilde{\psi}\phi_1}}\epsilon_{\alpha\beta\tilde{\psi}\phi_1}\mathcal{D}_{\tilde{\psi}}\mathcal{A}_{\phi_1}
-\sqrt{\frac{V_3}{C_1}}\sum_{k=1}^3\sqrt{b_{\tilde{\psi} k}}\epsilon_{\alpha\beta\tilde{\psi} k}\mathcal{D}_{\tilde{\psi}}
\varphi_k, \nonumber \\
i\tau^{(2)}\equiv& \sqrt{\frac{C_1a_1}{V_3}}[\mathcal{A}_r,\mathcal{A}_{\phi_1}]
+\sum_{k,l=1}^3\left(\sqrt{c_{rk}}[\mathcal{A}_r,\varphi_k]+\sqrt{c_{\phi_1k}}[\mathcal{A}_{\phi_1},\varphi_k]
+\sqrt{d_{kl}}[\varphi_k,\varphi_l]\right), \label{tau123def} \\
\tau^{(3)}\equiv&\sqrt{c_{12}}\mathcal{F}_{\alpha\psi}
-\sqrt{c_{\beta r}}\epsilon_{\alpha\tilde{\psi} \beta r}\mathcal{D}_\beta\mathcal{A}_r
-\sqrt{\tilde{c}_{\beta \phi_1}}\epsilon_{\alpha\tilde{\psi}\beta \phi_1}\mathcal{D}_\beta\mathcal{A}_{\phi_1}
-\sqrt{\frac{V_3}{C_1}}\sum_{k=1}^3\sqrt{b_{\beta k}}\epsilon_{\alpha\tilde{\psi}\beta  k}\mathcal{D}_\beta\varphi_k. \nonumber
\end{align}
Putting everything together as in (2.4) in~\cite{Lee:2006gqa}, we obtain the Hamiltonian associated to the action
(\ref{totaction}) (with $c_2=0$) to be given by
\begin{align}
H= \hspace*{-0.2cm}\int \hspace*{-0.1cm}d^4x \,\textrm{Tr}\Bigg\{\sum_{i=1}^3\mathcal{T}_i
+\frac{1}{2}\sum_{\alpha,\beta=1}^2\Big(\sqrt{\frac{C_1}{V_3}}\tau^{(1)} +\tau^{(2)}\Big)^2
+\frac{1}{2}\sum_{\alpha=1}^2\Big(\sqrt{\frac{C_1}{V_3}}\tau^{(3)}+\tau^{(2)}\Big)^2\Bigg\}
+Q_{EM}, \label{Hamiltonianinitial}
\end{align}
where $Q_{EM}$ denotes the sum of electric and magnetic charges in the theory.
As is well-known (see (2.5) in~\cite{Lee:2006gqa}), these charges are boundary terms.
We will study these boundary terms in exquisite detail in section \ref{fsbndsec} (for the case where $c_2\neq0$ in
(\ref{totaction}) only).
Hence, for the time being, we shall not make them precise and focus instead on the bulk terms.
Also, this Hamiltonian incorporates the Gauss law in it, as explained in~\cite{Lee:2006gqa}.
Consequently, there are no constraints on the gauge and scalar fields of our theory imposed by the Gauss
law\footnote{The skeptical reader can alternatively be convinced of this last statement
by the combination of (\ref{idenkor}) and our later choice (\ref{A3zeroch}).}.

According to the plan of action described in the beginning of this section, having obtained the Hamiltonian
for our gauge theory, we should now proceed to minimize it. It turns out, however, that the minimization
process simplifies considerably if we first rewrite (\ref{Hamiltonianinitial}) in a certain manner.
(Further, in section \ref{loceqsec} we shall obtain important results from this rewriting!) Thus, we will now simply
rewrite the Hamiltonian (\ref{Hamiltonianinitial}) in a more convenient form and postpone the minimization problem
to section \ref{minimsec}.

The rewriting we will carry out consists on
introducing new, arbitrary coefficients in some of the terms inside the sums of squares of (\ref{Hamiltonianinitial})
and, at the same time, summing new terms to the Hamiltonian so that there is no change in its quadratic components.
We shall not yet make precise the additional crossed terms produced in this manner.
But the reader should not worry, the crossed terms will be determined meticulously in section \ref{loceqsec}
(In fact, their study leads to the important results we were anticipating a little before.)
Perhaps a toy model will make the rewriting we intend to perform most transparent.
Consider the Hamiltonian
\begin{align}
H^{(1)}=(\mathbb{A}+\mathbb{B})^2+\mathbb{C}.
\end{align}
Introducing the arbitrary parameters $(\hat{x},\,\hat{y})$, the above can be rewritten as
\begin{align}
H^{(1)}=(\mathbb{A}+\hat{x}\mathbb{B})^2+\hat{y}\mathbb{B}^2+\tilde{\mathbb{C}},
\end{align}
as long as the constraints 
\begin{align}
\hat{x}^2+\hat{y} =1, \quad\quad
\tilde{\mathbb{C}}=\mathbb{C}+2\mathbb{AB}(1-\hat{x}),
\end{align}
are enforced. Written in this language, our earlier statement of ignoring the ``additional crossed
terms'' simply means that the second constraint above shall not be studied presently, but rather
in section \ref{loceqsec}.

Actually, we shall only rewrite the term $\mathcal{T}_4$
and leave $(\mathcal{T}_1,\,\mathcal{T}_2,\,\mathcal{T}_3)$ as they are.
We do so piecewise and
first focus on the first three terms of $\mathcal{T}_4^{(1)}$ in (\ref{mathcalt4}):
\begin{align}
\frac{1}{2}\sum_{\alpha,\beta=1}^2\Big(
\sqrt{\frac{C_1c_{11}}{V_3}}\mathcal{F}_{\alpha\beta}
-\sqrt{\frac{C_1c_{\tilde{\psi} r}}{V_3}}\epsilon_{\alpha\beta\tilde{\psi} r}\mathcal{D}_{\tilde{\psi}}\mathcal{A}_r
-\sqrt{\frac{C_1\tilde{c}_{\tilde{\psi}\phi_1}}{V_3}}\epsilon_{\alpha\beta\tilde{\psi}\phi_1}\mathcal{D}_{\tilde{\psi}}
\mathcal{A}_{\phi_1}+\ldots\Big)^2. \label{t41terms}
\end{align}
In the above, we introduce arbitrary coefficients in the second and third terms, which depend on $(\alpha,\,\beta)$. Clearly,
these must be anti-symmetric in the mentioned indices, so as not to yield zero
due to the present epsilon tensors. We absorb the minus signs in the coefficients and also transfer
the factor of $(1/2)$ inside the square. All in all, we rewrite the above as
\begin{align}
\nonumber
&\displaystyle\sum_{\alpha,\beta=1}^2\Big(
\sqrt{\frac{C_1c_{11}}{2V_3}}\mathcal{F}_{\alpha\beta}
+\sqrt{\frac{C_1c_{\tilde{\psi} r}}{V_3}}s_{\alpha\beta}^{(1)}\epsilon_{\alpha \beta\tilde{\psi} r}\mathcal{D}_{\tilde{\psi}}
\mathcal{A}_r+\sqrt{\frac{C_1\tilde{c}_{\tilde{\psi} \phi_1}}{V_3}}s_{\alpha\beta}^{(2)}\epsilon_{\alpha\beta\tilde{\psi}\phi_1}
\mathcal{D}_{\tilde{\psi}}\mathcal{A}_{\phi_1}+\ldots\Big)^2\\
&\displaystyle+\frac{C_1c_{\tilde{\psi} r}}{V_3}s^{(1)}(\mathcal{D}_{\tilde{\psi}}\mathcal{A}_r)^2
+\frac{C_1\tilde{c}_{\tilde{\psi} \phi_1}}{V_3}s^{(2)}(\mathcal{D}_{\tilde{\psi}}\mathcal{A}_{\phi_1})^2+\chi_s,
\label{terms}
\end{align}
where $\chi_s$ contains the additional crossed terms created by the inclusion of the $(s_{\alpha\beta}^{(1)},\,
s_{\alpha\beta}^{(2)})$ coefficients and we demand the constraints
\begin{eqnarray}
\label{conss}
2(s_{12}^{(i)})^2+s^{(i)}=1, \quad\quad \forall i=1,2
\end{eqnarray}
hold true, so as to ensure the quadratic pieces remain the same.
In exactly the same way, the first three terms of $\mathcal{T}_4^{(2)}$ in (\ref{mathcalt4}), namely
\begin{align}
\frac{1}{2}\sum_{\alpha=1}^2\Big(
\sqrt{\frac{C_1c_{12}}{V_3}}\mathcal{F}_{\alpha\tilde{\psi}}
-\sqrt{\frac{C_1c_{\beta r}}{V_3}}\epsilon_{\alpha\tilde{\psi} \beta r}\mathcal{D}_\beta\mathcal{A}_r
-\sqrt{\frac{C_1\tilde{c}_{\beta \phi_1}}{V_3}}\epsilon_{\alpha\tilde{\psi}\beta \phi_1}\mathcal{D}_\beta\mathcal{A}_{\phi_1}
+\ldots\Big)^2,
\end{align}
can be rewritten as
\begin{align}
\nonumber
&\sum_{\alpha=1}^2\Big(
\sqrt{\frac{C_1c_{12}}{2V_3}}\mathcal{F}_{\alpha\tilde{\psi}}
+\sqrt{\frac{C_1c_{\beta r}}{V_3}}t_{\alpha}^{(1)}\epsilon_{\alpha\tilde{\psi} \beta r}\mathcal{D}_\beta\mathcal{A}_r
+\sqrt{\frac{C_1\tilde{c}_{\beta \phi_1}}{V_3}}t_{\alpha}^{(2)}\epsilon_{\alpha\tilde{\psi}\beta \phi_1}
\mathcal{D}_\beta\mathcal{A}_{\phi_1}
+\ldots\Big)^2 \\
&+\frac{C_1c_{\beta r}}{V_3}t^{(1)}(\mathcal{D}_\beta\mathcal{A}_r)^2
+\frac{C_1\tilde{c}_{\beta \phi_1}}{V_3}t^{(2)}(\mathcal{D}_\beta\mathcal{A}_{\phi_1})^2+\chi_t,\label{termt}
\end{align}
where $\chi_t$ takes into account the additional crossed terms created by the inclusion of
$(t_{\alpha}^{(1)},\,t_{\alpha}^{(2)})$ and we impose the constraints
\begin{eqnarray}
\label{const}
\sum_{\alpha=1}^2(t_{\alpha}^{(i)})^2+t^{(i)}=1, \quad\quad \forall i=1,2,
\end{eqnarray}
which guarantee the squared terms are not affected in the rewriting.

With the very same idea in mind, we look at the fifth terms in both $\mathcal{T}_4^{(1)}$ and $\mathcal{T}_4^{(2)}$
next:
\begin{align}
\frac{1}{2}\sum_{\alpha,\beta=1}^2\Big(\ldots 
-i\sqrt{\frac{C_1a_1}{V_3}}[\mathcal{A}_r,\mathcal{A}_{\phi_1}]+\ldots\Big)^2+
\frac{1}{2}\sum_{\alpha=1}^2\Big(\ldots 
-i\sqrt{\frac{C_1a_1}{V_3}}[\mathcal{A}_r,\mathcal{A}_{\phi_1}]+\ldots \Big)^2.
\end{align}
We introduce antisymmetric (in their indices) coefficients in both the two terms, add squared terms
that make sure we do not alter that part and encompass the new crossed terms in $\chi_4$, which we do not
presently determine. We also pull in the factor of $(1/2)$, as before. Explicitly, the above becomes
\begin{align}
\nonumber
&\sum_{\alpha,\beta=1}^2\Big(\ldots 
-ig_{\alpha\beta}^{(4)}\sqrt{\frac{C_1a_1}{V_3}}[\mathcal{A}_r,\mathcal{A}_{\phi_1}]+\ldots\Big)^2+
\sum_{\alpha=1}^2\Big(\ldots 
-ih_{\alpha\tilde{\psi}}^{(4)}\sqrt{\frac{C_1a_1}{V_3}}[\mathcal{A}_r,\mathcal{A}_{\phi_1}]+\ldots \Big)^2 \\
& +\frac{C_1a_1}{V_3}q^{(4)}[\mathcal{A}_r,\mathcal{A}_{\phi_1}]^2+\chi_4, \label{term4}
\end{align}
where we require that the following must be satisfied\footnote{
Note that the relative difference in signs between (\ref{cons4}) and the previous constraints (\ref{conss}) and (\ref{const})
is a consequence of the overall factors of $(-i)$ in the terms of the action being considered.}:
\begin{align}
\label{cons4}
2(g_{12}^{(4)})^2+\sum_{\alpha=1}^2(h_{\alpha\tilde{\psi}}^{(4)})^2-q^{(4)}=1.
\end{align}
Similarly, the last terms in $\mathcal{T}_4^{(1)}$ and $\mathcal{T}_4^{(2)}$,
\begin{align}
\frac{1}{2}\sum_{\alpha,\beta=1}^2\Big(\ldots 
-i\sum_{k,l=1}^3\sqrt{d_{kl}}[\varphi_k,\varphi_l]\Big)^2+
\frac{1}{2}\sum_{\alpha=1}^2\Big(\ldots 
-i\sum_{k,l=1}^3\sqrt{d_{kl}}[\varphi_k,\varphi_l]\Big)^2 \label{commrewr},
\end{align}
are rewritten in the form
\begin{align}
\nonumber
&\sum_{\alpha,\beta=1}^2\Big(\ldots 
-i\sum_{k,l=1}^3g_{\alpha\beta kl}^{(1)}\sqrt{d_{kl}}[\varphi_k,\varphi_l]\Big)^2+
\sum_{\alpha=1}^2\Big(\ldots 
-i\sum_{k,l=1}^3h_{\alpha\tilde{\psi} kl}^{(1)}\sqrt{d_{kl}}[\varphi_k,\varphi_l]\Big)^2 \\
&+\sum_{k,l=1}^3q_{kl}^{(1)} d_{kl}[\varphi_k,\varphi_l]^2+\chi_1, \label{term1}
\end{align}
with the constraint
\begin{align}
\label{cons1}
2(g_{12kl}^{(1)})^2+\sum_{\alpha=1}^2(h_{\alpha\tilde{\psi} kl}^{(1)})^2-q_{kl}^{(1)}=1, \quad\quad \forall k,l=1,2,3,
\end{align}
where $g_{\alpha\beta kl}^{(1)}$ has been defined to be antisymmetric in $(\alpha,\,\beta)$
and in $(k,\,l)$. Analogously, $h_{\alpha\tilde{\psi} kl}^{(1)}$ is antisymmetric in ($\alpha,\,\tilde{\psi})$
and in $(k,\,l)$ by definition.
We do an identical rewriting of the sixth and seventh terms of $\mathcal{T}_4^{(1)}$ and $\mathcal{T}_4^{(2)}$
too. That is, we rewrite the mentioned terms (whose original form can be directly read from (\ref{mathcalt4}) and
(\ref{tau123def}) or even simply inferred from the subsequent equation)
in the more convenient form
\begin{align}
\nonumber
&\sum_{\alpha,\beta=1}^2\Big(\ldots-i\sum_{k=1}^3g_{\alpha\beta k}^{(2)}\sqrt{c_{rk}}[\mathcal{A}_r,\varphi_k] 
-i\sum_{k=1}^3g_{\alpha\beta k}^{(3)}\sqrt{c_{\phi_1k}}[\mathcal{A}_{\phi_1},\varphi_k]+\ldots\Big)^2\\
\nonumber
&+\sum_{\alpha=1}^2\Big(\ldots -i\sum_{k=1}^3h_{\alpha\tilde{\psi} k}^{(2)}\sqrt{c_{rk}}[\mathcal{A}_r,\varphi_k]
-i\sum_{k=1}^3h_{\alpha\tilde{\psi} k}^{(3)}\sqrt{c_{\phi_1k}}[\mathcal{A}_{\phi_1},\varphi_k]+\ldots\Big)^2 \\
&+\sum_{k=1}^3 q_k^{(2)}c_{rk}[\mathcal{A}_r,\varphi_k]^2
+\sum_{k=1}^3 q_k^{(3)}c_{\phi_1k}[\mathcal{A}_{\phi_1},\varphi_k]^2+\chi_2+\chi_3. \label{termn}
\end{align}
We also demand the following constraints
\begin{align}
\label{consn}
2(g_{12k}^{(i)})^2+\sum_{\alpha=1}^2(h_{\alpha\tilde{\psi} k}^{(i)})^2-q_k^{(i)}=1, \quad\quad \forall i=1,2,
\quad\quad \forall k=1,2,3.
\end{align}
Here, $g_{\alpha\beta k}^{(i)}$ has been defined to be antisymmetric in ($\alpha,\,\beta)$
and $h_{\alpha\tilde{\psi} k}^{(i)}$ in ($\alpha,\,\tilde{\psi})$, for both $i=2,\,3$.

The only two terms left, fourth terms of $\mathcal{T}_4^{(1)}$ and $\mathcal{T}_4^{(2)}$ in (\ref{mathcalt4}),
will be rewritten in a slightly trickier way. Essentially, we will first ``mix'' them and then
multiply those mixed terms with new coefficients. Again, we will make sure that the squared terms
are not affected in the rewriting by subjecting the coefficients introduced to constraint equations.
For the time being, we will not determine the additional
crossed terms thus produced. To make the idea more precise, let us
first consider a toy model to illustrate how we will proceed. Consider the Hamiltonian
\begin{align}
H^{(2)}=
\frac{1}{2}(\hat{\mathbb{A}}+\hat{\mathbb{B}})^2+
\frac{1}{2}(\hat{\mathbb{C}}+\hat{\mathbb{D}})^2=
\frac{1}{2}(\hat{\mathbb{A}}^2+\hat{\mathbb{B}}^2+\hat{\mathbb{C}}^2+\hat{\mathbb{D}}^2)+\textrm{ crossed terms}.
\label{quadratic1}
\end{align}
We will ``mix'' the terms $(\hat{\mathbb{B}},\,\hat{\mathbb{D}})$ in the above.
To this aim, we define $\hat{\mathbb{E}}\equiv \hat{\mathbb{B}}+\hat{\mathbb{D}}$.
Next, we insert inside the squares the factors of $(1/2)$
and introduce the arbitrary coefficients ($\hat{u},\,\hat{v}$). All these changes allow us to rewrite the toy Hamiltonian as
\begin{align}
H^{(2)}=(\frac{\hat{\mathbb{A}}}{\sqrt{2}}+\hat{u}\hat{\mathbb{E}})^2
+(\frac{\hat{\mathbb{C}}}{\sqrt{2}}+\hat{v}\hat{\mathbb{E}})^2
=\frac{1}{2}(\hat{\mathbb{A}}^2+\hat{\mathbb{C}})^2+(\hat{u}^2+\hat{v}^2)(\hat{\mathbb{B}}^2+\hat{\mathbb{D}}^2)
+\textrm{ crossed terms}. \label{quadratic2}
\end{align}
If we demand that the squared terms in (\ref{quadratic1}) and (\ref{quadratic2}) match, then it is clear that
$(\hat{u},\,\hat{v})$ must satisfy the following constraint:
\begin{align}
\hat{u}^2+\hat{v}^2=\frac{1}{2}.
\end{align}

Coming back to the fourth terms in
$\mathcal{T}_4^{(1)}$ and $\mathcal{T}_4^{(2)}$ that motivated the just explained toy model, these are given by
\begin{align}
\frac{1}{2}\sum_{\alpha,\beta=1}^2\Big(\ldots
-\sum_{k=1}^3\sqrt{b_{\tilde{\psi} k}}\epsilon_{\alpha\beta\tilde{\psi} k}\mathcal{D}_{\tilde{\psi}}\varphi_k+\ldots\Big)^2 
+\frac{1}{2}\sum_{\alpha=1}^2\Big(
\ldots-\sum_{k=1}^3\sqrt{b_{\beta k}}\epsilon_{\alpha\tilde{\psi}\beta  k}\mathcal{D}_\beta\varphi_k+\ldots\Big)^2.
\label{terms4}
\end{align}
Following the logic above exposed, we introduce $\delta\equiv(\alpha,\,\tilde{\psi})$ and rewrite (\ref{terms4}) as
\begin{align}
\sum_{\alpha,\beta=1}^2\hspace*{-0.2cm}\Big(\ldots
+\hspace*{-0.2cm}\sum_{\delta,k=1}^3\sqrt{b_{\delta k}}\epsilon_{\alpha\beta}\cdot m_{\delta k}^{(1)}
\mathcal{D}_\delta\varphi_k+\ldots\Big)^2 
+\sum_{\alpha=1}^2\hspace*{-0.1cm}\Big(
\ldots
+\hspace*{-0.2cm}\sum_{\delta,k=1}^3\sqrt{b_{\delta k}}\epsilon_{\alpha\tilde{\psi}}\cdot m_{\delta k}^{(2)}
\mathcal{D}_\delta\varphi_k+\ldots\Big)^2, \label{termm}
\end{align}
plus some extra crossed terms which we shall refer to symbolically as
$\chi_m$. The dot products appearing above will be made precise soon enough, in section \ref{minimsec}.
The new coefficients above must satisfy
\begin{align}
\label{consm}
\sum_{i=1}^2(m_{\delta k}^{(i)})^2=\frac{1}{2}, \quad\quad \forall \delta,k=1,2,3,
\end{align}
which makes sure the quadratic terms have not been changed during the rewriting. Note that
there is no antisymmetry relating the indices of these coefficients, unlike in previous cases.

We are now ready to collect results and present the Hamiltonian following from the action (\ref{totaction})
(with $c_2=0$) in the most convenient form for our subsequent investigations.
Appropriately summing (\ref{terms}), (\ref{termt}),
(\ref{term4}), (\ref{term1}), (\ref{termn}) and (\ref{termm}) we obtain the desired rewriting of $\mathcal{T}_4$ in
(\ref{mathcalt4}). Further adding $(\mathcal{T}_1,\,\mathcal{T}_2,\,\mathcal{T}_3)$ as given in (\ref{mathcalt1}),
the Hamiltonian in (\ref{Hamiltonianinitial}) can be rewritten as in (3.158) in~\cite{Dasgupta:2016rhc}:
\begin{align}
\begin{array}{lll}
\label{3158}
H=&\displaystyle\int d^4x\,\,\textrm{Tr}\Bigg\{
\frac{C_1}{V_3}\Bigg[\sum_{\alpha=1}^2(\sqrt{c_{11}}\mathcal{F}_{\alpha0}-\sqrt{c_{\alpha \tilde{3}}}
\mathcal{D}_\alpha\mathcal{A}_{\tilde{3}})^2+(\sqrt{c_{12}}\mathcal{F}_{\tilde{\psi}0}-
\sqrt{c_{\tilde{\psi}\tilde{3}}}\mathcal{D}_{\tilde{\psi}}\mathcal{A}_{\tilde{3}})^2 \\
&\displaystyle+(\sqrt{c_{0r}}\mathcal{D}_0\mathcal{A}_r-i\sqrt{a_2}
[\mathcal{A}_{\tilde{3}},\mathcal{A}_r])^2+
(\sqrt{\tilde{c}_{0\phi_1}}\mathcal{D}_0\mathcal{A}_{\phi_1}-i\sqrt{a_4}[\mathcal{A}_{\tilde{3}},\mathcal{A}_{\phi_1}])^2 
+c_{0\tilde{3}}(\mathcal{D}_0\mathcal{A}_{\tilde{3}})^2\Bigg] \\
&\displaystyle+\sum_{k,l=1}^3\Bigg[
(\sqrt{b_{0k}}\mathcal{D}_0\varphi_k-i\sqrt{c_{\tilde{3}k}}[\mathcal{A}_{\tilde{3}},\varphi_k])^2
+q_{kl}^{(1)} d_{kl}[\varphi_k,\varphi_l]^2
+\sum_{\gamma=2}^3q_k^{(\gamma)}c_{y_\gamma k}[\mathcal{A}_{y_\gamma},\varphi_k]^2\Bigg]\\
&\displaystyle+\sum_{\alpha,\beta=1}^2\Big(
\sqrt{\frac{C_1c_{11}}{2V_3}}\mathcal{F}_{\alpha\beta}
+\sqrt{\frac{C_1c_{\tilde{\psi} r}}{V_3}}s_{\alpha\beta}^{(1)}\epsilon_{\alpha \beta\tilde{\psi} r}
\mathcal{D}_{\tilde{\psi}}\mathcal{A}_r
+\sqrt{\frac{C_1\tilde{c}_{\tilde{\psi} \phi_1}}{V_3}}s_{\alpha\beta}^{(2)}\epsilon_{\alpha\beta\tilde{\psi} \phi_1}
\mathcal{D}_{\tilde{\psi}}\mathcal{A}_{\phi_1}\\
&\displaystyle-ig_{\alpha\beta}^{(4)}\sqrt{\frac{C_1a_1}{V_3}}[\mathcal{A}_r,\mathcal{A}_{\phi_1}]
-i\sum_{k,l=1}^3g_{\alpha\beta kl}^{(1)}\sqrt{d_{kl}}[\varphi_k,\varphi_l]
-i\sum_{k=1}^3\sum_{\gamma=2}^3g_{\alpha\beta k}^{(\gamma)}\sqrt{c_{y_\gamma k}}[\mathcal{A}_{y_\gamma},\varphi_k] \\
&\displaystyle
+\sum_{\delta,k=1}^3\sqrt{b_{\delta k}}\epsilon_{\alpha\beta}\cdot m_{\delta k}^{(1)}
\mathcal{D}_\delta\varphi_k
\Big)^2+\sum_{\alpha=1}^2\Big(
\sqrt{\frac{C_1c_{12}}{2V_3}}\mathcal{F}_{\alpha\tilde{\psi}}
+\sqrt{\frac{C_1c_{\beta r}}{V_3}}t_{\alpha}^{(1)}\epsilon_{\alpha\tilde{\psi} \beta r}\mathcal{D}_\beta\mathcal{A}_r\\
&\displaystyle
+\sqrt{\frac{C_1\tilde{c}_{\beta \phi_1}}{V_3}}t_{\alpha}^{(2)}\epsilon_{\alpha\tilde{\psi}\beta \phi_1}
\mathcal{D}_\beta\mathcal{A}_{\phi_1}
-ih_{\alpha\tilde{\psi}}^{(4)}\sqrt{\frac{C_1a_1}{V_3}}[\mathcal{A}_r,\mathcal{A}_{\phi_1}]
-i\sum_{k,l=1}^3h_{\alpha\tilde{\psi} kl}^{(1)}\sqrt{d_{kl}}[\varphi_k,\varphi_l]\\
&\displaystyle
-i\sum_{k=1}^3\sum_{\gamma=2}^3h_{\alpha\tilde{\psi} k}^{(\gamma)}\sqrt{c_{y_\gamma k}}[\mathcal{A}_{y_\gamma},\varphi_k]
+\sum_{\delta,k=1}^3\sqrt{b_{\delta k}}\epsilon_{\alpha\tilde{\psi}}\cdot m_{\delta k}^{(2)}
\mathcal{D}_\delta\varphi_k
\Big)^2
+\frac{C_1}{V_3}\Bigg[c_{\tilde{\psi} r}s^{(1)}(\mathcal{D}_{\tilde{\psi}}\mathcal{A}_r)^2\\
&\displaystyle
+\tilde{c}_{\tilde{\psi} \phi_1}s^{(2)}(\mathcal{D}_{\tilde{\psi}}\mathcal{A}_{\phi_1})^2
+c_{\beta r}t^{(1)}(\mathcal{D}_\beta\mathcal{A}_r)^2 
+\tilde{c}_{\beta \phi_1}t^{(2)}(\mathcal{D}_\beta\mathcal{A}_{\phi_1})^2
+a_1q^{(4)}[\mathcal{A}_r,\mathcal{A}_{\phi_1}]^2\Bigg]
+\chi_T\Bigg\}\\
&\displaystyle+Q_{EM},
\end{array}
\end{align}
where we have defined $(y_2,\,y_3)\equiv(r,\,\phi_1)$ (as a short-hand notation) and 
\begin{align}
\chi_T\equiv \chi_s+\chi_t+\chi_4+\chi_1+\chi_2+\chi_3 +\chi_m.
\end{align}
That is, $\chi_T$ accounts for all crossed terms produced when rewriting $\mathcal{T}_4$ as just explained.
$\chi_T$ will be the main object of study of section \ref{loceqsec}, but presently we shall not shed light into it.

We remind the reader that most of the notation used above was introduced in section \ref{actionsec}.
In particular, table \ref{table1} provides a quick guide to find the explicit form of the prefactors that have a supergravity
interpretation in terms of the warp factors in (\ref{easychoice}) and (\ref{tildeF2}), the deformation parameter $\theta_{nc}$
in (\ref{ncdeformation}) and the leading term of the dilaton in (\ref{consdil}).
For clarity and completeness, we include table \ref{table2}, which summarizes the form and properties of the new
coefficients introduced in going form (\ref{totaction}) to (\ref{3158}). Note that these coefficients
do {\it not} have a supergravity interpretation. Instead, the constraint relations we demanded in this section that
they should satisfy should be regarded as their {\it defining} equations.
These are (\ref{conss}), (\ref{const}), (\ref{cons4}), (\ref{cons1}), (\ref{consn}) and (\ref{consm}), which put together
recover (3.160) in~\cite{Dasgupta:2016rhc}.

\begin{table}[t]
\begin{tabular}{| c||c |}
\hline
Coefficient & Given in  \\  \hline  \hline 
$s_{\alpha\beta}^{(i)},\,s^{(i)}$ & (\ref{conss}) \\  \hline
$t_\alpha^{(i)},\,t^{(i)}$ & (\ref{const}) \\ \hline
\smash{\raisebox{30pt}{\rlap{\hspace{5cm} \, with: \, $\alpha,\,\beta,\,i=1,\,2$ and $k,\,l,\,\delta=1,\,2,\,3$,}}} 
\smash{\raisebox{5pt}{\rlap{\hspace{6cm}  \, $(s_{\alpha\beta}^{(i)},\,g_{\alpha\beta}^{(4)},\,g_{\alpha\beta kl}^{(1)},\,
g_{\alpha\beta k}^{(i)})$ antisymmetric in $(\alpha,\,\beta)$,}}} 
\smash{\raisebox{-20pt}{\rlap{\hspace{6.1cm}   $(h_{\alpha\tilde{\psi}}^{(4)},\,h_{\alpha\tilde{\psi}kl}^{(1)},\,
h_{\alpha\tilde{\psi} k}^{(i)})$ antisymmetric in $(\alpha,\,\tilde{\psi})$ and}}} 
\smash{\raisebox{-45pt}{\rlap{\hspace{6cm}   $(g_{\alpha\beta kl}^{(1)},\,h_{\alpha\tilde{\psi}kl}^{(1)},\,
q_{kl}^{(1)})$ antisymmetric in $(k,\,l)$.}}} 
$g_{\alpha\beta}^{(4)},\,h_{\alpha\tilde{\psi}}^{(4)},\,q^{(4)}$ & (\ref{cons4}) \\ \hline
$g_{\alpha\beta kl}^{(1)},\,h_{\alpha\tilde{\psi}kl}^{(1)},\,q_{kl}^{(1)}$ & (\ref{cons1}) \\  \hline
$g_{\alpha\beta k}^{(i)},\,h_{\alpha\tilde{\psi}k}^{(i)},\,q_k^{(i)}$ & (\ref{consn}) \\  \hline
$m_{\delta k}^{(i)}$ & (\ref{consm}) \\ \hline
\end{tabular}
\caption{List of coefficients appearing in the Hamiltonian (\ref{3158}) that do not have a supergravity
interpretation, the equation numbers of their defining relations and their antisymmetry properties.
Note that $m_{\delta k}^{(i)}$'s are {\it not} constrained by antisymmetry. These coefficients
are introduced while rewriting the Hamiltonian (\ref{Hamiltonianinitial}) as (\ref{3158}).}
\label{table2}
\end{table}

\FloatBarrier

\subsubsection{Minimization of the Hamiltonian \label{minimsec}}

Having written the Hamiltonian of our theory as (\ref{3158}), we now make the following crucial observation:
this is a sum of squared terms, plus boundary terms $Q_{EM}$ and ``crossed terms'' $\chi_T$.
Ignoring momentarily $(Q_{EM},\,\chi_T)$, it is clear that in order to minimize the energy of the system
each such squared term must vanish separately. In this section we enforce the just described minimization and thus
obtain the (bulk) equations of motion for the $SU(N)$ gauge theory in the four-dimensional space $X_4$ parametrized by
$(t,\,x_1,\,x_2,\,\tilde{\psi})$.

Let us start by setting to zero the first six squared terms in (\ref{3158}). (These are the terms stemming from 
$(\mathcal{T}_1,\,\mathcal{T}_2,\,\mathcal{T}_3)$ in (\ref{mathcalt1}).)
Since we wish our discussion to be as general as possible, we assume that the coefficients $C_1/V_3$
and $c_{0\tilde{3}}$ do not vanish. Then, we obtain the following:
\begin{align}
\begin{array}{lll}
&\displaystyle (\sqrt{c_{11}}\mathcal{F}_{\alpha0}-\sqrt{c_{\alpha\tilde{3}}}\mathcal{D}_\alpha
\mathcal{A}_{\tilde{3}})^2=0, &\displaystyle \quad\quad
(\sqrt{c_{12}}\mathcal{F}_{\tilde{\psi}0}-\sqrt{c_{\tilde{\psi}\tilde{3}}}\mathcal{D}_{\tilde{\psi}}\mathcal{A}_{\tilde{3}})^2
=0, \\
&\displaystyle (\sqrt{c_{0r}}\mathcal{D}_0\mathcal{A}_r-i\sqrt{a_2}[\mathcal{A}_{\tilde{3}},\mathcal{A}_r])^2=0,
&\displaystyle \quad\quad
(\sqrt{\tilde{c}_{0\phi_1}}\mathcal{D}_0\mathcal{A}_{\phi_1}-i\sqrt{a_4}[\mathcal{A}_{\tilde{3}},\mathcal{A}_{\phi_1}])^2=0, \\
&\displaystyle (\sqrt{b_{0k}}\mathcal{D}_0\varphi_k-i\sqrt{c_{\tilde{3}k}}[\mathcal{A}_{\tilde{3}},\varphi_k])^2=0,
&\displaystyle \quad\quad
\mathcal{D}_0\mathcal{A}_{\tilde{3}}=0, \label{D0A30}
\end{array}
\end{align}
which should hold true $\forall \alpha=1,\,2$ and $\forall k=1,\,2,\,3$.
Recall now that both the gauge fields $(\mathcal{A}_a,\,\mathcal{A}_{\tilde{\psi}})$ (with $a=0,\,1,\,2$)
and the real scalars $(\mathcal{A}_{\tilde{3}},\,\mathcal{A}_{\phi_1},\,\mathcal{A}_r)$ (in the adjoint representation
of $SU(N)$) depend only on the coordinates $(t,\,x_1,\,x_2,\,\tilde{\psi})$.
As we pointed out in the beginning of section \ref{hamilsec}, not only are we interested in obtaining the minimum
energy configuration for the aforementioned fields, but we also want them to satisfy the BPS conditions.
Hence, we search for static solutions to (\ref{D0A30}).
This implies we will consider in the ongoing that the fields only depend on $(x_1,\,x_2,\,\tilde{\psi})$
and thus, using (\ref{covder}), the above reduces to
\begin{align}
\begin{array}{llll}
&\displaystyle (\sqrt{c_{11}}\mathcal{D}_\alpha \mathcal{A}_0
-\sqrt{c_{\alpha\tilde{3}}}\mathcal{D}_\alpha\mathcal{A}_{\tilde{3}})^2=0, &\displaystyle \quad\quad 
(\sqrt{c_{12}}\mathcal{D}_{\tilde{\psi}}\mathcal{A}_0
-\sqrt{c_{\tilde{\psi}\tilde{3}}}\mathcal{D}_{\tilde{\psi}}\mathcal{A}_{\tilde{3}})^2=0, \\
&\displaystyle (\sqrt{c_{0r}}[\mathcal{A}_0,\mathcal{A}_r]
-\sqrt{a_2}[\mathcal{A}_{\tilde{3}},\mathcal{A}_r])^2=0, &\displaystyle \quad\quad
(\sqrt{\tilde{c}_{0\phi_1}}[\mathcal{A}_0,\mathcal{A}_{\phi_1}]
-\sqrt{a_4}[\mathcal{A}_{\tilde{3}},\mathcal{A}_{\phi_1}])^2=0, \\
&\displaystyle (\sqrt{b_{0k}}[\mathcal{A}_0,\varphi_k]
-\sqrt{c_{\tilde{3}k}}[\mathcal{A}_{\tilde{3}},\varphi_k])^2=0, &\displaystyle \quad\quad 
[\mathcal{A}_0,\mathcal{A}_{\tilde{3}}]=0, \label{6firsts}
\end{array}
\end{align}
valid again $\forall \alpha=1,\,2$ and $\forall k=1,\,2,\,3$.

To proceed further, we need to choose a gauge.
We make the gauge choice in (3.161) in~\cite{Dasgupta:2016rhc}:
\begin{align}
\mathcal{A}_0=\mathcal{A}_{\tilde{3}}. \label{3161}
\end{align}
This follows from our earlier identifications in (\ref{idenkor}), where the scalar field $\mathcal{A}_{\tilde{3}}$
was singled out from the other two scalars $(\mathcal{A}_{\phi_1},\,\mathcal{A}_r)$. One could certainly single out
$\mathcal{A}_{\phi_1}$ or $\mathcal{A}_r$ instead and appropriately modify the above gauge choice.
We will not entertain these options in the present work, as they do not lead to further physical insight.
However, the interested reader
can find enough detail on the $\mathcal{A}_0=\mathcal{A}_r$ gauge choice in (3.178)-(3.182) in~\cite{Dasgupta:2016rhc}.
With the choice (\ref{3161}), the set of equations in (\ref{6firsts}) reduces to (3.162) in~\cite{Dasgupta:2016rhc}:
\begin{align}
\begin{array}{llll}
&\displaystyle (\sqrt{c_{11}}
-\sqrt{c_{\alpha\tilde{3}}})^2(\mathcal{D}_\alpha\mathcal{A}_{\tilde{3}})^2=0, &\displaystyle \quad\quad
(\sqrt{c_{12}}
-\sqrt{c_{\tilde{\psi}\tilde{3}}})^2(\mathcal{D}_{\tilde{\psi}}\mathcal{A}_{\tilde{3}})^2=0, \\
&\displaystyle (\sqrt{c_{0r}}
-\sqrt{a_2})^2[\mathcal{A}_{\tilde{3}},\mathcal{A}_r]^2=0,  &\displaystyle \quad\quad
(\sqrt{\tilde{c}_{0\phi_1}}
-\sqrt{a_4})^2[\mathcal{A}_{\tilde{3}},\mathcal{A}_{\phi_1}]^2=0, \\ &\displaystyle  
(\sqrt{b_{0k}}
-\sqrt{c_{\tilde{3}k}})^2[\mathcal{A}_{\tilde{3}},\varphi_k]^2=0, &\displaystyle \quad\quad 
\forall\alpha=1,2, \quad \forall k=1,2,3. \label{3162}
\end{array}
\end{align}
Note that the last equation in (\ref{6firsts}) does not appear above, since it is trivially satisfied
by our gauge choice.

The above has the trivial solution $\mathcal{A}_{\tilde{3}}=0$.
Another possible solution would be to simultaneously satisfy 
\begin{align}
\label{eq1coe}
c_{11}=c_{\alpha\tilde{3}}, \quad c_{12}=c_{\tilde{\psi}\tilde{3}}, \quad c_{0r}=a_2,\quad
\tilde{c}_{0\phi_1}=a_4, \quad b_{0k}=c_{\tilde{3}k}, \quad \forall\alpha=1,2, \,\, \forall k=1,2,3.
\end{align}
Let us explore this option by using the explicit form of the above coefficients, summarized previously in
table \ref{table1}.
From (\ref{c11final}), (\ref{tildea1a2}) and (\ref{csfinal}), we immediately see that 
the first equation will be satisfied if and only if 
\begin{align}
\cos^2\theta_{nc}+F_2\sin^2\theta_{nc}=1. \label{Fthcons}
\end{align}
Similarly, using (\ref{379}), (\ref{c12final}), (\ref{tildeb2}) and (\ref{cpsis}) in the second equation,
one can right away conclude (\ref{Fthcons}) is required so that $c_{12}=c_{\tilde{\psi}\tilde{3}}$.
The same deduction follows from introducing (\ref{a1a2a4}), (\ref{tildea1a2}) and (\ref{csfinal})
in $c_{0r}=a_2$. On the other hand, using these same
results in $\tilde{c}_{0\phi_1}=a_4$, one finds that, besides (\ref{Fthcons}),
it is also necessary to impose
\begin{align}
\frac{(\tilde{F}_2\tan\theta_{nc})^2}{\tilde{F}_2-F_3}(1+F_2\tan^2\theta_{nc})=0. \label{othcon}
\end{align}
Finally, from (\ref{bakfinal}) and (\ref{cpsis}) it follows that $b_{0k}=c_{\tilde{3}k}$
iff we demand (\ref{Fthcons}). Summing up, to ensure (\ref{eq1coe})
we must enforce both (\ref{Fthcons}) and (\ref{othcon}). But in doing so,
we do not wish to constraint our set up by choosing a particular form for the warp factors.
(We want to keep our M-theory configuration (M, 1) of part \ref{parta} as general as possible.)
Hence, we conclude that the second possible solution to (\ref{3162}) is given by $\theta_{nc}=0$.

Between $\mathcal{A}_{\tilde{3}}=0$ and $\theta_{nc}=0$, there is a preferred solution to (\ref{3162}).
Recall section \ref{ncsect}: $\theta_{nc}$ was introduced as an alternative and computationally simpler
way to account for the axionic background of~\cite{Witten:2011zz}, which was there shown to be
an essential ingredient to study knots using the D3-NS5 system. In our approach too
(as we will show in section (\ref{twistsec})), $\theta_{nc}$ shall play a key role and allow us to construct
a three-dimensional space capable of supporting knots.
Accordingly, we set to zero the first six squared terms in the Hamiltonian (\ref{3158}) via
\begin{align}
\mathcal{A}_{\tilde{3}}=0, \label{A3zeroch}
\end{align}
along with the gauge choice in (\ref{3161})\footnote{This
implies $\mathcal{A}_0=0$, known as the Weyl gauge or also as the
axial gauge.}. Also, bear in mind all fields are time-independent now. 

Let us next turn our attention to the final five terms, as well as the last two terms in the third line of
the Hamiltonian (\ref{3158}). (These are the squared terms we introduced to make sure that
while rewriting the Hamiltonian (\ref{Hamiltonianinitial})
as (\ref{3158}) all quadratic terms remain unaffected.) Minimization of the energy requires them all to vanish which,
for $(C_1/V_3)\neq0$, means that
\begin{align}
\begin{array}{lllll}
&s^{(1)}(\mathcal{D}_{\tilde{\psi}}\mathcal{A}_r)^2=0, &\,\,
s^{(2)}(\mathcal{D}_{\tilde{\psi}}\mathcal{A}_{\phi_1})^2=0, &\,\,
t^{(1)}(\mathcal{D}_\beta\mathcal{A}_r)^2=0,  &\,\,
t^{(2)}(\mathcal{D}_\beta\mathcal{A}_{\phi_1})^2=0, \\
&a_1q^{(4)}[\mathcal{A}_r,\mathcal{A}_{\phi_1}]^2=0, &\,\,
q_{kl}^{(1)} d_{kl}[\varphi_k,\varphi_l]^2=0, &\,\,
q_k^{(\gamma)}c_{y_\gamma k}[\mathcal{A}_{y_\gamma},\varphi_k]^2=0,  \label{eom27}
\end{array}
\end{align}
for all $\beta=1,\,2$, $k,\,l=1,\,2,\,3$ and $\gamma=2,\,3$.
If we consider that, generically, all the coefficients
$(s^{(1)},\,s^{(2)},\,t^{(1)},\,t^{(2)},\,a_1,\,q^{(4)},\,q_k^{(\gamma)},\,c_{y_\gamma k})$ 
are not zero, then satisfying (\ref{eom27}) implies (3.167) and (3.169) in~\cite{Dasgupta:2016rhc}:
\begin{align}
\label{detazero}
\mathcal{D}_\eta\mathcal{A}_r=\mathcal{D}_\eta\mathcal{A}_{\phi_1}=[\mathcal{A}_r,\mathcal{A}_{\phi_1}]=
[\mathcal{A}_r,\varphi_k]=[\mathcal{A}_{\phi_1},\varphi_k]=0,
\quad\quad \forall\eta=1,2,\tilde{\psi}, \quad \forall k=1,2,3.
\end{align}
On the other hand, if we do not wish to trivialize the system, we cannot conclude that
most generically all $q_{kl}^{(1)}$'s are non-zero. (Note that this would imply $[\varphi_k,\varphi_l]=0$
for all $(k,\,l)$.) Hence, as the simplest non-trivial case, we will consider only one such
(independent) coefficient vanishes. Following~\cite{Dasgupta:2016rhc}, we choose $q_{12}^{(1)}=0$.
Then, to fulfill (\ref{eom27}), we must impose (3.171) in~\cite{Dasgupta:2016rhc} too:
\begin{align}
\label{3132zero}
[\varphi_1,\varphi_2]\neq0, \quad\quad
[\varphi_1,\varphi_3]=[\varphi_2,\varphi_3]=0.
\end{align}
In this manner, we have enforced (\ref{eom27}).

In our minimization of the Hamiltonian (\ref{3158}), we now focus on the squared term between the fourth and sixth lines
and demand its vanishing:
\begin{align}
\nonumber
&\sqrt{\frac{C_1}{V_3}}\left(
\sqrt{\frac{c_{11}}{2}}\mathcal{F}_{\alpha\beta}
+\sqrt{c_{\tilde{\psi} r}}s_{\alpha\beta}^{(1)}\epsilon_{\alpha \beta\tilde{\psi}r}
\mathcal{D}_{\tilde{\psi}}\mathcal{A}_r
+\sqrt{\tilde{c}_{\tilde{\psi} \phi_1}}s_{\alpha\beta}^{(2)}\epsilon_{\alpha\beta\tilde{\psi} \phi_1}
\mathcal{D}_{\tilde{\psi}}\mathcal{A}_{\phi_1}
-ig_{\alpha\beta}^{(4)}\sqrt{a_1}[\mathcal{A}_r,\mathcal{A}_{\phi_1}]\right)\\
& \label{bhn12sqterm}
-i\sum_{\delta,k,l=1}^3\left(g_{\alpha\beta kl}^{(1)}\sqrt{d_{kl}}[\varphi_k,\varphi_l]
+\sum_{\gamma=2}^3g_{\alpha\beta k}^{(\gamma)}\sqrt{c_{y_\gamma k}}[\mathcal{A}_{y_\gamma},\varphi_k]
+i\sqrt{b_{\delta k}}\epsilon_{\alpha\beta}\cdot m_{\delta k}^{(1)}
\mathcal{D}_\delta\varphi_k\right)=0,
\end{align}
which should be true for all $\alpha,\,\beta=1,\,2$. Needless to say, minimization of the energy requires all
squared terms to vanish simultaneously. This implies the choices previously made to set to zero other squared terms
must now be enforced as well. Thus, inserting (\ref{detazero}) and (\ref{3132zero}) in the above, our equations
reduce to
\begin{align}
\sqrt{\frac{C_1c_{11}}{2V_3}}\mathcal{F}_{\alpha\beta}
-2ig_{\alpha\beta 12}^{(1)}\sqrt{d_{12}}[\varphi_1,\varphi_2]
+\sum_{\delta,k=1}^3\sqrt{b_{\delta k}}\epsilon_{\alpha\beta}\cdot m_{\delta k}^{(1)}
\mathcal{D}_\delta\varphi_k=0, \quad\quad \forall\alpha,\beta=1,2, \label{bhn1ontw}
\end{align}
where we have used the fact that $g_{\alpha\beta 12}^{(1)}=-g_{\alpha\beta 21}^{(1)}$ by definition and $d_{12}=d_{21}$,
as can be seen from (\ref{dklfinal}). Since (\ref{bhn1ontw}) is antisymmetric in $(\alpha,\,\beta)$,
we can focus on the case $\alpha=1$ and $\beta=2$. With the convention that $\epsilon_{12}=1$, noting that (\ref{bakfinal})
tells us that
$b_{12}=b_{21}$ and choosing
coefficients as in (3.173) in~\cite{Dasgupta:2016rhc}; namely
\begin{align}
g_{1212}^{(1)}=m_{\tilde{\psi}3}^{(1)}=m_{12}^{(1)}=-m_{21}^{(1)}=\frac{1}{\sqrt{2}}, \quad\quad
m_{11}^{(1)}=m_{22}^{(1)}=m_{13}^{(1)}=m_{23}^{(1)}=m_{\tilde{\psi}1}^{(1)}=m_{\tilde{\psi}2}^{(1)}=0, \label{choosingmgs}
\end{align}
it is a matter of minor algebra to obtain (3.172) in~\cite{Dasgupta:2016rhc}:
\begin{align}
\mathcal{F}_{12}
+\sqrt{\frac{V_3}{C_1c_{11}}}\left[-2i\sqrt{d_{12}}[\varphi_1,\varphi_2]
+\sqrt{b_{12}}(\mathcal{D}_1\varphi_2-\mathcal{D}_2\varphi_1)
+\sqrt{b_{\tilde{\psi}3}}\mathcal{D}_{\tilde{\psi}}\varphi_3\right]=0.\label{3172}
\end{align}
Note that the dot product in (\ref{bhn1ontw}) has been interpreted as a usual scalar product in this case.

This is the first non-trivial equation of motion following from the minimization of the energy of the Hamiltonian (\ref{3158}).
Further, since all fields appearing in it are static, the above is a BPS condition.
Notice now that, schematically, our BPS condition is of the form
\begin{align}
\mathcal{F}+\mathcal{D}\varphi+[\varphi,\varphi]=0.
\end{align}
The well-versed reader will of course be familiar with the Bogomolny, Hitchin and Nahm equations, which we can sketch
as follows:
\begin{align}
\textrm{Bogomolny: }\,\, \mathcal{F}+\mathcal{D}\varphi=0, \quad\quad 
\textrm{Hitchin: }\,\, \mathcal{F}+[\varphi,\varphi]=0, \quad\quad 
\textrm{Nahm: }\,\, \mathcal{D}\varphi+[\varphi,\varphi]=0. \label{explbhn}
\end{align}
Written in this manner, it is evident that our BPS condition is just a combination of all these
Bogomolny, Hitchin and Nahm equations. We will thus refer to (\ref{3172}) as the first BHN equation.

Before proceeding further, let us pause for a moment and study what are the consequences of the choices of coefficients
made so far. These choices are $q_{12}^{(1)}=0$ and (\ref{choosingmgs}). As can be checked in table \ref{table2},
these coefficients are required to satisfy the constraint equations (\ref{cons1}) and (\ref{consm}).
So, combining our choices and the constraints, we are led to conclude that
\begin{align}
\nonumber
&2\left(g_{12kl}^{(1)}\right)^2+\sum_{\alpha=1}^2\left(h_{\alpha\tilde{\psi}kl}^{(1)}\right)^2-q_{kl}^{(1)}=1
\quad \forall k,l=2,3, \quad\quad h_{\alpha\tilde{\psi}12}^{(1)}=-h_{\alpha\tilde{\psi}21}^{(1)}=0 \quad \forall \alpha=1,2, \\
\label{m2zeroes}
&m_{\psi3}^{(2)}=m_{12}^{(2)}=m_{21}^{(2)}=0, \quad\quad 
m_{11}^{(2)},m_{22}^{(2)},m_{13}^{(2)},m_{23}^{(2)},m_{\tilde{\psi}1}^{(2)},m_{\tilde{\psi}2}^{(2)}=\pm\frac{1}{\sqrt{2}}
\end{align}
must hold true in the following.

The last step in the minimization of the energy of our system with Hamiltonian (\ref{3158}) is to demand
the vanishing of the squared term between the sixth and the eighth lines in that same equation.
This must be done in a consistent manner to all previous choices made in this section.
The necessary vanishing we just mentioned is
\begin{align}
\nonumber
&\sqrt{\frac{C_1}{V_3}}\left(\sqrt{\frac{c_{12}}{2}}\mathcal{F}_{\alpha\tilde{\psi}}
+\sqrt{c_{\beta r}}t_{\alpha}^{(1)}\epsilon_{\alpha\tilde{\psi} \beta r}\mathcal{D}_\beta\mathcal{A}_r
+\sqrt{\tilde{c}_{\beta \phi_1}}t_{\alpha}^{(2)}\epsilon_{\alpha\tilde{\psi}\beta \phi_1}
\mathcal{D}_\beta\mathcal{A}_{\phi_1}
-ih_{\alpha\tilde{\psi}}^{(4)}\sqrt{a_1}[\mathcal{A}_r,\mathcal{A}_{\phi_1}]\right) \\
&-i\sum_{\delta,k,l=1}^3\left(h_{\alpha\tilde{\psi} kl}^{(1)}\sqrt{d_{kl}}[\varphi_k,\varphi_l]
+\sum_{\gamma=2}^3h_{\alpha\tilde{\psi} k}^{(\gamma)}\sqrt{c_{y_\gamma k}}[\mathcal{A}_{y_\gamma},\varphi_k]
+i\sqrt{b_{\delta k}}\epsilon_{\alpha\tilde{\psi}}\cdot m_{\delta k}^{(2)}
\mathcal{D}_\delta\varphi_k\right)=0, \label{bhnalpsc0}
\end{align}
for all $\alpha,\,\beta=1,\,2$.
Using (\ref{detazero}), (\ref{3132zero}) and (\ref{m2zeroes}) in the above, we have that
\begin{align}
\sqrt{\frac{C_1c_{12}}{2V_3}}\mathcal{F}_{\alpha\tilde{\psi}}
+\sum_{\delta,k=1}^3\sqrt{b_{\delta k}}\epsilon_{\alpha\tilde{\psi}}\cdot m_{\delta k}^{(2)}
\mathcal{D}_\delta\varphi_k=0 \quad\quad \forall\alpha=1,2. \label{bhn2almm}
\end{align}
Here, $\delta=3$ should be understood as making reference to the $\tilde{\psi}$ direction.
Without loss of generality, we take the definition of the dot product above to be
\begin{align}
\sum_{\delta,k=1}^3\sqrt{b_{\delta k}}\epsilon_{\alpha\tilde{\psi}}\cdot m_{\delta k}^{(2)}\mathcal{D}_\delta\varphi_k\equiv
-6\sum_{\delta,k=1}^3\sqrt{b_{\delta k}}\epsilon_{[\alpha\tilde{\psi}}m_{\delta k]}^{(2)}\mathcal{D}_\delta\varphi_k
+\sqrt{b_{\tilde{\psi}\alpha}}\epsilon_{\alpha\tilde{\psi}}m_{\tilde{\psi}\alpha}^{(2)}\mathcal{D}_{\tilde{\psi}}\varphi_\alpha,
\label{dotm2def}
\end{align}
with the indices of the first term on the right-hand side necessarily different from each other.
This seemingly involved term is not so complicated and, upon using the antisymmetry of the epsilon tensors,
is explicitly given by
\begin{align}
&-\frac{1}{2}\sum_{\delta,k=1}^3\sqrt{b_{\delta k}}\left[\epsilon_{\alpha{\tilde{\psi}}}(m_{\delta k}^{(2)}-m_{k\delta}^{(2)})
+\epsilon_{\delta{\tilde{\psi}}}(m_{\alpha k}^{(2)}-m_{k\alpha}^{(2)})
+\epsilon_{\delta k}(m_{\alpha3}^{(2)}-m_{{\tilde{\psi}}\alpha}^{(2)}) \right. \nonumber \\
&\left.+\epsilon_{\alpha k}(m_{\delta3}^{(2)}-m_{{\tilde{\psi}}\delta}^{(2)})
+\epsilon_{\alpha\delta}(m_{k3}^{(2)}-m_{{\tilde{\psi}} k}^{(2)})
+\epsilon_{k{\tilde{\psi}}}(m_{\alpha\delta}^{(2)}-m_{\delta\alpha}^{(2)})\right]\mathcal{D}_\delta\varphi_k.
\end{align}
In good agreement with (\ref{m2zeroes}), we now implement the second line there, choosing the plus sign
for all the $m^{(2)}$ coefficients in the last equality.
In this manner, the above reduces considerably to
\begin{align}
-\frac{1}{2}\sum_{\delta,k=1}^3\sqrt{b_{\delta k}}[\epsilon_{\delta k}m_{{\tilde{\psi}}\alpha}^{(2)}
+\epsilon_{\alpha k}m_{{\tilde{\psi}}\delta}^{(2)}+\epsilon_{\alpha\delta}m_{{\tilde{\psi}} k}^{(2)}]\mathcal{D}_\delta\varphi_k.
\end{align}
As we said, the dot product is taken by definition such that
all indices in this term should be different from each other. In other words, $\delta=1(2)$ if $\alpha=2(1)$ and $k=3$.
This leads to, for $\alpha,\,\beta=1,\,2$ with $\alpha\neq\beta$,
\begin{align}
\label{dcott1res}
-\frac{1}{2}\sqrt{b_{\beta 3}}[\epsilon_{\beta \tilde{\psi}}m_{{\tilde{\psi}}\alpha}^{(2)}
+\epsilon_{\alpha \tilde{\psi}}m_{{\tilde{\psi}}\beta}^{(2)}+\epsilon_{\alpha\beta}m_{{\tilde{\psi}} 3}^{(2)}]
\mathcal{D}_\beta\varphi_3=
\begin{cases}
\displaystyle\sqrt{\frac{b_{23}}{2}}\mathcal{D}_2\varphi_3 &\quad \textrm{if }\alpha=1,\,\,\beta=2, \\
\displaystyle\sqrt{\frac{b_{13}}{2}}\mathcal{D}_1\varphi_3 &\quad \textrm{if }\alpha=2, \,\,\beta=1,
\end{cases}
\end{align}
where the normalization convention used is $\epsilon_{1\tilde{\psi}}=\epsilon_{2\tilde{\psi}}=1$. 
Finally, using the above in (\ref{bhn2almm}) and with minor algebra,
we obtain the remaining two BHN equations, as in (3.177) in~\cite{Dasgupta:2016rhc}:
\begin{align}
\begin{array}{llll}
\label{3177one}
&\displaystyle\mathcal{F}_{1{\tilde{\psi}}}+\sqrt{\frac{V_3}{C_1c_{12}}}
(\sqrt{b_{{\tilde{\psi}}1}}\mathcal{D}_{\tilde{\psi}}\varphi_1
+\sqrt{b_{23}}\mathcal{D}_2\varphi_3)=0, \vspace*{0.2cm}\\
\vspace*{0.2cm}
&\displaystyle\mathcal{F}_{2{\tilde{\psi}}}+\sqrt{\frac{V_3}{C_1c_{12}}}
(\sqrt{b_{{\tilde{\psi}}2}}\mathcal{D}_{\tilde{\psi}}\varphi_2+\sqrt{b_{13}}\mathcal{D}_1\varphi_3 )=0.
\end{array}
\end{align}

Collecting thoughts, in this section we have shown that the vanishing of the different squared terms in the Hamiltonian
(\ref{3158}) for static configurations leads to the BHN equations (\ref{3172}) and (\ref{3177one}).
The name BHN simply denotes that
these are a combination of the well-known Bogomolny, Hitchin and Nahm equations.
In obtaining such BHN equations, we chose the gauge (\ref{3161}) and further found that the gauge and scalar fields
in the bosonic sector of the theory should also satisfy (\ref{A3zeroch}), (\ref{detazero}) and (\ref{3132zero}).
Additionally, we made the coefficient choices $q_{12}^{(1)}=0$, (\ref{choosingmgs}) and (\ref{m2zeroes}), with
the plus sign in all cases of the last equality there. One can easily check that all our choices
respect the defining equations of the coefficients, summarized previously in table \ref{table2}.
However, this analysis completely ignored the  $(Q_{EM},\,\chi_T)$ terms in (\ref{3158}). In the next section,
we start to shed light in this direction by studying $\chi_T$.

\subsubsection{Consistency requirements and advantage of rewriting
(\ref{Hamiltonianinitial}) as (\ref{3158})\label{loceqsec}}

We already pointed out the crucial fact that the electric and magnetic charges $Q_{EM}$ in the Hamiltonian (\ref{3158})
are (not yet specified) boundary terms. That is, the Hamiltonian as a whole is defined in the $X_4$ space (the bulk)
but the terms $Q_{EM}$
are defined solely in $X_3$ (the boundary). (We remind the reader that the spaces $X_4$ and $X_3$ were defined in (\ref{decom}).)
The goal in this section is to ensure that $\chi_T$ in (\ref{3158})
does {\it not} contribute to the boundary terms $Q_{EM}$.
Further, we want to ensure that $\chi_T$ is in good agreement with the bulk energy minimization performed in the previous
section. Anticipating events, we will see that such consistency leads to new constraints on the scalar fields
of our gauge theory.
In this manner, we shall be able to focus on the study of the boundary theory only, since the bulk theory will by then be
set to zero by requiring that the fields satisfy (\ref{A3zeroch}), (\ref{detazero}) and (\ref{3132zero}),
together with the BHN equations (\ref{3172}) and (\ref{3177one}) and the new constraints we shall presently find.

But let us take a step back first: what is $\chi_T$ to begin with?
In order to determine $\chi_T$ precisely we will compare the Hamiltonians (\ref{Hamiltonianinitial})
and (\ref{3158}), i.e. the Hamiltonians before and after the inclusion of the coefficients in table \ref{table2}.
By definition, $\chi_T$ is simply the collection of all crossed terms produced during this rewriting. To make our task
computationally easier, we will make use of all the equations above mentioned, which guarantee that the bulk theory is minimized.

Explicitly, using (\ref{A3zeroch}), (\ref{detazero}) and (\ref{3132zero}) in (\ref{Hamiltonianinitial}), the Hamiltonian
before the rewriting is given by
\begin{align}
\nonumber
H=&\int d^4x\,\,\textrm{Tr}\Bigg[
\frac{1}{2}\sum_{\alpha,\beta=1}^2\Big(
\sqrt{\frac{C_1c_{11}}{V_3}}\mathcal{F}_{\alpha\beta}
-\sum_{k=1}^3\sqrt{b_{\tilde{\psi} k}}\epsilon_{\alpha\beta\tilde{\psi} k}\mathcal{D}_{\tilde{\psi}}\varphi_k
-i\sum_{k,l=1}^2\sqrt{d_{kl}}[\varphi_k,\varphi_l]
\Big)^2\\
&+\frac{1}{2}\sum_{\alpha=1}^2\Big(
\sqrt{\frac{C_1c_{12}}{V_3}}\mathcal{F}_{\alpha\tilde{\psi}}
-\sum_{k=1}^3\sqrt{b_{\beta k}}\epsilon_{\alpha\tilde{\psi}\beta k}\mathcal{D}_\beta\varphi_k
-i\sum_{k,l=1}^2\sqrt{d_{kl}}[\varphi_k,\varphi_l]\Big)^2+Q_{EM}\Bigg]. \label{fixedh}
\end{align}
Let us for the time being ignore $Q_{EM}$.
We already said and it can be clearly seen from (\ref{dklfinal}) too, that $d_{12}=d_{21}$. However, $[\varphi_1,\varphi_2]=-[\varphi_2,\varphi_1]$.
Hence, when summing over $k,\,l=1,\,2$ in the pertinent terms above, these will vanish unless they are squared. In other words, the non-zero 
crossed terms in our Hamiltonian (\ref{fixedh}) are just two:
\begin{align}
\begin{array}{llll}
\label{zetasdef}
&\displaystyle\zeta_1\equiv-\frac{1}{2}\sqrt{\frac{C_1c_{11}}{V_3}}\sum_{\alpha,\beta=1}^2\sum_{k=1}^3
\sqrt{b_{\tilde{\psi} k}}\epsilon_{\alpha\beta\tilde{\psi} k}\textrm{Tr}\{\mathcal{F}_{\alpha\beta},\mathcal{D}_{\tilde{\psi}}\varphi_k\}, \\
&\displaystyle\zeta_2\equiv-\frac{1}{2}\sqrt{\frac{C_1c_{12}}{V_3}}\sum_{\alpha=1}^2\sum_{k=1}^3
\sqrt{b_{\beta k}}\epsilon_{\alpha\tilde{\psi}\beta k}\textrm{Tr}\{\mathcal{F}_{\alpha\tilde{\psi}},\mathcal{D}_\beta\varphi_k\}.
\end{array}
\end{align}
Simply carrying out the sums above and noting that (\ref{bakfinal}) implies that
$b_{\tilde{\psi}k}$ and $b_{ak}$ are the same for all
values of $a=1,\,2$ and $k=1,\,2,\,3$ (yet not equal to each other), we get
\begin{align}
\begin{array}{llll}
\label{summedzetas}
\zeta_1=&\displaystyle \sqrt{\frac{C_1c_{11}b_{\tilde{\psi}3}}{V_3}}\,\,\textrm{Tr}\left\{\mathcal{F}_{12},
\mathcal{D}_{\tilde{\psi}}(\varphi_1+\varphi_2+\varphi_3)\right\}, \vspace*{0.2cm} \\
\zeta_2=&\displaystyle -\frac{1}{2}\sqrt{\frac{C_1c_{12}b_{12}}{V_3}}\textrm{Tr}\left[
\left\{\mathcal{F}_{2\tilde{\psi}},\mathcal{D}_1(\varphi_1+\varphi_2+\varphi_3)\right\}
-\left\{\mathcal{F}_{1\tilde{\psi}},\mathcal{D}_2(\varphi_1+\varphi_2+\varphi_3)\right\}\right],
\end{array}
\end{align}
with the normalization convention $\epsilon_{12k\tilde{\psi}}=1$ for all $k=1,\,2,\,3$.
On the other hand, using (\ref{A3zeroch}), (\ref{detazero}), (\ref{3132zero}) and the choices $q_{12}^{(1)},\,h_{\alpha\tilde{\psi}12}^{(1)}=0$ (for all $\alpha=1,\,2$)
in (\ref{3158}), we obtain the Hamiltonian after the rewriting as
\begin{align}
\nonumber
H=\hspace*{-0.2cm}&\int d^4x\,\,\textrm{Tr}\Bigg[
\sum_{\alpha,\beta=1}^2\Big(
\sqrt{\frac{C_1c_{11}}{2V_3}}\mathcal{F}_{\alpha\beta}
+\sum_{\delta,k=1}^3\sqrt{b_{\delta k}}\epsilon_{\alpha\beta}\cdot m_{\delta k}^{(1)}\mathcal{D}_\delta\varphi_k
-i\sum_{k,l=1}^2g_{\alpha\beta kl}^{(1)}\sqrt{d_{kl}}[\varphi_k,\varphi_l]\Big)^2\\
&+\sum_{\alpha=1}^2\Big(
\sqrt{\frac{C_1c_{12}}{2V_3}}\mathcal{F}_{\alpha\tilde{\psi}}
+\sum_{\delta,k=1}^3\sqrt{b_{\delta k}}\epsilon_{\alpha\tilde{\psi}}\cdot m_{\delta k}^{(2)}\mathcal{D}_\delta\varphi_k
\Big)^2 +\chi_T\Bigg]+Q_{EM}.
\label{fixedh2}
\end{align}
We know that the squared terms of this and the previous Hamiltonian are the same (provided the coefficients above
satisfy the constraints in table \ref{table2}, as already discussed in the previous section). Hence, let us just focus on the crossed terms.
There are four of them:
\begin{align}
\begin{array}{llll}
\label{zetaprimesdef}
&\displaystyle\zeta_1^\prime\equiv-i\sqrt{\frac{2C_1c_{11}d_{12}}{V_3}}\sum_{\alpha,\beta=1}^2g_{\alpha\beta 12}^{(1)}
\,\,\textrm{Tr}\{\mathcal{F}_{\alpha\beta},[\varphi_1,\varphi_2]\}, \\
&\displaystyle\zeta_2^\prime\equiv\sqrt{\frac{C_1c_{11}}{2V_3}}\sum_{\alpha,\beta=1}^2\sum_{\delta,k=1}^3
\sqrt{b_{\delta k}}\epsilon_{\alpha\beta}\cdot m_{\delta k}^{(1)}\,\,\textrm{Tr}\{\mathcal{F}_{\alpha\beta},\mathcal{D}_\delta\varphi_k\}, \\
&\displaystyle\zeta_3^\prime\equiv-2i\sqrt{d_{12}}\sum_{\alpha,\beta=1}^2\sum_{\delta,m=1}^3g_{\alpha\beta 12}^{(1)}
\sqrt{b_{\delta m}}\epsilon_{\alpha\beta}\cdot m_{\delta m}^{(1)}\,\,\textrm{Tr}
\{[\varphi_1,\varphi_2],\mathcal{D}_\delta\varphi_m\}, \\
&\displaystyle\zeta_4^\prime\equiv\sqrt{\frac{C_1c_{12}}{2V_3}}\sum_{\alpha=1}^2\sum_{\delta,k=1}^3
\sqrt{b_{\delta k}}\epsilon_{\alpha\tilde{\psi}}\cdot m_{\delta k}^{(2)}\,\,\textrm{Tr}\{\mathcal{F}_{\alpha\tilde{\psi}},\mathcal{D}_\delta\varphi_k\},
\end{array}
\end{align}
where we have used the (anti)symmetry properties $d_{12}=d_{21}$ and $g_{\alpha\beta 12}^{(1)}=-g_{\alpha\beta 21}^{(1)}$ to carry out the sums over
$k,\,l$ in the first and third terms.
In this language, $\chi_T$ is 
\begin{align}
\label{chitdef}
\chi_T=\sum_{i=1}^2\zeta_i-\sum_{i=1}^4\zeta_i^\prime.
\end{align}

In our way to determine $\chi_T$, let us first focus on $\zeta_4^\prime$. Using the coefficient choices in (\ref{m2zeroes}) for the plus
sign in all cases, the dot product definition in (\ref{dotm2def}) and the result (\ref{dcott1res}) and further summing over $\alpha$, it is easy to 
see that
\begin{align}
\zeta_4^\prime=\frac{1}{2}\sqrt{\frac{C_1c_{12}}{V_3}}\,\,\textrm{Tr}\left(
\{\mathcal{F}_{1\tilde{\psi}},\sqrt{b_{\tilde{\psi}1}}\mathcal{D}_{\tilde{\psi}}\varphi_1+\sqrt{b_{23}}\mathcal{D}_2\varphi_3\}+
\{\mathcal{F}_{2\tilde{\psi}},\sqrt{b_{\tilde{\psi}2}}\mathcal{D}_{\tilde{\psi}}\varphi_2+\sqrt{b_{13}}\mathcal{D}_1\varphi_3\}
\right),
\end{align}
where the normalization convention employed is once again $\epsilon_{13}=\epsilon_{23}=1$. With the aid of the BHN equations in (\ref{3177one}),
$\zeta_4^\prime$ is seen to be a squared (and not a crossed) term:
\begin{align}
\zeta_4^\prime =-\frac{C_1c_{12}}{2V_3}\sum_{\alpha=1}^2\textrm{Tr}(\mathcal{F}_{\alpha\tilde{\psi}})^2.
\end{align}
The conclusion that $\zeta_4^\prime$ is not a crossed term of course implies that it does not contribute to $Q_{EM}$, as we wished in the first place.
Further, since $\zeta_4^\prime$ is a squared term, it can be absorbed by an appropriate relabeling of the coefficients in table \ref{table2},
where the defining equations remain unaltered. Consequently, $\zeta_4^\prime$ does not contribute to $\chi_T$ and
we need not worry over it in the ongoing.

We turn our attention to $\zeta_1^\prime$, $\zeta_2^\prime$ and $\zeta_3^\prime$ next.
As before, we interpret the dot product in  $\zeta_2^\prime$ and $\zeta_3^\prime$ as a regular scalar product, we
use our coefficient choices in (\ref{choosingmgs}) and sum over $\alpha,\,\beta$ in (\ref{zetaprimesdef}). In the process,
one must not forget the antisymmetric properties of the coefficients summarized in table \ref{table2}. The described
computation is not hard and yields
\begin{align}
\begin{array}{lllll}
\zeta_1^\prime=&\displaystyle
2i\sqrt{\frac{C_1c_{11}d_{12}}{V_3}}\,\,\textrm{Tr}\Big\{\mathcal{F}_{12},[\varphi_1,\varphi_2]\Big\}, \vspace*{0.2cm}\\
\zeta_2^\prime=&\displaystyle \sqrt{\frac{C_1c_{11}}{V_3}}\,\,\textrm{Tr}\Big\{\mathcal{F}_{12},\sqrt{b_{12}}
(\mathcal{D}_1\varphi_2-\mathcal{D}_2
\varphi_1)+\sqrt{b_{\tilde{\psi}3}}\mathcal{D}_{\tilde{\psi}}\varphi_3\Big\}, \vspace*{0.2cm} \\
\zeta_3^\prime=&\displaystyle
2i\sqrt{d_{12}}\,\,\textrm{Tr}\{[\varphi_1,\varphi_2],\sqrt{b_{12}}(\mathcal{D}_1\varphi_2-\mathcal{D}_2\varphi_1)
+\sqrt{b_{\tilde{\psi}3}}\mathcal{D}_{\tilde{\psi}}\varphi_3\}.
\end{array}
\end{align}
It can be easily checked that, further introducing the first BHN equation (\ref{3172}) in the above,
the following is true:
\begin{align}
\zeta_1^\prime+\zeta_2^\prime= -\frac{2C_1c_{11}}{V_3}\,\,\textrm{Tr}(\mathcal{F}_{12})^2, \quad\quad
\zeta_3^\prime= 8d_{12}\,\,\textrm{Tr}[\varphi_1,\varphi_2]^2
-2i\sqrt{\frac{C_1c_{11}d_{12}}{V_3}}\,\,\textrm{Tr}\{[\varphi_1,\varphi_2],\mathcal{F}_{12}\}.
\end{align}
The same observation we made for $\zeta_4^\prime$ should be invoked presently too:
the squared terms can be absorbed by a relabeling of the coefficients in table \ref{table2}.
They do not contribute to $Q_{EM}$ and do not affect the bulk minimization of section \ref{minimsec}.
In other words, we can consistently conclude that they do not contribute to $\chi_T$ and simply ignore them
in the following. The only term which contributes to $\chi_T$ from the above is
\begin{align}
\zeta_3^\prime=
-2i\sqrt{\frac{C_1c_{11}d_{12}}{V_3}}\,\,\textrm{Tr}\{[\varphi_1,\varphi_2],\mathcal{F}_{12}\}.
\label{zetap3}
\end{align}

Putting everything together, we say that
\begin{align}
\chi_T=\zeta_1+\zeta_2-\zeta_3^\prime,
\end{align}
which must either be reduced to a sum of squared terms (that would then be accounted for by an inconsequential
redefinition of the coefficients in table \ref{table2}) or be set to zero.
In this manner, the Hamiltonian (\ref{3158}) will lead to
a boundary theory determined by $Q_{EM}$ solely, while a consistent bulk energy minimization is ensured via BHN and other
constraining equations on the gauge and scalar fields.
What is more,
it is evident that $\zeta_1-\zeta_3^\prime$ and $\zeta_2$ will have to satisfy this condition separately, as the
BHN equations (\ref{3172}) and (\ref{3177one}) do not mix $\mathcal{F}_{12}$ with $(\mathcal{F}_{1\tilde{\psi}},\,
\mathcal{F}_{2\tilde{\psi}})$. For this very same reason, we must demand right away
\begin{align}
\label{consist1}
\mathcal{D}_{\tilde{\psi}}\varphi_1=\mathcal{D}_{\tilde{\psi}}\varphi_2=\mathcal{D}_1\varphi_3=\mathcal{D}_2\varphi_3=0.
\end{align}
We will refer to these as the first set of consistency requirements we mentioned in the title of the present section.
Implementing the above and using (\ref{3172}), $\zeta_1$ in (\ref{summedzetas}) and $\zeta_3^\prime$ in (\ref{zetap3})
combine to give
\begin{align}
\zeta_1-\zeta_3^\prime=-\frac{2C_1c_{11}}{V_3}\,\,\textrm{Tr}(\mathcal{F}_{12})^2
-\sqrt{\frac{C_1c_{11}b_{12}}{V_3}}\,\,\textrm{Tr}\{\mathcal{F}_{12},\mathcal{D}_1\varphi_2-\mathcal{D}_2\varphi_1\}.
\end{align}
It goes without saying that the first term on the right-hand side above is squared and thus does not contribute to $\chi_T$.
That is not the case with the second term, though. To make it vanish, we will demand
\begin{align}
\label{consist2}
\mathcal{D}_1\varphi_2-\mathcal{D}_2\varphi_1=0,
\end{align}
another consistency requirement. The attentive reader won't take long staring at $\zeta_2$ in (\ref{summedzetas})
in combination with the two relevant BHN equations in (\ref{3177one}) to realize that yet another (and last)
consistency requirement is that in (3.174) in~\cite{Dasgupta:2016rhc}:
\begin{align}
\label{consist3}
\mathcal{D}_1\varphi_1+\mathcal{D}_2\varphi_2=0.
\end{align}
Then, $\zeta_2$ simplifies to
\begin{align}
\zeta_2=\frac{1}{2}\sqrt{\frac{C_1c_{12}b_{12}}{V_3}}\,\,\textrm{Tr}\left[\left\{\mathcal{F}_{2\tilde{\psi}},
\mathcal{D}_1(\varphi_1+\varphi_2)\right\}+\left\{\mathcal{F}_{1\tilde{\psi}},\mathcal{D}_1(\varphi_1-\varphi_2)\right\}\right].
\end{align}
We cannot make squares of the above, so it better vanish. Indeed it is zero, as can be seen from combining
the requirements (\ref{consist1}) and the BHN equations (\ref{3177one}), leading to
\begin{align}
\mathcal{F}_{1\tilde{\psi}}=\mathcal{F}_{2\tilde{\psi}}=0. \label{falphapsi0}
\end{align}
The other BHN equation, namely (\ref{3172}), also reduces in view of our consistency requirements and is now
given by
\begin{align}
\label{fbhn12}
\mathcal{F}_{12}+\sqrt{\frac{V_3}{C_1c_{11}}}\left(
2i\sqrt{d_{12}}[\varphi_1,\varphi_2]
+\sqrt{b_{\tilde{\psi}3}}\mathcal{D}_{\tilde{\psi}}\varphi_3\right)=0.
\end{align}
Finally, we note that $\chi_T$ has by now been converted to some sum of squared terms which does not affect our analysis
and definitely does not contribute to $Q_{EM}$, as was our goal in the beginning of this section.

In conclusion, for the gauge choice (\ref{3161}), the energy of the Hamiltonian (\ref{3158}) is minimized when all 
(\ref{A3zeroch}), (\ref{detazero}), (\ref{3132zero}), (\ref{consist1}), (\ref{consist2}) and (\ref{consist3})
are satisfied, together with the BHN equations (\ref{falphapsi0}) and (\ref{fbhn12}).
In this case, $\chi_T$ is zero (or, more precisely, is absorbed by an immaterial redefinition of coefficients,
as already explained)
and we are only left with the boundary terms $Q_{EM}$ to be considered.

To finish this section, let us clarify what is the advantage of rewriting the Hamiltonian
(\ref{Hamiltonianinitial}) as (\ref{3158}). The so called consistency requirements
(\ref{consist1}), (\ref{consist2}) and (\ref{consist3}) that we obtained in this section
to ensure no crossed terms were produced in the aforementioned rewriting are actually vital results in our analysis.
They simplify the BHN equations, which are conjectured to be directly related to knot invariants
(for example, see section 3.2 in~\cite{Witten:2011zz}).
But their simplifying power goes well beyond the BHN equations.

In~\cite{Dasgupta:2016rhc}, these consistency requirements are obtained in an altogether different manner:
after generalizing to the $c_2\neq0$ case and
by comparing our gauge theory to the twisted gauge theory\footnote{The reader
should not worry at this time over terminology.
We shall introduce the concept of topological twist and twist our own theory in due time, in
section \ref{twistsec}.} in~\cite{Witten:2011zz} and~\cite{Gaiotto:2011nm}.
More precisely, our consistency requirements in (\ref{consist1}) are equal to (3.218) and (3.220) in~\cite{Dasgupta:2016rhc},
(\ref{consist2}) is the same as (3.207) (albeit all three equations are expressed in the twisted language there)
and (\ref{consist3}) is exactly (3.174). Among all the necessary constraints in our set up, (\ref{consist2}) is particularly
useful. Unlike in the present work and in~\cite{Dasgupta:2016rhc}, in both~\cite{Witten:2011zz} and~\cite{Gaiotto:2011nm}
this constraint is not a consistency requirement of the
twisted gauge theory. This term simply does not vanish and hence is part of one of the twisted BHN equations.
However, this term greatly adds to the computational difficulties. Hence, to keep things as simple as possible,
in~\cite{Witten:2011zz} the prefactor for this term is made to vanish, via an S-duality. Then, the quite involved generalization to the case where the prefactor
does not vanish is studied in~\cite{Gaiotto:2011nm}.
The fact that (\ref{consist2}) is true in our construction thus avoids us the subtleties and struggles related to
having to consider the S-dual picture first and mimic the extension in~\cite{Gaiotto:2011nm} afterwards! 

Although the S-dual picture is not required in our analysis, for completeness and to provide a
transparent comparison to the well-known analysis in~\cite{Witten:2011zz}, this has been fully worked out around (3.252)-(3.275)
in~\cite{Dasgupta:2016rhc}. We thus refer the reader seeking an M-theory realization of the S-dual picture,
as well as quantitative details on its relation to the configuration (M, 1) in section \ref{ns5d3sect1}, to
the cited work. Here, we will take full advantage of having (\ref{consist2}) as part of our
gauge theory and rid ourselves of further complications along this direction.
Instead, we will now look at the generalization of all the results so far in section \ref{hamilsec} to the case that really
concerns us, where $c_2\neq 0$ in (\ref{totaction}). This will in turn directly lead us to the study of the corresponding
boundary theory in section (\ref{bndsec}).

\subsection{Generalization to the case where $c_2\neq 0$ in (\ref{totaction}) \label{c2not0sec}}

We have by now gained considerable insight into the bulk physics of the theory with action (\ref{totaction})
but with no topological term (i.e. $c_2=0$ there). The inclusion of this topological term is, however,
far from trivial, both conceptually and computationally.
To relax a bit the computational difficulties, we will begin this section by doing
the following approximation: we will in the ongoing consider that
\begin{align}
c_{11}=c_{12} \label{c1112e}
\end{align}
in (\ref{totaction}).
Looking at the definitions of these coefficients in (\ref{intc11}), we see that this 
amounts to requiring that $e^{2\phi_0}H_4=1$. Further using (\ref{Hs}), our simplification
reduces to a constraint equation on the so far completely arbitrary warp factors (\ref{easychoice})
and (\ref{tildeF2}) and constant leading value of the dilaton in (\ref{consdil})\footnote{We remind the reader that
any specific choice of these warp factors and dilaton should be checked to preserve $\mathcal{N}=2$ supersymmetry.
This idea will be made precise in section \ref{bcsec}.}:
\begin{align}
\label{consc1s}
\frac{e^{2\phi_0}\tilde{F}_2F_3\sec^2\theta_{nc}\sin^2\theta_1}{\tilde{F}_2\cos^2\theta_1+F_3\sin^2\theta_1}=1.
\end{align}
Clearly, this is not too stringent a constraint, as there is ample freedom of choice to satisfy it.
For a physical interpretation of our assumption, one should 
look at the metric of the M-theory configuration (M, 1) in (\ref{340}). We then see that (\ref{c1112e})
implies that $(t,\,x_1,\,x_2,\,\tilde{\psi})$ are now Lorentz invariant directions. In other words,
our approximation leads to a restoration of the Lorentz symmetry along $\tilde{\psi}$ in the subspace $X_4$
that we defined in (\ref{decom}).

Having made this simplification, we proceed to show an intermediate result, which will immediately prove useful
in deriving the Hamiltonian following from the action (\ref{totaction}) with $c_2\neq0$. This consists on
working out a convenient component form of the integrand of this topological term in the action:
\begin{align}
\mathcal{F}^{(X_4)}\wedge \mathcal{F}^{(X_4)}\equiv \sum_{\substack{\mu<\nu \\ \rho<\lambda}}
\mathcal{F}_{\mu\nu}\mathcal{F}_{\rho\lambda} dx_\mu\wedge dx_\nu\wedge dx_\rho\wedge dx_\lambda
=d^4x\,\sum_{\mu<\nu}\mathcal{F}_{\mu\nu}\ast\mathcal{F}^{\mu\nu}, \label{FstF}
\end{align}
where, as usual, the Hodge dual of the field strength is defined as
\begin{align}
\ast\mathcal{F}^{\mu\nu}\equiv\frac{1}{2}\sum_{\rho,\lambda}\epsilon^{\mu\nu\rho\lambda}\mathcal{F}_{\rho\lambda},
\end{align}
$d^4x$ is the volume element of the now Minkowskian spacetime $X_4$ and $x_\mu$ refers collectively
to its coordinates $(t,\,x_1,\,x_2,\,\tilde{\psi})$.

Using the approximation (\ref{c1112e}), (\ref{FstF}) and recalling (\ref{c2final}), we are ready to write the first line in the action (\ref{totaction})
of our theory (which we denote as $S_{L1}$) in the following suitable manner:
\begin{align}
S_{L1}= \int d^4x\,\,\textrm{Tr}\sum_{\mu<\nu}\left(\frac{C_1c_{11}}{V_3}\mathcal{F}_{\mu\nu}\mathcal{F}^{\mu\nu}
+C_1\sin\theta_{nc}q(\theta_{nc})\mathcal{F}_{\mu\nu}\ast\mathcal{F}^{\mu\nu}\right).
\end{align}
The reader will of course right away notice that $S_{L1}$ is precisely Maxwell's action with a
$\Theta$-term (see, for example, in (2.1) in~\cite{Leao:2001gz}). The correlation becomes fully apparent
once we identify our coefficients (which only depend on supergravity variables)
with the Yang-Mills coupling and gauge theory $\Theta$-parameter as
\begin{align}
\frac{C_1c_{11}}{V_3}\equiv\frac{4\pi}{g_{YM}^2}, \quad\quad C_1\sin\theta_{nc}q(\theta_{nc})\equiv\frac{\Theta}{2\pi}.
\label{gaugeparam}
\end{align}
The above makes concrete the long standing promise of section \ref{ncsect}. There, we claimed that introducing
the non-commutative deformation labeled by the parameter $\theta_{nc}$ would lead to a $\Theta$-term
in the four-dimensional gauge theory associated to the M-theory configuration (M, 1). From (\ref{gaugeparam})
it is clear that $\theta_{nc}=0$ would lead to no $\Theta$-term in the gauge theory, so the deformation
is indeed successful in replacing the axionic background of~\cite{Witten:2011zz} to source this topological term.
(Later on, in section (\ref{twistsec}), we shall see that this topological term is a fundamental ingredient to convert
the boundary $X_3$ of $X_4$ into a suitable space for the embedding of knots.
This is because such term allows us to define a topological theory in $X_3$.)
It is standard to combine the Yang-Mills coupling and the $\Theta$-parameter into a single
{\it complex} coupling constant $\tau$ as
\begin{align}
\tau\equiv \frac{\Theta}{2\pi}+i\frac{4\pi}{g_{YM}^2}=C_1\left(\sin\theta_{nc}q(\theta_{nc})+i\frac{c_{11}}{V_3}\right),
\label{comtaudef}
\end{align}
where the last equality follows from our prior identification (\ref{gaugeparam}) and reproduces (3.183)
in~\cite{Dasgupta:2016rhc}.

The Hamiltonian associated to $S_{L1}$ can be directly read from (2.2) in~\cite{Leao:2001gz}.
Note however that we must do an overall sign change (we work in the opposite Minkowski signature convention)
and account for the different overall normalization too. Explicitly, we obtain
\begin{align}
H_{L1}=&\int d^4x\,\,\textrm{Tr}\left(\frac{2i}{\tau-\bar{\tau}}\Pi^i\Pi_i
+i\frac{\tau+\bar{\tau}}{\tau-\bar{\tau}}\Pi^iB_i
+\frac{i}{2}\frac{\tau\bar{\tau}}{\tau-\bar{\tau}}B^iB_i\right) \nonumber \\
=&\frac{2i}{\tau-\bar{\tau}}\int d^4x\,\,\textrm{Tr}\left(\Pi^i+\frac{\tau}{2}B^i\right)
\left(\Pi_i+\frac{\bar{\tau}}{2}B_i\right),
\end{align}
where $i=(x_1,\,x_2,\,\tilde{\psi})$ spans the spatial coordinates of $X_4$ and the canonical momenta and
magnetic field in our case are given by
\begin{align}
\Pi^i=\frac{C_1c_{11}}{V_3}\mathcal{F}^{0i}, \quad\quad B^i=2\epsilon^{ijk}\mathcal{F}_{jk}.
\end{align}
This is the same Hamiltonian that appears in (3.187) in~\cite{Dasgupta:2016rhc} too:
\begin{align}
H_{L1}=\frac{2i}{\tau-\bar{\tau}}\int d^4x\,\,
\textrm{Tr}\left(\frac{C_1c_{11}}{V_3}\mathcal{F}^{0i}+\tau
\epsilon^{ijk}\mathcal{F}_{jk}\right)\left(\frac{C_1c_{11}}{V_3}\mathcal{F}_{0i}+\bar{\tau}
\epsilon_{ilm}\mathcal{F}^{lm}\right), \label{3182}
\end{align}
where $\bar{\tau}$ denotes the complex conjugate of $\tau$.
An uncomplicated yet very useful rewriting of this Hamiltonian in terms of only the complex coupling $\tau$ and
the field strengths is the following:
\begin{align}
\nonumber
H_{L1}=&\int d^4x\,\,
\textrm{Tr}\Bigg[\frac{\tau-\bar{\tau}}{2i}\sum_{i=1}^3(\mathcal{F}_{0i}\mathcal{F}^{0i})+
\frac{4i|\tau|^2}{\tau-\bar{\tau}}\sum_{\alpha,\beta=1}^2(\mathcal{F}_{\alpha\beta}\mathcal{F}^{\alpha\beta})+
\frac{8i|\tau|^2}{\tau-\bar{\tau}}\sum_{\alpha=1}^2(\mathcal{F}_{\alpha\tilde{\psi}}\mathcal{F}^{\alpha\tilde{\psi}})\\
&+(\tau+\bar{\tau})\sum_{i,j,k=1}^3\epsilon_{0ijk}(\mathcal{F}^{0i}\mathcal{F}^{jk})\Bigg],
\label{gaugeH3}
\end{align}
which the reader may verify quite effortlessly.

At this point, we are ready to write the full Hamiltonian following from (\ref{totaction}), topological piece included.
All that is left to do is couple the Hamiltonian (\ref{gaugeH3}) to the real scalar fields
$\mathcal{A}_r,\,\mathcal{A}_{\phi_1},\,\mathcal{A}_{\tilde{3}}$ and $\varphi_k$'s (with $k=1,\,2,\,3$).
Our prior meticulous analysis of the $c_2=0$ case makes this task almost trivial.
Keeping the last term in (\ref{gaugeH3}) separate, we can couple the scalar fields as in (\ref{3158}).
The only difference is that, now, the prefactors for the terms involving field strengths
will be different, matching the ones in (\ref{gaugeH3}).
Of course, the coefficients that do not have a supergravity interpretation remain constrained
as summarized in table \ref{table2}. Explicitly, the full Hamiltonian is
\newpage
\begin{align}
\begin{array}{lll}
\label{c2notzerohamtot}
H= &\displaystyle\hspace*{-0.2cm}\int d^4x\,\,\textrm{Tr}\Bigg\{
\sum_{\alpha=1}^2\Big(\sqrt{\frac{\tau-\bar{\tau}}{2i}}\mathcal{F}_{\alpha0}-\sqrt{\frac{C_1c_{\alpha \tilde{3}}}{V_3}}
\mathcal{D}_\alpha\mathcal{A}_{\tilde{3}}\Big)^2+\Big(\sqrt{\frac{\tau-\bar{\tau}}{2i}}\mathcal{F}_{\tilde{\psi}0}-
\sqrt{\frac{C_1c_{\tilde{\psi}\tilde{3}}}{V_3}}\mathcal{D}_{\tilde{\psi}}\mathcal{A}_{\tilde{3}}\Big)^2 \\
&\displaystyle\hspace*{-0.2cm}+\frac{C_1}{V_3}\Big[(\sqrt{c_{0r}}\mathcal{D}_0\mathcal{A}_r-i\sqrt{a_2}
[\mathcal{A}_{\tilde{3}},\mathcal{A}_r])^2+
(\sqrt{\tilde{c}_{0\phi_1}}\mathcal{D}_0\mathcal{A}_{\phi_1}-i\sqrt{a_4}[\mathcal{A}_{\tilde{3}},\mathcal{A}_{\phi_1}])^2 
+c_{0\tilde{3}}(\mathcal{D}_0\mathcal{A}_{\tilde{3}})^2\Big] \\
&\displaystyle\hspace*{-0.2cm}+\sum_{k,l=1}^3\Big[
(\sqrt{b_{0k}}\mathcal{D}_0\varphi_k-i\sqrt{c_{\tilde{3}k}}[\mathcal{A}_{\tilde{3}},\varphi_k])^2
+q_{kl}^{(1)} d_{kl}[\varphi_k,\varphi_l]^2
+\sum_{\gamma=2}^3q_k^{(\gamma)}c_{y_\gamma k}[\mathcal{A}_{y_\gamma},\varphi_k]^2\Big]\\
&\displaystyle\hspace*{-0.2cm}+\sum_{\alpha,\beta=1}^2\Big(
\sqrt{\frac{2i|\tau|^2}{\tau-\bar{\tau}}}\mathcal{F}_{\alpha\beta}
+\sqrt{\frac{C_1c_{\tilde{\psi} r}}{V_3}}s_{\alpha\beta}^{(1)}\epsilon_{\alpha \beta\tilde{\psi} r}
\mathcal{D}_{\tilde{\psi}}\mathcal{A}_r
+\sqrt{\frac{C_1\tilde{c}_{\tilde{\psi} \phi_1}}{V_3}}s_{\alpha\beta}^{(2)}\epsilon_{\alpha\beta\tilde{\psi} \phi_1}
\mathcal{D}_{\tilde{\psi}}\mathcal{A}_{\phi_1}\\
&\displaystyle\hspace*{-0.2cm}-ig_{\alpha\beta}^{(4)}\sqrt{\frac{C_1a_1}{V_3}}[\mathcal{A}_r,\mathcal{A}_{\phi_1}]
-i\sum_{k,l=1}^3g_{\alpha\beta kl}^{(1)}\sqrt{d_{kl}}[\varphi_k,\varphi_l]
-i\sum_{k=1}^3\sum_{\gamma=2}^3g_{\alpha\beta k}^{(\gamma)}\sqrt{c_{y_\gamma k}}[\mathcal{A}_{y_\gamma},\varphi_k] \\
&\displaystyle\hspace*{-0.2cm}
+\sum_{\delta,k=1}^3\sqrt{b_{\delta k}}\epsilon_{\alpha\beta}\cdot m_{\delta k}^{(1)}
\mathcal{D}_\delta\varphi_k
\Big)^2+\sum_{\alpha=1}^2\Big(
\sqrt{\frac{4i|\tau|^2}{\tau-\bar{\tau}}}\mathcal{F}_{\alpha\tilde{\psi}}
+\sqrt{\frac{C_1c_{\beta r}}{V_3}}t_{\alpha}^{(1)}\epsilon_{\alpha\tilde{\psi} \beta r}\mathcal{D}_\beta\mathcal{A}_r\\
&\displaystyle \hspace*{-0.2cm}
+\sqrt{\frac{C_1\tilde{c}_{\beta \phi_1}}{V_3}}t_{\alpha}^{(2)}\epsilon_{\alpha\tilde{\psi}\beta \phi_1}
\mathcal{D}_\beta\mathcal{A}_{\phi_1}
-ih_{\alpha\tilde{\psi}}^{(4)}\sqrt{\frac{C_1a_1}{V_3}}[\mathcal{A}_r,\mathcal{A}_{\phi_1}]
-i\sum_{k,l=1}^3h_{\alpha\tilde{\psi} kl}^{(1)}\sqrt{d_{kl}}[\varphi_k,\varphi_l]\\
&\displaystyle\hspace*{-0.2cm}
-i\sum_{k=1}^3\sum_{\gamma=2}^3h_{\alpha\tilde{\psi} k}^{(\gamma)}\sqrt{c_{y_\gamma k}}[\mathcal{A}_{y_\gamma},\varphi_k]
+\sum_{\delta,k=1}^3\sqrt{b_{\delta k}}\epsilon_{\alpha\tilde{\psi}}\cdot m_{\delta k}^{(2)}
\mathcal{D}_\delta\varphi_k
\Big)^2
+\frac{C_1}{V_3}\Big[c_{\tilde{\psi} r}s^{(1)}(\mathcal{D}_{\tilde{\psi}}\mathcal{A}_r)^2\\
&\displaystyle \hspace*{-0.2cm}
+\tilde{c}_{\tilde{\psi} \phi_1}s^{(2)}(\mathcal{D}_{\tilde{\psi}}\mathcal{A}_{\phi_1})^2
+c_{\beta r}t^{(1)}(\mathcal{D}_\beta\mathcal{A}_r)^2 
+\tilde{c}_{\beta \phi_1}t^{(2)}(\mathcal{D}_\beta\mathcal{A}_{\phi_1})^2
+a_1q^{(4)}[\mathcal{A}_r,\mathcal{A}_{\phi_1}]^2\Big]
+\tilde{\chi}_T\\
&\displaystyle\hspace*{-0.2cm}+(\tau-\bar{\tau})\sum_{i,j,k=1}^3\epsilon_{0ijk}\mathcal{F}^{0i}\mathcal{F}^{jk}\Bigg\}
+\tilde{Q}_{EM}.
\end{array}
\end{align}
Note that the terms $(\tilde{\chi}_T,\,\tilde{Q}_{EM})$ are now written with a tilde to denote they are not the same
as those appearing in (\ref{3158}), although they still stand for the crossed terms related to the coefficients
of table \ref{table2} and the electric and magnetic charges in the theory, respectively.
Note the close resemblance between the above and the Hamiltonian for the $c_2=0$ case in (\ref{3158}).
Essentially, they are the same up to prefactors in the terms containing field strengths, but there is an all important
additional term now (appearing in the last line in (\ref{c2notzerohamtot})).

This similarity between the $c_2=0$ Hamiltonian and the $c_2\neq0$ one 
allows us to easily generalize the results in section \ref{c20sec} to the present and relevant case.
In particular, it is remarkably simple to minimize the energy of (\ref{c2notzerohamtot}) for static configurations.
That is, to find the BPS conditions for our gauge and scalar fields. Let us nevertheless show a few steps in the process
in the following for clarity, since we will not minimize the energy in exactly the same way.

As before, we choose to work in the gauge (\ref{3161}) and demand that (\ref{A3zeroch}) and (\ref{detazero}) hold true.
This time, instead of ensuring the vanishing of the seventh squared term via (\ref{3132zero}), we will choose
\begin{align}
q_{kl}^{(1)}=0, \quad\quad \forall k,l=1,2,3. \label{qkl10}
\end{align}
This choice leads to a more rich dynamics of the $\varphi_k$ scalar fields
(than that we considered in the $c_2=0$ case), which, as we shall see,
will play a role in the study of the boundary theory in section (\ref{twistsec}) later on. For the time being, the mentioned
choices reduce the Hamiltonian to (3.225) in~\cite{Dasgupta:2016rhc}:
\begin{align}
\begin{array}{llll}
H =&\displaystyle \hspace*{-0.2cm}\int \hspace*{-0.1cm}d^4x \,\,\textrm{Tr}\Bigg\{
\sum_{\alpha=1}^2\Big(
\sqrt{\frac{4i|\tau|^2}{\tau-\bar{\tau}}}\mathcal{F}_{\alpha\tilde{\psi}}
-i\sum_{k,l=1}^3h_{\alpha\tilde{\psi} kl}^{(1)}\sqrt{d_{kl}}[\varphi_k,\varphi_l]
+\sum_{\delta,k=1}^3\sqrt{b_{\delta k}}\epsilon_{\alpha\tilde{\psi}}\cdot m_{\delta k}^{(2)}
\mathcal{D}_\delta\varphi_k\Big)^2 \\
&\displaystyle\hspace*{-0.2cm}+\sum_{\alpha,\beta=1}^2\Big(
\sqrt{\frac{2i|\tau|^2}{\tau-\bar{\tau}}}\mathcal{F}_{\alpha\beta}
-i\sum_{k,l=1}^3g_{\alpha\beta kl}^{(1)}\sqrt{d_{kl}}[\varphi_k,\varphi_l]
+\sum_{\delta,k=1}^3\sqrt{b_{\delta k}}\epsilon_{\alpha\beta}\cdot m_{\delta k}^{(1)}
\mathcal{D}_\delta\varphi_k\Big)^2 \\
&\displaystyle\hspace*{-0.2cm}
+(\tau+\bar{\tau})\sum_{i,j,k=1}^3\epsilon_{0ijk}\mathcal{F}^{0i}\mathcal{F}^{jk}+\tilde{\chi}_T\Bigg\}+\tilde{Q}_{EM}.
\label{3218}
\end{array}
\end{align}

In section \ref{c20sec}, we did many coefficient choices to simplify the computation as much as possible.
On this occasion, we wish to keep our coefficients arbitrary for as long as possible (this freedom of choice
will be beneficial once we look at the boundary theory). Consequently, we will take as our BHN equations
the following:
\begin{align}
\begin{array}{llll}
&\displaystyle\sqrt{\frac{4i|\tau|^2}{\tau-\bar{\tau}}}\mathcal{F}_{\alpha\tilde{\psi}}
-i\sum_{k,l=1}^3h_{\alpha\tilde{\psi} kl}^{(1)}\sqrt{d_{kl}}[\varphi_k,\varphi_l]
+\sum_{\delta,k=1}^3\sqrt{b_{\delta k}}\epsilon_{\alpha\tilde{\psi}}\cdot m_{\delta k}^{(2)}
\mathcal{D}_\delta\varphi_k=0, \\
&\displaystyle\sqrt{\frac{2i|\tau|^2}{\tau-\bar{\tau}}}\mathcal{F}_{\alpha\beta}
-i\sum_{k,l=1}^3g_{\alpha\beta kl}^{(1)}\sqrt{d_{kl}}[\varphi_k,\varphi_l]
+\sum_{\delta,k=1}^3\sqrt{b_{\delta k}}\epsilon_{\alpha\beta}\cdot m_{\delta k}^{(1)}
\mathcal{D}_\delta\varphi_k=0,
\label{finalbhns}
\end{array}
\end{align}
for all $\alpha,\,\beta=1,\,2$. In view of the detailed computation in section \ref{loceqsec},
it is not hard to infer that on this occasion too we will be able to absorb $\tilde{X}_T$ 
through a meaningless renaming of coefficients by imposing
certain consistency requirements to our scalar fields $\varphi_k$'s. The conditions there derived,
namely (\ref{consist1}), (\ref{consist2}) and (\ref{consist3}),
are completely independent of the prefactors in the various terms of the Hamiltonian.
Hence, the only alteration needed in that calculation consists on accommodating
the choice (\ref{qkl10}) instead of (\ref{3132zero}). The attentive reader will surely be easily convinced that
the consistency requirements generalize to
\begin{align}
\mathcal{D}_1\varphi_2-\mathcal{D}_2\varphi_1=\mathcal{D}_1\varphi_3-\mathcal{D}_{\tilde{\psi}}\varphi_1=
\mathcal{D}_2\varphi_3-\mathcal{D}_{\tilde{\psi}}\varphi_2=\mathcal{D}_1\varphi_1+\mathcal{D}_2\varphi_2+
\mathcal{D}_{\tilde{\psi}}\varphi_3=0 \label{finalvars}
\end{align}
in the present case. Once the energy has thus been minimized, the Hamiltonian reduces to
\begin{align}
H =(\tau+\bar{\tau})\int d^4x 
\sum_{i,j,k=1}^3\epsilon_{0ijk}\textrm{Tr}(\mathcal{F}^{0i}\mathcal{F}^{jk})+\tilde{Q}_{EM}.
\label{321888}
\end{align}
In the following section, we will devote quite some effort to the study of the above Hamiltonian.
But before jumping into the pertinent details, let us briefly review
the main contents of the present section.

We have shown that the action (\ref{totaction}) is associated to the Hamiltonian (\ref{c2notzerohamtot}).
Both of them are defined in the space $X_4$.
A consistent minimization of the energy of (\ref{c2notzerohamtot}) for static configurations of the fields, working in the gauge (\ref{3161}), 
is obtained by imposing the constraints (\ref{A3zeroch}), (\ref{detazero}) and (\ref{finalvars}). We also require that
the BHN equations
in (\ref{finalbhns}) be satisfied. In this energy minimization process, the coefficients of table \ref{table2}
remain mostly arbitrary.
The only choice made is that in (\ref{qkl10}). The Hamiltonian then reduces to (\ref{321888}).

\section{The boundary theory \label{bndsec}}

As we just mentioned, the minimization of the energy of the Hamiltonian stemming from the M-theory configuration (M, 1)
presented in section \ref{c2not0sec} leads to (\ref{321888}). In the present section, we will first show that
(\ref{321888}) is defined only in $X_3$, the boundary of $X_4$. 

This realization then requires us to find suitable boundary conditions for all the fields in the gauge theory. 
Of course, we are referring to half-BPS boundary conditions: ones that break the $\mathcal{N}=4$ supersymmetry of the theory to
$\mathcal{N}=2$. Although so far we have insisted that by construction the configuration (M, 1) is $\mathcal{N}=2$ supersymmetric,
it is only at this stage that we shall be able to make this claim fully precise.
Indeed, as we shall see, this desired amount of supersymmetry requires of no constraint on the parameters
that characterize (M, 1) (those summarized in table \ref{table1}) and is enforced by appropriate boundary conditions only.

Finally, we shall note that, if the configuration (M, 1) is to be useful for the study of knots and their invariants,
the theory in $X_3$ better be topological. In this manner, it will be possible to embed
the knots (which are topological objects) in $X_3$ consistently. To this aim, we will present the notion of
topological twist and show that, upon twisting, our gauge theory indeed becomes a suitable framework for the realization of
knots.

A graphical summary of the main results of section \ref{bndsec} is as shown in blue in figure \ref{fig12}.
From this schematic point of view, section \ref{fsbndsec} can be understood as the derivation of (\ref{nottopbnd}).
Similarly, section \ref{bcsec} contains the details on (\ref{bcond0})-(\ref{bcond2}) and sections \ref{twistsec}
and \ref{twistbulksec} deal with the technicalities involved in topological twisting all previously cited results.

\subsection{First steps towards determining the boundary theory \label{fsbndsec}}

In this section, we have one very concrete goal: to rewrite the Hamiltonian of our gauge theory after its energy has been
minimized (this is given by (\ref{321888})) as an integral over $X_3$ instead of $X_4$.
(Once more, we remind the reader that these spaces were defined and described around (\ref{decom}).)
In other words, we want to show that, for the gauge choice (\ref{3161}) and
after imposing the BPS conditions  (\ref{A3zeroch}), (\ref{detazero}), (\ref{finalvars})
and (\ref{finalbhns}), the total Hamiltonian (\ref{c2notzerohamtot}) reduces to a boundary Hamiltonian.
As a matter of a fact, this does not involve any conceptual hurdle, so let us jump into computation right away.

After having left the electric and magnetic charges $\tilde{Q}_{EM}$
unspecified for the whole of section \ref{hamilsec}, we finally take it upon us to specify them.
As we already hinted previously, we will do so by comparing our Hamiltonian (\ref{c2notzerohamtot})
to that in (2.4) in~\cite{Lee:2006gqa} and then inferring $\tilde{Q}_{EM}$ from (2.5) in that same reference.
Obviously, one could do the computation explicitly. However, this won't give us any further insight into our theory
and so we do not attempt such approach here.
From our identifications in (\ref{idenkor}) and our choice (\ref{A3zeroch}),
it is clear that the electric charge vanishes in our case:
\begin{align}
\tilde{Q}_{EM}\equiv \tilde{Q}_E+\tilde{Q}_M, \quad\quad \tilde{Q}_E=0.
\end{align}
It is also easy to see that the magnetic charge is of the form
\begin{align}
\label{qmbnd}
\tilde{Q}_M=\int d^4x \,\,\partial_{\tilde{\psi}} q_M=\int d^3x\,\,q_M, \quad\quad d^3x\equiv dtdx_1dx_2,
\end{align}
where we have ignored terms which are total derivatives along the unbounded directions $(t,\,x_1,\,x_2)$, since they
do not affect the physics of our theory and where we have rewritten $\tilde{Q}_M$ as a boundary term, defined in $X_3$
instead of the whole $X_4$.
Of course, this comes as no surprise: we have long been anticipating that the electric and magnetic charges would
be restricted to $X_3$ only.
Further using (\ref{idenkor}) and noting that (\ref{c2notzerohamtot}) is exactly  (2.4) in~\cite{Lee:2006gqa} up to
prefactors, it is clear that $q_M$ is given by
\begin{align}
q_M=\sum_{k,l,m=1}^3\textrm{Tr}\Bigg[
\sum_{\alpha,\beta=1}^2d_1\epsilon_{ k\alpha\beta}
\varphi_k\mathcal{F}_{\alpha\beta}
+\epsilon_{ klm}
\Big(\frac{id_2}{3}\varphi_k[\varphi_l,\varphi_m]
+d_3\varphi_k\mathcal{D}_l\varphi_m\Big)\Bigg], \label{hatqm}
\end{align}
where 
$(d_1,\,d_2,\,d_3)$ are coefficients that account for the difference of prefactors between our Hamiltonian
and that in~\cite{Lee:2006gqa}. Their determination is not straightforward, so let us work them out in details.

Simply looking at our Hamiltonian (\ref{c2notzerohamtot}), it is evident that the field strength $\mathcal{F}_{\alpha\beta}$
picks up the additional prefactor $\sqrt{2i|\tau|^2(\tau-\bar{\tau})^{-1}}$ for all $\alpha,\,\beta=1,\,2$, as compared
to~\cite{Lee:2006gqa}. Similarly, for fixed values of $(l,\,m)$, it follows that to $\mathcal{D}_l\varphi_m$
we must associate the prefactor $\sqrt{b_{lm}}m_{lm}^{(1)}$\footnote{To fully understand this 
prefactor, the reader may find it useful to recall that the dot product appearing in the relevant
term of the Hamiltonian was taken to be the usual scalar product around (\ref{3172}).}.
Actually, the only non-trivial prefactors are those that we should attach to $\varphi_k$ and $[\varphi_l,\varphi_m]$.
To establish what they are, we first note that
\begin{align}
\sum_{\alpha,\beta=1}^2\sum_{k,l=1}^3g_{\alpha\beta kl}^{(1)}\sqrt{d_{kl}}[\varphi_k,\varphi_l]=
4\sqrt{d_{12}}\left( g_{1212}^{(1)}[\varphi_1,\varphi_2]
+ g_{1213}^{(1)}[\varphi_1,\varphi_3]
+ g_{1223}^{(1)}[\varphi_2,\varphi_3]\right), \label{sumsvarcoms}
\end{align}
where we have used the fact that $g_{\alpha\beta kl}^{(1)}$ is antisymmetric in $(\alpha,\,\beta)$ and in $(k,\,l)$
by definition (see table \ref{table2}) and $d_{kl}$ is independent of $(k,\,l)$ (see (\ref{dklfinal})).
From the above it follows that to the $[\varphi_l,\varphi_m]$ term we must associate the factor $4\sqrt{d_{lm}}g_{12lm}^{(1)}$.
Let us denote as $(y_1,\,y_2,\,y_3)$ the prefactors that we need to associate to $(\varphi_1,\,\varphi_2,\,\varphi_3)$, respectively.
From (\ref{sumsvarcoms}), we also have that
\begin{align}
y_1y_2=4\sqrt{d_{12}}g_{1212}^{(1)}, \quad\quad 
y_1y_3=4\sqrt{d_{12}}g_{1213}^{(1)}, \quad\quad
y_2y_3=4\sqrt{d_{12}}g_{1223}^{(1)}.
\end{align}
This can be easily solved to yield
\begin{align}
\label{ysdef}
y_1=2d_{12}^{1/4}\sqrt{\frac{g_{1212}^{(1)}g_{1213}^{(1)}}{g_{1223}^{(1)}}}, \quad\quad
y_2=2d_{12}^{1/4}\sqrt{\frac{g_{1223}^{(1)}g_{1212}^{(1)}}{g_{1213}^{(1)}}}, \quad\quad
y_3=2d_{12}^{1/4}\sqrt{\frac{g_{1223}^{(1)}g_{1213}^{(1)}}{g_{1212}^{(1)}}}.
\end{align}
Putting all our observations on the prefactors together, our discussion implies
\begin{align}
d_1=y_k \sqrt{\frac{2i|\tau|^2}{\tau-\bar{\tau}}}, \quad\quad
d_2=4 y_k\sqrt{d_{lm}}g_{12 lm}^{(1)}, \quad\quad
d_3=y_k\sqrt{b_{lm}}m_{lm}^{(1)}, \label{finalds}
\end{align}
which fully specifies the magnetic charge in our theory.
Note that the indices of these coefficients are to be contracted with the appropriate terms in (\ref{hatqm}).
Note also that (\ref{finalds}) agrees with (3.233) in~\cite{Dasgupta:2016rhc}, after appropriately
summing over the free index $k$.

Once we have the explicit form of $\tilde{Q}_{EM}$ in (\ref{321888}), we can focus on the only other term
in this Hamiltonian, namely
\begin{align}
H_{top}\equiv (\tau+\bar{\tau})\int d^4x 
\sum_{i,j,k=1}^3\epsilon_{0ijk}\textrm{Tr}(\mathcal{F}^{0i}\mathcal{F}^{jk}).
\end{align}
Recall that $(i,\,j,\,k)$ stand for the spatial directions of $X_4$: $(x_1,\,x_2,\,\tilde{\psi})$.
Recall also that, after our simplifying assumption in (\ref{c1112e}), $X_4$ is now a Lorentz-invariant space.
A quick exercise of opening indices in both (\ref{FstF}) and the above allows us to rewrite $H_{top}$ as
\begin{align}
H_{top}=(\tau+\bar{\tau})\int_{X_4}\textrm{Tr}\left(\mathcal{F}^{(X_4)}\wedge\mathcal{F}^{(X_4)}\right).
\end{align}
It is well-known that the above can be rewritten as a Chern-Simons type of boundary integral,
\begin{align}
S_{top}=(\tau+\bar{\tau})\int_{X_3}\textrm{Tr}\left(\mathcal{A}\wedge d\mathcal{A}+\frac{2i}{3}\mathcal{A}\wedge\mathcal{A}
\wedge\mathcal{A}\right),
\end{align}
which is gauge-invariant iff $(\tau+\bar{\tau})$ is an integer multiple of $2\pi$. We will discuss this subtlety
shortly, in section \ref{twistsec}. For the time being, however, we will just collect our results so far.
Using (\ref{qmbnd}) and  $H_{top}$ in (\ref{321888}), we can indeed write the Hamiltonian of our theory, after
its bulk energy has been minimized, as a boundary action, the way we wanted:
\begin{align}
S_{bnd}\equiv \tilde{Q}_M+S_{top}=\int d^3x\,\,q_M
+(\tau+\bar{\tau})\int_{X_3}\textrm{Tr}\left(\mathcal{A}\wedge d\mathcal{A}+\frac{2i}{3}\mathcal{A}\wedge\mathcal{A}
\wedge\mathcal{A}\right), \label{nottopbnd}
\end{align}
with $q_M$ as in (\ref{hatqm}) and the gauge and scalar fields in the theory satisfying the constraint and BHN equations
mentioned at the end of the previous section.

At this stage, we have been able to minimize the energy of the four-dimensional gauge theory
defined in $X_4$ that follows from the M-theory
configuration (M, 1) of part \ref{parta}. By construction, this bulk theory has $\mathcal{N}=4$ supersymmetry.
After such minimization, we have just found out that we are left with
a theory whose action is given by (\ref{nottopbnd}). That is, we have a theory defined on the three-dimensional boundary
$X_3$ of $X_4$. All through parts \ref{parta} and \ref{partb}, we have insisted that the presence of this boundary
provides a half-BPS condition to the full four-dimensional theory, thus reducing the amount of supersymmetry to
$\mathcal{N}=2$. But, of course, this does not happen naturally: in general, arbitrary boundary conditions on the fields
break all supersymmetry. In the next section, we derive the constraints required to ensure the desired maximally supersymmetric
boundary conditions. In this way, we will finally make precise what we mean when we say that the
warp factors in (\ref{easychoice}) and (\ref{tildeF2}) and the dilaton in (\ref{consdil}) should be chosen
such that $\mathcal{N}=2$ supersymmetry is ensured\footnote{We remind the reader that, presently, the choice is constrained by
(\ref{consc1s}), owing to our simplifying assumption in (\ref{c1112e}).}. 

\subsection{Ensuring maximally supersymmetric boundary conditions \label{bcsec}}

Whether boundary conditions that preserve some amount of supersymmetry are possible in a four-dimensional, $\mathcal{N}=4$
Yang-Mills theory coupled to matter and, if so, what these look like are fundamental questions that were answered
in~\cite{Gaiotto:2008sa}. In this section, we review the relevant results of this work
and adapt them to our own theory. As we shall see,
ensuring that the boundary theory (\ref{nottopbnd}) previously derived has $\mathcal{N}=2$ supersymmetry is indeed possible
and only requires a mild constraint be satisfied by our supergravity parameters.

As a first step towards obtaining the much desired $\mathcal{N}=2$ boundary conditions, we must first understand
the symmetries of our M-theory configuration (M, 1). As was explained in section \ref{ns5d3sect1}
and as sketched in figure \ref{fig8}, (M, 1) is dual to the D3-NS5 system in type IIB.
The non-abelian enhanced scenario amounts to considering $N$ superposed D3-branes, as argued in section \ref{nonabsec}.
In the following, we will use this duality to our advantage and discuss the spacetime symmetries of (M, 1), in its non-abelian
version, in the simpler scenario of the multiple D3's ending on an NS5 system.
We remind the reader that the underlying metric and orientations of both the
multiple D3-branes and the single NS5-brane in this set up were introduced right at the beginning of section \ref{ns5d3sect1}
and are graphically summarized in figure \ref{fig1}{\bf A}.
It is also worth bearing in mind that, upon dimensional reduction,
the four-dimensional gauge theory on the world-volume of the D3-branes has $SU(N)$ as its gauge group and
$\mathcal{N}=4$ supersymmetry.
Having refreshed a bit our memory, it is easy enough to argue what symmetries are present in the D3-NS5 system.

Consider the usual type IIB superstring theory. This is defined in $\mathbb{R}^{1,9}$.
We will label the corresponding coordinates as $x_I$, with $I=0,\,1,\,\ldots,\,9$.
The associated metric is simply $\eta_{IJ}=\textrm{diag}(-1,\,1,\,\dots,\,1)$.
Hence, the spacetime symmetry group is $SO(1,9)$.
As is well-known, $SO(1,9)$ is generated by Gamma matrices $\Gamma_I$, which satisfy the usual Clifford algebra
\begin{align}
\{\Gamma_I,\Gamma_J\}=2\eta_{IJ}, \label{cliff}
\end{align}
and has $\boldsymbol{16}$ as is its irrep. Here, we consider a ten-dimensional gauge field
and Majorana-Weyl fermion, related to each other by their supersymmetry transformations.
We denote as $\varepsilon$ the supersymmetry generator, a Majorana-Weyl spinor satisfying
\begin{eqnarray}
\bar{\Gamma}\varepsilon=\varepsilon, \quad\quad \bar{\Gamma}\equiv \Gamma_0\Gamma_1\ldots\Gamma_9.
\label{susygen}
\end{eqnarray}
and thus transforming in the $\boldsymbol{16}$ of $SO(1,9)$. Here, $\Gamma_0\Gamma_1\ldots\Gamma_9$
stands for the antisymmetrized product of $(\Gamma_0,\,\Gamma_1,\,\ldots,\,\Gamma_9)$.

The inclusion of multiple, coincident D3-branes breaks $SO(1,9)$ to $SO(1,3)\times SO(6)$, the $SO(1,3)$
oriented along the same directions as the D3's. The NS5-brane further breaks the symmetry group to
(3.243) in~\cite{Dasgupta:2016rhc}:
\begin{align}
\mathcal{U}\equiv SO(1,2)\times SO(3)\times SO(3). \label{symmgroup}
\end{align}
This is most easily understood in two steps. First, the NS5-brane restricts one of the spatial
coordinates of the D3-branes to take only non-negative values. (In our notation, $\psi\geq0$,
as can be seen in figure \ref{fig1}{\bf A}.)
Demanding that Lorentz transformations leave the boundary ($\psi=0$) invariant, $SO(1,3)$ breaks to $SO(1,2)$.
On the other hand, the NS5-brane also breaks $SO(6)$ to $SO(3)\times SO(3)$.
One of these $SO(3)$'s acts on the three-dimensional subspace spanned by the NS5-brane
which is orthogonal to the directions shared with the D3's. (In the language of figure \ref{fig1}{\bf A},
along $(x_3,\,x_8,\,x_9)$.) The other $SO(3)$ then acts on the remaining spacetime directions.
(These are $(\theta_1,\,\phi_1,\,r)$, suppressed in figure \ref{fig1}{\bf A}.)
We denote as $\boldsymbol{V_8}$ the irrep of $\mathcal{U}$: the $(\boldsymbol{2},\boldsymbol{2},\boldsymbol{2})$
tensor product.

Having established $\mathcal{U}$ in (\ref{symmgroup}) as the symmetry group of the D3-NS5 system,
it follows that $\mathcal{U}$ is the symmetry of the configuration (M, 1) too.
However, caution is needed: some of the dualities required to obtain (M, 1) from the D3-NS5 system are non-trivial
(for example, the T-duality in figure \ref{fig1}{\bf C} to \ref{fig1}{\bf D}).
Consequently, for our coming analysis to hold true,
any specific choice of the warp factors (\ref{easychoice}) and (\ref{tildeF2}) and dilaton (\ref{consdil}),
with the constraint (\ref{consc1s}), that one may wish to consider in the metric of (M, 1) (\ref{340})
should be checked to be $\mathcal{U}$-invariant.

Focusing on the case where (M, 1) is indeed $\mathcal{U}$-invariant,
we can precisely reproduce the results in~\cite{Witten:2011zz}.
Let us see how.
As we saw in section \ref{actionsec},
the scalar fields associated to the directions on which the $SO(3)$'s of $\mathcal{U}$ act
are $(\mathcal{A}_{\tilde{3}},\,\varphi_1,\,\varphi_2)$ and $(\varphi_3,\,\mathcal{A}_{\phi_1},\,\mathcal{A}_r)$, respectively.
In the language of~\cite{Witten:2011zz,Gaiotto:2008sa}, these are collectively referred to as $\vec{X}$ and $\vec{Y}$.
This identification is the same as in (3.155) in~\cite{Dasgupta:2016rhc}:
\begin{align}
\vec{X}\equiv (\mathcal{A}_{\tilde{3}},\varphi_1,\varphi_2), \quad\quad \vec{Y}\equiv
(\varphi_3,\mathcal{A}_{\phi_1},\mathcal{A}_r) \label{idenxy}
\end{align}
and will soon prove useful to us.

Let us make yet one more observation before we determine the desired half-BPS boundary conditions.
We note that the $\boldsymbol{16}$ of $SO(1,9)$ decomposes as
\begin{align}
\boldsymbol{16}=\boldsymbol{V_8}\otimes\boldsymbol{V_2},
\end{align}
where $\boldsymbol{V_2}$ is a 2-dimensional real vector space. The natural elements that act on $\boldsymbol{V_2}$
are the even elements of the $SO(1,9)$ Clifford algebra that commute with $\mathcal{U}$.
It follows then that the supersymmetry generator $\varepsilon$ can be decomposed as
\begin{align}
\varepsilon=\varepsilon_8\otimes\varepsilon_2, \quad\quad \varepsilon_8\in\boldsymbol{V_8}, \quad\quad
\varepsilon_2\in\boldsymbol{V_2}.
\label{varepdecom}
\end{align}
In order for $\varepsilon$ to be $\mathcal{U}$-invariant, $\varepsilon_2$ must be a non-zero, fixed element of
$\boldsymbol{V_2}$ ($\varepsilon_8$ is just some arbitrary element of $\boldsymbol{V_8}$).
Again following~\cite{Witten:2011zz,Gaiotto:2008sa}, we choose
\begin{eqnarray}
\varepsilon_2=\left(
\begin{array}{ccc}
-a \\ 1
\end{array}
\right), \label{varepsilon0}
\end{eqnarray}
with $a$ a real parameter.
The above is precisely the last ingredient we need to finally discuss half-BPS boundary conditions in the four-dimensional
gauge theory following from (M, 1).

It is well established (for example, see~\cite{Brink}) that boundary conditions preserve some degree of supersymmetry
iff they ensure that the normal
(to the boundary) component of the corresponding supercurrent vanishes. This in turn constrains the associated
supersymmetry generator too. 
Thanks to the above discussion and, in particular, to our identifications (\ref{idenxy}),
we can directly read off from~\cite{Witten:2011zz,Gaiotto:2008sa} the boundary conditions
and constraint on $\varepsilon_2$ thus obtained.
We refer the interested reader to~\cite{Gaiotto:2008sa} for a detailed derivation of the results
we now quote. The boundary conditions on the fields are as follows.
The scalar fields $(\varphi_3,\,\mathcal{A}_{\phi_1},\,\mathcal{A}_r)$ must all vanish
at $\tilde{\psi}=0$:
\begin{align}
\varphi_3=\mathcal{A}_{\phi_1}=\mathcal{A}_r=0. \label{bcond0}
\end{align}
The remaining scalar fields must satisfy
\begin{align}
\mathcal{D}_{\tilde{\psi}}\mathcal{A}_{\tilde{3}}-\frac{2a}{1+a^2}[\varphi_1,\varphi_2]=0, \quad
\mathcal{D}_{\tilde{\psi}}\varphi_1-\frac{2a}{1+a^2}[\varphi_2,\mathcal{A}_{\tilde{3}}]=0, \quad
\mathcal{D}_{\tilde{\psi}}\varphi_2-\frac{2a}{1+a^2}[\mathcal{A}_{\tilde{3}},\varphi_1]=0
\end{align}
at the boundary. Due to our choice (\ref{A3zeroch}), the above further simplifies to
\begin{align}
[\varphi_1,\varphi_2]=\mathcal{D}_{\tilde{\psi}}\varphi_1=\mathcal{D}_{\tilde{\psi}}\varphi_2=0,
\label{bcond1}
\end{align}
for a general value of the parameter $a$.
At $\tilde{\psi}=0$, the gauge fields are required to obey
\begin{align}
\mathcal{F}_{\tilde{\psi}\mu}+\frac{a}{1-a^2}\epsilon_{\mu\nu\lambda}\mathcal{F}^{\nu\lambda}=0,
\quad\quad \forall\mu,
\label{bcond2}
\end{align}
where $(\mu,\,\nu,\,\lambda)$ label the spacetime directions $(t,\,x_1,\,x_2,\,\tilde{\psi})$.
As for the constraint on the supersymmetry generator, it relates the parameter $a$ in (\ref{varepsilon0})
to the Yang-Mills coupling and gauge theory $\Theta$-parameter as
\begin{align}
\frac{\Theta/(2\pi)}{4\pi/g_{YM}^2}=\frac{2a}{1-a^2}. \label{defa}
\end{align}
Owing to our prior identifications (\ref{gaugeparam})
of these two parameters to coefficients in our four-dimensional gauge theory,
we can give a supergravity interpretation of $a$ also:
\begin{align}
\frac{V_3\sin\theta_{nc}q(\theta_{nc})}{c_{11}}=\frac{2a}{1-a^2} \quad\rightarrow \quad a=\sqrt{1+
\left(\frac{c_{11}}{V_3\sin\theta_{nc}q(\theta_{nc})}\right)^2}-\frac{c_{11}}{V_3\sin\theta_{nc}q(\theta_{nc})}.
\end{align}
This is exactly what is suggested in (3.222) and (3.223) in~\cite{Dasgupta:2016rhc}.
Yet another way to express the same relation follows from using 
(\ref{c2final}) and (\ref{gaugeparam}) in (\ref{defa}), which reproduces (3.251) in~\cite{Dasgupta:2016rhc}:
\begin{align}
c_2=\frac{4\pi}{g_{YM}^2}\frac{2a}{1-a^2}.
\end{align}

Now that our boundary theory in (\ref{nottopbnd}) is $\mathcal{N}=2$-supersymmetric, we need to still overcome
one more difficulty.
If our M-theory configuration (M, 1) and the four-dimensional gauge theory stemming from it through dimensional reduction
are to be of use in the study of knots and their invariants: what is the three-dimensional space where knots should be realized?
Undoubtedly, $X_3$ spanned by $(t,\,x_2,\,x_2)$. Or more precisely, its Euclidean version.
Now, since knots are topological objects, it is clear that the theory in $X_3$ ought to be topological too.
(At least, this should be the case for our construction to be an appropriate framework to support knots.)
However, a quick look at our action (\ref{nottopbnd}) immediately tells us that this is not the case in our set up.
The second, Chern-Simons term in the boundary action is indeed topological, but the presence of the magnetic charge adds a
non-topological contribution
that naively seems undesirable from our point of view.
The resolution to this puzzle was first worked out in the well-known work~\cite{Kapustin:2006pk}
and it consists on performing a so-called topological twist to our four-dimensional gauge theory.
In the following, we summarize the basics of this technique and apply it to our own theory.

\subsection{Obtaining a Chern-Simons boundary action: topological twist \label{twistsec}}

We begin this section by introducing the concept of topological twist.
Following which, we shall show that topologically twisting our gauge theory, its corresponding boundary action
is Chern-Simons-like.

If we momentarily ignore the fact that $\tilde{\psi}\geq0$,
then the symmetry of our M-theory configuration (M, 1) is as in (\ref{symmgroup}), but with
$SO(1,2)$ replaced by $SO(1,3)$.
In this case,
the topological twist consists on extending the Lorentz symmetry $SO(1,3)$ acting
along $(t,\,x_1,\,x_2,\,\tilde{\psi})$
to a new symmetry $S^\prime$. 
$S^\prime$ rotates the $(t,\,x_1,\,x_2,\,\tilde{\psi})$ subspace and, simultaneously,
the $(\tilde{x}_3,\,\theta_1,\,x_8,\,x_9)$ subspace too. It is not hard to see that
this new symmetry necessarily leads to the reinterpretation of the scalar fields
$(\mathcal{A}_{\tilde{3}},\,\varphi_1,\,\varphi_2,\,\varphi_3)$
associated to the new rotation directions as a one-form:
\begin{align}
\Phi=\sum_{\mu}\Phi_\mu dx^\mu, \quad\quad
(\Phi_0,\Phi_1,\Phi_2,\Phi_3)=i(\varphi_3,\varphi_1,\varphi_2,\mathcal{A}_{\tilde{3}}).
\label{twsc}
\end{align}
There should be no confusion regarding notation. As introduced in (\ref{FstF}) and used through all the previous section,
$x_\mu$ refers to the spacetime coordinates $(t,\,x_1,\,x_2,\,\tilde{\psi})$.
The precise identification between the components of this one-form and our scalars suggested above
is such that we match the notation in~\cite{Witten:2011zz}.
It also matches (3.156) in~\cite{Dasgupta:2016rhc}.
However, other identifications could also be entertained. In fact, we will do so later on, in section \ref{twistbulksec}.

As a short aside, it will soon prove useful to introduce some notation.
Following both~\cite{Witten:2011zz} and (3.157) in~\cite{Dasgupta:2016rhc},
we combine the scalar fields $(\mathcal{A}_{\phi_1},\,\mathcal{A}_r)$
associated to the directions $(\phi_1,\,r)$ not affected by $S^\prime$ into a complex scalar field:
\begin{align}
\sigma\equiv\mathcal{A}_r+i\mathcal{A}_{\phi_1}, \quad\quad \bar{\sigma}=\mathcal{A}_r-i\mathcal{A}_{\phi_1}.
\label{sigdefs}
\end{align}
In the same spirit of using the same notation as in~\cite{Witten:2011zz}, we shall rescale
our gauge fields as in (3.191) in~\cite{Dasgupta:2016rhc}:
\begin{align}
A=\sum_{\mu}A_\mu dx^\mu, \quad\quad A_\mu=i\mathcal{A}_\mu, \quad\quad \forall \mu. \label{twAs}
\end{align}
The corresponding field strengths are then
\begin{align}
F=dA+A\wedge A=\sum_{\mu,\nu}F_{\mu\nu}dx^\mu\wedge dx^\nu,
\quad\quad F_{\mu\nu}=\partial_\mu A_\nu-\partial_\nu A_\mu+[A_\mu,A_\nu]. \label{fullfst}
\end{align}
Clearly, this leads us to introduce new covariant derivatives, which match the ones used so far  (introduced earlier in
(\ref{covder})):
\begin{align}
D_\mu\equiv\partial_\mu+[A_\mu,\quad]=\partial_\mu+i[\mathcal{A}_\mu,\quad]\equiv\mathcal{D}_\mu, \quad\quad \forall\mu.
\label{twder}
\end{align}

Of course, the above topological twist must be made compatible with the fact that $\tilde{\psi}\geq0$ in our set up,
before we can apply it to our four-dimensional gauge theory.
What is more, it must also be made compatible with having $\mathcal{N}=2$ supersymmetric boundary conditions on the fields.
In other words, before proceeding further, all the results in section \ref{bcsec} must be extended to the case
where the gauge theory is twisted. Such generalization was first done in~\cite{Witten:2011zz,Kapustin:2006pk},
where the reader may find all the computational details. In the following, we simply review the main pertinent results in
these works, while adapting them to our present construction.

We begin by making the supersymmetry generator $\varepsilon$ in (\ref{susygen}) compatible with the new symmetry $S^\prime$.
That is, we demand
\begin{align}
(\Gamma_{\mu\nu}+\Gamma_{\tilde{\mu}\tilde{\nu}})\varepsilon=0, \quad\quad \forall\mu,\nu=t,x_1,x_2,\tilde{\psi},
\quad\quad \forall\tilde{\mu},\tilde{\nu}=\tilde{x}_3,\,\theta_1,\,x_8,\,x_9,
\end{align}
so that $\varepsilon$ is $S^\prime$-invariant. This condition has a two-dimensional space of solutions.
If we denote as $(\varepsilon_l,\,\varepsilon_r)$ the basis of solutions, then the supersymmetry generator
can be written as a linear combination of them both:
\begin{align}
\varepsilon=\varepsilon_l+\hat{t}\varepsilon_r, \quad\quad \hat{t}\in\mathbb{C}, \label{twisvar}
\end{align}
where the hat on $\hat{t}$ is meant to differentiate the above complex variable from the time coordinate $t$.
At this point, one repeats the same procedure as in the previous section:
one requires that the component of the supercurrent associated to $\varepsilon$ above that is normal to the
$\tilde{\psi}=0$ boundary vanishes.
In this manner, we reproduce the same boundary conditions as before
(these are (\ref{bcond0})-(\ref{bcond2})), but in the twisted case:
\begin{align}
\sigma=\bar{\sigma}=\Phi_0=[\Phi_1,\Phi_2]=D_{\tilde{\psi}}\Phi_1=D_{\tilde{\psi}}\Phi_2=F_{\tilde{\psi}\mu}
-\frac{i}{2}\frac{\hat{t}^2+1}{\hat{t}^2-1}\epsilon_{\mu\nu\lambda}F^{\nu\lambda}=0, \quad\quad \forall\mu.
\label{twbctot}
\end{align}
Comparing the last boundary condition above with its untwisted counterpart in (\ref{bcond2}), it follows that
the parameters $a$ and $\hat{t}$ are related to each other. Since $a$ is additionally related to
the gauge theory parameters $(g_{YM}^2,\,\Theta)$, so must $\hat{t}$ be.
These relationships also follow from studying the constraint imposed on the supersymmetry generator
by demanding the vanishing of the normal component of its supercurrent.
In this latter approach, as shown in~\cite{Witten:2011zz}, the constraint that $\varepsilon$ in 
(\ref{twisvar}) must satisfy turns out to be the exact same constraint that $\varepsilon_2$ in (\ref{varepsilon0})
has to satisfy in the untwisted case, which then led us to (\ref{defa}).
Either of the two approaches yields (3.224) and (3.246) in~\cite{Dasgupta:2016rhc}:
\begin{align}
\hat{t}=-i\frac{1+ia}{1-ia}.
\end{align}

The above can be rewritten in many interesting ways. For example,
using (\ref{defa}), we can write $\hat{t}$ as a function of the Yang-Mills coupling and $\Theta$-parameter
of our gauge theory: $\hat{t}=\hat{t}(g_{YM}^2,\Theta)$. Further using (\ref{gaugeparam}),
we can express $\hat{t}$ in terms of supergravity parameters of our M-theory configuration (M, 1):
$\hat{t}=\hat{t}(c_{11},\,V_3,\,\theta_{nc})$.
A particularly neat result follows from considering (\ref{comtaudef}) as well:
\begin{align}
\hat{t}=\pm\frac{|\tau|}{\tau}, \label{hatttau}
\end{align}
which the reader can verify without excessive algebraic effort and which is (3.184) in~\cite{Dasgupta:2016rhc}.
This is interesting because it is not obvious a priori that the two complex parameters $(\tau,\,\hat{t})$
that characterize the twisted gauge theory should be related to one another. Additionally, it is surprising 
that they should have such a mathematically simple relation.

Having introduced the topological twist and verified its consistency with all the (super)symmetries in our set up,
we can proceed to twist the boundary action (\ref{nottopbnd}).
As anticipated, this will give rise to a topological theory in $X_3$. Let us see how exactly.

Using (\ref{twsc})-(\ref{twder}) in (\ref{nottopbnd}), we see that the boundary theory after twisting becomes
\begin{align}
S_{bnd}^{(t)}=-\int d^3x\,\, q_M^{(t)}-(\tau+\bar{\tau})\int_{X_3}\textrm{Tr}\big(A\wedge dA+\frac{2}{3}A\wedge A\wedge A\big).
\label{bndal}
\end{align}
From (\ref{hatqm}), the twisted magnetic charge density $q_M^{(t)}$ can  easily be seen to be
\begin{align}
q_M^{(t)}=\sum_{a,b,c=0}^2\textrm{Tr}\Big[\sum_{\alpha,\beta=1}^2d_1\epsilon_{a\alpha\beta}\Phi_a
F_{\alpha\beta}+\epsilon_{abc}\Big(\frac{d_2}{3}\Phi_a[\Phi_b,\Phi_c]+d_3\Phi_aD_b\Phi_c\Big)\Big],
\label{twqm}
\end{align}
with $(d_1,\,d_2,\,d_3)$ as in (\ref{finalds}), albeit the indices there need to be appropriately reinterpreted.
As we will soon open up all indices and make explicit their meaning,
the reader should not worry too much over notation at this stage.
It is perhaps worth mentioning that, in the last term, $D_{\tilde{\psi}}$ does not appear, unlike in the untwisted case
(\ref{hatqm}).
This is simply because the boundary conditions (\ref{twbctot}) guarantee no such contribution occurs.
On the other hand, although (\ref{3161}) and (\ref{A3zeroch}) also force $D_0\Phi=0$, we shall carry these vanishing terms
around because they will make the coming derivation of the topological boundary action more transparent.
It goes without saying that one can do the same calculation without them too.

It turns out, however, that (\ref{bndal}) is not quite the correct twisted boundary theory.
One more term, proportional to the Chern-Simons term in (\ref{bndal}), must be added to the above:
\begin{align}
S_{bnd,tot}^{(t)}=S_{bnd}^{(t)}+b_2\int_{X_3}\textrm{Tr}\big(A\wedge dA+\frac{2}{3}A\wedge A\wedge A\big),
\quad\quad b_2\in\mathbb{C}. \label{b2top}
\end{align}
This additional term is required to ensure that all observables and states on the twisted gauge theory are
invariant under the supersymmetry generated by $\varepsilon$ in (\ref{twisvar}).
Upon including such term, one more striking observation can be made:
not only are $\tau$ and $\hat{t}$ related to each other, but also all physics of the twisted theory depends solely
on a particular combination of the two parameters:
\begin{align}
\Psi\equiv \frac{\tau+\bar{\tau}}{2}+\frac{\tau-\bar{\tau}}{2}\frac{\hat{t}-\hat{t}^{-1}}{\hat{t}+\hat{t}^{-1}}.
\label{psideff}
\end{align}
$\Psi$ is usually referred to as ``canonical parameter''
and it appears in the correct boundary theory as
\begin{align}
S_{bnd,tot}^{(t)}=-\int d^3x\,\, q_M^{(t)}+i\Psi\int_{X_3}\textrm{Tr}\big(A\wedge dA+\frac{2}{3}A\wedge A\wedge A\big).
\label{bndtotac}
\end{align}
Note that this allows us to determine the value of $b_2$, the coefficient of the required extra piece in the boundary action,
since
\begin{align}
-(\tau+\bar{\tau})+b_2=i\Psi\quad\implies\quad b_2=
\frac{\tau+\bar{\tau}}{2}(2+i)+i\frac{\tau-\bar{\tau}}{2}\frac{\hat{t}-\hat{t}^{-1}}{\hat{t}+\hat{t}^{-1}}.
\end{align}

Of course, none of the statements in the above paragraph are obvious.
Their proofs were worked out in exquisite detail in sections 3.4 and 3.5 in~\cite{Kapustin:2006pk}.
Unfortunately, a review of these derivations is beyond the scope of the present work.
Nonetheless, the reader should find no difficulty going through the cited reference, as we have carefully made our notation
coincident with the one there used. 

Having established (\ref{bndtotac}) as the twisted boundary action, 
showing its topological nature amounts to appropriately rewriting it.
We will do so in a few steps, the first consisting on expressing the twisted magnetic charge density $q_M^{(t)}$
in differential geometry language. To this aim, let us first introduce the exterior covariant derivative
of the twisted scalar fields (\ref{twsc}):
\begin{align}
d_A\Phi\equiv d\Phi+[A,\Phi]. \label{extcovder}
\end{align}
If we restrict $d_A\Phi$ to $X_3$ (where $\tilde{\psi}=0$ and thus $d\tilde{\psi}=0$ too) and
since $\Phi_3=0$ due to (\ref{A3zeroch}) and (\ref{twsc}), the above can be explicitly written as
\begin{align}
\label{bnextcovder}
d_A\Phi&=\sum_{a,b=0}^2\Big(\frac{\partial\Phi_b}{\partial x^a}dx^a\wedge dx^b+[A_adx^a,\Phi_b dx^b]\Big) \\&=
(D_0\Phi_1-D_1\Phi_0)dt\wedge dx_1
+(D_0\Phi_2-D_2\Phi_0)dt\wedge dx_2
+(D_1\Phi_2-D_2\Phi_1)dx_1\wedge dx_2.
\nonumber
\end{align}
Then, we can use (\ref{bnextcovder}) to introduce three more quantities, defined
in $X_3$, that will soon become relevant to us:
\begin{align}
\nonumber
\Phi\wedge F&=\Big(\sum_{a=0}^2\Phi_a dx^a\Big)\wedge \Big(\sum_{\alpha,\beta=1}^2F_{\alpha\beta}dx^\alpha\wedge dx^\beta\Big)
=2\Phi_0F_{12}d^3x, \\
\Phi\wedge\Phi\wedge\Phi&=
(\Phi_0[\Phi_1,\Phi_2]-\Phi_1[\Phi_0,\Phi_2]+\Phi_2[\Phi_0,\Phi_1])d^3x,  \label{aux3qt}\\
\Phi\wedge d_A\Phi&=
[\Phi_0(D_1\Phi_2-D_2\Phi_1)-\Phi_1(D_0\phi_2-D_2\Phi_0)+\Phi_2(D_0\Phi_1-D_1\Phi_0)]d^3x.\nonumber
\end{align}
(We remind the reader that $d^3x=dt\wedge dx_1\wedge dx_2$ is the normalized volume element of $X_3$.)
Note that, in the above, we did not take into account the whole twisted field strength introduced in (\ref{fullfst}).
The reasons are similar to those which led us to (\ref{bnextcovder}). Specifically,
$F_{0\mu}=0$ for all $\mu$, due to the constraint (\ref{3161}) and our gauge choice (\ref{A3zeroch}).
Also, $\tilde{\psi}=0$ at the three-dimensional boundary $X_3$ of our spacetime $X_4$,
implying $d\tilde{\psi}=0$ there and thus no field strength stretching along this direction.

To appreciate the benefit of having calculated (\ref{aux3qt}),
let us now carry out the sums in (\ref{twqm}).
In doing so, we shall use (\ref{finalds}) and, through explicit computation,
clear any doubt regarding index notation, as previously promised. The first sum can be easily seen to yield
\begin{align}
\sum_{a,b,c=0}^2\sum_{\alpha,\beta=1}^2d_1\epsilon_{a\alpha\beta}\Phi_a
F_{\alpha\beta}=2y_3\sqrt{\frac{2i|\tau|^2}{\tau-\bar{\tau}}}\Phi_0F_{12}, 
\label{sum1res}
\end{align}
with the normalization convention $\epsilon_{012}=1$ and $y_3$ given by (\ref{ysdef}).
The second sum gives
\begin{align}
\sum_{a,b,c=0}^2d_2\epsilon_{abc}\Phi_a[\Phi_b,\Phi_c]=
8\sqrt{d_{12}}y_3g_{1212}^{(1)}(\Phi_0[\Phi_1,\Phi_2]-\Phi_1[\Phi_0,\Phi_2]+\Phi_2[\Phi_0,\Phi_1]),
\label{sum2res}
\end{align}
where we have used the fact that $d_{kl}$ is independent of $(k,\,l)$ (see (\ref{dklfinal}))
to take $d_{12}$ as common factor and also the equalities
\begin{align}
y_1g_{1232}^{(1)}=y_2g_{1231}^{(1)}=y_3g_{1212}^{(1)},
\end{align}
which follow readily from (\ref{ysdef}). The third and last sum appearing in the twisted magnetic charge density
is
\begin{align}
\nonumber
\sum_{a,b,c=0}^2\epsilon_{abc}d_3\Phi_aD_b\Phi_c=&
y_3\Phi_0\Big(\sqrt{b_{12}}m_{12}^{(1)}D_1\Phi_2-\sqrt{b_{21}}m_{21}^{(1)}D_2\Phi_1\Big)
-y_1\Phi_1\Big(\sqrt{b_{\tilde{\psi}2}}m_{\tilde{\psi}2}^{(1)}D_0\Phi_2 \nonumber \\
&-\sqrt{b_{23}}m_{23}^{(1)}D_2\Phi_0\Big)
+y_2\Phi_2\Big(\sqrt{b_{\tilde{\psi}1}}m_{\tilde{\psi}1}^{(1)}D_0\Phi_1-\sqrt{b_{13}}m_{13}^{(1)}D_1\Phi_0\Big).
\label{3sumqmt}
\end{align}
Recall that, so far, we have only made the choice of coefficients in (\ref{qkl10}). We shall now make further choices.
In particular, we want to impose
\begin{align}
y_3\sqrt{b_{12}}m_{12}^{(1)}=y_3\sqrt{b_{21}}m_{21}^{(1)}=y_1\sqrt{b_{\tilde{\psi} 2}}m_{\tilde{\psi} 2}^{(1)}
=y_1\sqrt{b_{23}}m_{23}^{(1)}=y_2\sqrt{b_{\tilde{\psi} 1}}m_{\tilde{\psi}1}^{(1)}=y_2\sqrt{b_{13}}m_{13}^{(1)}.
\end{align}
Since $b_{12}=b_{21}=b_{23}$ and $b_{\tilde{\psi} 1}=b_{\tilde{\psi} 2}$ from (\ref{bakfinal}), the above 
(together with (\ref{ysdef})) implies
choosing our coefficients $(m^{(1)},\,g^{(1)})$ such that
\begin{eqnarray}
\begin{array}{lll}
&\displaystyle m_{12}^{(1)}=\pm m_{21}^{(1)},
&\quad\quad \displaystyle \sqrt{b_{\tilde{\psi} 1}}m_{\tilde{\psi} 2}^{(1)}=\sqrt{b_{12}}m_{23}^{(1)}, \\ 
&\displaystyle \sqrt{b_{\tilde{\psi} 1}}m_{\tilde{\psi} 1}^{(1)}=\sqrt{b_{12}}m_{13}^{(1)},
&\quad\quad \displaystyle y_3m_{12}^{(1)}=y_1m_{23}^{(1)}=y_2m_{13}^{(1)}.
\end{array}
\end{eqnarray}
A concrete such choice is to fix $(m_{13}^{(1)},\,m_{21}^{(1)},\,m_{23}^{(1)},\,m_{\tilde{\psi}1}^{(1)},\,m_{\tilde{\psi}2}
^{(1)})$ to
\begin{align}
\frac{g_{1212}^{(1)}}{g_{1213}^{(1)}}m_{13}^{(1)}=-m_{21}^{(1)}=\frac{g_{1212}^{(1)}}{g_{1223}^{(1)}}m_{23}^{(1)}=-
\left|\sqrt{\frac{b_{\tilde{\psi}1}}{b_{12}}}\right|\frac{g_{1212}^{(1)}}{g_{1213}^{(1)}}m_{\tilde{\psi}1}^{(1)}=-
\left|\sqrt{\frac{b_{\tilde{\psi}1}}{b_{12}}}\right|\frac{g_{1212}^{(1)}}{g_{1223}^{(1)}}m_{\tilde{\psi}2}^{(1)}=m_{12}^{(1)},
\label{choosem1s}
\end{align}
with $m_{12}^{(1)}$ not yet fixed to any particular value.
It is important to note that our choices are in good agreement with the defining relation
(\ref{consm}), since we have the full spectrum of $m^{(2)}$'s unfixed to satisfy those
equalities. In this case, the sum in (\ref{3sumqmt}) gives
\begin{align}
y_3\sqrt{b_{12}}m_{12}^{(1)}[\Phi_0(D_1\Phi_2-D_2\Phi_1)-\Phi_1(D_0\Phi_2-D_2\Phi_0)+\Phi_2(D_0\Phi_1-D_1\Phi_0)].
\label{sum3res}
\end{align}
Now, comparing our prior auxiliary quantities in (\ref{aux3qt}) with the sums (\ref{sum1res}), (\ref{sum2res}) and
(\ref{sum3res}), it follows that $q_M^{(t)}$ in (\ref{twqm}) can be written in the very convenient form
suggested in (3.232) in~\cite{Dasgupta:2016rhc}:
\begin{align}
\int d^3x\,\,q_M^{(t)}=-\int_{X_3}\textrm{Tr}\Big(2D_1\Phi\wedge F
+\frac{2}{3}D_2\Phi\wedge\Phi\wedge\Phi+D_3\Phi\wedge d_A\Phi\Big),
\end{align}
where we have defined the coefficients $(D_1,\,D_2,\,D_3)$ as
\begin{align}
D_1\equiv-y_3\sqrt{\frac{i|\tau|^2}{2(\tau-\bar{\tau})}}, \quad\quad D_2\equiv-4y_3\sqrt{d_{12}}g_{1212}^{(1)}, \quad\quad
D_3\equiv-y_3\sqrt{b_{12}}m_{12}^{(1)}. \label{Dsdef}
\end{align}
Using the above in our boundary action (\ref{bndtotac}), we obtain (3.236) in~\cite{Dasgupta:2016rhc}:
\begin{align}
S_{bnd,tot}^{(t)}=&\int_{X_3}\textrm{Tr}\Big(2D_1\Phi\wedge F
+\frac{2}{3}D_2\Phi\wedge\Phi\wedge\Phi+D_3\Phi\wedge d_A\Phi\Big) \nonumber \\
&+i\Psi\int_{X_3}\textrm{Tr}\big(A\wedge dA+\frac{2}{3}A\wedge A\wedge A\big). \label{bndactaltop}
\end{align}

The second step required to rewrite (\ref{bndactaltop}) as a topological action consists on suitably
fixing $(D_2,\,D_3)$. Specifically, we require that (3.237)
in~\cite{Dasgupta:2016rhc} holds true:
\begin{align}
D_2=\frac{D_1^3}{(i\Psi)^2}, \quad\quad D_3=\frac{D_1^2}{i\Psi}.
\end{align}
From (\ref{ysdef}) and (\ref{Dsdef}) it follows that, in terms of the coefficients of tables \ref{table1} and \ref{table2}
(the first ones having a supergravity interpretation), the above constraints are given by
\begin{align}
1=\frac{g_{1213}^{(1)}g_{1223}^{(1)}}{(i\Psi)^2(g_{1212}^{(1)})^2}\left(\frac{i|\tau|^2}{2(\tau-\bar{\tau})}\right)^{3/2},
\quad\quad
1=\frac{-2d_{12}^{1/4}}{(i\Psi)m_{12}^{(1)}}\sqrt{\frac{g_{1213}^{(1)}g_{1223}^{(1)}}{g_{1212}^{(1)}}}
\left(\frac{i|\tau|^2}{2(\tau-\bar{\tau})}\right). \label{gsconstt}
\end{align}
These constraints can be easily satisfied: the coefficients appearing here must fulfill (\ref{cons1}) and (\ref{consm}),
where we have already chosen (\ref{qkl10}) and (\ref{choosingmgs}). Clearly, there is still ample freedom of choice
left for us. Hence, we choose to fix $(g_{1213}^{(1)},\,g_{1223}^{(1)})$ such that the above holds true.
Then, easy algebra yields
\begin{align}
S_{bnd,tot}^{(t)}=&i\Psi\int_{X_3}\textrm{Tr}\Big(A\wedge dA+\frac{2}{3}A\wedge A\wedge A+2\tilde{\Phi}\wedge dA+2\tilde{\Phi}
\wedge A\wedge A \nonumber \\
&+\frac{2}{3}\tilde{\Phi}\wedge\tilde{\Phi}\wedge\tilde{\Phi}+\tilde{\Phi}\wedge d\tilde{\Phi}
+\tilde{\Phi}\wedge [A,\tilde{\Phi}]\Big), \label{bndactal}
\end{align}
where we have used (\ref{fullfst}) and(\ref{extcovder})
and where $\tilde{\Phi}$ is just the one-form $\Phi$ in (\ref{twsc}) rescaled in the following manner:
\begin{align}
\tilde{\Phi}\equiv\frac{D_1}{i\Psi}\Phi. \label{tildepsideff}
\end{align}
A couple of trace identities allow us to further rewrite the boundary theory in what will soon become
a particularly enlightening form. The identities in question are
\begin{align}
\textrm{Tr}(\tilde{\Phi}\wedge[A,\tilde{\Phi}])=2\textrm{Tr}(\tilde{\Phi}\wedge A\wedge\tilde{\Phi}), \quad\quad
\textrm{Tr}(A\wedge d\tilde{\Phi})=\textrm{Tr}(\tilde{\Phi}\wedge dA),
\label{doubletrace}
\end{align}
which the reader may easily verify through explicit computation with the aid of (\ref{3161}), (\ref{A3zeroch}), (\ref{twsc}),
(\ref{twAs}) and (\ref{fullfst}). The second identity holds up to a total derivative only. However, since
these terms are defined in $X_3$, the three-dimensional space labeled by the unbounded directions $(t,\,x_1,\,x_2)$,
the total derivative term does not affect the physics following from $S_{bnd,tot}^{(t)}$ and so we ignore it in the
ongoing. Combining (\ref{bndactal}) and (\ref{doubletrace}), we obtain
\begin{align}
S_{bnd,tot}^{(t)}=&i\Psi\int_{X_3}\textrm{Tr}\Big(A\wedge dA+\frac{2}{3}A\wedge A\wedge A+2A\wedge d\tilde{\Phi}+2\tilde{\Phi}
\wedge A\wedge A \nonumber \\
&+\frac{2}{3}\tilde{\Phi}\wedge\tilde{\Phi}\wedge\tilde{\Phi}+\tilde{\Phi}\wedge d\tilde{\Phi}
+2\tilde{\Phi}\wedge A\wedge \tilde{\Phi}\Big). \label{acbndtth}
\end{align}

The third and last step on our way to a topological boundary theory consists on defining a {\it modified}
gauge field, analogous to that in (3.240) in~\cite{Dasgupta:2016rhc}, which is a linear combination
of the twisted gauge and scalar fields (\ref{twsc}) and (\ref{twAs}):
\begin{align}
A_D\equiv A+\tilde{\Phi}. \label{modgadef}
\end{align}
It is a matter of simple algebra to check that
\begin{align}
\nonumber
A_D\wedge dA_D=&A\wedge dA+\tilde{\Phi}\wedge dA+A\wedge d\tilde{\Phi}+\tilde{\Phi}\wedge d\tilde{\Phi}, \\
A_D\wedge A_D\wedge A_D=&A\wedge A\wedge A+A\wedge \tilde{\Phi}\wedge \tilde{\Phi}+\tilde{\Phi}\wedge A\wedge A+
\tilde{\Phi}\wedge \tilde{\Phi}\wedge \tilde{\Phi} \nonumber \\
&+A\wedge A\wedge \tilde{\Phi}+A\wedge \tilde{\Phi}\wedge A 
+\tilde{\Phi}\wedge A\wedge \tilde{\Phi}+\tilde{\Phi}\wedge \tilde{\Phi}\wedge A.
\end{align}
Since the trace of a product is invariant under cyclic permutations of the terms in that product
and also due to (\ref{doubletrace}), it
is easy to see that, as promised, indeed (\ref{acbndtth}) defines a topological field theory in $X_3$,
albeit in terms of the just introduced modified gauge field $A_D$:
\begin{align}
S_{bnd,tot}^{(t)}=&i\Psi\int_{X_3} \textrm{Tr}(A_D\wedge dA_D+\frac{2}{3}A_D\wedge A_D\wedge A_D).\label{CSbounda}
\end{align}
The above Chern-Simons action is that in (3.241) in~\cite{Dasgupta:2016rhc} as well.
Needless to say, this satisfies the goal stated at the beginning of the present section.
Yet, before proceeding ahead, there are a couple of issues worth mentioning.

First, we note that in (\ref{CSbounda}) there is still one free parameter: $D_1$.
Recall that $\Psi$ is given by (\ref{psideff}). Hence, it depends only on $(\tau,\,\hat{t})$. These two parameters
have an interpretation in terms of our supergravity parameters (the warp factors and dilaton of the M-theory configuration
(M, 1)). As such, they are fixed when a specific model (M, 1) is considered. It turns out $D_1$ can also be fixed.
As argued in~\cite{Witten:2011zz}, supersymmetric Wilson loop operators can be associated to the boundary theory
with action (\ref{CSbounda}) iff the Chern-Simons gauge field $A_D$ is invariant under the supersymmetry generated by
$\varepsilon$ in (\ref{twisvar}). Schematically, we can express this as (3.242) in~\cite{Dasgupta:2016rhc}:
\begin{align}
\delta A_D=\delta(A+\tilde{\Phi})=\delta(A+\frac{D_1}{i\Psi}\Phi)=0,
\end{align}
where we have made use of (\ref{tildepsideff}) and (\ref{modgadef}).
As our notation is now such that it precisely matches the one used in~\cite{Witten:2011zz},
the interested reader should have no difficulty in following the discussion in
section 2.2.4 of that same reference. In it, the reader shall find the proof that the above constraint
sets the value of $D_1$ to
\begin{align}
D_1=i\Psi\frac{t-t^{-1}}{2}=\frac{i}{4}(t-t^{-1})\Big[\tau+\bar{\tau}
+(\tau-\bar{\tau})\frac{\hat{t}-\hat{t}^{-1}}{\hat{t}+\hat{t}^{-1}}\Big],
\end{align}
where the second equality follows from (\ref{psideff}). As we just said, $(\tau,\,\hat{t})$
are fixed for a given model (M, 1). However, from (\ref{ysdef}) and (\ref{Dsdef}), we see that $D_1$
depends on various coefficients: $(d_{12},\,b_{12},\,m_{12}^{(1)},\,g_{1212}^{(1)},\,g_{1213}^{(1)},\,g_{1223}^{(1)})$.
As given by (\ref{bakfinal}) and (\ref{dklfinal}), $(d_{12},\,b_{12})$ are also fixed once a particular model (M, 1)
is chosen via warp factors and constant dilaton. We remind the reader
that $(g_{1213}^{(1)},\,g_{1223}^{(1)})$ were already fixed
in demanding that (\ref{gsconstt}) be satisfied. Consequently, on this occasion we choose $g_{1212}^{(1)}$
such that the above holds true and keep $m_{12}^{(1)}$ arbitrary.
Of course, this new choice is still in good agreement with the constraints summarized in table \ref{table2}:
the still unspecified coefficients $(m^{(2)},\,h^{(1)})$ allow us to enforce all required equalities.
Specifically, (\ref{cons1}) may be satisfied by appropriately fixing $h_{1\tilde{\psi}kl}^{(1)}$
for all $(k,\,l=1,\,2,\,3)$, while maintaining $h_{2\tilde{\psi}kl}^{(1)}$ arbitrary.
Similarly, enforcing (\ref{consm}) implies all $(m_{13}^{(2)},\,m_{21}^{(2)},\,m_{23}^{(2)},\,m_{\tilde{\psi}1}^{(2)},
\,m_{\tilde{\psi}2}^{(2)})$ are already determined.

Second, we must refer to the point already mentioned in passing in section \ref{fsbndsec}.
Namely, the fact that the non-abelian Chern-Simons theory (\ref{CSbounda}) is gauge-invariant iff
$(i\Psi)$ is an integer multiple of $2\pi$\footnote{As the lucid work~\cite{Witten:2010cx} shows,
an appropriate analytical continuation of (\ref{CSbounda}) would allow for a path integral formalism
in case that such requirement is not met. This is hard to realize in our M-theory construction of model (M, 1),
since it would require a (to date) nonexistent formalism:
{\it topological M-theory}.
Needless to say, a careful study of such scenario is beyond the scope of the present work and we shall not
proceed in this direction. The interested reader can gain more insight on this topic from the discussion
between (3.346) and (3.350) in~\cite{Dasgupta:2016rhc}.}.
In other words, a path integral formalism associated to the action (\ref{CSbounda})
is only well defined for
\begin{align}
\frac{i\Psi}{2\pi}\in\mathbb{Z}. \label{psiconstr}
\end{align}
From its very definition in (\ref{psideff}), we see that
$\Psi$ does not necessarily satisfy such a property.
Perhaps this observation is even more evident from (\ref{comtaudef}) and (\ref{hatttau}),
expressing $\Psi$ only
in terms of coefficients with a supergravity interpretation, which depend only on the specific choice of M-theory model (M, 1):
\begin{align}
\Psi=C_1\sin\theta_{nc}q(\theta_{nc})-\frac{C_1 c_{11}^2}{V_3\sin\theta_{nc}q(\theta_{nc})}
\frac{V_3\sin\theta_{nc}q(\theta_{nc})-ic_{11}}{V_3\sin\theta_{nc}q(\theta_{nc})+ic_{11}}. \label{psisugra}
\end{align}
The conclusion from both perspectives is one and the same: we must impose some constraints
on the warp factors (\ref{easychoice}) and (\ref{tildeF2}) dilaton in (\ref{consdil}) if our topological boundary
is to have a path integral representation. (See table \ref{table1} for a guide to the equations
linking the coefficients in (\ref{psisugra}) and the just mentioned warp factors and dilaton.)
Given that in the present work we wish not study a concrete model (M, 1), we will not elaborate on the required
constraints here. However, our analysis is only valid for the subset of M-theory configurations (M, 1)
that satisfy (\ref{psiconstr}).

\subsubsection{Twisting the bulk \label{twistbulksec}}

Let us  briefly refresh our memory.
In part \ref{parta}, we constructed the M-theory model (M, 1).
In this part \ref{partb}, we derived the 
Hamiltonian (\ref{c2notzerohamtot}), defined in $X_4$ (the bulk) and associated to (M, 1). Then,
a consistent minimization of its energy, for static configurations of the fields,
led to the Hamiltonian (\ref{321888}). We further rewrote this as the action (\ref{nottopbnd}), which is
defined in $X_3$: the boundary of $X_4$. 
Upon topologically twisting (\ref{nottopbnd}), we obtained the Chern-Simons action (\ref{CSbounda}):
a suitable framework for the realization of knots in our set up.
Quite evidently, our analysis shall be consistent only when we also topologically twist the bulk
energy minimization equations that allowed us to obtain (\ref{nottopbnd}) to begin with.
Doing so is the aim of the present section.

The set of energy minimization equations we must twist are, as already pointed out at the very end of section \ref{c2not0sec}:
(\ref{A3zeroch}), (\ref{detazero}), (\ref{finalbhns}) and (\ref{finalvars}).
Before twisting, however, we make the following observation:
the various coefficient choices made so far in order to obtain a topological boundary theory considerably
simplify the BHN equations (\ref{finalbhns}).

To be precise, consider the third term
in the second BHN equation for $\alpha=1$ and $\beta=2$ and interpret the dot product there appearing as
a usual scalar product, in the same spirit as we did earlier in (\ref{bhn1ontw}). Once more, we work with the normalization
convention that $\epsilon_{12}=1$. Then, this term can be written as
\begin{align}
\sum_{\delta,k=1}^3\sqrt{b_{\delta k}}m_{\delta k}^{(1)}\mathcal{D}_\delta\varphi_k=
\sqrt{b_{12}}\left(\sum_{\alpha=1}^2\sum_{k=1}^3m_{\alpha k}^{(1)}\mathcal{D}_\alpha\varphi_k\right)
+\sqrt{b_{\tilde{\psi}1}}\left(\sum_{k=1}^3m_{\tilde{\psi}k}^{(1)}\mathcal{D}_{\tilde{\psi}}\varphi_k\right),
\end{align}
where we have used the fact that $b_{1k}=b_{2k}$ for all $k=1,\,2,\,3$ and the same is true for $b_{\tilde{\psi}k}$,
as can be seen from (\ref{bakfinal}). If we now insert in the above our coefficient choices in (\ref{choosem1s})
and further set the till now arbitrary parameters $(m_{11}^{(1)},\,m_{22}^{(1)},\,m_{\tilde{\psi}3}^{(1)})$ to
\begin{align}
m_{11}^{(1)}=m_{22}^{(1)}=m_{\tilde{\psi}3}^{(1)}\sqrt{\frac{b_{\tilde{\psi}1}}{b_{12}}}=m_{12}^{(1)},
\label{chooselastms}
\end{align}
then we obtain
\begin{align}
\sum_{\delta,k=1}^3\sqrt{b_{\delta k}}m_{\delta k}^{(1)}\mathcal{D}_\delta\varphi_k=&
\sqrt{b_{12}}m_{12}^{(1)}\Bigg[\Big(\mathcal{D}_1\varphi_1+\mathcal{D}_2\varphi_2+\mathcal{D}_{\tilde{\psi}}\varphi_3\Big)
+\Big(\mathcal{D}_1\varphi_2-\mathcal{D}_2\varphi_1\Big)\nonumber \\
&+\frac{g_{1213}^{(1)}}{g_{1212}^{(1)}}\Big(\mathcal{D}_1\varphi_3-\mathcal{D}_{\tilde{\psi}}\varphi_1\Big)
+\frac{g_{1223}^{(1)}}{g_{1212}^{(1)}}\Big(\mathcal{D}_2\varphi_3-\mathcal{D}_{\tilde{\psi}}\varphi_2\Big)\Bigg].
\end{align}
Written in this manner, it is straightforward to see that the consistency requirements (\ref{finalvars})
set to zero each term between brackets on the right-hand side above. Further,
since the BHN equation of which this term is part of is antisymmetric under the exchange of $(\alpha,\,\beta)$,
the above holds true for all allowed values of these indices. That is,
\begin{align}
\sum_{\delta,k=1}^3\sqrt{b_{\delta k}}\epsilon_{\alpha\beta}\cdot m_{\delta k}^{(1)}\mathcal{D}_\delta\varphi_k=0,
\quad\quad \forall \alpha,\beta=1,2.
\end{align}

In much the same way, one can show that the third term
in the first BHN equation (\ref{finalbhns}) also vanishes:
\begin{align}
\sum_{\delta,k=1}^3\sqrt{b_{\delta k}}\epsilon_{\alpha\tilde{\psi}}\cdot m_{\delta k}^{(2)}
\mathcal{D}_\delta\varphi_k=0, \quad\quad \forall \alpha=1,2. \label{zero3sumalps}
\end{align}
If one interprets the dot product above as the usual scalar product, the proof is exactly as
before. In more details, one must obtain the values of the $m^{(2)}$ coefficients from
(\ref{consm}), (\ref{choosem1s}) and (\ref{chooselastms}). Also, one must
realize that $b_{12}=b_{\tilde{\psi}1}$ owing to our approximation (\ref{c1112e}),
which implies $e^{2\phi_0}H_4=1$ in (\ref{bssdeff}). However, if one would like to consider the more general scenario
where (\ref{c1112e}) is not imposed, (\ref{zero3sumalps}) can still be enforced by simply entertaining more
elaborated interpretations of the dot product, in the vein of (\ref{dotm2def}) earlier on.

All in all, the conclusion is that our choices of the coefficients in table \ref{table2}
reduce the BHN equations in (\ref{finalbhns}) to
\begin{align}
\sqrt{\frac{4i|\tau|^2}{\tau-\bar{\tau}}}\mathcal{F}_{\alpha\tilde{\psi}}
-i\sum_{k,l=1}^3h_{\alpha\tilde{\psi} kl}^{(1)}\sqrt{d_{kl}}[\varphi_k,\varphi_l]=0, \quad
\sqrt{\frac{2i|\tau|^2}{\tau-\bar{\tau}}}\mathcal{F}_{\alpha\beta}
-i\sum_{k,l=1}^3g_{\alpha\beta kl}^{(1)}\sqrt{d_{kl}}[\varphi_k,\varphi_l]=0, \label{finbhneqs}
\end{align}
for all $\alpha,\,\beta=1,\,2$.
As explained around (\ref{explbhn}), these are just Hitchin equations! This is a remarkable result:
in our set up,
the BHN equations {\it naturally} decouple to Hitchin equations and a set of constraint equations on the scalar
fields there appearing. Such result becomes even more relevant in view that
Hitchin equations are precisely the starting point in the study of knots and their invariants in~\cite{Gukov:2007ck}.
The very same Hitchin equations are also related to a number of other interesting topics, such as
the Geometric Langlands Program~\cite{Gukov:2006jk}.

However exciting these directions may be,
let us get back on track: currently, our aim is to twist all energy minimization equations.
To this aim and as already anticipated in section \ref{twistsec}, it is convenient to consider
a different mapping between our scalar fields and their twisted one-form counterpart.
In particular, instead of (\ref{twsc}), we would like to consider the identification in (3.282) in~\cite{Dasgupta:2016rhc}:
\begin{align}
\Lambda=\sum_{\mu}\Lambda_\mu dx^\mu, \quad\quad
(\Lambda_0,\Lambda_1,\Lambda_2,\Lambda_{\tilde{\psi}})=i(\mathcal{A}_{\tilde{3}},\varphi_1,\varphi_2,\varphi_3).
\label{twscnew}
\end{align}
All other twisted fields remain as previously explained in (\ref{sigdefs})-(\ref{twder}).
In this manner, the twisted version of (\ref{A3zeroch}) and (\ref{detazero}) is
\begin{align}
\Lambda_0=D_\eta\sigma=D_\eta\bar{\sigma}=[\sigma,\bar{\sigma}]=[\sigma,\Lambda_k]=[\bar{\sigma},\Lambda_k]=0,
\quad\quad \forall\eta= x_1,x_2,\tilde{\psi}, \quad \forall k=1,2,3. \label{setofcot}
\end{align}
Similarly, the twisted version of the Hitchin equations in (\ref{finbhneqs}) is given by
\begin{align}
F_{\alpha\tilde{\psi}}
-\frac{\aleph}{\sqrt{2}}\sum_{k,l=1}^3h_{\alpha\tilde{\psi} kl}^{(1)}[\Lambda_k,\Lambda_l]=0, \quad\quad
F_{\alpha\beta}
-\aleph\sum_{k,l=1}^3g_{\alpha\beta kl}^{(1)}[\varphi_k,\varphi_l]=0, \quad\quad \forall\alpha,\beta=1,2.
\label{finhittw}
\end{align}
where we have defined $\aleph$ as the following constant:
\begin{align}
\aleph\equiv \sqrt{\frac{d_{12}(\tau-\bar{\tau})}{2i|\tau|^2}}.
\end{align}
The above definition uses the fact that, as can be seen from (\ref{dklfinal}), all $d_{kl}$ coefficients
have the same value. Note that, from (\ref{comtaudef}) and the equations mentioned in table \ref{table1},
it follows that $\aleph$ depends entirely on supergravity parameters only. That is, parameters that
characterize the M-theory model (M, 1).

At this stage, the only equations left to be twisted are those in (\ref{finalvars}). 
These become
\begin{align}
D_1\Lambda_2-D_2\Lambda_1=D_1\Lambda_{\tilde{\psi}}-D_{\tilde{\psi}}\Lambda_1=
D_2\Lambda_{\tilde{\psi}}-D_{\tilde{\psi}}\Lambda_2=D_1\Lambda_1+D_2\Lambda_2+D_{\tilde{\psi}}\Lambda_{\tilde{\psi}}=0.
\label{twfinvar}
\end{align}
Our identifications (\ref{twscnew}) allow us to further rewrite the above in a very concise manner
in a differential geometry language. To do so, we first compute a few auxiliary quantities. We begin with the Hodge
dual of $\Lambda$. Since (\ref{setofcot}) sets the time component of this one-form to zero,
we can carry out this computation in the three-dimensional subspace spanned by $(x_1,\,x_2,\,\tilde{\psi})$.
As we already explained, the simplifying assumption (\ref{c1112e}) converts this to a Euclidean space.
Consequently, the calculation is trivial and yields
\begin{align}
\ast \Lambda=\Lambda_1 dx_2\wedge d\tilde{\psi}-\Lambda_2 dx_1\wedge d\tilde{\psi}+\Lambda_{\tilde{\psi}}dx_1\wedge dx_2.
\end{align}
Making use of the exterior covariant derivative introduced in (\ref{extcovder})
and in much the same way as earlier in (\ref{bnextcovder}), it is easy to see that
\begin{align}
\begin{array}{llll}
d_A\Lambda&\displaystyle=(D_1\Lambda_2-D_2\Lambda_1)dx_1\wedge dx_2
+\sum_{\alpha=1}^2(D_\alpha\Lambda_{\tilde{\psi}}-D_{\tilde{\psi}}\Lambda_\alpha)dx_\alpha\wedge d\tilde{\psi}, \\
d_A\ast\Lambda&=(D_1\Lambda_1+D_2\Lambda_2+D_{\tilde{\psi}}\Lambda_{\tilde{\psi}})dx_1\wedge dx_2\wedge d\tilde{\psi}.
\end{array}
\end{align}
Upon comparing the above with (\ref{twfinvar}), it is clear that this last set of constraint equations can 
be written as in (3.287) in~\cite{Dasgupta:2016rhc}:
\begin{align}
d_A\Lambda=0=d_A\ast\Lambda,
\end{align}
which completes the twisting of all energy minimization equations in $X_4$.

Hereupon, we have gathered a good amount of knowledge about the four-dimensional gauge theory following from
the M-theory configuration (M, 1), dual to the model in~\cite{Witten:2011zz}.
In the following, we rephrase our findings in such a way that their merit is made most visible.

Appropriately compactifying (M, 1), we have obtained its associated four-dimensional action (\ref{totaction}),
defined in the space $X_4$.
Then, we have derived the corresponding Hamiltonian and written it in the particularly convenient form (\ref{3158}).
Clearly, the coefficients appearing in the Hamiltonian are expressed only in terms
of supergravity parameters of (M, 1), by construction.
Minimization of the energy of this Hamiltonian for static configurations of the fields
led to a series of constraint equations (BPS conditions) on these gauge and scalar fields.
For the gauge choice (\ref{3161}), they are given by
(\ref{A3zeroch}), (\ref{detazero}), (\ref{finalbhns}) and (\ref{finalvars}).
It turns out that all these are the same equations mentioned in~\cite{Witten:2011zz}
and derived using localization techniques for path integrals in~\cite{Kapustin:2006pk}.
Consequently, we have reproduced the results of~\cite{Witten:2011zz},
but we have done so in the well-known, conceptually simple classical Hamiltonian formalism.
In the process, we have established a precise mapping between
the usual gauge theory parameters $(g_{YM},\,\Theta,\,\tau)$ and the parameters that
characterize model (M, 1): (\ref{gaugeparam}) and (\ref{comtaudef}).
In other words, we have given a concrete, {\it simple} procedure to reproduce~\cite{Witten:2011zz}
and simultaneously provided a {\it supergravity interpretation} for it.

After the minimization process above described, the non-vanishing part of the Hamiltonian was
rewritten as the action in (\ref{nottopbnd}). This is defined in the three-dimensional space $X_3$,
the boundary of $X_4$.
Of course, if our construction is to be a suitable framework for the study of knot invariants,
knots should be embedded in $X_3$. Hence, the boundary action should be topological for our goals.
Upon a topological twist, this was proven to be indeed the case: (\ref{nottopbnd}) converts to the
Chern-Simons action (\ref{CSbounda}). Note that the Chern-Simons gauge field is a linear combination
of the twisted gauge and scalar fields, as given by (\ref{modgadef}).
Further, $\mathcal{N}=2$ supersymmetry was made compatible with this construction, requiring only
appropriate boundary conditions for the twisted fields, stated in (\ref{twbctot}).

The careful analysis of the theory in $X_3$ showed that it indeed has all required features to host knots.
What is more, additional support to this claim followed from this very same analysis in the following manner. 
Overall coherence required us to twist the energy minimization conditions in the bulk if we were
to focus on the twisted boundary theory. We then noted that,
in obtaining (\ref{CSbounda}), we were forced to make certain choices for the coefficients
summarized in table \ref{table2}. Aptly translating such choices to our BPS conditions
revealed that these were simplified to precisely the set of equations that are the starting point
for the study of knots and their invariants in~\cite{Gukov:2007ck}!
For completeness, we remind the reader that the twisted BPS equations are those in (\ref{setofcot})-(\ref{twfinvar}).

\section{Summary, conclusions and outlook \label{finalsect}}

In the first part of this work (sections \ref{ns5d3sect1} and \ref{ns5d3sect2}), we have constructed two M-theory configurations: (M, 1) and (M, 5).
They have both been obtained from the type IIB D3-NS5 system of~\cite{Witten:2011zz}
by means of a well defined series of dualities and modifications.
As depicted in figure \ref{fig8}, (M, 1) has been proven to be dual to the aforementioned model in~\cite{Witten:2011zz},
while (M, 5) has been argued to be dual to the resolved conifold with fluxes in~\cite{Ooguri:1999bv}.
An apparent indication of the seeming unrelatedness between (M, 1) and (M, 5)
(and hence between the models in~\cite{Witten:2011zz} and~\cite{Ooguri:1999bv}) is their supersymmetry:
$\mathcal{N}=2$ and $\mathcal{N}=1$, respectively. However, we have been able to trace 
their dissimilarities to a difference in the orientation of branes in a dual type IIB picture:
compare figures \ref{fig1}{\bf B} and \ref{fig3}{\bf B}. We have thus showed that, although distinct,
\cite{Witten:2011zz} and~\cite{Ooguri:1999bv} are intimately related.
So much so, that they constitute one and the same physics approach to the study of knots,
albeit in different frameworks, each suitable to address specific knots invariants.

In the second part, we have derived and studied in depth the four-dimensional gauge theory following from
the configuration (M, 1). This gauge theory is defined in a space that we have named $X_4$.
In sections \ref{actionsec} and \ref{hamilsec},
we have obtained its action and written the associated Hamiltonian in a particularly
enlightening form: a sum of squared terms, plus contributions from the three-dimensional
boundary $X_3$ of $X_4$. Energy minimization then sets each such squared term to zero independently
and, for static configurations of the fields, leads to various BPS conditions.
These are precisely the ``localization equations'' of~\cite{Witten:2011zz,Gaiotto:2011nm,Kapustin:2006pk},
obtained via elaborate techniques of localization of certain path integrals.
This correspondence implies that our approach reproduces all the results in~\cite{Witten:2011zz},
but in a much simpler formalism. Further, due to our careful deduction of the Hamiltonian of the gauge theory
directly from (M, 1), we have been able to map all parameters in~\cite{Witten:2011zz} to
variables of the M-theory model (M, 1). In this manner, we have been able to give a precise supergravity interpretation
to all the findings in~\cite{Witten:2011zz}.

Finally, in section \ref{bndsec}, we have focused on the boundary theory.
We have shown that, upon a topological twist, a Chern-Simons action captures the physics in $X_3$.
Remarkably, the Chern-Simons gauge field is a particular linear combination of the twisted gauge and scalar
fields of the gauge theory in $X_4$, exactly as in~\cite{Witten:2011zz}.
Additionally, we have obtained the appropriate half-BPS boundary conditions for all the fields, which ensure that
the theory in $X_4$ is indeed $\mathcal{N}=2$ supersymmetric.
It follows that the space $X_3$ has all required features to host knots.
In other words, after Euclideanization, knots can consistently be embedded in $X_3$ and studied in the framework of
the previously described four-dimensional gauge theory.

The details regarding such embedding of knots, as well as the study of their linking number, can be found in section
3.3 of~\cite{Dasgupta:2016rhc}. In fact, this is a coherent and natural follow up to the present paper.
Let us briefly summarize its contents. The key observation there is as follows:
the inclusion of certain M2-branes in the model (M, 1) can 
simultaneously account for the correct insertion of knots in $X_3$ and source related changes in the BPS conditions in $X_4$.
Such M2-branes make it intuitive and natural to explain why four-dimensional techniques may be useful for the study
of knots and their invariants. What is more,
the modifications thus sourced to the BPS conditions are accurately those
identified as surface operators in~\cite{Witten:2011zz,Gukov:2008sn,Gaiotto:2011nm,Gukov:2007ck,Gukov:2014gja}.
And so,~\cite{Dasgupta:2016rhc} is able to give a supergravity interpretation to these operators as M2-brane states.
Finally, restriction to the abelian case, along with the implementation of Heegard splitting, monodromy identification and
the two strands braid group action in terms of $2\times 2$ matrices whose components are evolution operators,
allow for the computation of the linking number for any arbitrary knot.

There are many interesting future directions. In fact,
both the present paper and~\cite{Dasgupta:2016rhc} form the first volume in a series of papers to appear
that will attempt to cover a good deal of them. On the one hand, we have not yet exploited most of the immense potential
of model (M, 1) and its four-dimensional gauge theory.
For example, a non-abelian extension of the construction in section 3.3 of~\cite{Dasgupta:2016rhc}
should readily reproduce the all-famous Jones polynomial and its generalizations, as suggested by~\cite{Witten:2011zz}.
Another exciting connection is to Khovanov homology: finite-dimensional vector spaces associated to knots.
Khovanov homology arises naturally from a four-dimensional gauge theory in the presence of surface operators,
just like ours.
The puzzle of why the coefficients of the Jones and related polynomials should be integers was resolved in~\cite{Khovanov},
in terms of Khovanov homology.
What is more, Khovanov's invariants are stronger than those of Jones (for instance, see~\cite{BarNatan}).

On the other hand, turning our attention to model (M, 5), we see that most of the analysis is pending.
Most notoriously,
the details on its connection to~\cite{Ooguri:1999bv} through a flop transition, the derivation of
its pertinent four-dimensional gauge theory and the suitable embedding of knots in it. Once this is done,
a wide range of possibilities unfolds.
Two such are the computation of HOMFLY-PT polynomials, along the lines of~\cite{Labastida:2000zp} 
and the study of A-polynomials, as in~\cite{Aganagic:2012jb}.

\acknowledgments{
I am specially indebted to Keshav Dasgupta, P. Ramadevi and Radu Tatar for many enlightening discussions, pointing out
useful references and their observations on preliminary versions of the present paper.
I am grateful to Maxim Emelin and Evan McDonough for their clarifying explanations of various subtleties
in string theory.
The figures in this work could not have been possible without the help of Jatin Panwar.

This work was supported in part by the National Science and
Engineering Research Council of Canada, Grant No. 210381.}

\end{document}